\documentclass[a4paper,11pt,twoside,notitlepage]{report}
\usepackage[]{graphicx}
\usepackage{amssymb}
\usepackage{amsmath}
\usepackage{feynmf}
\usepackage{geometry}
\usepackage{fancyhdr}
\usepackage{appendix}
\usepackage{setspace}
\def\lsim{\mathrel{\rlap{\lower4pt\hbox{\hskip1pt$\sim$}}
    \raise1pt\hbox{$<$}}}                
\def\spa#1{\phantom{\fbox{\rule[-#1cm]{0cm}{0cm}}}} 
\def\bes{\begin{equation*}}
\def\beas{\begin{eqnarray*}}
\def\ees{\end{equation*}}
\def\eeas{\end{eqnarray*}}
\def\be{\begin{equation}}
\def\bea{\begin{eqnarray}}
\def\ee{\end{equation}}
\def\eea{\end{eqnarray}}
\def\im{\mathcal{I} m }
\def\re{\Re e}

\def\ocortado{\lower20pt\hbox{
\begin{fmfchar*}(15,15)
\fmfleft{i}
\fmfright{o}
\fmf{phantom}{i,v1}
\fmf{phantom}{v2,o}
\fmf{plain,left,tension=0.01}{v1,v2,v1}
\fmf{plain,tension=0.01 }{v1,v2}
 \end{fmfchar*}
}}

\def\mercedes{\lower20pt\hbox{
\begin{fmfchar*}(15,15)
\fmfleft{i1,i2,i3}
\fmfright{o1,o2,o3}
\fmf{phantom}{i1,v1}
\fmf{phantom}{i2,v2}
\fmf{phantom}{i3,v3}
\fmf{phantom}{v4,o1}
\fmf{phantom}{v5,o2}
\fmf{phantom}{v6,o3}
\fmf{plain,left,tension=0.01}{v2,v5,v2}
\fmf{phantom,left,tension=0.01}{v1,v2,v3,v6,v5,v4,v1}
\fmf{plain,tension=0.01}{v2,v7}
\fmf{plain,tension=.35}{v4,v7}
\fmf{plain,tension=.35}{v6,v7}
\fmf{phantom,tension=.35}{v1,v7}
\fmf{phantom,tension=.35}{v3,v7}
\fmf{phantom,tension=0.01}{v5,v7}
 \fmfdot{v7}
 \end{fmfchar*}
}}

\def\fourpoint{\lower20pt\hbox{
\begin{fmfchar*}(15,15)
\fmfleft{i1,i2,i3}
\fmfright{o1,o2,o3}
\fmf{phantom}{i1,v1}
\fmf{phantom}{i2,v2}
\fmf{phantom}{i3,v3}
\fmf{phantom}{v4,o1}
\fmf{phantom}{v5,o2}
\fmf{phantom}{v6,o3}
\fmf{plain,left,tension=0.01}{v2,v5,v2}
\fmf{phantom,left,tension=0.01}{v1,v2,v3,v6,v5,v4,v1}
\fmf{phantom,tension=0.01}{v2,v7}
\fmf{plain,tension=.45}{v4,v8}
\fmf{plain,tension=.45}{v6,v8}
\fmf{plain,tension=.45}{v1,v7}
\fmf{plain,tension=.45}{v3,v7}
\fmf{phantom,tension=0.01}{v5,v8}
 \fmf{plain,tension=.45}{v7,v8}
\fmfdot{v7,v8}
 \end{fmfchar*}
}}

\def\fivepoint{\lower20pt\hbox{
\begin{fmfchar*}(15,15)
\fmfleft{i1,i2,i3}
\fmfright{o1,o2,o3}
\fmf{phantom}{i1,v1}
\fmf{phantom}{i2,v2}
\fmf{phantom}{i3,v3}
\fmf{phantom}{v4,o1}
\fmf{phantom}{v5,o2}
\fmf{phantom}{v6,o3}
\fmf{plain,left,tension=0.01}{v2,v5,v2}
\fmf{phantom,left,tension=0.01}{v1,v2,v3,vu,v6,v5,v4,vd,v1}
\fmffixed{(0,.95*h)}{vd,vu}
\fmf{plain,tension=0.01}{v2,v7}
\fmf{phantom,tension=.5}{v3,v7}
\fmf{plain,tension=.5}{v1,v7}
\fmf{plain,tension=.01}{v5,v8}
\fmf{phantom,tension=.5}{v6,v8}
\fmf{plain,tension=0.5}{v4,v8}
 \fmf{plain,tension=.3}{v7,v9}
 \fmf{plain,tension=.3}{v8,v9}
 \fmf{phantom,tension=.3}{v7,v10}
 \fmf{phantom,tension=.3}{v8,v10}
\fmf{plain,tension=.3}{vu,v9}
\fmf{phantom,tension=.3}{vd,v10}
\fmfdot{v7,v8,v9}
 \end{fmfchar*}
}}

\begin{document}
\setlength{\unitlength}{1mm}
\begin{fmffile}{Wittendiagrams}


\thispagestyle{empty}


\vspace{-4.9cm}

\begin{center}
\includegraphics[scale=0.6]{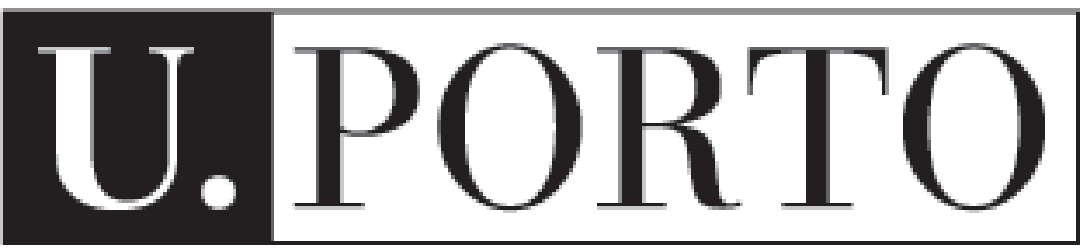}
\par\end{center}

\vspace{2cm}

\begin{center}
\textbf{\Huge High Energy Scattering in the}\\
\vspace{.4cm}
\textbf{\Huge   AdS/CFT Correspondence}
\par\end{center}{\LARGE \par}

\vspace{2cm}

\begin{center}
{\Large Jo\~ao Miguel Augusto Penedones Fernandes}\\
{\Large ~}\\
{\Large ~}
\par\end{center}{\Large \par}

\vspace{1cm}

\begin{center}
\includegraphics[scale=0.25]{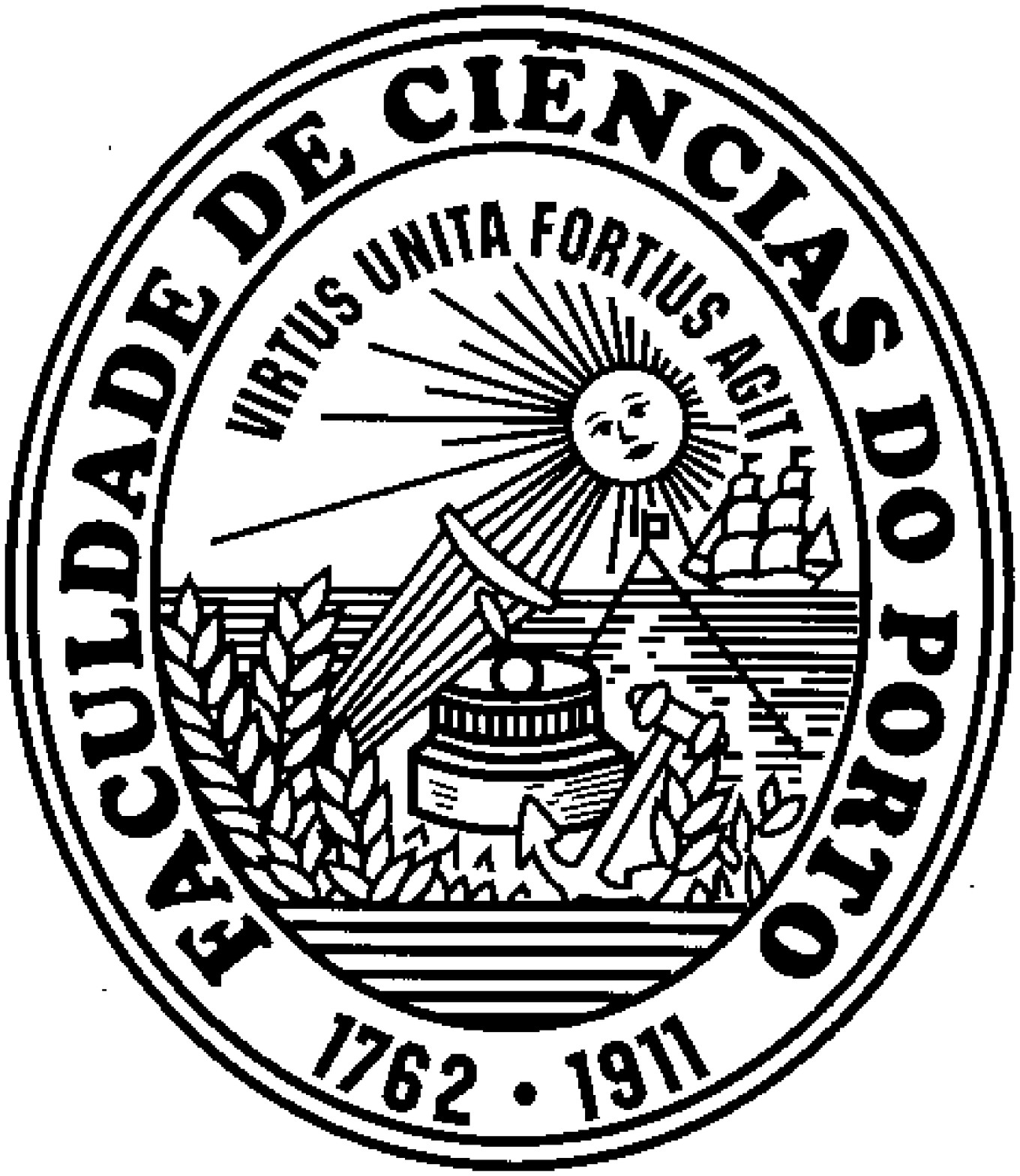}
\par\end{center}

\vspace{3cm}

\begin{center}
{\Large Departamento de F\'isica}\\ \vspace{.2cm}
{\Large Faculdade de Ci\^encias da Universidade do Porto}\\ \vspace{.2cm}
{\Large July 2007}
\par\end{center}

\thispagestyle{empty}

\newpage\thispagestyle{empty}~\newpage\thispagestyle{empty}

\vspace{-2cm}

\begin{center}
\includegraphics[scale=0.6]{uporto_pos_opaco}
\par\end{center}

\vspace{2cm}

\begin{center}
\textbf{\Huge High Energy Scattering in the}\\
\vspace{.4cm}
\textbf{\Huge  AdS/CFT Correspondence}
\par\end{center}{\LARGE \par}

\vspace{2cm}

\begin{center}
{\Large Jo\~ao Miguel Augusto Penedones Fernandes}\\
{\Large ~}\\
{\Large PhD Thesis supervised by Prof. Miguel Sousa da Costa}
\par\end{center}{\Large \par}

\vspace{1cm}

\begin{center}
\includegraphics[scale=0.25]{fc_logo}
\par\end{center}

\vspace{3cm}

\begin{center}
{\Large Departamento de F\'isica}\\ \vspace{.2cm}
{\Large Faculdade de Ci\^encias da Universidade do Porto}\\ \vspace{.2cm}
{\Large July 2007}
\par\end{center}

\newpage\thispagestyle{empty}~\newpage


\pagenumbering{roman}

\addcontentsline{toc}{chapter}{Abstracts}

\chapter*{Abstract}
This work explores the celebrated AdS/CFT correspondence  \cite{Malda} in the regime of
high energy scattering in Anti--de Sitter (AdS) spacetime.
In particular, we develop the eikonal approximation
to high energy scattering in AdS and 
 explore its consequences  for
the dual Conformal Field Theory (CFT).

Using position space Feynman rules,
we rederive the eikonal approximation for high energy scattering in flat space.
Following this intuitive position space perspective, we then generalize the eikonal approximation 
for high energy scattering in AdS and other spacetimes.
Remarkably, we are able to resum, in terms of a generalized phase shift, ladder and cross ladder
Witten diagrams associated to the exchange of an AdS spin $j$ field, to all orders in the coupling constant. 
In addition, we confirm our results with an alternative derivation of the eikonal approximation in AdS,
based on gravitational shock waves. 
 
By the AdS/CFT correspondence, the eikonal amplitude in AdS is related to the four point function
of CFT primary operators in the regime of large 't Hooft coupling $\lambda$,
including all terms of the $1/N$ expansion.
We then show that the eikonal amplitude determines the behavior of 
the CFT four point function 
for small values of the cross ratios in a Lorentzian regime 
and  that this controls 
its  high spin and dimension conformal partial wave decomposition.
These results allow us to determine the anomalous dimension 
of high spin and dimension double trace primary operators, by relating it to
the AdS eikonal phase shift.
Finally we find that, at large energies and large impact parameters in AdS, 
the gravitational interaction dominates all other interactions, as in flat space.
Therefore, the anomalous dimension of double trace operators, associated to graviton exchange in AdS,
yields a universal prediction for CFT's with AdS gravitational duals.


\newpage\thispagestyle{empty}~\newpage

\chapter*{Resumo} 
Este trabalho explora a correspond\^encia AdS/CFT   \cite{Malda} no regime de difus\~ao a alta energia
no espa\c co--tempo de Anti--de Sitter (AdS).
Em particular, desenvolvemos a aproxima\c c\~ao eikonal para difus\~ao a alta energia em AdS e 
exploramos as suas consequ\^encias para a Teoria de Campo Conforme (CFT) dual.

Usando regras de Feynman no espa\c co das posi\c c\~oes,  
rederivamos a aproxima\c c\~ao eikonal  para difus\~ao a alta energia em espa\c co plano.
De seguida, seguindo esta perspectiva intuitiva no  espa\c co das posi\c c\~oes,
generalizamos a aproxima\c c\~ao eikonal  para difus\~ao a alta energia em AdS e outros espa\c cos--tempo.
Desta forma, conseguimos somar, em termos de uma diferen\c ca de fase generalizada, os digramas de Witten
do tipo escada e escada cruzada associados \`a troca de um campo com spin $j$ em AdS, incluindo todas as 
ordens na constante de acoplamento.
Por fim, confirmamos os nossos resultados com uma deriva\c c\~ao alternativa da  aproxima\c c\~ao eikonal
em AdS, baseada em ondas de choque gravitacionais.

A correspond\^encia AdS/CFT relaciona a amplitude eikonal em AdS com a fun\c c\~ao de correla\c c\~ao 
de quatro operadores prim\'arios da CFT dual calculada no regime de forte acoplamento de 't Hooft, 
incluindo todos os termos da expans\~ao em $1/N$.
Em primeiro lugar, mostramos que  a amplitude eikonal determina o comportamento desta 
fun\c c\~ao de correla\c c\~ao a quatro pontos para pequenos valores dos cross ratios num regime
Lorentziano.
De seguida, mostramos que este comportamento da  fun\c c\~ao de correla\c c\~ao 
controla a sua decomposi\c c\~ao em ondas parciais conformes  com  spin e dimens\~ao elevados.
Estes resultados permitem--nos determinar as dimens\~oes an\'omalas de operadores prim\'arios de duplo tra\c co 
com spin e dimens\~ao elevados, relacionando-as com  a diferen\c ca de fase eikonal em AdS.
Finalmente, concluimos que em processos de difus\~ao 
a alta energia e grande par\^ametro de impacto 
a interac\c c\~ao gravitational
domina sobre todas as outras interac\c c\~oes,tal como em espa\c co plano.
Desta forma, as dimens\~oes an\'omalas de operadores de duplo tra\c co, associadas \`a troca de gavit\~oes em AdS,
\'e uma previs\~ao universal para CFT's duais a teorias gravitacionais em AdS.


\newpage\thispagestyle{empty}~\newpage

\chapter*{R\'esum\'e}

Ce travail explore la correspondance  AdS/CFT    \cite{Malda}  dans le r\'egime de diffusion \`a haute \'energie 
dans l'espace--temps de Anti--de Sitter.
En particulier, nous d\'eveloppons l'approximation eikonal pour la difusion a haute \'energie dans AdS et
explorons ses cons\'equences pour la Th\'eorie des Champ Conforme (CFT) duale.
 
Employant les r\`egles de Feynman dans le espace des positions, nous rederivons la approximation eikonal pour la 
diffusion aux hautes \'energies dans l'espace plain.
Apr\`es cette perspective intuitive de l'espace des positions, nous g\'en\'eralisons la approximation eikonal
pour la diffusion \`a haute \'energie dans AdS et dans d'autres espaces--temps.
De cette fa\c on, nous pouvons somer, en termes d'une diff\'erence de phase g\'en\'eralis\'e,  les diagrammes de Witten 
de \'echelle et d' \'echelle crois\'e associ\'es \`a l'\'echange de un champ avec spin $j$ en AdS,
\`a tous les ordres dans la constante de couplage. 
Finalement, nous confirmons nos r\'esultats avec une d\'erivation alternative 
de la approximation eikonal en AdS, bas\'ee sur les ondes de choc gravitationnelles. 

Par la correspondance AdS/CFT, l'amplitude eikonale en AdS est li\'ee \`a la fonction de correlacion des quatre 
op\'erateurs primaires
de la th\'eorie duale dans le r\'egime du grand couplage de 't Hooft, avec tous les termes de l'expansion en $1/N$. 
Nous prouvons que l'amplitude eikonal d\'etermine le comportement de cette fonction de corr\'elation
pour des petites valeurs des cross ratios dans un r\'egime Lorentzian. 
Ensuite, nous prouvons  que ce  comportement de la fonction de corr\'elation contr\^ole 
sa d\'ecomposition en ondes partielles conformes avec spin et dimension \'elev\'es.
Ces r\'esultats nous permettent de d\'eterminer la dimension anomale des op\'erateurs primaires  de double trace
avec spin et dimension \'elev\'es, en la reliant \`a la d\'ephasage eikonal  en AdS.
 Enfin nous constatons que dans des processus de diffusion \`a haute 
energie et avec des grands param\`etres d'impacte en AdS, 
l'interaction gravitationnelle domine toutes les autres interactions, pareil que dans l'espace plain.
Par cons\'equent, la dimension anomale des op\'erateurs de double trace, associ\'ee \`a l'\'echange des gravitons dans AdS, 
est une pr\'evision universelle pour des CFT's duaux a theories gravitationnelles en AdS.


\newpage\thispagestyle{empty}~\newpage

\addcontentsline{toc}{chapter}{Acknowledgements}
\chapter*{Acknowledgements}

Fortunately, as usual in my life, I have many reasons to be grateful for the privileged experience that my PhD was.
I am specially grateful to my advisor, Miguel Costa, for his constant support and generosity of time, during the last four years.
His determination and high scientific standards were always a great inspiration for me. 
I am also very grateful to Lorenzo Cornalba for having taught me many of the subjects in this thesis.
Collaborating with Miguel and Lorenzo was both an exceptional intellectual privilege as well as
a genuine pleasure.
I would like to thank my co--advisor, Costas Bachas, for pleasant discussions and for 
his hospitality at \'Ecole Normale Sup\'erieure of Paris,
where I stayed for some nice and fruitful periods.
I am also thankful to my collaborators and other colleagues for many stimulating discussions about String Theory
and other subjects. They created the perfect environment for my scientific research and education.

For financial support, I thank Funda\c c\~ao para a Ci\^encia e a Tecnologia for a research fellowship
and Centro de F\'isica do Porto for travel expenses.
For that and for all my academic education, I am indebted to  the Portuguese (and European) people.

I must also thank  my numerous family and friends for their permanent support and friendship.
Indeed, their presence makes me feel deeply fortunate and happy.

Finally, I thank my parents for always encouraging my curiosity.

\newpage\thispagestyle{empty}~\newpage

\addcontentsline{toc}{chapter}{Contents}
\tableofcontents
\newpage

\newpage

\pagenumbering{arabic}

\chapter{Introduction}
\label{ch:intro}

Quantum Chromodynamics (QCD) is the fundamental theory ruling the hadronic world.
QCD is a non--abelian gauge theory with two types of fundamental particles: quarks and gluons.
The quarks are Dirac fermions transforming in the fundamental representation 
of the gauge group  $SU(3)$, they are the building blocks for hadrons.
The gluons are massless vector bosons mediating the strong force that bounds the quarks inside hadrons.    

Historically, the scientific triumph of QCD was supported by two main types of experimental results.
The first was the group theoretical structure organizing the known hadrons into the {\em eightfold way},
which seeded the Quark model. 
The second relies on the peculiar property of {\em asymptotic freedom} that QCD enjoys.
This property means that the quarks are weakly coupled at short distances, making it possible 
to accurately describe high energy phenomena using standard perturbative techniques.
Dynamical predictions, such as Bjorken scaling (and deviations from it) in deep inelastic scattering 
and jet production in high energy scattering, have been spectacularly verified in several particle accelerators 
around the world, over the past four decades \cite{QuarkHistory}.

\begin{figure}
\begin{center}
\includegraphics[width=6cm]
{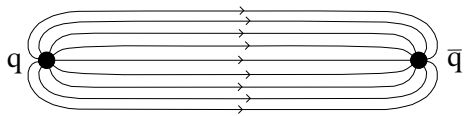}
\caption{The flux lines of the  chromoelectric field between quark and anti--quark.}
\label{fluxlines}
\end{center}
\end{figure}

In spite of these great successes, there are still many features in the phenomenology of mesons and baryons that 
cannot be derived from QCD.
The problem arises from the other side of asymptotic freedom, the low energy {\em confinement} of quarks 
inside colorless hadrons. 
In many aspects, the confined phase of QCD suggests an underlying string description.
The intuitive idea is that the flux lines of the chromoelectric field do not spread in space as in electromagnetism.
Instead, they organize in flux tubes due to the non--linear structure of non--abelian gauge theories.
This picture immediately leads to a long range force between quark and anti--quark, given that
the energy of the system grows linearly with the distance between them (figure \ref{fluxlines}).

String Theory was discovered forty years ago as an attempt to understand  hadronic physics.
By that time, QCD and String Theory competed as models of the strong force.
Of course, this QCD/String dispute was decided long ago in favor of QCD.
However, the modern viewpoint replaces {\em dispute} by {\em duality}, and rephrases the 
main question: {\em Is QCD a String Theory?}


\section{Hadronic Spectrum \& Strings}


Although the fundamental particles of QCD are quarks and gluons, the confinement mechanism disallows 
their direct observation. Instead, the observed spectrum is characterized by a long list of 
colorless bound states of the fundamental particles. Most of these bound states are unstable and
are found as resonances in scattering experiments.
At the present day, we are still unable to accurately predict the observed hadronic spectrum directly from the 
QCD dynamics \footnote{See  \cite{Juge} and  \cite{deTeramond} and references therein for attempts using the lattice formulation of QCD
and the AdS/CFT correspondence.}.
Nevertheless, from a phenomenological perspective, the hadronic spectrum has several inspiring features.

\begin{figure}
\begin{center}
\includegraphics[width=8cm]{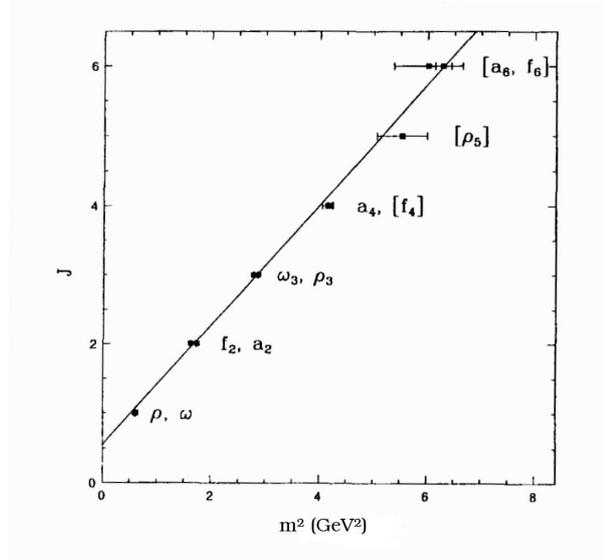}
\caption{  The Chew--Frautschi plot.  
Spin $J$ of the isospin $I=1$ even parity mesons against their mass squared.
(From reference  \cite{pomeronbook})}
\label{regge}
\end{center}
\end{figure}

In figure \ref{regge} we plot the spin $J$ of the lighter mesons against their mass squared $m^2$.
The result is well modeled by a linear Regge trajectory
$$
J=\alpha\left(m^2\right)=\alpha(0)+\alpha' m^2\ ,
$$
where $\alpha(0)$ and $\alpha'$ are known as the intercept and the Regge slope, respectively.
In fact, most of the hadronic resonances fall on approximately linear Regge trajectories with slopes
around $1 (GeV)^{-2}$ and different intercepts.
A linear relation between spin and mass squared suggests a description of the bound states as 
string like objects rotating at relativistic speeds. Indeed, the spin of a classical open string with tension $T$,
rotating as a straight line segment with endpoints traveling at the speed of light, is given by
$\alpha'=(2\pi T)^{-1}$ times its energy squared \footnote{See section 2.1.3 of  \cite{GSW} for details.}.

\begin{figure}
\begin{center}
\includegraphics[width=6cm]{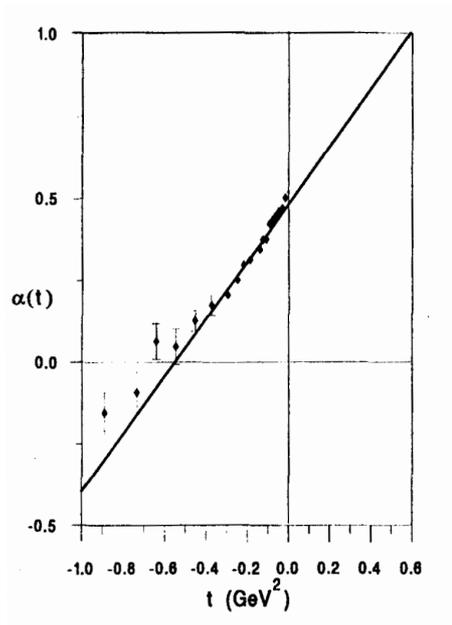}
\caption{Regge trajectory determined from the  large energy (20--200 $GeV$) behavior of the differential 
cross section of the process $\pi^- + p \to \pi^0 + n$.
The straight line is obtained by extrapolating the trajectory in figure \ref{regge}.
 (From reference  \cite{pomeronbook})}
\label{scatregge}
\end{center}
\end{figure}

\begin{figure}
\begin{center}
\includegraphics[width=14cm]
{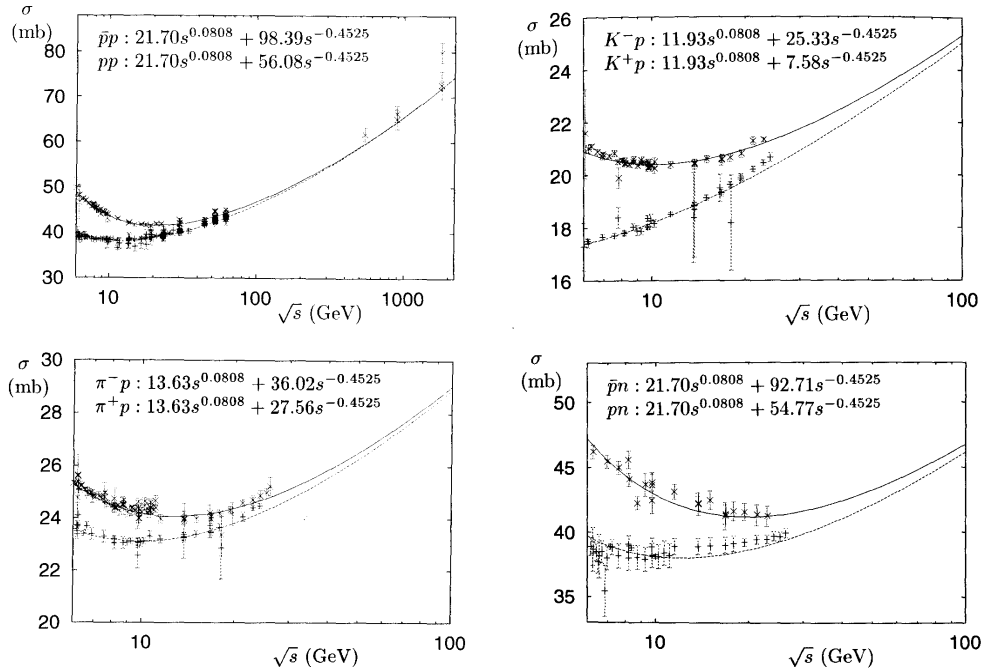}
\caption{Total cross sections for elastic scattering at high energy.
The cross sections rise slowly due to pomeron exchange. (From reference  \cite{pomeronQCDbook})}
\label{pp}
\end{center}
\end{figure}

A related stringy feature of QCD is the high energy behavior of scattering amplitudes.
Experimentally, at large center--of--mass energy $\sqrt{s}$, the hadronic scattering amplitudes show Regge behavior
$$
A(s,t) \sim \beta(t) s^{\alpha(t)}\ ,
$$
where $t$ is the square of the momentum transferred.
The appropriate Regge trajectory $\alpha(t)$ that dominates a given scattering process is selected by
the exchanged quantum numbers. For example, the process 
$$
\pi^- + p \to \pi^0 + n
$$
is dominated by the exchange of isospin $I=1$ even parity mesons, i. e. the  Regge trajectory in figure \ref{regge}.
In figure \ref{scatregge} we plot the Regge trajectory obtained from the behavior of the differential 
cross section at large $s$. 
Elastic scattering is characterized by the exchange of the vacuum quantum numbers.
In this case the scattering amplitude is dominated by the {\em Pomeron} trajectory \cite{pomeronQCDbook,pomeronbook}
$$
\alpha_P(t)\simeq 1,08 + 0,25\,t\ , \ \ \ \ \ \ \ \ \ \ ( GeV \ {\rm units})\ .
$$
There is some evidence from lattice simulations that there are glueball states
lying on this trajectory starting from spin $J=2$  \cite{Meyer, Boschi-Filho}.
Furthermore, an even glueball state with spin 2 lying on the pomeron trajectory
seems to have been found in experiments  \cite{Barberis}. 
However, in real QCD, glueball states mix with mesons and their identification is not clear  \cite{pomeronQCDbook}.
An important consequence of the pomeron intercept being larger than 1, is that hadrons effectively expand
at high energies. More precisely, the total cross section for
elastic processes in QCD grows with center--of--mass energy,
$$
\sigma \sim s^{\alpha_P(0)-1} \sim  s^{0.08}\ ,
$$
as can be seen in figure \ref{pp}.
This expansion with energy reinforces the picture of hadrons as stringlike objects.
It is well known  \cite{Karliner} that the average size of a fundamental string is given by the divergent sum,
$$
<R^2>\sim \alpha' \sum_{n=1}^{\infty}\frac{1}{n}\ ,
$$
coming from the contributions of zero point fluctuations of each string mode.
However, in a scattering experiment, only the modes with frequency smaller than
the energy $\sqrt{s}$ can be resolved. This effectively cuts off the sum to $n \lsim \sqrt{\alpha's}$,
and leads to a finite string size increasing logarithmically with energy 
$$
<R^2>\sim\alpha' \log(s)\ .
$$
This rough estimate is consistent with a form factor analysis of the Regge behaved scattering amplitude.
Expanding around the graviton \emph{T}--channel pole,
$$
A(s,t=-q^2)\sim \frac{s^{2+\alpha't}}{-t} \sim  \frac{s^2}{q^2}  e^{ -\alpha'\log(s) q^2}\ ,
$$ 
we obtain a gaussian form factor 
$$
F(x)\sim\int dq \, e^{\,iq\cdot x}   e^{ -\alpha'\log(s) q^2} \sim  e^{-\frac{x^2}{4 \alpha'\log(s)}}\ ,
$$
with the estimated width.

\begin{figure}
\begin{center}
\includegraphics[width=10cm]{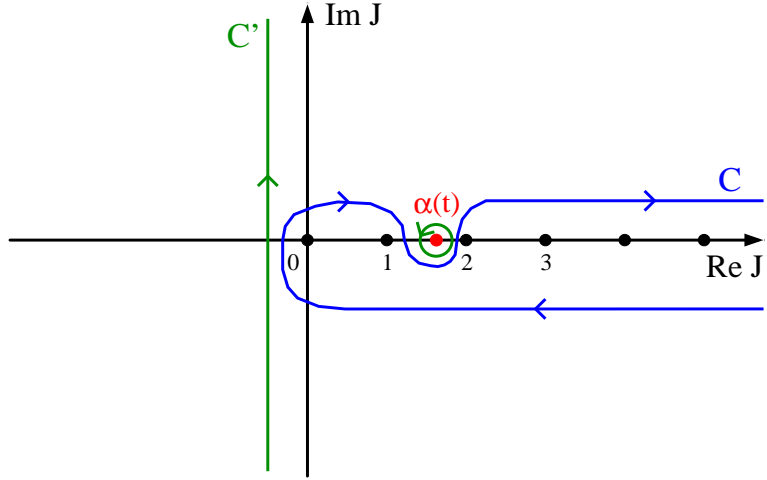}
\caption{Integration contours in the complex $J$ plane.}
\label{contour}
\end{center}
\end{figure}

Regge theory  \cite{Regge,Regge2} is the natural framework to understand the connection between the two mentioned manifestations of
 Regge trajectories.
In the Regge limit $s\gg t$, we expect the scattering amplitude, due to \emph{T}--channel exchange
of a full Regge trajectory $\alpha(t)$, to have the form
\begin{equation}
A(s,t)\sim \sum_J \frac{a_J }{\alpha(t)-J }s^J \ ,
\label{Reggepoles}
\end{equation}
where $a_J$ encodes the coupling between the external particles and the exchanged spin $J$ particle.
The last expression can also be thought of as the Regge limit of the  \emph{T}--channel partial wave expansion
of the scattering amplitude.
Here, we have written explicitly the physical $t$ poles of the partial wave coefficients.
The basic idea of Regge theory is to analytically continue the functions of $J$ from the integers 
to the complex plane and write the sum over $J$
as a contour integral \footnote{We shall focus on the main idea of  Regge theory and
disregard many important details of the rigorous treatment  \cite{pomeronQCDbook}.}
$$
A(s,t)\sim \int_C dJ \frac{a(J) }{\sin(\pi J) }\frac{s^J }{\alpha(t)-J }\ .
$$ 
The contour $C$ encircles the non--negative integers as in figure \ref{contour}.
Under rather generic assumptions, the function $a(J)$ decays rapidly enough as $|J| \to \infty$
in the region $\re(J)>0$, so that 
we can deform the contour $C$ in  figure \ref{contour},  to the contour $C'$ over the line $\re (J)=-1/2$,
plus the contribution from the Regge pole at $J=\alpha(t)$ and from other possible singularities of $a(J)$. 
Finally, the contribution from  $C'$ is negligible for large $s$ and  we obtain Regge behavior
\begin{equation}
A(s,t)\sim \frac{a(\alpha(t) )  }{\sin(\pi \alpha(t)) } s^{\alpha(t)}\ ,\ \ \ \ \ \ \ \ \ 
(s\to\infty \ ,\  t \ {\rm fixed})\ ,
\label{Reggebehavior}
\end{equation}
assuming that the function $a(J)$ is analytic for $\re(J)> \re(\alpha(t))$.

Regge behavior is hard to understand from a standard quantum field theory perspective.
 \emph{T}--channel exchange of a spin $J$ particle leads to a high energy behavior
$A\propto s^J$ so that the scattering amplitude in the Regge limit should be dominated by the
highest spin particle. The way out, suggested by the hadronic spectrum, is to consider particles with unbounded spin.
However, interacting higher spin theories are very difficult to construct and, to date, all known theories including
fundamental particles with  unbounded spin are string theories.

Consider the open bosonic string as a concrete and instructive example.
Its spectrum organizes in parallel Regge trajectories with slope $\alpha'$ and different intercepts,
with the leading Regge trajectory given by
$$
J=\alpha(m^2)=1+\alpha' m^2 \ , \ \ \ \ \ \ \ \ \ \ \ \ J=0,1,2,\cdots\ .
$$
Scattering of open string states exhibits Regge behavior at high energy 
$$
A(s,t)\sim \Gamma(-\alpha(t) )  s^{\alpha(t) } \ ,\ \ \ \ \ \ \ \ \ 
(s\to\infty \ ,\  t \ {\rm fixed})\ .
$$
Matching with (\ref{Reggebehavior}) we find that open bosonic string  corresponds to the specific choice 
of coupling $a_J=1/J!$ in (\ref{Reggepoles}).

In summary, the Regge trajectories characteristic of the strong interaction indicate a stringy
structure of hadrons.
This motivated the discovery of the fascinating Veneziano's amplitude and the subsequent development
of String Theory, which has become an outstanding conceptual paradigm in Theoretical Physics.
However, in their original goal of describing  hadronic physics, standard flat space
string theories failed miserably.
Indeed, they have too many shortcomings as models of the hadronic world: 
extra dimensions; either tachyons or supersymmetry; 
a massless spectrum containing spin 0 and spin 2 particles; 
exponential falloff of  amplitudes in the hard scattering limit $s\to \infty $ with $t/s$ fixed; etc.
Amusingly, these annoying features of String Theory were turned into virtues by the more ambitious perspective of being the 
{\em theory of everything}.
More recently, the connection between QCD and String Theory has reborn in a surprising way.
As we shall see below, we expect the string dual to QCD to live in a curved background with one extra dimension.



\section{'t Hooft Limit}


Another important route to gauge/string duality is the parametric limit found by 't Hooft in 1974  \cite{hooftlimit}.
His idea was to study the large $N$ limit of  $SU(N)$ gauge theory.
If we can understand the theory in this limit, then perhaps real QCD will have similar properties or
we can try to approach the physical value $N=3$ with a perturbative expansion in $1/N$. 

\begin{figure}
\begin{center}
\includegraphics[width=10cm]
{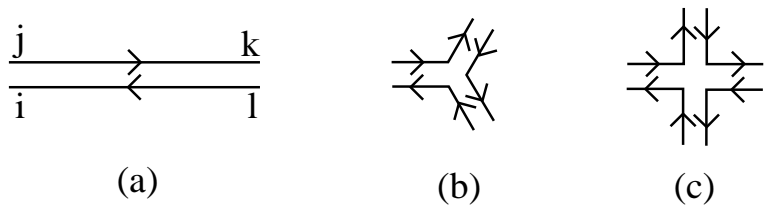}
\caption{ Double line elements for perturbative computations in gauge theory:
(a) propagator, (b) cubic vertex and (c) quartic vertex.}
\label{propver}
\end{center}
\end{figure}

\begin{figure}
\begin{center}
\includegraphics[width=8cm]
{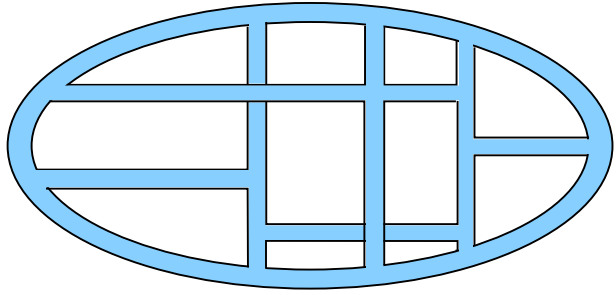}
\caption{ Generic Feynman diagram in the double line notation defining a non--planar surface with $F=6$ faces, $E=23$ edges 
and $V=15$ vertices.}
\label{graph}
\end{center}
\end{figure}

Consider, for simplicity,  pure $U(N)$ Yang--Mills theory with the standard lagrangian density
$$
\frac{1}{4g_{{\rm YM}}^2} {\rm Tr}\, dA^2\ .
$$
The gauge bosons \footnote{After gauge fixing there are also Fadeev--Popov ghosts, but from the color point of view
these work exactly as gluons. Actually, the discussion also applies to any other field transforming in the adjoint
representation of the gauge group.} transform in the adjoint representation and therefore carry an index 
$a=1,2,\cdots, N^2$. Using the fundamental representation of the group generators $T_a$, we can write
$$
A_i^j = \sum_a A^a [T_a]_i^j\ ,
$$
and think of the gauge field $A^a$ as carrying two indices $i,j=1,2,\cdots,N$, one fundamental and other anti--fundamental.
Then, the gluon propagator  satisfies
$$
<A_i^j A_k^l> \propto \delta_i^l \delta_k^j\ ,
$$
therefore it can be represented by two parallel lines as in figure \ref{propver}(a).
Furthermore, the interaction vertices can also be thickened as in figure  \ref{propver}(b) and \ref{propver}(c),
in order to glue with the double line propagators.
In this language, a generic vacuum Feynman diagram like the one in figure \ref{graph} defines a two dimensional surface
with $F$ faces (color loops), $E$ edges (propagators) and $V$ vertices.
The trace over the color index yields a factor of $N$ for each color loop; each propagator gives a factor of $g_{{\rm YM}}^2$
and each vertex
yields a factor of $1/g_{{\rm YM}}^2$. Thus, a generic Feynman diagram  is proportional to
$$
N^F g_{{\rm YM}}^{2(E-V)}= (Ng_{{\rm YM}}^2)^{F}  g_{YM}^{2(E-V-F)} = \left(g_{{\rm YM}}^2\right)^{-\chi}\lambda^{F}\ ,
$$ 
where we have introduced the 't Hooft coupling $\lambda\equiv Ng_{{\rm YM}}^2$ and the Euler characteristic $\chi=F+V-E$
of the  two dimensional surface. 
The Euler characteristic is completely determined by the genus  $g$ (number of handles) of the surface, $\chi=2-2g$.
In addition, the number of holes $h$ is just the number of faces $F$.
The perturbation series of gauge theory can then be organized in the double expansion
\begin{equation}
\sum_{g=0}^{\infty} \left(g_{{\rm YM}}^2\right) ^{2g-2} \sum_{h=2}^{\infty} C_{g,h}\lambda^h\ ,
\label{openstring}
\end{equation}
which strongly resembles the perturbative expansion of an open string theory with string coupling 
$$
g_s\sim g_{{\rm YM}}^2\ .
$$
Moreover, if we sum over the number of holes,
$$
\sum_{g=0}^{\infty} g_s ^{2g-2}  C_{g}(\lambda)\ ,
$$
the expansion looks like perturbative closed string theory. 

The 't Hooft limit is the limit $N\to\infty$ with 't Hooft coupling $\lambda = Ng_{{\rm YM}}^2$ fixed.
In this limit, only {\em planar} Feynman diagrams (with $g=0$) contribute and the putative closed string theory
becomes free. The 't Hooft coupling should be regarded either as a  coupling constant of the 
 worldsheet sigma--model, describing the dynamics of the dual free closed string; or, equivalently,  
as a geometric modulus characterizing the dual closed string theory background.

The planar limit easily generalizes to theories with quarks and scalar fields. 
For instance, quark loops are suppressed by a factor of $1/N$. This just reflects the fact that 
there are $N$ times more gluons than  quarks to sum over.
Moreover, the decay widths of mesons and glueballs, as well as  their mixing, are also $1/N$ effects.
We conclude that, in the planar limit, all resonances become stable and correspond to the dual closed string spectrum.
Then, at finite $N$, strings start to interact and heavy states become unstable.
Unfortunately, it is very hard to find the precise string theory dual to a given gauge theory.
In fact, it took more than twenty years to find a concrete four dimensional realization of
the gauge/string duality.

\section{Open/Closed Duality}          \label{openclosed}

\begin{figure}
\begin{center}
\includegraphics[width=8cm]
{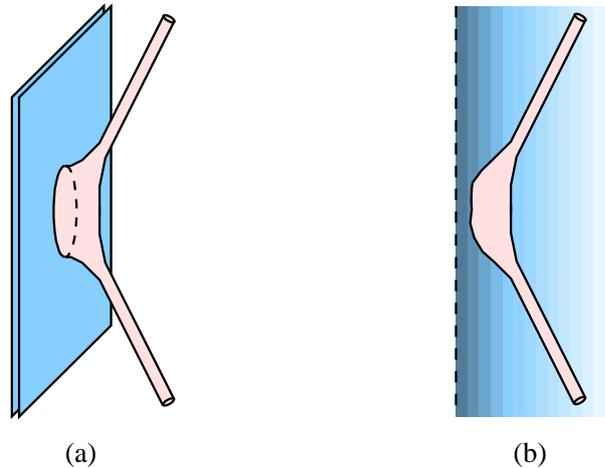}
\caption{ The interaction of a closed string with $N$ D--branes (a) is dual to propagation of the
same closed string in a deformed background due to the D--branes tension and charge (b).}
\label{stringbrane}
\end{center}
\end{figure}

The modern viewpoint over the gauge/string duality suggested by 't Hooft's $1/N$ expansion, is to embed it into the broader,
but perhaps more intuitive, duality between  open and closed string theories.
First, one interprets expression (\ref{openstring})  literally  as an open string perturbative expansion 
(or as an appropriate limit of it, usually the low energy limit).  
Then, the boundary conditions at the end--points of the open string will define D--branes,
which correspond to the space where the original gauge theory is defined.
Notice, though, that the open string is, in general, allowed to move in a larger space.
The last step is to identify the dual closed string theory. 
The intuitive idea is that the closed string moves in the same target space as the open string
but deformed by the presence of the D--branes. 
In other words, the sum over holes in diagrams describing a closed string
interacting with  D--branes  (figure \ref{stringbrane}(a)) 
is traded by a deformation of the closed string worldsheet theory (figure \ref{stringbrane}(b)).
This is then the desired dual closed string theory.

As just described, the derivation of a gauge/string duality might seem a simple algorithmic task.
Of course, this is not the case and, up to now, nobody was able to apply this program to QCD, 
with the exception of the 2--dimensional case  \cite{GrossTaylor}.
Nevertheless, there has been important progress in understanding with more detail some of the  steps involved.
In particular, how to write gauge theory Feynman diagrams as integrals over the moduli space of Riemann surfaces  \cite{GopakumarI}.
However, so far these direct approaches starting directly from the gauge theory 
side of the duality have not been very successful (see  
 \cite{Thorn, ThornYM, Gopakumar, GopakumarII, GopakumarIII, GopakumarIV}
for interesting attempts in this direction).

Presently, there are a few inspiring models of lower dimensional gauge theories where the program can be made rigorous:
the duality between 3--dimensional Chern--Simons theory and the A--model topological closed string
theory on the $S^2$ resolved conifold  \cite{GopakumarVafa}; and the duality  between 
double--scaled matrix models and minimal closed bosonic string theories  \cite{GaiottoRastelli}.
Usually, the string theories involved in these toy models are, in some sense, topological so that the open string theories are
precisely equivalent to gauge theories. 
In contrast, the  celebrated AdS/CFT correspondence  \cite{Malda, WittenAdS, GubserKlebanovPolyakov} was obtained as the low energy
limit of an underlying open/closed duality.


\section{AdS/CFT Correspondence}


Let us now briefly describe the reasoning that lead to the prototypical example of AdS/CFT correspondence.
The basic idea is to apply the picture in figure \ref{stringbrane} to the system
of $N$ coincident D3--branes of type IIB string theory placed in 10--dimensional Minkowski space
and then focus on its low energy dynamics.
More precisely, we shall consider the limit of vanishing string length, $l_s\equiv \sqrt{\alpha'} \to 0$, keeping
the string coupling $g_s$, the number of branes $N$ and the energy fixed.
On one hand, the system of figure \ref{stringbrane}(a) has closed and open string degrees of freedom, 
describing bulk and brane excitations, respectively.
As $\alpha'\to 0$, all massive string modes disappear and we are left with two decoupled systems: free gravity in 
$\mathbb{M}^{10}$ and  ${\cal N}=4$ supersymmetric $SU(N)$ Yang--Mills theory  \footnote{
In this discussion we neglect the decoupled $U(1)$
factor associated to the freedom to move the center--of--mass of the brane system.}
in the 4--dimensional brane worldvolume.
The Yang--Mills coupling constant is related to the string coupling by $g_{{\rm YM}}^2=2\pi g_s$,
in agreement with the generic expectations of 't Hooft's $1/N$ expansion. 
On the other hand, the equivalent description of figure  \ref{stringbrane}(b) corresponds to a closed string background
which is asymptotically $\mathbb{M}^{10}$, but is deformed by the D3--branes tension and charge. 
In the supergravity approximation, this is given by the metric \footnote{
The solution has also non-vanishing RR 5--form but this is not important for the argument.}
\begin{equation}
ds^2=h^{-1/2}\,ds^2_{\mathbb{M}^{4}} + h^{1/2}(dr^2 + r^2d\Omega^2_5) \ ,
\label{metric}
\end{equation}
where $d\Omega^2_5$ is the metric on the 5--sphere, the $\mathbb{M}^{4}$ corresponds to the branes worldvolume and
$$
h=1+\frac{\ell^{\,4}}{r^4}\ ,\ \ \ \ \ \ \ \ \ \ \ \ \ \ 
\ell^{\,4}=4\pi g_s N \alpha'^{2} \ .
$$
The low energy limit is slightly more subtle in this description. The limit $\alpha' \to 0$ naively applied
to the closed string background (\ref{metric}) yields just free gravity in $\mathbb{M}^{10}$. 
However, this is not the full result.
The vanishing of the metric time component at the location of the original branes ($r=0$) 
means that low--energy fluctuations from the point of view of an observer at infinity
can be very energetic local excitations in the near horizon region $r\ll \ell$.
This is just the usual red--shift phenomena associated with black hole horizons.
To appropriately determine the low energy limit of the near horizon region,  we introduce
a new coordinate $U\equiv r/\ell^{\,2}$ which we keep fixed as  $\alpha' \to 0$.
Then the near horizon geometry becomes
\begin{equation}
ds^2=\ell^{\,2} \left( \frac{dU^2}{U^2} + U^2\, ds^2_{\mathbb{M}^{4}}\right)  +\ell^{\,2} \, d\Omega^2_5 \ ,
\label{AdS5S5metric}
\end{equation}
corresponding to the product space  AdS$_5\times S^5$ both with radius $\ell$.
We conclude that the low energy limit of the dual closed string description of the brane+bulk system yields 
again two decoupled systems: type IIB closed strings in  AdS$_5\times S^5$ and, as before, free gravity in $\mathbb{M}^{10}$.
This naturally leads to the fascinating conjectured duality between 4--dimensional 
${\cal N}=4$ supersymmetric Yang--Mills (SYM) theory and 
type IIB string theory in AdS$_5\times S^5$.

It is interesting to see that this duality follows precisely the pattern suggested by 't Hooft's $1/N$ expansion.
Indeed, the 't Hooft coupling of the gauge theory determines the worldsheet coupling constant,
$$
\frac{\alpha'}{\ell^{\,2}}=\frac{1}{\sqrt{4\pi g_s N}}=\frac{1}{\sqrt{2 \lambda}}\ ,
$$
controlling string effects.
Clearly, the closed string worldsheet theory is weakly coupled when the dual gauge theory is strongly coupled.
This is a generic feature of gauge/string dualities that makes them so powerful, but also so hard to prove.
When the radius $\ell$ of AdS is much larger than the string length $\ell_s$ we can, 
in first approximation, analyze the dynamics of the low energy gravitational theory for  string massless modes.
Compactifying on the 5--sphere, we obtain a gravitational theory on AdS$_5$ whose Newton constant $G$, 
in units of the AdS radius, is
$$
G 
= \frac{\pi}{2 N^2}\ .
$$
Therefore, the $1/N$ expansion of the gauge theory correspond to the perturbative treatment of 
gravitational interactions between  string massless modes.

By now, there is a plethora of examples of the AdS/CFT correspondence. These are usually obtained 
by applying similar reasoning to other brane configurations in string theory.
The known examples have an impressive variety, differing in spacetime dimensionality, gauge symmetry,
global symmetry (like the amount of supersymmetry),
particle content (including fundamental fermions), $\beta$--function (including confining theories), etc.
However, the crucial closed string theory dual to physical QCD remains unknown.


The use of the AdS/CFT correspondence as a tool to describe gauge theory phenomena is  limited
by our poor understanding of String Theory.
Usually, one is restricted to the large $N$ and large $\lambda$ regime, where the string theory reduces
to a classical gravitational theory.
In the past years, a great effort has been made to understand the planar limit at generic values of 't Hooft coupling $\lambda$.
This amounts to the full comprehension of the string worldsheet theory, including the strong coupling regime.
The study of the extrapolation between large $\lambda $, where string theory is under control, and small   $\lambda $,
where gauge theory perturbative computations are reliable, has led to many interesting developments and ideas.
From these, the most prominent is perhaps the  surprising hypothesis of integrability of planar $\mathcal{N}=4$ SYM 
 \cite{MZ,Bei,BS,BSLR,Pol,Kaza,BHL,BES}.


\section{Outline of the Thesis}


In this thesis we shall concentrate  on $1/N$ effects, keeping the 't Hooft coupling $\lambda$ large.
As explained above, this regime is dual to perturbative quantum gravity in AdS.
Unfortunately, the loop expansion in the gravitational coupling $G$ is plagued with the usual ultra--violet (UV) divergences
present also in flat space when the regulator length $\ell_s$ is neglected.
Indeed, in most circumstances, we are forced to restrict our attention to tree level gravitational interactions,
which, in AdS, are already very complex.
Our strategy to go beyond tree level will be to focus on a specific kinematical limit of AdS scattering amplitudes.
It is well known that, in flat space, the quantum effects of various types of interactions can be reliably re--summed
to all orders in the relevant coupling constant, in specific kinematical regimes
  \cite{LevySucher, Abarbanel,tHoofteik,Kabat,ACV,tHoofteik3d,Deser88, Deser93}.
In particular, the amplitude for the scattering of two particles can be approximately computed in the 
{\em eikonal} limit of small momentum transfer compared to the center--of--mass energy.
In this limit, even the gravitational interaction can be approximately evaluated to \textit{all} orders in $G$,
and the usual perturbative UV problems are rendered harmless by the resummation process.
Moreover, at large energies, the gravitational interaction dominates all other interactions,
quite independently of the underlying theory  \cite{tHoofteik}. 

The main result of this thesis is the generalization of the eikonal approximation to high energy scattering in AdS
and its interpretation from the point of view of the dual CFT.
We remark that, although our results will include all terms of the $1/N$ expansion,
there are still finite $N$ effects that are not captured by our computations.
This is the case of  instanton effects which give rise to the usual 
non--perturbative factor $e^{-{\cal O}(1/g_s)} \sim e^{-{\cal O}(N/\lambda)}$.
Therefore we shall always consider $N\gg \lambda $, 
corresponding to small string coupling $g_s\ll 1$, therefore
ensuring the smallness of non--perturbative effects.

As usual in investigations of the AdS/CFT correspondence, 
this thesis is naturally divided in two main parts:
chapter \ref{ch:eikonal} is mainly devoted to computations in Anti--de Sitter spacetime;
while chapters  \ref{ch:cpw} and  \ref{ch:eikAdSCFT} concern the study of the dual CFT.

We start chapter \ref{ch:eikonal} by reviewing the eikonal approximation for potential scattering
in Quantum Mechanics.
We then rederive the standard eikonal
approximation to ladder and cross ladder diagrams in flat space  \cite{LevySucher, Abarbanel}, 
using Feynman rules in position space. This derivation makes the physical meaning of the eikonal
approximation transparent. Each particle follows a null
geodesic corresponding to its classical trajectory, insensitive to
the presence of the other. The leading effect of the interaction,
at large energy, is then just a phase $e^{\,I/4}$ determined by the tree level
interaction between the null geodesics ${\bf x}(\lambda)$ and ${\bf \bar{x}}(\bar{\lambda})$
of the incoming particles,
\begin{equation}
I= (-ig)^2 \int_{-\infty}^{\infty}d\lambda d\bar{\lambda} \,
\Pi^{(j)}\left({\bf x}(\lambda),{\bf \bar{x}}(\bar{\lambda})\right)\ ,
\label{geoint}
\end{equation}
where $g$ is the coupling and $\Pi^{(j)}$ is the propagator for the exchanged spin 
$j$ particle contracted with the external momenta. We shall see in section \ref{eikAdS} that this
intuitive description generalizes to AdS, therefore resumming ladder
and cross ladder Witten diagrams.
Moreover, we are able to formulate the eikonal approximation for high energy scattering
in a general spacetime, emphasizing its essential properties.
Closing chapter \ref{ch:eikonal}, we present an alternative derivation of 
the eikonal approximation in AdS,
based on gravitational shock waves. 
 
In chapter  \ref{ch:cpw}, we prepare the path for the 
CFT interpretation of the eikonal approximation in AdS.
Firstly, we review the conformal partial wave decomposition of the CFT four point correlator
\be
\hat{A}\left( \mathbf{p}_{1},\cdots ,\mathbf{p}_{4}\right) =\left\langle
\mathcal{O}_{1}\left( \mathbf{p}_{1}\right) \mathcal{O}_{2}\left( \mathbf{p}%
_{2}\right) \mathcal{O}_{1}\left( \mathbf{p}_{3}\right) \mathcal{O}%
_{2}\left( \mathbf{p}_{4}\right) \right\rangle \ , \label{corre}
\ee
of primary operators $\mathcal{O}_{1}$ and $\mathcal{O}_{2}$.
In particular, we establish the \emph{S}--channel ($\mathbf{p}_{1}\rightarrow \mathbf{p}_{2}$)
partial wave expansion of the disconnected graph 
$$
\hat{A}\left( \mathbf{p}_{1},\cdots ,\mathbf{p}_{4}\right) =\left\langle
\mathcal{O}_{1}\left( \mathbf{p}_{1}\right)
\mathcal{O}_{1}\left( \mathbf{p}_{3}\right) \right\rangle\,\left\langle
 \mathcal{O}_{2}\left( \mathbf{p}_{2}\right) 
\mathcal{O}_{2}\left( \mathbf{p}_{4}\right) \right\rangle \ ,
$$
as an exchange of an infinite tower of primary composite operators 
$\mathcal{O}_{1}\partial \cdots \partial \mathcal{O}_{2}$ of increasing dimension and spin.
Secondly, we introduce an impact parameter representation for the conformal partial waves appropriate
to describe the eikonal kinematical regime.

In chapter \ref{ch:eikAdSCFT}, we explore the consequences of the eikonal approximation
in AdS for the CFT four point correlator (\ref{corre}).
Using the eikonal approximation in AdS, we establish the behavior
of $\hat{A} $ in the limit of $\mathbf{p}_{1}\sim \mathbf{p}_{3}$. 
The relevant limit is not controlled by the
standard OPE, since the eikonal kinematics is intrinsically
Lorentzian. Nonetheless, the amplitude $\hat{A}$ is related to the
usual Euclidean correlator $A$ by analytic continuation and can be
easily expressed in terms of the impact parameter representation 
introduced in chapter  \ref{ch:cpw}.

\begin{figure}
[ptb]
\begin{center}
\includegraphics[width=12cm]{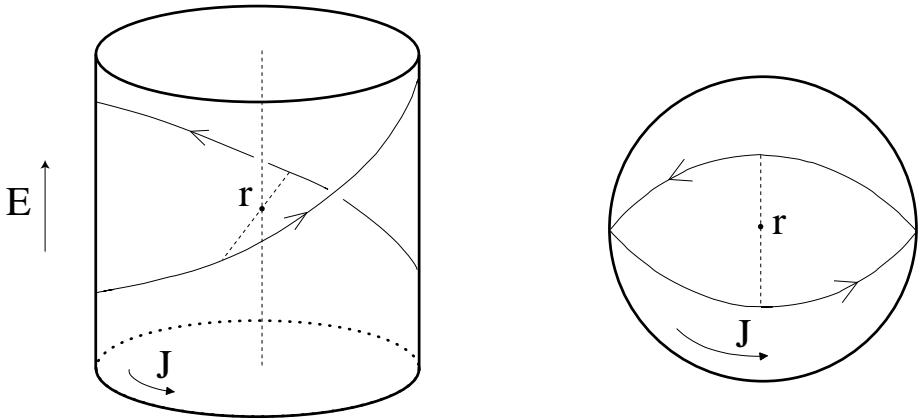}
\caption{Classical null trajectories of two  particles
moving in AdS$_{d+1}$ with total energy $E$ and
relative angular momentum $J$.
Their impact parameter $r$ is given by $\tanh\left( r/2\right)=J/E$.}
\label{nullgeodincylinder}
\end{center}
\end{figure}

In the eikonal regime, the \emph{S}--channel partial wave decomposition 
of the amplitude $\hat{A}$ is dominated, 
as in flat space, by composite states
of the two incoming particles. These are dual to the composite primary
operators $\mathcal{O}_{1}\partial \cdots \partial \mathcal{O}_{2}$ of
classical dimension $E$ and spin $J$,  already present at zero order. 
We then find that the eikonal approximation to $\hat{A}$
controls the anomalous dimension $2\Gamma \left( E,J\right) $ of these
intermediate two--particle states, in the limit of large $E$ and $J$. 

Heuristically,
the basic idea can be summarized in two steps. Firstly, the two incoming particles
approximately follow two null geodesics in AdS$_{d+1}$ with total energy $E$ and
relative angular momentum $J$, as described by Figure \ref{nullgeodincylinder}. 
The corresponding  $(d-1)$--dimensional impact parameter space is the  transverse hyperboloid $H_{d-1}$
and the minimal geodesic distance $r$ between the null geodesics is given by
$$
\tanh \left( \frac{r}{2}\right) =\frac{J}{E}\ .
$$
Then, the eikonal approximation determines the phase  $e^{-2\pi i \Gamma}$
due to the exchange of a particle of spin $j$ and dimension $\Delta$ in AdS.
As described above, this phase shift is determined by the interaction between
the two null geodesics. We shall see that computing (\ref{geoint}) in AdS gives
\begin{equation}
2\,\Gamma (E,J)\simeq 
-\frac{g^{2}}{2\pi }\,(E^{2}-J^{2})^{\,j-1}\,\Pi_{\perp }\left( r\right) \ 
\ \ \ \ \ \ \ \ \ \ \ \ \ \left( E\sim J\rightarrow \infty \right)   \ ,
\label{finalres}
\end{equation}
where $g$ is the coupling in AdS and $\Pi _{\perp }$ is the Euclidean scalar
propagator of dimension $\Delta -1$ in the transverse space $H_{d-1}$.
Secondly, the phase shift is related to the anomalous dimension by the following 
argument. Recall that  \cite{AdSreview}, due to the
conformal structure of AdS, wave functions have
discrete allowed frequencies. More precisely, a state of
dimension $\delta$ with only positive frequencies will be almost periodic in global time $\tau$, 
acquiring only a phase $e^{-2\pi i \delta }$ as $\tau \rightarrow \tau +2\pi $. 
Since the interaction between the two particles occurs in a global time span of
$\pi$, we conclude that the full dimension of the composite state is
$\delta= E + 2\Gamma \left( E,J\right) $.

We end chapter  \ref{ch:eikAdSCFT} focusing on the tree level 
contribution for the eikonal amplitude, to study the 
 \emph{T}--channel ($\mathbf{p}_{1}\rightarrow \mathbf{p}_{3}$)
partial wave expansion of an AdS tree level exchange graph in the same channel. 
We establish that a spin $j$ AdS exchange  has a partial wave expansion
of bounded spin $J\le j$ and that the eikonal approximation fixes the coefficients of the
intermediate primaries of highest spin $J=j$.
Armed with this result, we deduce that the
leading contribution to $\Gamma $, for $E\sim J\rightarrow \infty $, is determined
completely by the tree level \emph{T}--channel exchange, so that (\ref{finalres}) is exact
to all orders in the coupling $g$. Moreover, in gravitational theories, the
leading contribution to $\Gamma $ comes from the graviton  \cite{tHoofteik}, with $j=2$ and 
$\Delta =d$. The result (\ref{finalres}) is then valid to all orders in the
gravitational coupling $G=g^{2}/8\pi$. 
This provides a powerful prediction for
the duality between strings on AdS$_5\times S^5$ and four dimensional ${\cal N}=4$ SYM:
the anomalous dimension of the above double trace operators is
\begin{equation}
2\,\Gamma (E,J)\simeq - \frac{1}{4N^2}\,\frac{(E-J)^4}{EJ} \ 
\ \ \ \ \ \ \ \ \ \ \ \ \ \left( E\sim J\rightarrow \infty \right) \ ,
\label{resSYM4}
\end{equation}
for $E-J \ll J$, so that the impact parameter $r$ is much larger than the $S^5$ radius $\ell=1$
and the effects of massive KK modes are negligible.

Finally, we present our conclusions and comments on open questions in chapter \ref{ch:conc}.


\section{Preliminaries \& Notation}


This section contains the indispensable information on AdS/CFT 
necessary for the considerations in the subsequent chapters.
We shall avoid the use of specific realizations of the  
AdS/CFT correspondence by considering a generic formulation
summarizing its essential properties.
We emphasize the importance of this section given that it
introduces the powerful and infrequent notation based on the embedding of AdS space 
into flat space, which will be extensively used in this thesis.


\subsection{AdS Geometry} \label{AdSgeom}


We start by recalling that AdS$_{d+1}$ space (of dimension $d+1$) can be defined as 
a pseudo--sphere in the embedding space $\mathbb{R}^{2,d}$.
More precisely, AdS space of radius $\ell$ is given by the set of points
\begin{equation}
{\bf x} \in \mathbb{R}^{2,d}\ ,\ \ \ \ \ \ \ \ 
\mathbf{x}^{2}=-\ell^{\,2} \ .\label{pseudoS}
\end{equation}
Throughout this thesis, within the context of AdS/CFT, all points, vectors and scalar products are 
 taken in the embedding space  $\mathbb{R}^{2,d}$.
Also, from now on we choose units such that $\ell=1$.
It is clear from the above definition that the AdS isometry group is the Lorentz group $SO(2,d)$.
Rigorously, AdS$_{d+1}$ is the universal covering space of the pseudo--sphere (\ref{pseudoS}), obtained
by decompactifying along the non--contractible timelike cycle associated with the action of $SO(2) \subset SO(2,d)$.

\begin{figure}
\begin{center}
\includegraphics[width=6cm]{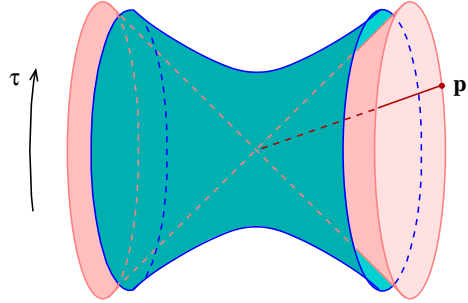}
\caption{Embedding of AdS$_2$ in $\mathbb{R}^{2,1}$. 
A point  $\mathbf{p}$ in the boundary of AdS$_2$ is a null ray in  $\mathbb{R}^{2,1}$.
Rotations along the non--contractible timelike cycle are global time translations.}
\label{AdSTwo}
\end{center}
\end{figure}

We shall define the holographic boundary of AdS$_{d+1}$ as the set of outward null rays 
from the origin of the embedding space $\mathbb{R}^{2,d}$,
i.e. a point in the boundary of AdS is given by 
\begin{equation}
{\bf p} \in \mathbb{R}^{2,d} \ ,\ \ \ \ \ \ \ \ 
{\bf p}^2=0 \ ,\ \ \ \ \ \ \ \ 
{\bf p}\sim \lambda {\bf p}\ ,\ \ \ \ \ \ \ \ (\lambda>0)\ .
\label{AdSboundary}
\end{equation}
Again, the boundary is  the universal covering space of (\ref{AdSboundary}).
In figure \ref{AdSTwo} the embedding of the AdS geometry is represented for the AdS$_2$ case. 
Usually, the AdS boundary is identified with a specific section $\Sigma$ of the light cone 
in  $\mathbb{R}^{2,d}$, containing one representative point for each null ray. 
Then, the embedding space naturally induces a metric on $\Sigma$.
The metrics on two different light cone sections, $\Sigma$ and $\Sigma'$, are related by a Weyl transformation,
since any  ${\bf p} \in \Sigma$ can be rescaled to ${\bf p'}=\Lambda({\bf p}) {\bf p} \in\Sigma' $,
yielding
$$
ds_{\Sigma'}^2= d{\bf p'}^2=\left({\bf p} d \Lambda + \Lambda d{\bf p}\right)^2=  \Lambda^2 d{\bf p}^2 =  \Lambda^2 ds_{\Sigma}^2\ .
$$
Furthermore, the natural action of the Lorentz group $SO(2,d)$ on the space of null rays defines
the action of the conformal group on  $\Sigma$  \footnote{This was explained long  ago by Dirac  \cite{Dirac}.}.
However, Lorentz transformations preserve the light cone but not the section  $\Sigma$.
Thus, their action ${\bf p} \to {\bf p}' $ on a point ${\bf p} \in \Sigma $ 
needs to be compensated with an appropriate re--scaling
 ${\bf p}'\to \lambda {\bf p}' $, so that the final point $\lambda {\bf p}'$ belongs to $\Sigma $.
This makes the action of the conformal group less transparent.
Therefore, in this thesis, we shall use the coordinate--invariant definition  (\ref{AdSboundary}) of the
AdS boundary.
Nevertheless, to develop some intuition on this invariant language, and to make contact with other studies on AdS/CFT, 
we shall now briefly review the two most common choices of light cone sections or, equivalently,
representations of the AdS boundary.

\begin{figure}
\begin{center}
\includegraphics[width=12cm]{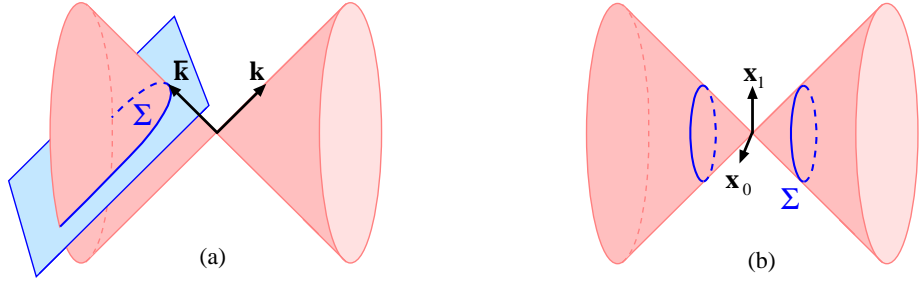}
\caption{Two choices of sections $\Sigma$ of the light cone of  $\mathbb{R}^{2,1}$ 
representing the  AdS$_2$ boundary: (a) 1--dimensional Minkowski space; (b) $S^0 \times \mathbb{R}_{\tau} $ 
(after decompactifying the global time circle).}
\label{AdS2boundary}
\end{center}
\end{figure}

The first choice of light cone section  $\Sigma$ is associated with the familiar 
statement that the boundary of AdS$_{d+1}$ is $d$--dimensional Minkowski space.
In this case,  $\Sigma$ is defined by the set of points  ${\bf p}$ satisfying
\begin{equation}
{\bf p} \in \mathbb{R}^{2,d} \ ,\ \ \ \ \ \ \ \ 
{\bf p}^2=0 \ ,\ \ \ \ \ \ \ \ 
-2 {\bf p} \cdot {\bf k}=1\ ,
\label{AdSboundaryPoincare}
\end{equation}
with ${\bf k}$ a fixed null vector in  $\mathbb{R}^{2,d}$.
This choice is usually associated with AdS written in Poincar\'e coordinates.
These are obtained by splitting the embedding space $\mathbb{R}^{2,d}$ as the direct product
$\mathbb{M}^2 \times \mathbb{M}^d$, with the $\mathbb{M}^2$
containing the null vector ${\bf k}$.
Poincar\'e coordinates $\{y,{\bf y} \}$, where $y\in \mathbb{R}^+$, ${\bf y} \in \mathbb{M}^d$, are then defined by
\begin{equation}
{\bf x}= \frac{1}{y} \left( {\bf \bar{k}} + (y^2+{\bf y}^2) {\bf k} + {\bf y} \right)\ ,
\label{Poincare}
\end{equation}
with ${\bf \bar{k}}$ the null vector in $\mathbb{M}^2$ satisfying  $ -2{\bf k} \cdot {\bf \bar{k}}=1$.
In this expression, and often in the rest of the thesis, the point  ${\bf y} \in \mathbb{M}^d$ represents a point
in the embedding space $\mathbb{R}^{2,d}=\mathbb{M}^2 \times \mathbb{M}^d$ at the origin of the $\mathbb{M}^2$ factor.
In other words, ${\bf y} $ parametrizes the subspace of  $\mathbb{R}^{2,d}$ orthogonal to  ${\bf k}$  and ${\bf \bar{k}}$.
It is now trivial to compute the AdS metric in Poincar\'e coordinates,
$$
ds^2=d{\bf x}^2= \frac{1}{y^2} \left( dy^2+ d{\bf y}^2 \right)\ .
$$
Similarly, one can parametrize the section $\Sigma$ using ${\bf y} \in \mathbb{M}^d$, by writing 
\begin{equation}
{\bf p} = {\bf \bar{k}} + {\bf y}^2 {\bf k} + {\bf y}\ .
\label{kofy}
\end{equation}
Then, the induced metric on $\Sigma$ is indeed the Minkowski one.
The fact that $y \,{\bf x} \to {\bf p}$ as $y\to 0$ leads to the usual statement that the AdS boundary
is at $y=0$.
In figure \ref{AdS2boundary}(a) we plot this choice of section for the  AdS$_2$ case. 

\begin{figure}
\begin{center}
\includegraphics[keepaspectratio,height=6cm]{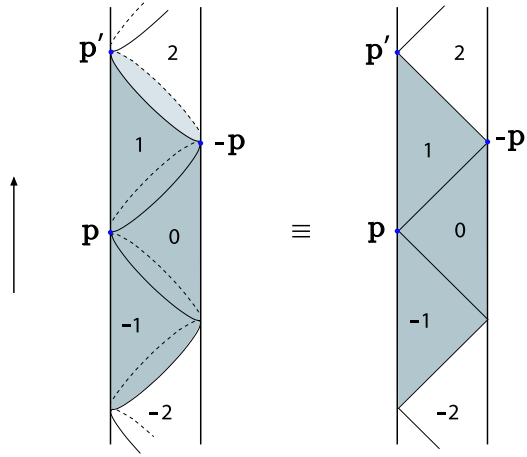}
\caption{Poincar\'e patches of an arbitrary boundary point $\mathbf{p}$,
separated by the null surfaces $\mathbf{x\cdot p}=0$. 
Here, AdS is represented as a cylinder with boundary $\mathbb{R}\times {S}^{d-1}$. 
Throughout this thesis we shall mostly use a two--dimensional simplification of this picture, 
as shown in the figure. The point $-\mathbf{p}$ and an image point $\mathbf{p}'$ of $\mathbf{p}$ are also shown.}
\label{DecompAdS}
\end{center}
\end{figure}

The coordinate system (\ref{Poincare}) only covers part of AdS and, similarly, the section $\Sigma$
is not the entire AdS boundary.
To see this, we use the boundary point $\mathbf{p}$ to divide AdS in an infinite sequence of Poincar\'{e}
patches, separated by the null surfaces $\mathbf{x\cdot p}=0$ and labeled by integers $n$ increasing 
as we move forward in global time. 
The  Poincar\'e coordinates  (\ref{Poincare}) cover only the $n=0$ patch, which is the one 
spacelike related to the boundary point $\mathbf{p}$, as shown in figure \ref{DecompAdS}.

The splitting $\mathbb{M}^2 \times \mathbb{M}^d$ of the embedding space $\mathbb{R}^{2,d}$ establishes a direct
map between the usual generators $\{\mathcal{D}, \mathcal{P}_a, \mathcal{K}_a, \mathcal{J}_{ab}\} $ 
of the conformal group acting on $\mathbb{M}^d$ and the Lorentz generators $ {\bf J}_{\mu\nu} $ of $SO(2,d)$,
$$
\mathcal{D} = 2{\bf k}^{\mu} {\bf \bar{k}}^{\nu} {\bf J}_{\mu\nu} \ ,\ \ \ \ \ \ \  
\mathcal{P}_a = -2 {\bf k}^{\mu}  {\bf J}_{\mu a} \ ,\ \ \ \ \ \ \ 
\mathcal{K}_a = -2{\bf \bar{k}}^{\mu} {\bf J}_{\mu a} \ ,\ \ \ \ \ \ \ 
\mathcal{J}_{ab} =  {\bf J}_{ab}\ ,
$$
with the indices $\mu,\nu$  varying over all  $\mathbb{R}^{2,d}$ directions and with the indices $a,b$
restricted to the   $\mathbb{M}^d$ factor.
Notice that conformal transformations that leave the $d$--dimensional metric invariant (the Poincar\'e group)
are, in this language, the transformations that preserve the section $\Sigma$. 
Indeed, acting on ${\bf p}({\bf y})$, as defined in (\ref{kofy}), we obtain
$$
e^{ -2 {\bf k}^{\mu} {\bf a}^{\nu} {\bf J}_{\mu \nu} }\, {\bf p}({\bf y}) = {\bf p}\left( {\bf y} + {\bf a} \right) \ ,\ \ \ \ \ \ \  \ \ 
e^{\,\omega^{ab}{\bf J}_{ab} } \,{\bf p}({\bf y}) =  {\bf p}\left( e^{\,\omega^{ab}{\bf J}_{ab} } \,{\bf y} \right)\ ,
$$
where  ${\bf a} \in \mathbb{M}^d \subset \mathbb{R}^{2,d} $.
In contrast, the transformations that 
do not preserve the  section $\Sigma$, sending ${\bf p}({\bf y})\in \Sigma $ to $ \Lambda({\bf y}){\bf p}({\bf y'})$,
induce a conformal scaling of the $d$--dimensional metric 
$$
ds_{\Sigma}^2=d{\bf y}^2 \to  d{\bf y'}^2 =\Lambda^{-2}({\bf y})d{\bf y}^2\ .
$$ 
Dilatations,
$$
e^{ \,2\alpha {\bf k}^{\mu} {\bf \bar{k}}^{\nu} {\bf J}_{\mu\nu} } \,{\bf p}({\bf y}) = 
e^{-\alpha}\,{\bf p}\left( e^{\,\alpha} \,{\bf y} \right) \ ,
$$
and special conformal transformations,
$$
e^{-2{\bf \bar{k}}^{\mu} {\bf a}^{\nu} {\bf J}_{\mu \nu} }\, {\bf p}({\bf y}) =  
\left( 1+ 2{\bf a}\cdot {\bf y} + {\bf a}^2 {\bf y}^2\right)
{\bf p}\left( \frac{ {\bf y} +{\bf y}^2 {\bf a} }{1+ 2{\bf a}\cdot {\bf y} + {\bf a}^2 {\bf y}^2}  \right)\ ,
$$
belong to this class.

The other frequent face of the AdS boundary is $\mathbb{R}\times S^{d-1}$, with $\mathbb{R}$ representing
the time direction.
This is usually  associated with AdS written in global coordinates.
Here, one splits the embedding space $\mathbb{R}^{2,d}$ as the product of a timelike $\mathbb{R}^2$
by an Euclidean $\mathbb{R}^{d}$. 
It is convenient to define the  timelike plane as the space spanned by two orthogonal and normalized timelike vectors
in $\mathbb{R}^{2,d}$, that is two points ${\bf x}_0,{\bf x}_1$ in AdS$_{d+1}$ satisfying    ${\bf x}_0 \cdot {\bf x}_1 =0$.
One can then introduce global coordinates $\{\tau,\rho,{\bf n} \}$ for AdS as follows
\begin{equation}
{\bf x} = \cosh(\rho) \left[ \cos(\tau) \,{\bf x}_0 + \sin(\tau) \, {\bf x}_1 \right]+ 
\sinh(\rho) \,{\bf n}\ ,
\label{Global}
\end{equation}
where the decompactified angle $\tau$ is the global time, $\rho>0$ is a radial coordinate 
and the vector ${\bf n} $ parametrizes
the unit sphere $S^{d-1} \subset \mathbb{R}^{d}$  orthogonal to  ${\bf x}_0$ and ${\bf x}_1$.
A natural  choice of light cone section $\Sigma$ is  the set of null vectors in   $\mathbb{R}^{2,d}$ with unit
projection onto the chosen timelike plane, i.e. the set of points
\begin{equation}
{\bf p} = \cos(\tau) \,{\bf x}_0 + \sin(\tau) \, {\bf x}_1 +{\bf n}\ .
\label{AdSboundaryGlobal}
\end{equation}
This corresponds to the intuitive idea that one approaches the AdS boundary as $\rho \to \infty$.
Indeed,  $2e^{-\rho} {\bf x}\to {\bf p} $ as  $\rho \to \infty$.
In figure \ref{AdS2boundary}(b) we plot this choice of  light cone section for the  AdS$_2$ case. 
Finally, we point out that one must be cautious when working in embedding coordinates since two 
general bulk points $\mathbf{x}$ and $\mathbf{x'}$, or two boundary points $\mathbf{p}$ and $\mathbf{p'}$, 
related by a global time translation of integer multiples of $2\pi$, have the same embedding in  $\mathbb{R}^{2,d}$.

In the sequel, we will find the transverse hyperbolic space $H_{d-1}\subset\mathbb{M}^{d}$,
given by the upper mass--shell
$$
{\bf y}^{2}=-1\ ,
$$
where $ {\bf y}\in\mathbb{M}^{d}$ is future directed. 
We shall denote with $\mathrm{M}\subset\mathbb{M}^{d}$ the future Milne wedge given by ${\bf y}^{2}\leq0$,
with $ {\bf y}\in\mathbb{M}^{d}$ future directed.
Similarly, we shall use  $-\mathrm{M}$ and  $-H_{d-1}$ to designate the past Milne wedge and
the associated transverse hyperbolic space. 
Finally, we represent by
\begin{eqnarray*}
\int_{{\rm AdS}} d\mathbf{x} \, f( {\bf x} ) & =& \int_{\mathbb{R}^{2,d} } d\mathbf{x}~2\,\delta\left(  \mathbf{x}
^{2}+1\right)  f( {\bf x} ) ,\ \ \ \ \ \ \ \ \ \ \ \\
\int_{H_{d-1}} d\mathbf{y} \, f( {\bf y} ) & =& \int_{{\rm M} } d\mathbf{y}~2\,\delta\left(  \mathbf{y}
^{2}+1\right)  f( {\bf y} ) ,
\end{eqnarray*}
a generic integral on AdS$_{d+1}$ and $H_{d-1}$, respectively. 
With this notation, we have
\[
\int_{\mathrm{M}}d {\bf y}   =
\int_{0}^{\infty}r^{d-1}dr\int_{H_{d-1}} d{\bf w} \ ,
\]
where $ {\bf y} =r  {\bf w}$ and $ {\bf w} \in H_{d-1}$.


\subsection{AdS Dynamics}   \label{AdSdyn}


Particle propagation simplifies considerably in the embedding space language.
For example, the timelike geodesic corresponding to a particle stopped at $\rho=0$ in global coordinates
(\ref{Global}), is simply the intersection, in the embedding space $\mathbb{R}^{2,d}$,
of the timelike plane generated by  ${\bf x}_0$ and  ${\bf x}_1$,
with the pseudo--sphere defining AdS.
Then, by Lorentz transformations, we conclude that any 
timelike geodesic is given by the intersection of a timelike plane through the origin of $\mathbb{R}^{2,d}$
with the pseudo--sphere defining AdS. Moreover, the energy of this particle, as measured by an observer
following another timelike geodesic, is just a measure of the intersection angle between the two 
timelike planes defining the geodesics. 
Most importantly for our purposes are null geodesics in AdS.
These are also geodesics in the embedding space $\mathbb{R}^{2,d}$, 
$$
{\bf x}(\lambda)={\bf x}_0 + \lambda {\bf k}\ ,
$$ 
with ${\bf x}_0^2=-1$,  ${\bf k}^2=0$ and  ${\bf x}_0 \cdot{\bf k}=0$.
As it is well known, lightlike geodesics reach the AdS boundary in finite global time.
More precisely, the above null geodesic starts at the point $- {\bf k}$ of the AdS boundary,
travels during a global time interval $\Delta \tau =\pi$, and 
ends up again in the AdS boundary  at the  point ${\bf k}$.
On the other hand, if one tries to reach a null geodesic by taking the limit of a timelike geodesic with infinite energy,
one obtains an infinite sequence of null geodesics bouncing back and forth on the AdS boundary
between consecutive images of ${\bf k}$ and $-{\bf k}$.

The embedding space is also very useful to describe field dynamics in AdS.
The d'Alembertian  operator in AdS can be  written using just partial derivatives
in the flat embedding space,
\begin{equation}
\Box_{\rm AdS} = \boldsymbol{\partial}^2+ {\bf x}\cdot \boldsymbol{ \partial}(d+{\bf x}\cdot \boldsymbol{\partial}) \ .
\label{box}
\end{equation}
Rigorously, in order to take partial derivatives $\boldsymbol{ \partial}$,
it would be necessary to extend the domain of the fields from AdS to the full embedding space.
However, the action of $\Box_{\rm AdS}$ is independent of this extension so that we can use
this convenient  notation without reference to any extension of the domain.
On the other hand, this shows that a  scalar field $ \psi $ satisfying the massless Klein--Gordon equation in the embedding space,
$$
 \boldsymbol{\partial}^2 \, \psi =0\ ,
$$
and the scaling relation
$$
\psi(\lambda{\bf x} )= \lambda^{-\Delta}\psi({\bf x} )\ ,
$$
obeys the massive Klein--Gordon equation in AdS,
$$
\Box_{\rm AdS}\, \psi = m^2 \,\psi\ ,
$$
with
$$
m^2= \Delta(\Delta-d)\ .
$$

The wave equation in AdS requires the specification of boundary conditions on the AdS timelike boundary.
Firstly, one needs to determine the behavior near the AdS boundary of solutions of the  Klein--Gordon equation in AdS.
For this purpose, it is convenient
to introduce generic coordinates for AdS, related with a given light cone section $\Sigma$ parametrized by ${\bf y} $,
\begin{equation}
{\bf x}= \frac{1}{y} \,{\bf p}({\bf y})  + y \,{\bf \bar{p}}({\bf y})\ ,
\label{ySigma}
\end{equation}
where $y$ is the inverse of the radial coordinate along the light cone so that $y=0$ corresponds to the boundary at infinity.
The null vectors $ {\bf p}, {\bf \bar{p}}$ satisfy $ -2{\bf p}\cdot {\bf \bar{p}}=1 $.
Notice that the Poincar\'e coordinates (\ref{Poincare}) are precisely of this form with ${\bf p}({\bf y})$ given
in (\ref{kofy}) and ${\bf \bar{p}}({\bf y})={\bf \bar{k}}$ a fixed null vector.
Similarly, the AdS global coordinates (\ref{Global}) also have the generic form  (\ref{ySigma})
with $2e^{-\rho}$ playing the role of $y$, the boundary coordinates ${\bf y}=\{\tau, {\bf n} \} $ and   
\begin{align*}
{\bf p}(\tau, {\bf n})& =  \cos(\tau) \,{\bf x}_0 + \sin(\tau) \, {\bf x}_1 + {\bf n}\ ,\\
4\,{\bf \bar{p}}(\tau, {\bf n})& =  \cos(\tau) \,{\bf x}_0 + \sin(\tau) \, {\bf x}_1 - {\bf n} \ .
\end{align*}
In the coordinate system (\ref{ySigma}) the AdS metric near the boundary reads
$$
ds^2=d {\bf x}^2 = \frac{1}{y^2}\left[dy^2+ds^2_{\Sigma}+O(y) \right]\ .
$$
Therefore, a solution of the massive Klein--Gordon equation in AdS can have two leading behaviors as $y\to 0$,
\begin{equation}
  \psi ({\bf x}) \simeq \frac{1}{2\Delta-d} \, \phi({\bf y}) \,y^{d-\Delta}  +  \tilde{\phi}({\bf y})\, y^{\Delta}\ ,
\label{asymptoticKG}
\end{equation}
where we have normalized $\phi({\bf y})$ for later convenience.
This asymptotic behavior is crucial for the AdS/CFT correspondence.

The Feynman propagator in AdS satisfies
$$
\left[ \Box_{\rm AdS} -  \Delta(\Delta-d)\right] \Pi_{\Delta}({\bf x} ,{\bf \bar{x}} ) = i\delta ({\bf x} ,{\bf \bar{x}} )\ .
$$
As usual in quantum field theory, the Lorentzian propagators require an appropriate $i\epsilon$ prescription
as ${\bf x}$ crosses the light cone of ${\bf \bar{x}} $. 
The best way to accomplish this is to define the Lorentzian propagator by a Wick rotation of the 
Euclidean one. The latter satisfies the similar equation 
$$
\left[ \Box_{H_{d+1}} -  \Delta(\Delta-d)\right] \Pi_{\Delta}({\bf x} ,{\bf \bar{x}} ) = -\delta ({\bf x} ,{\bf \bar{x}} )\ ,
$$
but now on the hyperbolic space $H_{d+1}$,
which is the Euclidean version of  AdS$_{d+1}$.
The propagator can be explicitly written using the hypergeometric function,
\begin{equation}
\Pi_{\Delta} = \mathcal{C}_{\Delta} \, u^{-\Delta}\,
F  \left( \Delta, \frac{2\Delta-d+1}{2}, 2\Delta-d+1 , -\frac{4}{u}\right)~,~
\label{AdSprop}
\end{equation}
where
$$
\mathcal{C}_{\Delta}=\frac{1}{2 \, \pi ^{\frac{d}{2}} } 
\frac {\Gamma \left(  \Delta \right) }{ \Gamma \left(  \Delta-\frac{d}{2}+1 \right)}
$$
is a normalization constant and 
$$
u=({\bf x}- {\bf \bar{x}})^2
=-2(1+{\bf x} \cdot {\bf \bar{x}})
$$
is the Lorentz invariant chordal distance, which is always positive in the Euclidean regime.
In contrast, in the Lorentzian regime there are negative (timelike) chordal distances which require 
a prescription to pass through the branch points of $\Pi_{\Delta}(u)$ at $u=0$ and $u=-4$.
The correct prescription is obtained by following the complex path of $u$ as the AdS points are Wick rotated
from the Euclidean to the Lorentzian setting.
  \begin{figure}
\begin{center}
\includegraphics[width=11cm]{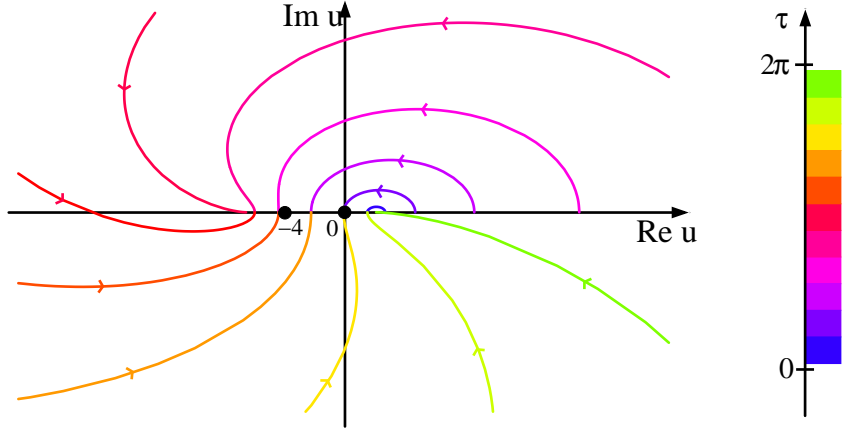}
\caption{Complex paths $u(\theta)$ in (\ref{wickrotation}) for several values of global time $\tau>0$.
All paths start in the positive real axis (Euclidean regime) and end up again somewhere in the real axis (Lorentzian regime).
 As $\tau$ increases, the paths spiral more and more around the branch point at $u=0$.}
\label{Wickspiral}
\end{center}
\end{figure}
To see the explicit result of this prescription, write ${\bf x} $ in global coordinates (\ref{Global}).
Choosing  ${\bf \bar{x}}={\bf x}_0$ for simplicity, we obtain
$$
u=-2+2\cosh(\rho)\cos(\tau)\ .
$$
Then, consider the standard Wick rotation parametrized by $0\leq \theta \leq 1$,
\begin{equation}
u(\theta)=-2+2\cosh(\rho)\cos\left(-i\tau e^{\frac{i\pi}{2}\theta}\right)\ ,
\label{wickrotation}
\end{equation}
with $\theta=0$ in the Euclidean setting and $\theta=1$ in the Lorentzian one.
The Lorentzian propagator is then given by the function $\Pi_{\Delta}(u)$ in the Riemann sheet 
containing the endpoint $u(1)$ of the complex path  $u(\theta)$, starting from the Euclidean point $u(0)$
 in the positive real axis. In figure \ref{Wickspiral}, we plot the  paths  $u(\theta)$ 
for several values of global time $\tau>0$, which is sufficient since the paths  $u(\theta)$ in (\ref{wickrotation}) are 
insensitive to the sign of $\tau$. Denoting with $|\tau| <\tau_*$, the region where  ${\bf x}$ and ${\bf \bar{x}}$
are spacelike separated, we have
$$
\cosh(\rho)\cos\left(\tau_*\right)=1\ .
$$
In the region   $|\tau-\pi| <\tau_*$ the point   ${\bf x}$ is spacelike separated from $-{\bf \bar{x}}$
(see figure \ref{xpatchs}). In general, for  
$$
|\tau- n\pi| <\tau_*
$$ 
we have $-4/u < 1$ and therefore there is no phase factor contribution from the branch cut 
of the hypergeometric function in the propagator expression. 
Following carefully the spiral structure of the complex paths in  figure \ref{Wickspiral}, we conclude that, in these regions,
the factor $u^{-\Delta} $ in the propagator yields a phase
$$
e^{-i \pi  \Delta|n|}\ .
$$
For $\cosh(\rho)\cos\left(\tau\right)<1$, the complex paths   $u(\theta)$ end on the branch cut of the hypergeometric function.
In this case, one should use  $-4/u(1-\epsilon)$  as the argument of the  hypergeometric function.
These comments are summarized in figure \ref{xpatchs}.

\begin{figure}
\begin{center}
\includegraphics[width=2.5cm]{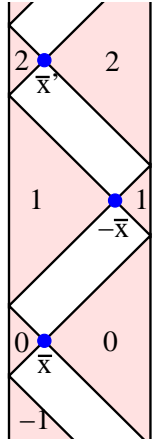}
\caption{The light cone of ${\bf \bar{x}}$ divides AdS in different patches where the propagator has different phases.
In the numbered shaded regions, the propagator has the phase $e^{-i \pi  \Delta|n|}$.
In the white regions of AdS, one needs to be careful with the branch cut of the hypergeometric function in (\ref{AdSprop}).
}
\label{xpatchs}
\end{center}
\end{figure}

In the AdS/CFT correspondence, there is also a bulk to boundary propagator.
This can be obtained as a limit of the bulk to bulk propagator described above.
To approach the boundary, we write a point in AdS in the form (\ref{ySigma}) and take the limit $y\to 0$,
\begin{equation}
K_{\Delta}({\bf p}, {\bf x})= \lim_{y \to 0} y^{-\Delta}\,\Pi_{\Delta}
\left(  \frac{1}{y}\,{\bf p} + y\,{\bf \bar{p}} \,  ,\,{\bf x}  \right)
= \mathcal{C}_{\Delta} \frac{ e^{-i \pi \Delta|n| } }{ |2 {\bf p}\cdot{\bf x} |^{\Delta}}
\ ,
\label{bblimitbb}
\end{equation}
where now $n$  numbers the Poincar\'e patches of ${\bf p}$ as in figure \ref{DecompAdS}.
We shall often use the $i\epsilon $ prescription
$$
K_{\Delta}({\bf p}, {\bf x})=  \frac{\mathcal{C}_{\Delta} }{ (-2 {\bf p}\cdot{\bf x} +i\epsilon)^{\Delta}}
\ ,
$$
which  works for ${\bf x}$ in the $n=-1,0,1 $ Poincar\'e patches of ${\bf p}$ (see figure \ref{DecompAdS}).
As a function of ${\bf x} \in$ AdS, the bulk to boundary propagator $K_{\Delta}({\bf p}, {\bf x})$ satisfies the
Klein--Gordon equation in AdS
with mass squared $\Delta(\Delta-d)$. Therefore, it must follow the general asymptotic behavior (\ref{asymptoticKG}).
Indeed, as  ${\bf x}$ approaches the boundary ($y\to 0$)
\begin{equation}
K_{\Delta}\left( {\bf p'}, {\bf x}=\frac{1}{y}\,{\bf p} + y\,{\bf \bar{p}}   \right) \simeq
y^{\Delta} \, K_{\Delta}\left( {\bf p'}, {\bf p}  \right) +  y^{d-\Delta} \,
\frac{1}{2\Delta-d} \,\delta_{\Sigma}\left( {\bf p'}, {\bf p}  \right)\ , 
\label{asymptoticprop}
\end{equation}
where 
$$
 K_{\Delta}\left( {\bf p'}, {\bf p}  \right)= 
\mathcal{C}_{\Delta} \frac{ e^{-i \pi \Delta|n| } }{ |2 {\bf p'}\cdot{\bf p} |^{\Delta}} 
$$ 
is the boundary to boundary propagator and $ \delta_{\Sigma}$ is the $d$--dimensional Dirac delta function on
the lightcone section $\Sigma$.
The standard normalization of the bulk to boundary propagator is  chosen such that there is no factor 
of $1/(2\Delta-d) $ multiplying the delta function in (\ref{asymptoticprop}).
As we shall see, our normalization, obtained from the limit of the bulk to bulk propagator, is
more convenient for the computation of correlation functions within the AdS/CFT correspondence  \cite{KlebanovWitten}.

Finally, we recall that the bulk to boundary propagator  can be used to write a generic solution of the Klein--Gordon equation
in AdS
$$
\psi({\bf x} )= \int_{\Sigma}d{\bf p}  \, \phi({\bf p})\, K_{\Delta}({\bf p}, {\bf x})\ .
$$
Invariance of this integral under a change of light cone section $\Sigma\to\Sigma'$ requires
the boundary wave function $ \phi({\bf p})$ to be an homogeneous function on the light cone,
$$
 \phi(\lambda {\bf p}) = \lambda^{\Delta-d} \phi({\bf p})\ .
$$

\subsection{CFT on the AdS Boundary}

We shall define the conformal field theory (CFT) on its natural habitat, the light cone of $\mathbb{R}^{2,d}$.
A CFT correlator of scalar primary operators located at points $\mathbf{p}_{1},\ldots,\mathbf{p}_{n}$, 
can then be conveniently described by an amplitude
\[
A\left(  \mathbf{p}_{1},\ldots,\mathbf{p}_{n}\right)
\]
invariant under $SO\left(  2,d\right)  $, and therefore only a function of the invariants
\[
\mathbf{p}_{ij}\equiv-2\mathbf{p}_{i}\cdot\mathbf{p}_{j}~.
\]
The amplitude $A$ is homogeneous in each entry
\[
A\left( \ldots, \lambda\mathbf{p}_{i}, \ldots \right) = \lambda^{-\Delta_{i}} A\left( \ldots, \mathbf{p}_{i}, \ldots \right)~,
\]
where $\Delta_{i}$ is the conformal dimension of the $i$--th scalar primary operator.
In this language, one recovers  standard results on $d$--dimensional CFT's by choosing a specific section $\Sigma$
of the light cone of $\mathbb{R}^{2,d}$. 
Correlations functions of primary operators defined on different light cone sections $\Sigma$ and $\Sigma'$ 
with $ds^2_{\Sigma'}=\Lambda^2 ds^2_{\Sigma} $, are then related by
$$
A_{\Sigma'}\left(  \ldots, \mathbf{p}'_{i}, \ldots \right)= A\left( \ldots, \Lambda(\mathbf{p}_{i}) \mathbf{p}_{i}, \ldots \right)= 
A_{\Sigma}\left(  \ldots, \mathbf{p}_{i} , \ldots\right)  \prod_i  \Lambda(\mathbf{p}_{i})^{-\Delta_i}\ ,
$$
as required for these CFT correlators.
The power of this notation is that it makes symmetry properties transparent.
For example, the two point function of two primary operators can only be a function of the invariant $\mathbf{p}_{12} $.
Moreover, the homogeneity condition implies the vanishing of the two point function for operators with different
conformal dimension and fixes
$$
A\left(  \mathbf{p}_{1},\mathbf{p}_{2}\right) \propto \mathbf{p}_{12}^{-\Delta}\ ,
$$
for operators of the same dimension.
Notice that in the light cone section (\ref{AdSboundaryPoincare}) we have ${\bf p}_{12}= ({\bf y}_1- {\bf y}_2)^2$,
which gives the standard flat space CFT two point function of primary operators.
In the AdS/CFT context, it is convenient to normalize CFT operators such that
the two point function is given by the boundary to boundary propagator defined in the previous section,
$$
< \mathcal{O}_{\Delta} (  {\bf p}_{1})  \mathcal{O}_{\Delta} (  {\bf p}_{2}) >=
K_{\Delta}\left(  {\bf p}_{1},{\bf p}_{2}\right)\ .
$$
The three point function 
$$
\left\langle 
\mathcal{O}_{1}\left( {\bf p}_{1}\right)  \mathcal{O}_{2}\left( {\bf p}_{2}\right)
 \mathcal{O}_{3}\left( {\bf p}_{3}\right) \right\rangle
$$
is also determined, up to a constant, by conformal symmetry, since
$$
 {\bf p}_{12}^{(\Delta_3- \Delta_1-\Delta_2)/2}
\, {\bf p}_{23}^{(\Delta_1- \Delta_2-\Delta_3)/2} \,{\bf p}_{13}^{(\Delta_2- \Delta_1-\Delta_3)/2}
$$
is the only possible combination of the  Lorentz invariants  ${\bf p}_{ij} $ with the required
scaling properties. 

In this thesis, we shall focus a great deal of our attention on four point amplitudes of scalar primary operators. 
More precisely, we shall consider correlators of the form
\[
A\left( {\bf p}_{1},{\bf p}_{2},{\bf p}_{3},{\bf p}_{4} \right) = \left\langle 
\mathcal{O}_{1}\left( {\bf p}_{1}\right)  \mathcal{O}_{2}\left( {\bf p}_{2}\right)
 \mathcal{O}_{1}\left( {\bf p}_{3}\right) \mathcal{O}_{2}\left( {\bf p}_{4}\right) \right\rangle
\]
where the scalar operators $\mathcal{O}_{1}$ and $\mathcal{O}_{2}$ have dimensions
$$
\Delta_{1}= \eta + \nu \ ,\ \ \ \ \ \ \ \ \ \ \  \Delta _{2}= \eta - \nu \ ,
$$
respectively. It is then convenient to write the four point amplitude $A$ as
\begin{equation}
A\left(  {\bf p}_{i}\right)  =K_{\Delta_1}\left(  {\bf p}_{1},{\bf p}_{3}\right)
K_{\Delta_2}\left(  {\bf p}_{2},{\bf p}_{4}\right)
\mathcal{A}\left(  z,\bar{z}\right)~,
\label{CFTamp}
\end{equation}
so that $\mathcal{A}$ is a generic function of the two cross--ratios $z$ and $\bar{z}$,
which we define, following  \cite{Osborn, Osborn22}, in terms of the kinematical invariants ${\bf p}_{ij}$ as 
 \footnote{Throughout the thesis, we shall consider barred and unbarred variables as independent, 
with complex conjugation denoted by $\star$. In general $\bar{z}=z^{\star}$ when considering the 
analytic continuation of the CFT$_{d}$ to Euclidean signature. For Lorentzian signature, either 
$\bar{z}=z^{\star}$ or both $z$ and $\bar{z}$ are real. These facts follow simply from solving the 
quadratic equations for $z$ and $\bar{z}$.}
\be
z\bar{z}  =\frac{{\bf p}_{13}{\bf p}_{24}}{{\bf p}_{12}
{\bf p}_{34}}\ ,\ \ \ \ \ \ \ \ \ \ \ \ \ \ \ \ \ 
\left(  1-z\right)  \left(  1-\bar{z}\right)    =\frac{{\bf p}
_{14}{\bf p}_{23}}{{\bf p}_{12}{\bf p}_{34}}\ .     \label{zzbar}
\ee
With this definition, the disconnected graph gives $\mathcal{A}=1$.

\subsection{AdS/CFT Correspondence}

We are now in position to state the generic conjectured correspondence between string theories in AdS$_{d+1}$
and conformal gauge theories defined on the AdS boundary. 
Since both theories have the same symmetry group $SO(2,d)$, it is not unreasonable to believe that there is 
a duality between them.
Our point of view will be to find specific CFT properties dictated by  the existence of a dual string theory on AdS.
Therefore, we consider a generic gauge theory with quantum fields $\Phi_i$ transforming in the adjoint representation of the gauge
group  \footnote{These include gauge fields or adjoint scalars. We will not consider the case of particles (like quarks) transforming
in the fundamental representation  since, from
the general discussion of section \ref{openclosed}, these are expected to be dual to open strings.}.
The basic gauge invariant local operators are single trace,
$$
\mathcal{O}_{\alpha}=\frac{1}{N} \,{\rm Tr} \,F_{\alpha}(\Phi_i)\ ,
$$ 
where $F_{\alpha}$ is an arbitrary function (usually, a polynomial) of the fields  $\Phi_i$  and their derivatives.
Theories with a large $N$ limit have an action functional of the form
$$
N^2\,\int_{\Sigma} d{\bf p}  \,  \mathcal{L}\left(\mathcal{O}_{\alpha}({\bf p})\right)\ ,
$$
where
$ \mathcal{L}$ is, in principle, an arbitrary function of the operators $\mathcal{O}_{\alpha}$, 
encoding the coupling constants of the theory (like the 't Hooft coupling in  Yang--Mills theory).
In most cases the action is single trace and $ \mathcal{L}$ is just a linear function.
Furthermore, we shall restrict our attention to conformal field theories, where the lagrangian density $ \mathcal{L}({\bf p})$ is
an exactly marginal operator of dimension $d$ (see  \cite{deBoer} for RG flows within AdS/CFT).
The AdS/CFT correspondence states that, for each single trace primary operator $\mathcal{O}_{\alpha}$ of the CFT,
there is a field $\psi_{\alpha}$ in AdS, corresponding to a particular string mode.
On the other hand, multiple trace local operators are dual to multistring states in AdS.
In a given realization of the AdS/CFT correspondence, this map can usually be made very precise.
Here, we restrict ourselves to generic features of the correspondence.
For instance, a scalar field $\psi$ in AdS with mass squared $m^2=\Delta(\Delta-d)>1-d^2/4$, 
is dual to a scalar primary operator $\mathcal{O}$ of dimension $\Delta> 1+ d/2$.

\begin{figure}
\begin{center}
\includegraphics[width=6cm]{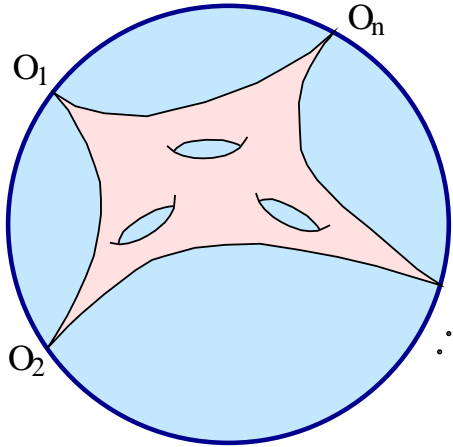}
\caption{A connected  diagram of the string perturbative expansion defining the CFT correlator 
$ \left\langle  \mathcal{O}_{1}\left( {\bf p}_{1}\right)  \cdots   \mathcal{O}_{n}\left( {\bf p}_{n}\right) \right\rangle$.
Here, the interior of the circle represents AdS spacetime and the circumference represents its boundary.}
\label{AdS/CFTcorrelations}
\end{center}
\end{figure}

The CFT dynamics  is defined by  the correlation functions of its gauge invariant local operators.
Furthermore, it is enough to consider single trace operators, since multiple trace operators
can be obtained from the operator product expansion (OPE) of the first  \footnote {See   \cite{WittenBC} 
for the relation between multiple trace deformations and AdS boundary conditions.}.
The dynamical information of the AdS/CFT correspondence can then be embodied in the relation
between the generating functional for CFT correlators and the string theory partition function with
appropriate boundary conditions on the AdS boundary.
In the case of the scalar field $\psi$ mentioned above, we have
\begin{equation}
\left\langle e^{i\int_{\Sigma} d{\bf p}  \, \phi({\bf p})\,\mathcal{O}({\bf p}) }\right\rangle_{{\rm CFT}}=
Z_{{\rm string}} \left[ \psi({\bf x}) \to \phi({\bf p})    \right]\ ,
\label{generatingfunction}
\end{equation}
where the AdS scalar field $\psi({\bf x})$ tends to the source $\phi({\bf p})$  at the AdS
boundary as defined by (\ref{asymptoticKG}).
In the Lorentzian version of AdS/CFT, this boundary condition and the equations of motion do not uniquely determine the field $\psi$
in the bulk. This freedom corresponds to different choices of initial and final states for the boundary correlators 
\cite{BB,LorentzianAdS}.
In this thesis, we shall always consider correlation functions in the vacuum state, by analytically continuing from the 
Euclidean theory.
CFT correlation functions are obtained by differentiating with respect to the source $\phi({\bf p})$.
On the string theory side, each differentiation sends into AdS a closed string in the state $\psi$.
In other words, a CFT $n$ point function is given by the string amplitude associated to the sum
over all worldsheets embedded in AdS with $n$ punctures fixed at the AdS boundary as in figure \ref{AdS/CFTcorrelations}.
Different CFT operators correspond to different boundary conditions on the worldsheet fields at the punctures.


When the AdS radius $\ell$ is much larger than the string length $\ell_s$, the low energy gravitational 
approximation, describing the string massless modes, is reliable
$$
Z_{{\rm string}} \simeq \int \mathcal{D}\psi 
e^{\,i S\left[ \psi   \right] }\ .
$$
Here, $\psi$ represents all massless fields in AdS.
Pictorically, the infinite string tension collapses the surface in  figure \ref{AdS/CFTcorrelations}
to Witten diagrams describing particle exchange in AdS  \cite{WittenAdS}.
In the case of $\mathcal{N}=4$ SYM, we have seen that this regime of small strings in AdS corresponds
to large 't Hooft coupling $\lambda$ in the gauge theory. 
In general, one expects the gravitational regime of the string theory
to be dual to some strongly coupled gauge theory.
In addition, the  perturbative expansion in the gravitational coupling $G$ corresponds
to the $1/N$ expansion in the CFT.
The Witten diagrams are drawn using the propagators and interaction vertices associated to
the effective action for the string lightest modes.
We shall concentrate on correlation functions 
$ \left\langle  \mathcal{O}_{1}\left( {\bf p}_{1}\right)  \cdots   \mathcal{O}_{n}\left( {\bf p}_{n}\right) \right\rangle$
of scalar primary operators.
These are obtained from Witten diagrams with $n$ bulk to boundary propagators, connected to the points $ {\bf p}_{i}$, 
glued to bulk to bulk propagators of the exchanged particles,
using the interaction vertices contained in the effective action. 

As a toy example, consider our scalar field $\psi$ with a cubic interaction in AdS,
$$
S=-\int_{{\rm AdS}} d{\bf x} \left[ \frac{1}{2}(\partial \psi)^2+ \frac{1}{2} m^2 \psi^2 +\frac{1}{3!} g \psi^3 \right] \ .
$$
A solution to the equation of motion $ ( \Box -m^2) \psi = g \psi^2 /2$, satisfying 
the required boundary conditions, can be obtained by iterating the following integral equation
\begin{equation}
 \psi( {\bf x}) = \int_{\Sigma} d{\bf p }\, K_{\Delta}({\bf p },{\bf x })\, \phi( {\bf p})
 -i\frac{g}{2} \int_{{\rm AdS}} d{\bf x'} \, \Pi_{\Delta}({\bf x },{\bf x' })\, \psi^2( {\bf x'}) \ .
\label{psiintegralequation}
\end{equation}
Then, the value of the action $S$ for this classical solution is given by the 
diagrammatic expansion
\begin{equation}
iS= \frac{1}{2} \ocortado + \frac{1}{6} \mercedes  +\frac{1}{8} \fourpoint  + \frac{1}{8} \fivepoint +\cdots \ ,
\label{action}
\end{equation}
where each Witten diagram is evaluated using the following rules:
each vertex yields a factor of $-ig$ and is integrated over AdS; internal lines represent bulk to bulk propagators;
lines connecting to the boundary represent bulk to boundary propagators; and the boundary points ${\bf p}$ 
are integrated over $\Sigma$ with weight $\phi( {\bf p}) $.
The correlation functions are obtained just by taking functional derivatives of (\ref{action}). 
For example, the two point function is 
$$
 \left\langle  \mathcal{O}\left( \mathbf{p}_{1}\right)  \mathcal{O}\left( \mathbf{p}_{2}\right) \right\rangle=
\left. \frac{\delta}{i\delta \phi( {\bf p}_1)}\frac{\delta}{i\delta \phi( {\bf p}_2)} e^{\,iS}\right|_{\phi=0}=
 K_{\Delta}({\bf p}_1,{\bf p}_2)
$$
and the three point function is
$$
 \left\langle  \mathcal{O}\left( \mathbf{p}_{1}\right)  \mathcal{O}\left( \mathbf{p}_{2}\right)
\mathcal{O}\left( \mathbf{p}_{3}\right) \right\rangle=g \int_{{\rm AdS}} d{\bf x}\,
 K_{\Delta}({\bf p}_1,{\bf x} ) K_{\Delta}({\bf p}_2,{\bf x} ) K_{\Delta}({\bf p}_3,{\bf x} )\ .
$$

So far we have considered the classical approximation $Z\simeq e^{\,iS}$.
Quantum effects in AdS correspond to Witten diagrams with internal loops.
Including these in (\ref{action}) defines the quantum effective action,
which becomes the new generating function for CFT correlators.

\chapter{Eikonal Approximation}
\label{ch:eikonal}

Geometric optics can be obtained as the infinite frequency limit of wave propagation.
In this limit, the phase function $S({\bf x})$ satisfies the eikonal equation
$$
(\boldsymbol{\nabla} S)^2=n^2({\bf x})\ ,
$$
where $n({\bf x})$ is the refraction index of the medium.
Light rays are then given by the integral curves of the gradient $\boldsymbol{\nabla} S$ of the phase function.
The function $S({\bf x})$ is also known as the lapse function. This terminology arises from 
the problem of finding the fastest light path from the boundary of a given compact space $\Omega$, 
with   refraction index $n({\bf x})$, to a point  ${\bf x} \in \Omega$.
In this case, the lapse function $S({\bf x})$ is the shortest time that light takes
to arrive at point ${\bf x}$ starting from the boundary of  $\Omega$.

In this chapter, we shall describe the eikonal approximation for particle scattering.
We start with the simpler case of potential scattering in non--relativistic quantum mechanics
since all important concepts are already present in this computation.
We then move to quantum field theory (QFT) on  Minkowski spacetime and rederive
the eikonal approximation using position space Feynman rules.
With this in mind, we are able to generalize the eikonal approximation to particle scattering
in AdS and other spacetimes.
Finally, we use gravitational shock wave techniques to establish an alternative, perhaps more physical,
derivation  of the AdS eikonal approximation.


\section{Potential Scattering in Quantum Mechanics}
\label{sec:QM}


Consider the Schr\"odinger equation
$$
\left[ -\frac{1}{2m}\boldsymbol{\nabla}^2 +V({\bf x}) \right] \psi({\bf x})= E  \psi({\bf x}) \ ,
$$
governing the wave function $\psi({\bf x})$ of a non--relativistic particle scattering in the potential $V({\bf x})$ at very
high energy $E=\omega^2/(2m)$. We write 
$$
\psi({\bf x})= (2\pi)^{-3/2} e^{\,i S({\bf x})}\ ,
$$
so that free propagation with momentum  ${\bf k}$ corresponds to a linear phase 
$$
S({\bf x})= {\bf k}\cdot {\bf x} \ ,
$$
with ${\bf k}^2=\omega^2 $.
With this definition, the standard scattering amplitude is given by  \cite{Sakurai}
$$
f( {\bf k'}, {\bf k})=-\frac{m}{2\pi} \int_{\mathbb{R}^3}  d {\bf x} \,e^{-i   {\bf k'}\cdot {\bf x}}
 \, V({\bf x})\,e^{\,i S({\bf x})}\ ,
$$
where $ {\bf k}$ and ${\bf k'}$ are the initial and final momenta, respectively, determining
the scattering angle through ${\bf k} \cdot {\bf k'} = \omega^2 \cos \theta$.

In the presence of a non--vanishing potential, the phase function $S({\bf x})$ obeys the 
eikonal like equation
$$
(\boldsymbol{\nabla} S)^2=\omega^2 - 2m V({\bf x}) -i  \boldsymbol{\nabla}^2 S \ .
$$
Then, the leading correction in $1/\omega$ is simply given by the accumulated phase shift along
the free particle's  trajectory 
$$
S({\bf x})={\bf k}\cdot {\bf x} - m \int_{-\infty}^{0} d\lambda V({\bf x}+ \lambda {\bf k} )
 + O\left( \omega^{-2}\right)\ .
$$
The eikonal approximation is valid at high energies and  fixed momentum transfer ${\bf q}= {\bf k'}-{\bf k}$.
In this regime, the scattering angle $\theta\simeq |{\bf q}|/\omega$  is very small and the momentum
transfer   ${\bf q}$ is essentially orthogonal to the propagation direction ${\bf k}\simeq{\bf k'}$.
Finally, the eikonal scattering amplitude reads
\begin{equation}
f( {\bf k'}, {\bf k})\simeq\frac{\omega}{2\pi i} \int_{\mathbb{R}^2}  d {\bf x}_{\perp} \,e^{-i   {\bf q}\cdot {\bf x}_{\perp}}
 \left[ e^{\,2i \delta({\bf x}_{\perp} )} -1 \right] \ ,
\label{QMeikonal}
\end{equation}
with
$$
\delta({\bf x}_{\perp} )= -\frac{m}{2}  \int_{-\infty}^{\infty} d\lambda V({\bf x}+ \lambda {\bf k} )\ .
$$
Notably, the eikonal approximation resums an infinite number of terms in the Born perturbative expansion, yielding
a scattering amplitude only valid in a specific kinematical regime, but including contributions of all orders 
in the scattering potential $V$.
Interestingly, the eikonal amplitude (\ref{QMeikonal}) satisfies the optical theorem  \cite{Gottfried}
$$
\sigma_{tot}=\int_{\mathbb{R}^2}  d {\bf q}\, |f( {\bf k'}, {\bf k}) |^2=
\frac{\omega}{\pi}  \int_{\mathbb{R}^2}  d {\bf x}_{\perp} \sin ^2  \delta({\bf x}_{\perp} )=
\frac{4\pi}{\omega} \im   \, f( {\bf k}, {\bf k})\ .
$$

For spherically symmetric potentials, one usually expands the scattering amplitude in 
partial waves
$$
f(\theta)=\frac{1}{2i \omega} \sum_{J\ge 0} (2J+1) {\rm P}_J(\cos \theta) \left[ e^{\,2i \delta_J} -1 \right]  \ ,
$$
where ${\rm P}_J$ are the Legendre polynomials and $\delta_J$ is the phase shift associated to
the spin $J$ partial wave. 
In the eikonal regime, one considers large energy and  small scattering angle, keeping
their product fixed by the momentum transfer $| {\bf q} |=q=\omega \theta$.
In this limit, the sum over spin $J$ can be replaced by an integral over the impact parameter $b=J/\omega$ and
the Legendre polynomials reduce to the zeroth order Bessel function ${\rm J}_0$, giving
the known impact parameter representation
$$
f(q) = -i\omega \int_0^{\infty}  db\, b \,{\rm J}_0(q b)
 \left[ e^{\,2i \delta_{J=b\omega} } -1 \right] \ .
$$
The eikonal amplitude (\ref{QMeikonal}) reduces exactly to the same expression after performing
the angular integral in transverse space and identifying $b=| {\bf x}_{\perp} |$
and $\delta_J\simeq\delta(b=J/\omega)$ for large $J$.
In other words, the eikonal approximation determines the leading behavior of the phase shift at large spin.


\section{Minkowski Spacetime} \label{flateikonal}


In relativistic QFT, all interactions are mediated by particle exchange and there is no direct
analogue of a static potential like in the previous section.
Therefore, the generalization of the semi--classical eikonal approximation is not obvious.
Nevertheless, this generalization  was found long time ago  \cite{LevySucher, Abarbanel}.
In this section we shall rederive the standard eikonal amplitude for high energy
scattering in  Minkowski spacetime from a position space perspective.
This will prove to be useful because the physical picture here developed will generalize to scattering in AdS.
We shall consider $(d+1)$--dimensional Minkowski space $\mathbb{M}^{d+1}$ in close analogy with AdS$_{d+1}$.
At high energies
$$
s=(2\omega)^2
$$
we can neglect the masses of the external particles and, for
simplicity, we shall consider first an interaction mediated by a
scalar field of mass $m$. In flat space we may choose the external
particle wave functions to be plane waves $\psi_i({\bf x})=e^{\,i\,
{\bf k}_i \cdot {\bf x}}\ (i=1,\cdots,4)$, so that the amplitude
is a function of the Mandelstam invariants
$$
s=-({\bf k}_1+{\bf k}_2)^2 \ , \ \ \ \ \ \ \ \ \ \ \
t =-({\bf k}_1+{\bf k}_3)^2  =  - {\bf q}^2 \ ,
$$
We then have
$$
-2{\bf k}_1\cdot {\bf k}_2=(2\omega)^2\, ,\ \ \ \ \ \ \ \ \ \ \ \ \ {\bf k}_i^{\,2}=0\, .
$$
The eikonal approximation is valid for $s \gg -t$, where the
momentum transferred ${\bf q}={\bf k}_1+{\bf k}_3$ is
approximately orthogonal to the external momenta.

The momenta of the incoming particles naturally decompose spacetime as $\mathbb{M}^2 \times \mathbb{R}^{d-1}$.
Using coordinates $\{u,v\}$ in  $\mathbb{M}^2$ and  ${\bf w}$ in the transverse space $\mathbb{R}^{d-1} $,
a generic point can be written using the exponential map
\begin{equation}
{\bf x}=e^{\,v  {\bf T}_2 + u  {\bf T}_1 }\,  {\bf w}= {\bf w} +  u\,{\bf T}_1 + v\,{\bf T}_2\ ,
\label{coord}
\end{equation}
where the vector fields ${\bf T}_1$ and  ${\bf T}_2$ are defined by
$$
{\bf T}_1 =  \frac{{\bf k}_1 }{2\omega }\ , \ \ \ \ \ \ \ \ \
{\bf T}_2 =  \frac{{\bf k}_2}{ 2\omega}\ .
$$
The incoming wave functions are then
$$
\psi_1({\bf x})=e^{-i\omega v} \ ,\ \ \ \ \ \ \ \
\psi_2({\bf x})=e^{-i\omega u}\ .
$$
The coordinate $u$ is an affine parameter along the null geodesics
describing the classical trajectories of particle 1. This set of
null geodesics, labeled by  $v$ and  ${\bf w}$, is then the
unique congruence associated with particle 1 trajectories. Since
${\bf T}_2 = \frac{d\,}{dv}$ is a Killing vector field, these
geodesics have a conserved charge $-{\bf T}_2 \cdot {\bf k}_1=\omega$. 
At the level of the wave function this charge
translates into the condition
$$
 \mathcal{L}_{{\bf T}_2 } \psi_1 = - i\omega \psi_1 \ .
$$
Notice also that the wave function $\psi_1$ is constant along each geodesic of the
null congruence,
$$
{\bf x}(\lambda) = {\bf y} + \lambda {\bf k}_1\ ,
$$
where ${\bf k}_1=2\omega\frac{d\,}{du}$ is the momentum vector field associated to particle 1 trajectories.
Hence
$$
\mathcal{L}_{{\bf k}_1 } \psi_1 = 0\ .
$$
Finally, the field equations imply that $\psi_1$ is
independent of the transverse space coordinate ${\bf w}$.
Similar comments apply to particle 2.

Neglecting terms of order $-t/s$, the outgoing wave functions
for particles 1 and 2 are still independent of the corresponding affine parameter,
but depend on the transverse coordinate  ${\bf w}$,
$$
\psi_3({\bf x})\simeq e^{\,i\omega v+i{\bf q}\cdot {\bf w} } \ ,\ \ \ \ \ \ \ \
\psi_4({\bf x})\simeq e^{\,i\omega u-i{\bf q}\cdot {\bf w} } \ .
$$
The dependence in transverse space is determined by the
transferred momentum ${\bf q}$. Physically, the transverse space
is the impact parameter space. In fact, for two null geodesics
associated to the external particles 1 and 2, labeled
respectively by $\{v,{\bf w}\}$ and $\{\bar{u},{\bf \bar{w}}\}$,
the classical impact parameter is given by the distance  $|{\bf
w}-{\bf \bar{w}}|$.

\begin{figure} 
[ptb]
\begin{center}
\includegraphics[width=11cm]{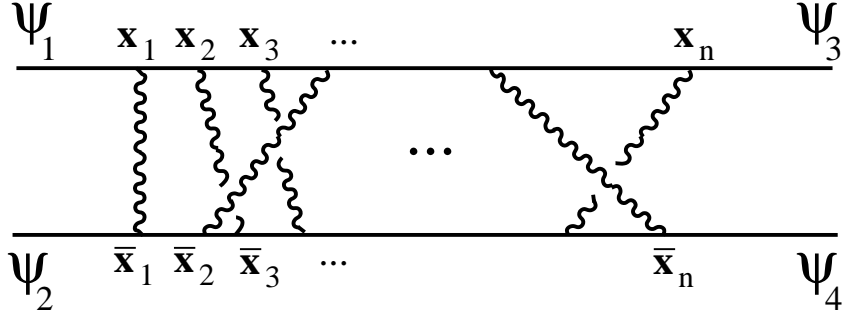}
\caption{  The crossed--ladder graphs describing  the \emph{T}--channel exchange of many soft particles
dominate the scattering amplitude in the eikonal regime. }
\label{fig1}
\end{center}
\end{figure}

The exchange of $n$ scalar particles described by figure \ref{fig1} gives the following contribution to
the scattering amplitude
\begin{eqnarray*}
&&\mathcal{A}_n = \frac{(-ig)^{2n}}{V}
\int_{\mathbb{M}^{d+1} } d{\bf x}_1 \cdots d{\bf x}_n
  d{\bf \bar{x}}_1 \cdots d{\bf \bar{x}}_n\,
\psi_3( {\bf x}_n )  \Delta({\bf x}_n-{\bf x}_{n-1})\cdots \Delta({\bf x}_2-{\bf x}_1)\psi_1( {\bf x}_1 )
\\
&&\ \ \ \ \ \ \psi_4( {\bf \bar{x}}_n ) \Delta({\bf \bar{x}}_n-{\bf \bar{x}}_{n-1})\cdots
\Delta({\bf \bar{x}}_2-{\bf \bar{x}}_1)\psi_2( {\bf \bar{x}}_1 )
\sum_{{\rm perm} \ {\sigma}}   \Delta_{m}({\bf x}_1-{\bf \bar{x}}_{\sigma_1})\cdots \Delta_{m}({\bf x}_n-{\bf \bar{x}}_{\sigma_n})\ ,
\end{eqnarray*}
where $V$ is the spacetime volume, $g$ is the coupling and where $\Delta({\bf x} )$ and  $\Delta_m({\bf x} )$ are,
respectively, the massless and massive Feynman propagators satisfying
$$
\left( \Box - m^2 \right) \Delta_m({\bf x} ) = i \delta({\bf x} )\ .
$$
The basic idea of the eikonal approximation is to put the horizontal propagators in figure  \ref{fig1}
almost on--shell. This is usually done in momentum space. For example, for the propagator between vertices
${\bf x}_j$ and ${\bf x}_{j+1}$, we approximate
$$
\frac{-i}{({\bf k}_1+{\bf K})^2-i\epsilon}\simeq \frac{-i}{2{\bf k}_1 \cdot {\bf  K} -i\epsilon}\ ,
$$
where ${\bf K}$ is the total momentum transferred up to the vertex
at ${\bf x}_j $. The physical meaning of this approximation
becomes clear in the coordinates (\ref{coord}),
\begin{eqnarray}
\Delta({\bf x}_{j+1}-{\bf x}_j)&\simeq&-i
 \int \frac{ d{\bf  K}}{ (2\pi)^{d+1}}
\frac{e^{\,i({\bf k}_1+ {\bf K})\cdot ({\bf x}_{j+1}-{\bf x}_j)} }{2{\bf k}_1 \cdot {\bf K} -i\epsilon}
\nonumber \\
&\simeq&\frac{1}{2\omega}\,\Theta(u_{j+1}-u_j)\,\delta(v_{j+1}-v_j)\,\delta^{d-1}({\bf w}_{j+1}-{\bf w}_j)\ .
\label{posprop}
\end{eqnarray}
In words, particle 1 can propagate from ${\bf x}_j$ to ${\bf x}_{j+1}$ only if ${\bf x}_{j+1}$ lies on the future
directed null geodesic that starts at ${\bf x}_j$ and has tangent vector ${\bf k}_1$.
This intuitive result can be derived directly in position space. In fact, in coordinates (\ref{coord}),
the propagator satisfies
$$
\Box \Delta(x)=\left(-4\partial_u \partial_v   + \partial^2_{{\bf w} } \right) \Delta(u,v,{\bf w} )
= 2i\delta(u)\delta(v)\delta^{d-1}({\bf w})\ .
$$
Since for particle 1 we have $ \partial_v =-i\omega$, for high energies
$\Box\simeq 4i\omega\partial_u$ and (\ref{posprop}) follows.

The eikonal approximation to the position space propagators greatly simplifies the scattering
amplitude for the exchange of $n$ scalar particles
\begin{eqnarray*}
V \mathcal{A}_n \simeq
\int_{\mathbb{M}^{d+1} } d{\bf x}_1  d{\bf \bar{x}}_1
\int_{u_1}^{\infty} du_2 \int_{u_2}^{\infty} du_3 \cdots \int_{u_{n-1}}^{\infty} du_n
\int_{\bar{v}_1}^{\infty} d\bar{v}_2 \int_{\bar{v}_2}^{\infty} d\bar{v}_3 \cdots
\int_{\bar{v}_{n-1}}^{\infty} d\bar{v}_n&&\\
(4\omega)^2\left(\frac{ig}{4\omega}\right)^{2n}
 e^{\, i{\bf q}\cdot {\bf w}}\,e^{-i{\bf q}\cdot{\bf \bar{w}}}
\sum_{{\rm perm}\ {\sigma}}   \Delta_{m}({\bf x}_1-{\bf \bar{x}}_{\sigma_1})\cdots \Delta_{m}({\bf x}_n-{\bf \bar{x}}_{\sigma_n})\ ,&&
\end{eqnarray*}
with
$$
{\bf x}_j={\bf w} + u_j \,{\bf T}_1 + v\,{\bf T}_2\ ,\ \ \ \ \ \ \ \ \ \
{\bf \bar{x}}_j={\bf \bar{w}}  + {\bar u} \,{\bf T}_1 + {\bar v}_j\,{\bf T}_2\ .
$$
Furthermore, the sum over permutations can be used to extend the integrals over the affine parameters
of external particle trajectories to the full real line,
\begin{equation*}
V\mathcal{A}_n \simeq \frac{(2\omega)^2}{n!}
\int_{-\infty}^{\infty}  dv d\bar{u}
\int_{\mathbb{R}^{d-1} } d{\bf w}  d{\bf \bar{w}}\, e^{\,i{\bf q}\cdot {\bf w}}\,e^{-i{\bf q}\cdot{\bf \bar{w}}}
\left( -\frac{g^2}{16\omega^2} \int_{-\infty}^{\infty} du d\bar{v}
\,\Delta_{m}\left({\bf x}-{\bf \bar{x}} \right)\right)^n \ ,
\end{equation*}
where
$$
{\bf x}  - {\bf \bar{x}} ={\bf w} + \frac{u-{\bar u}}{2\omega}\,{\bf k}_1
-{\bf \bar{w}} - \frac{ v- \bar{v}}{2\omega}\,{\bf k}_2\ .
$$
Summing over $n$, one obtains (the $n=0$ term corresponds to the
disconnected graph)
\begin{equation}
V\mathcal{A} \simeq (2\omega)^2
\int_{-\infty}^{\infty}  dv d\bar{u}
\int_{\mathbb{R}^{d-1} } d{\bf w}  d{\bf \bar{w}}\, e^{\,i{\bf q}\cdot {\bf w}}\,e^{-i{\bf q}\cdot{\bf \bar{w}}}
\,e^{\, I/4} \ .
\label{scalareik}
\end{equation}
The integral $I$ can be interpreted as the interaction
between two null geodesics of momentum ${\bf k}_1 $ and ${\bf k}_2$
describing the classical trajectories of the incoming particles.
In fact, using as integration variables the natural affine parameters $\lambda, \bar\lambda$ along the geodesics,
one has
$$
I=(-ig)^2\int_{-\infty}^{\infty}d\lambda d\bar{\lambda}
\,\Delta_{m}\left({\bf w}+ \lambda{\bf k}_1 -{\bf \bar{w}} - \bar{\lambda} {\bf k}_2\right)\ .
$$
Since the integral $I$ only depends on the impact parameter $|{\bf w}- {\bf \bar{w}}|$ of the null
geodesics, equation (\ref{scalareik}) simplifies to 
\begin{equation}
\mathcal{A} (s,t=-{\bf q}^2) \simeq 2s   
\int_{\mathbb{R}^{d-1} } d{\bf w}  \,
e^{\, i{\bf q} \cdot{\bf w}} e^{\,I/4}  \ .
\label{flateik}
\end{equation}
Explicit computation yields the Euclidean propagator of mass $m$ in the transverse 
$\mathbb{R}^{d-1}$ space
$$
I=\frac{-ig^2}{ {\bf k}_1\cdot  {\bf k}_2 }\int_{\mathbb{R}^{d-1} }\frac{d{\bf k}_{\perp}}{(2\pi)^{d-1}}\,
\frac{e^{\,i{\bf k}_{\perp}\cdot ({\bf w} -{\bf \bar{w}}) }}{{\bf k}^2_{\perp} +m^2}
=\frac{2ig^2}{s}\Delta_{\perp}( {\bf w} -{\bf \bar{w}})\ ,
$$
and we recover the well known eikonal amplitude.

The generalization of the above method to interactions mediated by a spin $j$ particle is now straightforward.
We only need to change the integral $I$ describing the scalar interaction between null geodesics.
The spin $j$ exchange alters the vertices of the local interaction, as well as the propagator
of the exchanged particles.
In general, the vertex includes $j$ momentum factors with a complicated
index structure. However, in the eikonal regime, the momenta entering the vertices
are approximately the incoming momenta ${\bf k}_1, {\bf k}_3$.
Therefore, the phase $I$ should be replaced  by
   \footnote{
The sign $(-)^j$ indicates that, for odd $j$, particles 1 and 2 have opposite charge
with respect to the spin $j$ interaction field. With this convention the interaction
is attractive, independently of $j$.}
\begin{eqnarray*}
I&=&-g^2(-2)^j\,({\bf k}_1)_{\alpha_1}\cdots ({\bf k}_1)_{\alpha_j}\,
({\bf k}_2)_{{\beta}_1}\cdots ({\bf k}_2)_{{\beta}_j}
\\
&&\int_{-\infty}^{\infty}d\lambda d\bar{\lambda}
\,\Delta_m^{\ \alpha_1 \cdots \alpha_j\,{\beta}_1 \cdots {\beta}_j }
\left({\bf w}+ \lambda{\bf k}_1 -{\bf \bar{w}} - \bar{\lambda} {\bf k}_2\right)\ ,
\end{eqnarray*}
where $\Delta_m^{\ \alpha_1 \cdots \alpha_j\,{\beta}_1 \cdots {\beta}_j }$
is the propagator of the massive spin $j$ field. Recall that the equations of motion for
a spin $j$ field $h^{\alpha_1 \cdots \alpha_j}$ imply that $h$ is symmetric, traceless
and transverse ($\partial_{\alpha_1}\,h^{\alpha_1 \cdots \alpha_j}=0$), together with the
mass--shell condition $\square=m^2$. Therefore, the relevant part of the propagator at high energies
is given by
$$
\eta^{(\alpha_1 \beta_1}\eta^{\alpha_2 \beta_2}\cdots \eta^{\alpha_j \beta_j)}\, \Delta_m\left({\bf x} -{\bf \bar{x}} \right)\ +\cdots,
$$
where the indices $\alpha_i$ and $\beta_i$ are separately symmetrized with weight $1$ and
$\Delta_m$ is the scalar propagator of mass $m$.
The neglected terms in $\cdots$ are trace terms, 
which vanish since ${\bf k}_i^{\,2}=0$, and derivative terms 
acting on $\Delta_m$, which vanish after integration along the two interacting 
geodesics. Compared to the scalar case, we have then an extra factor of $(-2{\bf k}_1\cdot{\bf k}_2)^{j}=s^j$, so that
$$
I = 2 i g^2\, s^{\, j-1} \,\Delta_{\perp}( {\bf w} -{\bf \bar{w}})\ .
$$
Note that we have normalized the coupling $g^2$ so that the leading behavior of the tree level amplitude at large $s$
is given by $-g^2s^j/t$. In the particular case of $j=2$ we have $g^2=8\pi G$,
where $G$ is the canonically normalized Newton constant.

It is known  \cite{Jackiw, KabatValEik} that the eikonal approximation is problematic for $j=0$ exchanges. In this case, the large
incoming momentum can be exchanged by the mediating particle, interchanging the role of $u,v$ in intermediate parts of the graph.
The eikonal approximation estimates correctly the large $s$ behavior of the amplitude at each order in perturbation theory, but
underestimates the relative coefficients, which do not resum to an exponential. Nonetheless, this is not problematic, since exactly in the
$j=0$ case the higher order terms are suppressed by powers of $s^{-1}$. For $j\geq 1$ the problematic
hard exchanges are suppressed at large energies and the eikonal approximation is valid. On the other hand,
for the QED case where $j=1$, there is a different set of graphs involving virtual fermions  \cite{ChengWu,KabatValEik}  that 
dominate the eikonal soft photons exchange. Therefore, also for $j=1$, the validity of the eikonal approximation 
is in question. None of these problems arise, though, for the most relevant case, the gravitational interaction with $j=2$.

\subsection{Partial Wave Expansion}          \label{flatPW}

\begin{figure}
[ptb]
\begin{center}
\includegraphics[width=10cm]{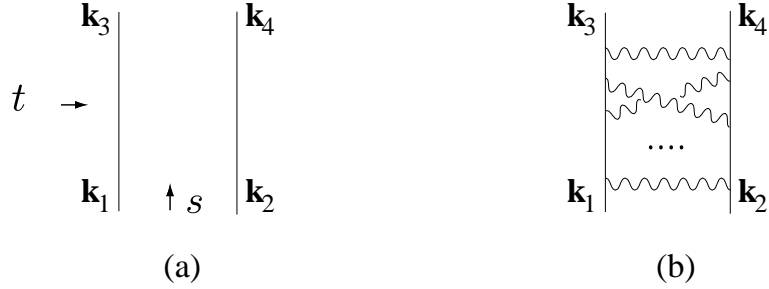}
\caption{In the eikonal regime, free propagation (a) is
modified primarily by interactions described by crossed--ladder graphs (b).}
\label{fig4}
\end{center}
\end{figure}

As in non-relativistic quantum mechanics, the Lorentz invariant scattering amplitude $\mathcal{A}(s,t)$
can be conveniently decomposed in \emph{S}--channel partial waves
\begin{equation}
\mathcal{A=}s^{\frac{3-d}{2}}~\sum_{J\geq 0}\mathcal{S}_{J}\left(
z\right) ~e^{-2\pi i\,\sigma _{J}\left( s\right) }~,\label{flatpwe}
\end{equation}%
with
\begin{equation*}
z=\sin ^{2}\left( \frac{\theta }{2}\right) =-\frac{t}{s}
\end{equation*}%
related to the scattering angle $\theta $ and with $\sigma
_{J}\left(
s\right) $ the phase shift for the spin $J$ partial wave \footnote{
We choose a non standard normalization and notation for the phase
shifts for later convenience. To revert to standard conventions, one must replace 
$-2\pi \sigma _{J}$ by $ 2\delta _{J}$.}. The angular functions $%
\mathcal{S}_{J}\left( z\right) $ are eigenfunctions of the
Laplacian on the sphere at infinity, with eigenvalue $-J\left(
J+d-2\right) $, and are polynomials in $z$ of order $J$, whose
explicit form depends on the dimension of spacetime. They can be
written as hypergeometric functions
\begin{equation*}
\mathcal{S}_{J}\left( z\right) =
\frac{2^{d} \pi^{\frac{d-1}{2}} }{\Gamma\left( \frac{d-1}{2}\right) }\,\frac{(2J+d-2)\Gamma(J+d-2) }
{  \Gamma(J+1)}\,
F\left( -J,J+d-2,\frac{d-1}{2}, z\right) 
\end{equation*}%
and they are normalized so that $\sigma =0$ corresponds to free
propagation with no interactions. Unitarity then implies
$\mathrm{Im}\, \sigma \leq 0$. The amplitude itself can be
computed in perturbation theory
\begin{equation*}
\mathcal{A}=\mathcal{A}_{0}+\mathcal{A}_{1}+\cdots ~,
\end{equation*}%
where
$\mathcal{A}_{0}=s^{\frac{3-d}{2}}\sum_{J}\mathcal{S}_{J}\left(
z\right) ~$\ corresponds to graph (a) in figure
\ref{fig4} describing free propagation in spacetime. All
\emph{S}--channel partial waves contribute to $\mathcal{A}_{0}$
with equal weight one, and vanishing phase shift.

In the eikonal regime we are interested in the limit $z\ll 1$ of
small scattering angle.
If the amplitude diverges at $z=0$, then its partial wave expansion contains
partial waves with unbounded intermediate spin $J$. In fact, one may then
replace the sum over $J$ with an integral over the impact
parameter $r$,
\begin{equation*}
\frac{r}{2}=\frac{J}{\sqrt{s}}~,
\end{equation*}%
denoting the phase shift $\sigma_J \left( s\right) $ by $\sigma \left( s,r\right) $ from now
on. More precisely, if we consider the double limit
\begin{equation}
z\rightarrow0\, ,~\ \ \ \ \ \ J\rightarrow\infty\, ,~\ \ \ \ \ \ \
\ \ \ \ \ \ \left( z\sim J^{-2}\right),  \label{limitflat}
\end{equation}
the angular functions $\mathcal{S}_{J}\left( z\right) $ become, in this limit, Bessel functions
defining the impact parameter partial waves $\mathcal{I}_{J}$,
\begin{equation}
\mathcal{I}_{J}=
2^{d+1} \pi^{\frac{d-1}{2}}  J^{d-2} \,\left(J\sqrt{z}\right)^{\frac{3-d}{2}}\,\mathbf{J}_{%
\frac{d-3}{2}}\left( 2J\sqrt{z}\right)  
=4\,s^{\frac{d-2}{2}} \int_{\mathbb{R}^{d-1}}~d{\bf w}~\delta \left(
|{\bf w}|-r\right) ~e^{\,i{\bf q} \cdot {\bf w}}
\ ,  \label{imprep2}
\end{equation}
with ${\bf q}^{2}=-t$.
The impact parameter limit corresponds to 
approximating the sphere at infinity by a transverse
Euclidean plane $\mathbb{R}^{d-1}$. This limit takes the sphere harmonic functions $\mathcal{S}_{J}$ 
into the plane harmonic functions  $\mathcal{I}_{J}$.
Returning to (\ref{flatpwe}) and replacing the sum over spins $J$ by the integral $\frac{\sqrt{s}}{2}\int dr$
over the impact parameter $r$, one  obtains the impact parameter representation of the amplitude
\begin{equation}
\mathcal{A}\,\simeq \,2s\int_{\mathbb{R}^{d-1}}~d{\bf w}\;e^{\,i{\bf q}\cdot
{\bf w}}\;e^{-2\pi i\,\sigma \left( s,r\right) }\,,
\label{eikonalre--sum}
\end{equation}
where $r=|{\bf w}|$.
Comparing with (\ref{flateik}), we find that the leading behavior of $\sigma \left( s,r\right)$ 
for large $r$, which controls small angle scattering, is given by the tree--level interaction $I$
between two null geodesics,
$$
\sigma\left( s,r\right) \simeq \frac{i}{8\pi}\,I \ .
$$
Indeed, a simple way of stating the eikonal approximation
is to say that  the leading behavior of the phase shift for large $r$,
is dominated by the leading tree--level amplitude
$\mathcal{A}_{1}$ and it is therefore determined by a simple
Fourier transform,
\begin{equation}
\mathcal{A}_{1}\,\simeq\,-4\pi
i~s\int_{\mathbb{R}^{d-1}}~d{\bf w}\;e^{\,i{\bf q}\cdot {\bf w}}\;\sigma(s,r)\,.
\label{imprep1}
\end{equation}
Moreover, the dominant interaction in $\mathcal{A}_{1}$ comes from
\emph{T}--channel exchanges of spin $j$ massless particles, so
that the full amplitude (\ref{eikonalre--sum}) approximately
resums the crossed--ladder graphs in figure \ref{fig4}(b). In
the limit of high energy $s$, the mediating particle with maximal
$j$ dominates the interaction. 
In theories of gravity, this particle is the graviton, with $j=2$  \cite{tHoofteik}.

\begin{figure}
[ptb]
\begin{center}
\includegraphics[width=14cm]{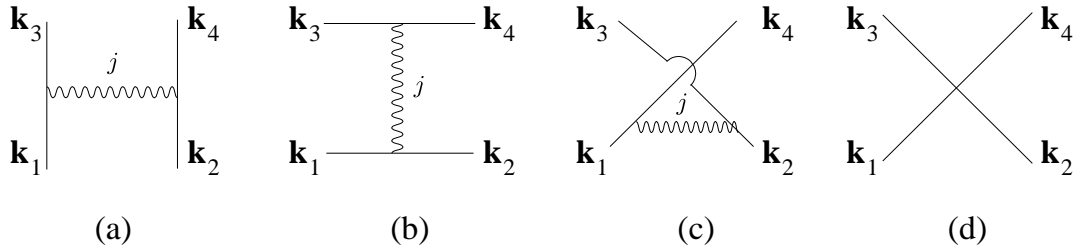} 
\caption{Some possible interactions at tree--level.
At high energies, graph  (a), with maximal spin
$j=2$, dominates the partial wave expansion at large intermediate spin.}
\label{fig2}
\end{center}
\end{figure}

To understand in more detail the generic behavior of amplitudes
for small scattering angle and large energies, consider some
sample interactions shown in figure \ref{fig2} contributing to
$\mathcal{A}_{1}$, where the exchanged particle is massless and has spin $j$.
The contribution of graph \ref{fig2}(a) to $\mathcal{A}_{1}$ has the generic form
\begin{equation}
g^2 \,\frac{s^{j}-c_{1}s^{j-1}t+\cdots}{-t} =
g^2 \,s^{j-1}\left( \frac{1}{z}+c_{1}+\cdots+c_{j}z^{j-1}\right) . \label{flatgenamp}
\end{equation}
The polynomial part in $z$ contributes to partial waves with spin $J<j$.
This can also be understood in the impact parameter representation (\ref{imprep1}), 
since polynomial terms in $z={\bf q}^{2}s$ give, after Fourier
transform, delta--function contributions to the phase shift localized at $r=0$. 
The universal term $s^{j-1}/z$ contributes, on the other
hand, to partial waves of all spins and gives a phase shift at
large $r$ given by
\begin{equation}
\sigma\left( s,r\right) \simeq \frac{i}{8\pi}\,I 
= -\frac{g^2}{4\pi}\, s^{\, j-1} \,\Delta_{\perp}(r) 
~,  \label{phaseshift}
\end{equation}
where $\Delta_{\perp}(r)$ is the scalar massless Euclidean
propagator in transverse space $\mathbb{R}^{d-1}$. At high
energies, the maximal $j$
dominates. Graph \ref{fig2}(b) is obtained by interchanging the role of $s$ and $t$ in (\ref{flatgenamp})
, and therefore is proportional to $s^{j-1}\left( z^{j}+c_{1}
z^{j-1}+\cdots+c_{j}\right)$. The corresponding intermediate
partial waves have spin $J\leq j$. Finally, graph \ref{fig2}(c) is
obtained from \ref{fig2}(a) by
replacing $t$ by $u=-\left( s+t\right) $ and therefore by sending 
$z\rightarrow 1-z$. Using the fact that $\mathcal{S}_{J}(1-z)=(-)^J\mathcal{S}_{J}(z)$, 
we can again expand the \emph{U}--channel exchange graph
\ref{fig2}(c) in
the form $s^{\frac{3-d}{2}}\sum_{J}\left( -\right) ^{J}\mathcal{S}%
_{J}~\sigma_{J}\left( s\right)$ where $\sigma_J(s)$ are the phase
shifts of
the \emph{T}--channel exchange, whose large spin behavior is given by (\ref%
{phaseshift}). At large spins, the contribution to the various
partial waves is alternating in sign and averages to a
sub--leading contribution. Finally, the contact graph \ref{fig2}(d) only
contains a spin zero
contribution. We conclude that, at large spins and energies, the \emph{T}--channel 
exchange of a graviton in graph \ref{fig2}(a) dominates all other interactions.


\section{Anti--de Sitter Spacetime} \label{eikonalAdS}


Let us now follow the intuitive picture developed in the previous section
to generalize the eikonal approximation  to hard scattering in Anti--de Sitter spacetime  \cite{Paper3}.
In this section we derive an eikonal approximation in AdS summing ladder and cross ladder
Witten diagrams. The corresponding partial wave analysis  is postponed to the next chapters.
As explained in section \ref{AdSgeom}, we define AdS$_{d+1}$ space, of dimension $d+1$, 
as the pseudo--sphere (\ref{pseudoS}) of radius $\ell = 1$ embedded in $\mathbb{R}^{2,d}$. 
Then, we denote a  point  ${\bf x} \in $ AdS by a point in the embedding space $\mathbb{R}^{2,d}$ 
obeying  ${\bf x}^2=-1$.

Consider  the Feynman graph in figure \ref{fig1}, but now  in AdS.
For simplicity, we consider the exchange of an AdS scalar field of dimension $\Delta$ and, for external fields,
we consider scalars of dimension $\Delta_1$ and $\Delta_2$. Then, the graph in figure \ref{fig1} evaluates to
\begin{equation}
\begin{array}{c}
\displaystyle{A_n = (ig)^{2n} \int_{{\rm AdS}}d{\bf x}_1 \cdots d{\bf x}_n d{\bf \bar{x}}_1 \cdots d{\bf \bar{x}}_n \,
\psi_3( {\bf x}_n )  \Pi_{\Delta_1}({\bf x}_n,{\bf x}_{n-1})\cdots \Pi_{\Delta_1}({\bf x}_2,{\bf x}_1)\psi_1( {\bf x}_1)}
\spa{0.5}
\\
\displaystyle{\psi_4( {\bf \bar{x}}_n ) \Pi_{\Delta_2}({\bf \bar{x}}_n,{\bf \bar{x}}_{n-1})\cdots
\Pi_{\Delta_2}({\bf \bar{x}}_2,{\bf \bar{x}}_1)\psi_2( {\bf \bar{x}}_1 )
\sum_{{\rm perm} \ \sigma}   \Pi_{\Delta}({\bf x}_1,{\bf \bar{x}}_{\sigma_1})
\cdots \Pi_{\Delta}({\bf x}_n,{\bf \bar{x}}_{\sigma_n})}\ ,\label{AdSamplitude}
\end{array}
\end{equation}
where $\Pi_{\Delta}({\bf x},{\bf \bar{x}})$ stands for the scalar propagator of mass $\Delta(\Delta-d)$ in AdS, satisfying
\begin{equation}
\Big[ \Box_{\rm AdS}-\Delta( \Delta-d ) \Big]\Pi_{\Delta}({\bf x},{\bf \bar{x}})=i\delta({\bf x},{\bf \bar{x}})\ .
\label{propeq}
\end{equation}
In general, this amplitude is very hard to compute. However, we expect some drastic simplifications
for specific external wave functions describing highly energetic particles scattering at
fixed impact parameters. In analogy with flat space, we expect the eikonal approximation to correspond
to the collapse of the propagators $\Pi_{\Delta_1}$ and $\Pi_{\Delta_2}$ into
null geodesics approximating classical trajectories of highly energetic particles.

\subsection{Null Congruences in AdS and Wave Functions}\label{NCWF}

\begin{figure}
\begin{center}
\includegraphics[width=9cm]{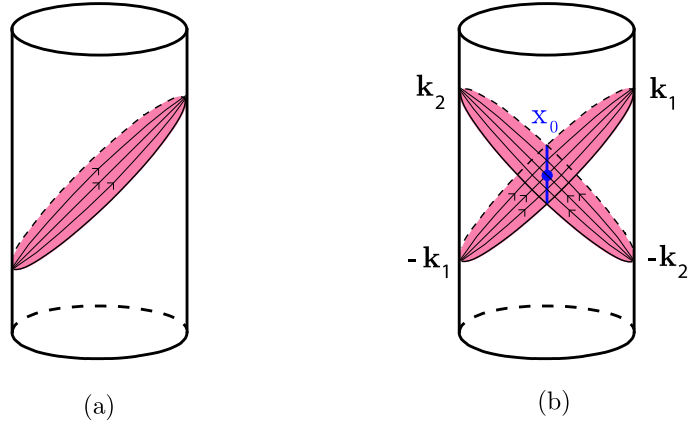}
\caption{  (a) A generic null hypersurface  ${\bf k}
\cdot{\bf y} =0$ in conformally compactified AdS. (b) The two null
hypersurfaces  ${\bf k}_1 \cdot{\bf y} =0$ and  ${\bf k}_2
\cdot{\bf y} =0$. Their intersection is the transverse hyperboloid
$H_{d-1}$ containing the reference point ${\bf x}_0$.
As we have seen in the introduction, the null vectors ${\bf k}_i$ and $-{\bf k}_i$
can be thought of as points in the AdS conformal boundary. }
\label{nullsurface}
\end{center}
\end{figure}

As mentioned in section \ref{AdSdyn}, a null geodesic in AdS is also a null geodesic in the embedding space
$$
{\bf x}(\lambda)={\bf y}+\lambda\, {\bf k} \ ,
$$
where ${\bf y}\in {\rm AdS}$ and the tangent vector ${\bf k}\in \mathbb{R}^{2,d}$ satisfies
$$
{\bf k}^2=0 \ ,\ \ \ \ \ \ \ \ \ \ \ \ \ \ {\bf k} \cdot{\bf y} =0 \ .
$$
We will follow the intuitive idea that the wave functions $\psi_1$ and $\psi_2$ correspond to the initial
states of highly energetic particles moving along two intersecting congruences of null geodesics.
As described in the previous section, in flat space there is a one--to--one correspondence between
null momenta (up to scaling) and congruences of null geodesics.
On the other hand, in AdS the situation is more complicated.
Given a null vector $ {\bf k}$ there is a natural set of null geodesics $ {\bf y}+\lambda {\bf k}$
passing through all points ${\bf y}\in {\rm AdS}$ belonging to the hypersurface ${\bf k} \cdot{\bf y} =0$,
as shown in figure \ref{nullsurface}(a). However,
to construct a congruence of null geodesics we need to extend this set to the full
AdS space. Contrary to flat space, in AdS this extension is not unique because
the spacetime conformal boundary is timelike. We will now describe how to construct
such a congruence in analogy with the construction presented for flat space.

We start with two null vectors ${\bf k}_1,\, {\bf k}_2$ associated with the incoming particles, 
as represented in  figure \ref{nullsurface}(b) and normalized as in flat space  
$$-2{\bf k}_1\cdot {\bf k}_2 = (2\omega)^2\  .$$
The transverse space is naturally defined as the intersection of the two null hypersurfaces associated to 
${\bf k}_1$ and ${\bf k}_2$.  It is the  hyperboloid $H_{d-1}$ defined by
$$
{\bf w} \in {\rm AdS} \ , \ \ \ \ \
{\bf k}_1 \cdot{\bf w} ={\bf k}_2 \cdot{\bf w} =0\ .
$$
In order to introduce coordinates in AdS$_{d+1}$ in analogy with (\ref{coord}), we choose an arbitrary
reference point ${\bf x}_0$ in this transverse space $H_{d-1}$. This allows us to define the vector fields
$$
{\bf T}_1 ({\bf x})  = 
\frac{({\bf k}_1\cdot {\bf x})\, {\bf x}_0 -  ({\bf x}_0 \cdot {\bf x})\, {\bf k}_1}{2\omega}
   \ , \ \ \ \ \ \ \ \ \
{\bf T}_2({\bf x})   = 
\frac{({\bf k}_2\cdot {\bf x})\, {\bf x}_0 -  ({\bf x}_0 \cdot {\bf x})\, {\bf k}_2}{2\omega}  \ ,
$$
which, from the embedding space perspective, are respectively the
generators of parabolic Lorentz transformations in the ${\bf
x}_0\,{\bf k}_1$ and   ${\bf x}_0\,{\bf k}_2$--plane. They
therefore generate AdS isometries. We may now introduce
coordinates $\{u,v,{\bf w}\}$ for ${\bf x} \in $ AdS$_{d+1}$ as
follows
\begin{equation}
\begin{array}{rcl}
{\bf x}&=& \displaystyle{e^{\,v  {\bf T}_2}\,e^{\,u  {\bf T}_1}\, {\bf w}}\spa{0.2}\\
&=& \displaystyle{{\bf w} -  u\, \frac{({\bf x}_0\cdot{\bf w}) \, {\bf k}_1}{2\omega} - v\,\frac{({\bf x}_0\cdot{\bf w}) \,{\bf k}_2}{2\omega}
+ uv\,\frac{({\bf x}_0\cdot{\bf w}) \,{\bf x}_0}{2} +  uv^2\,\frac{({\bf x}_0\cdot{\bf w}) \,{\bf k}_2}{8\omega}} \ ,
\end{array}
\label{coordAdS}
\end{equation}
where ${\bf w}\in H_{d-1}$.
It is important to realize that, contrary to the flat space case,  
$[{\bf T}_1, {\bf T}_2]\neq 0$ and therefore the order of the exponential maps in (\ref{coordAdS}) is important, as will become clear below.

As for flat space, the coordinate $u$ is an affine parameter along
null geodesics labeled by $v$ and ${\bf w}$, which form the
desired congruence for particle 1. In fact,  (\ref{coordAdS}) can
be written as
$$
{\bf x}= e^{\,v  {\bf T}_2}\, {\bf w} +u\, e^{\,v  {\bf T}_2} \, {\bf T}_1 ({\bf w} )\ .
$$
Hence, the geodesics in the null congruence associated to particle 1 are given by
\begin{equation}
{\bf x}={\bf y}+\lambda\, {\bf k}\, ,
\label{p}
\end{equation}
where
$$
{\bf y} = e^{\,v  {\bf T}_2}\, {\bf w} =
{\bf w}  - v\,\frac{({\bf x}_0\cdot{\bf w}) \,{\bf k}_2}{2\omega}\, .
$$
The normalization of the momentum  ${\bf k}$ and affine parameter $\lambda$ of the classical trajectories is fixed by
demanding, as in flat space, that the conserved charge 
$-{\bf T}_2\cdot {\bf k}=\omega$.
This gives
\begin{eqnarray*}
\lambda &=&  u\,\frac{({\bf x}_0\cdot{\bf w})^2}{2\omega}\, ,\\
{\bf k} &=& \frac{2\omega}{({\bf x}_0\cdot{\bf w})^2  } \,e^{\,v  {\bf T}_2}\, {\bf T}_1({\bf w} )
= \frac{2\omega}{({\bf x}_0\cdot{\bf w})^2}\,\frac{d\,}{du}
=- \frac{1}{{\bf x}_0\cdot{\bf w}  } \left(
{\bf k}_1-v\omega \,{\bf x}_0-\frac{v^2}{4}\,{\bf k}_2
\right)\ .
\end{eqnarray*}
Let us remark that different choices of ${\bf x}_0$ give different congruences, all containing
the null geodesics ${\bf w}+\lambda '\, {\bf k}_1$, which lay on
the  hypersurface ${\bf k}_1\cdot {\bf x} = 0$ at $v=0$. Starting from this hypersurface,
we  then constructed a congruence of null geodesics using
the AdS isometry generated by ${\bf T}_2$.

\begin{figure}
\begin{center}
\includegraphics[width=14cm]{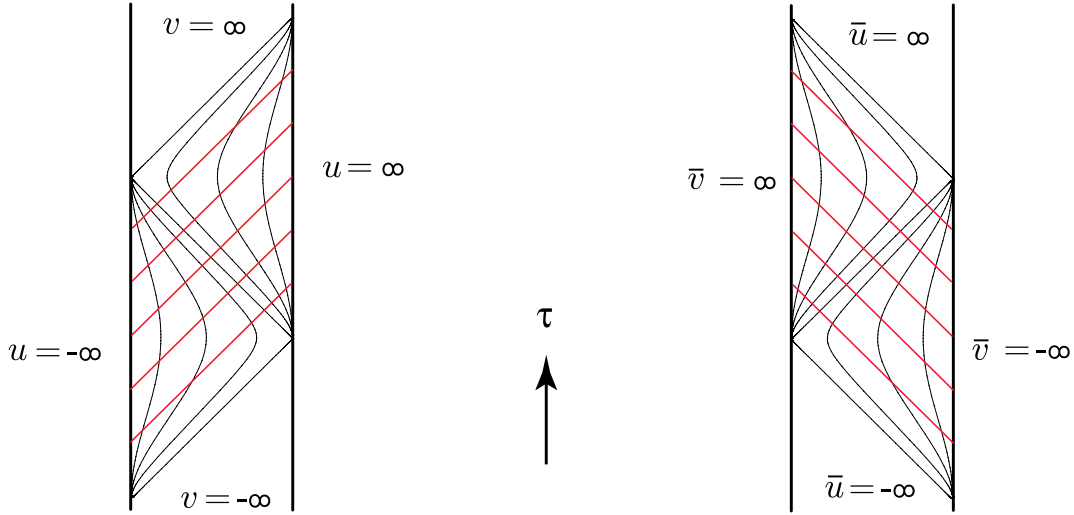}
\caption{  The coordinates $\{u,v\}$ and $\{\bar{u},\bar{v}\}$ for the simplest case of AdS$_2$.
In general, the wave function of particle 1 is independent of the coordinate $u$, while that of particle
2 is independent of the coordinate $\bar{v}$.}
\label{uvcoord}
\end{center}
\end{figure}

Contrary to flat space, the curves defined by constant $u$ and ${\bf w}$ in the coordinate system
(\ref{coordAdS}) are {\em not} null geodesics
(except for the curves on the surface $u=0$ which are null geodesics with affine parameter $v$).
These curves are the integral curves of the Killing vector field $ {\bf T}_2 = \frac{d\,}{dv}$.
In fact, these curves are not even null, as can be seen from the form of the AdS metric in these coordinates
\begin{equation}
ds^2=d {\bf w}^2- ({\bf x}_0\cdot{\bf w}  )^2 dudv - \frac{u^2}{4} ({\bf x}_0\cdot{\bf w}  )^2   dv^2\ ,
\label{AdSMetric}
\end{equation}
where $d {\bf w}^2$ is the metric on the hyperboloid $H_{d-1}$.
To construct the null congruence for particle 2, we introduce new coordinates
$\{\bar{u} ,   \bar{v},   {\bf \bar{w}} \}$ for ${\bf \bar{x}} \in $ AdS$_{d+1}$ as follows
\begin{equation}
{\bf \bar{x}}= e^{\,\bar{u}  {\bf T}_1}\,e^{\,\bar{v}  {\bf T}_2}\, {\bf \bar{w}}\ . \label{coordAdSbar}
\end{equation}
The two sets of coordinates are related by
\be
\bar{u}=u\left(1-\frac{uv}{4}\right)^{-1}\ , \ \ \ \ \ \ \ \ \ \ 
\bar{v}=v\left(1-\frac{uv}{4}\right)\ , \ \ \ \ \ \ \ \ \ \  
\bar{\bf w}={\bf w}\ .
\ee
The congruence associated with particle 2 is then the set of null geodesics
\begin{equation}
{\bf \bar{x}}={\bf \bar{y}}+\bar{\lambda}\, {\bf \bar{k}} \ ,
\label{pbar}
\end{equation}
with
\begin{eqnarray*}
&&{\bf \bar{y}}= e^{\,\bar{u}  {\bf T}_1}\, {\bf \bar{w}}=
{\bf \bar{w}}  - \bar{u}\,\frac{({\bf x}_0\cdot{\bf \bar{w}}) \,{\bf k}_1}{2\omega}\ ,\\
&&\bar{\lambda} =  \bar{v}\,\frac{({\bf x}_0\cdot{\bf \bar{w}})^2}{2\omega} \ ,\\
&&{\bf \bar{k}}= \frac{2\omega}{({\bf  x}_0\cdot{\bf  \bar{w}})^2 }\,
e^{\, \bar{u}  {\bf T}_1}\, {\bf T}_2({\bf  \bar{w}} )
=\frac{2\omega}{({\bf x}_0\cdot{\bf \bar{w}})^2}\,\frac{d\,}{d\bar{v}}
= - \frac{1}{{\bf x}_0\cdot {\bf  \bar{w}}  } \left(
{\bf k}_2- \bar{u}\omega \,{\bf x}_0-\frac{\bar{u}^2}{4}\,{\bf k}_1
\right)\ ,
\end{eqnarray*}
so that the conserved charge $-{\bf T}_1\cdot {\bf \bar{k}}=\omega$.
In figure \ref{uvcoord} we plot the curves of constant $u$ and $v$ (left) and of constant $\bar{u}$ and $\bar{v}$ (right)
in the simplest case of AdS$_2$.

As in flat space, the wave function describing particle 1 carries energy $\omega$ 
$$
\mathcal{L}_{{\bf T}_2} \psi_1 = \partial_v \psi_1 \simeq - i\omega \psi_1 \ .
$$
Therefore we choose
$$
\psi_1( {\bf x})= e^{-i\omega  v}F_1({\bf x} )\ ,
$$
where the function $F_1$ is approximately constant over the length scale $1/\omega$,
more precisely $|\partial F_1| \ll \omega |F_1|$.
The Klein--Gordon equation for the wave function $\psi_1$ implies
$$
\left[ \frac{4i\omega}{({\bf x}_0\cdot{\bf w})^2}\, \partial_u +\Box_{\rm AdS}- \Delta_1(\Delta_1-d) \right]F_1({\bf x} )=0\ ,
$$
since, as in flat space, the coordinate $v$ satisfies
$$
\square_{\rm AdS}\ v = (\nabla v)^2 = 0\, .
$$
The above equation can then be solved expanding $F_1$ in powers of $1/\omega$,
$$
F_1({\bf x} )=F_1(v,{\bf w})- \frac{({\bf x}_0\cdot{\bf w})^2}{4i\omega}
\int du \Big[\Box_{\rm AdS}-\Delta_1(\Delta_1-d) \Big]F_1(v,{\bf w})+\cdots
$$
Since the eikonal approximation gives only the leading behavior of the scattering amplitude at large
$\omega$, it is enough to consider only the first term
$F_1({\bf x} )\simeq F_1(v,{\bf w})$ so that,
to this order, we have
$$
\mathcal{L}_{\bf k} \psi_1 = 0\ ,
$$
as expected. We conclude that the function $F_1$ is a smooth transverse modulation independent of the affine parameter $\lambda$ of
the null geodesics associated with the classical trajectories of particle 1.
Similar reasoning applied to particle 2 leads to
$$
\psi_2( {\bf \bar{x}})\simeq e^{-i\omega \bar{u}}F_2(\bar{u},{\bf \bar{w}})\ .
$$

Finally, since in the eikonal regime the particles are only slightly deviated by the scattering process,
to leading order in $1/\omega$ the outgoing wave functions
are also independent of the corresponding affine parameters,
$$
\psi_3( {\bf x})\simeq e^{i\omega v}F_3(v, {\bf w})\ ,\ \ \ \ \ \ \ \
\psi_4( {\bf \bar{x}})\simeq e^{i\omega \bar{u}}F_4(\bar{u} , {\bf \bar{w}})\ ,
$$
with the same requirement $|\partial F| \ll \omega |F|$.

We have kept the discussion of this section completely coordinate
independent. On the other hand, given the choice of ${\bf k}_1$ and ${\bf k}_2$, the embedding
space ${\mathbb R}^{2,d}$ naturally splits into $\mathbb{M}^{2}\times \mathbb{M}^{d}$,
with $\mathbb{M}^{2}$ spanned by ${\bf k}_1$ and $ {\bf k}_2$ and with $\mathbb{M}^{d}$ its orthogonal complement.
Then, the points ${\bf x}_0$, $ {\bf w} $ and $ {\bf \bar{w}}$ are timelike points in  $H_{d-1}\subset \mathbb{M}^{d}$,
with the transverse space $H_{d-1}$ the upper unit mass--shell on  $\mathbb{M}^{d}$.
Finally, it is natural to introduce Poincar\'e coordinates $\{ y,{\bf y} \} $ similar to (\ref{Poincare}), 
$$
{\bf x} = \frac{1}{y}\left( \frac{ {\bf k}_1 }{2\omega }+ (y^2 + {\bf y}^2 ) \frac{ {\bf k}_2}{2\omega }  +{\bf y} \right)\ ,
$$ 
where $ {\bf y}$ parametrizes the $\mathbb{M}^{d}$ orthogonal to  ${\bf k}_1$ and ${\bf k}_2$ and
$y\in \mathbb{R}^+$ is a radial coordinate.
As explained in section \ref{AdSgeom}, in these coordinates, the action of $e^{\, \alpha {\bf T}_1}$ 
leaves the radial coordinate $y$ fixed and induces translations in $\mathbb{M}^{d}$,
$$
{\bf y} \to {\bf y}+\frac{\alpha}{ 2}\,   {\bf x}_0\ ,
$$ 
along the time direction indicated by ${\bf x}_0$.
Similar remarks apply to ${\bf T}_2$ with the roles of ${\bf k}_1$ and ${\bf k}_2$ interchanged.

\subsection{Eikonal Amplitude}

\begin{figure}
[ptb]
\begin{center}
\includegraphics[width=4cm]{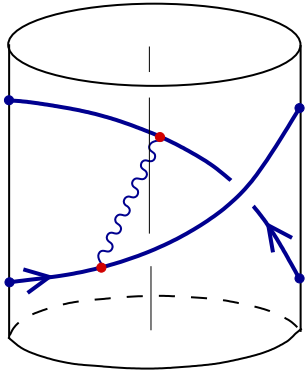}
\caption{The eikonal phase shift is simply the tree--level interaction between two null geodesics in AdS,
which we parametrize by $\lambda$ and $\bar{\lambda}$.}
\label{eikonalint}
\end{center}
\end{figure}

We are now in position to compute the leading behavior of the amplitude (\ref{AdSamplitude})
for the exchange of $n$ scalars in AdS at large $\omega$, using the techniques explained in section \ref{flateikonal}.
The first step is to obtain an approximation for the AdS propagator similar to (\ref{posprop}).
Since for particle 1 we have $\partial_v \simeq -i\omega $, we can approximate
$$
\Box_{\rm AdS} \simeq \frac{4i\omega}{({\bf x}_0\cdot{\bf w})^2} \,\partial_u\ ,
$$
in equation (\ref{propeq}) for the propagator of particle 1 between vertices ${\bf x}_j$ and  ${\bf x}_{j+1}$,
obtaining
$$
\frac{4i\omega}{({\bf x}_0\cdot{\bf w})^2} \,\partial_{u_j} \Pi_{\Delta_1}({\bf x}_j,{\bf x}_{j+1})=
\frac{2i}{({\bf x}_0\cdot{\bf w})^2}\,\delta(u_j-u_{j+1})\, \delta(v_j-v_{j+1})\,\delta_{H_{d-1}}({\bf w}_j,{\bf w}_{j+1})\ .
$$
The solution,
\be
\Pi_{\Delta_1}({\bf x}_j,{\bf x}_{j+1})\simeq \frac{1}{2\omega}\,\Theta(u_j-u_{j+1})\,
\delta(v_j-v_{j+1})\,\delta_{H_{d-1}}({\bf w}_j,{\bf w}_{j+1})\ ,   \label{AdSeikprop}
\ee
has the natural interpretation of propagation only along the particle classical trajectory and, in these
coordinates, takes almost exactly the same form as the corresponding propagator (\ref{posprop}) in flat space.
With this approximation to the propagator, the amplitude (\ref{AdSamplitude}) associated with
the exchange of $n$ scalar particles simplifies to
\begin{eqnarray*}
A_n &\simeq& (2\omega)^2 \int_{-\infty}^{\infty}dv d\bar{u} \int_{H_{d-1}} d{\bf w} d{\bf \bar{w}}
 F_1(v,{\bf w}) F_3(v,{\bf w})F_2(\bar{u} , {\bf \bar{w}})F_4(\bar{u} , {\bf \bar{w}})\\
&&
\int_{-\infty}^{\infty} du_1 \int_{u_1}^{\infty} du_2  \cdots \int_{u_{n-1}}^{\infty} du_n
\int_{-\infty}^{\infty} d\bar{v}_1\int_{\bar{v}_1}^{\infty} d\bar{v}_2  \cdots
\int_{\bar{v}_{n-1}}^{\infty} d\bar{v}_n\\
&&\left(\frac{ig \,({\bf x}_0\cdot{\bf w})( {\bf x}_0\cdot  {\bf \bar{w}}) }{4\omega}\right)^{2n}
\sum_{{\rm perm}\ \sigma}   \Pi_{\Delta}({\bf x}_1,{\bf \bar{x}}_{\sigma_1})\cdots
 \Pi_{\Delta}({\bf x}_n,{\bf \bar{x}}_{\sigma_n})\ ,
\end{eqnarray*}
where
$$
{\bf x}_j = e^{\,v   {\bf T}_2 }\,e^{\,u_j  {\bf T}_1}\, {\bf w} \ ,\ \ \ \ \ \ \ \
{\bf \bar{x}}_j= e^{\,\bar{u}  {\bf T}_1 }\,e^{\,\bar{v}_j  {\bf T}_2 }\, {\bf \bar{w}}\ .
$$
Notice that the extra powers of $({\bf x}_0\cdot{\bf w})^2/2$ and $( {\bf x}_0\cdot  {\bf \bar{w}})^2/2$ come from the integration
measure in the ${\bf x}_j$ and ${\bf \bar{x}}_j$ coordinates, respectively.
As for flat space, the integrals over the affine parameters can be extended
to the real line so that, after summing over $n$, we obtain
\begin{equation}
A \simeq (2\omega)^2 \int_{-\infty}^{\infty}dv d\bar{u} \int_{H_{d-1}} d{\bf w} d{\bf \bar{w}}
 F_1(v,{\bf w}) F_3(v,{\bf w})F_2(\bar{u} , {\bf \bar{w}})F_4(\bar{u} , {\bf \bar{w}})\, e^{\,I/4}\ ,
\label{eqFinal}
\end{equation}
with
$$
I= -\frac{g^2 \,({\bf x}_0\cdot{\bf w})^2 ( {\bf x}_0\cdot  {\bf \bar{w}})^2 }{(2\omega)^2}
\int_{-\infty}^{\infty} du d\bar{v} \, \Pi_{\Delta}\left({\bf x},{\bf \bar{x}}\right)\ .
$$
This can be rewritten as the tree--level interaction between two classical trajectories of the incoming particles
described by (\ref{p}) and (\ref{pbar}),
which are labeled respectively by ${\bf y}$  and ${\bf \bar{y}}$,
$$
I= (-ig)^2\,\int_{-\infty}^{\infty}d\lambda d\bar{\lambda}\,
\Pi_{\Delta}\left({\bf y}+ \lambda  \, {\bf k}({\bf y}) ,
{\bf \bar{y}} + \bar{\lambda} \, {\bf \bar{k}}({\bf \bar{y}}) \right)\ .
$$
Hence, the AdS eikonal amplitude just obtained is the direct analogue of the corresponding
flat space amplitude (\ref{scalareik}).
The eikonal interaction is depicted in figure \ref{eikonalint}.

The generalization of the above result to the case
of interactions mediated by a minimally coupled particle of spin $j$ is straightforward,
and we shall only give the relevant results. At high energies, the only
change concerns the propagator $\Pi_\Delta$, which now should be replaced by the propagator
of the spin--$j$ particle contracted with the null momenta of the geodesics
$$
\Pi^{(j)}_\Delta=(-2)^j\, {\bf k}_{\alpha_1}\cdots{\bf k}_{\alpha_j}\, {\bf \bar{k}}_{\beta_1}\cdots{\bf \bar{k}}_{\beta_j}
\, \Pi_\Delta^{\alpha_1,\cdots,\alpha_j,\beta_1,\cdots,\beta_j}\, ,
$$
where the indices $\alpha_i,\beta_j$ are tangent indices to AdS.
This follows immediately from the fact that, at high energies,
covariant derivatives $-i\nabla_\alpha$ in interaction
vertices can be replaced  by ${\bf k}_\alpha$ and ${\bf \bar k}_\alpha$ for
particle one and two, respectively. The spin--$j$ propagator is
totally symmetric and traceless in the indices $\alpha_1\dots
\alpha_j$ (and similarly in the indices $\beta_1\cdots \beta_j$),
it is divergenceless and satisfies
\begin{equation}
\Big[ \Box -\Delta( \Delta-d )+j \Big]
\Pi_\Delta^{\alpha_1,\cdots,\alpha_j,\beta_1,\cdots,\beta_j}({\bf x},{\bf \bar{x}})=
i\,g^{(\alpha_1\beta_1}g^{\alpha_2\beta_2}\cdots g^{\alpha_1\beta_1)}\,\delta({\bf x},{\bf \bar{x}})\,+\cdots ,
\label{propeqBIS}
\end{equation}
where the indices $\alpha_i$ and $\beta_i$ are separately symmetrized, and where the terms in $\cdots$  
contain derivatives of $\delta({\bf x},{\bf \bar{x}})$ and are not going to be of relevance to the discussion 
which follows, since they give subleading contributions at high energies. The eikonal expression 
(\ref{eqFinal}) is then valid in general, with the phase factor $I$ now replaced by
\begin{equation}
I= -g^2\,\int_{-\infty}^{\infty}d\lambda d\bar{\lambda}\,
\Pi^{(j)}_{\Delta}\left({\bf y}+ \lambda  \, {\bf k} ,
{\bf \bar{y}} + \bar{\lambda} \, {\bf \bar{k}} \right)\ .
\label{explicitgenphase}
\end{equation}
Note that we have normalized the interaction coupling as in flat space, where the tree level amplitude is given by $-g^2\, s^{j}/t$
at large $s$.

Finally, we remark that the explicit expression (\ref{explicitgenphase}) for the phase shift in (\ref{eqFinal}) is not valid
for generic values of the impact parameter  \cite{GG}. 
In the gravitational case, we expect the semi--classical eikonal approximation
to breakdown when the the impact parameter is smaller than (or of the order of) the Schwarschild radius
 $\left(G\sqrt{s}\right)^{\frac{1}{(d-2)}}$.
In fact, the regime of black hole formation in AdS has been related to the saturation of the 
Froissart bound in the dual gauge theory 
\cite{Giddings:2002cd,Kang:2004jd, Kang:2005bj, Nastase:2005bk}.
Furthermore, in string theories, string effects become relevant at small impact parameters, when the tidal forces
produced by one string are strong enough to change the internal state of the other.
We shall return to this issues with more detail in the concluding chapter.

\subsection{Transverse Propagator}\label{sectTP}

Now we compute the integral $I$. Its last expression shows that it is
a Lorentz invariant local function of ${\bf y},{\bf \bar{y}},{\bf k}$ and ${\bf \bar{k}}$. Moreover,
it is invariant under
$$
{\bf y}\to {\bf y}+\alpha\, {\bf k} \ ,\ \ \ \ \ \ \ \ \ \
{\bf \bar{y}} \to{\bf \bar{y}} + \bar{\alpha} \, {\bf \bar{k}}\ ,
$$
and it scales like $ I\to (\alpha\bar{\alpha})^{j-1}I $
when ${\bf k}\to \alpha\, {\bf k} $ and ${\bf \bar{k}} \to\bar{\alpha} \, {\bf \bar{k}} $.
Therefore, the integral $I$ is fixed up to  an undetermined function $G$,
\begin{eqnarray*}
I&=&2ig^2\,(-2{\bf k}\cdot{\bf \bar{k}})^{j-1}\,G\left({\bf y}\cdot {\bf \bar{y}}  -
\frac{({\bf k}\cdot{\bf \bar{y}}) \, ({\bf \bar{k}}\cdot{\bf y})}{{\bf k}\cdot   {\bf \bar{k}} } \right)\\
&=&2ig^2\,s^{j-1}
\,G(  {\bf w} \cdot {\bf \bar{w}} )
\ ,
\end{eqnarray*}
with $s$ defined in analogy with flat space
\begin{equation}
s=-2{\bf k}\cdot{\bf \bar{k}}=(2\omega)^{2}
\frac{(1+v\bar{u}/4)^2}{
({\bf x}_0\cdot{\bf w})\, ({\bf x}_0\cdot  {\bf \bar{w}})}\ .
\label{sAdS}
\end{equation}
To determine the function $G$ we use equation (\ref{propeqBIS}), contracting both sides with
$$
(-2)^j\, {\bf k}_{\alpha_1}\cdots{\bf k}_{\alpha_j}\, {\bf \bar{k}}_{\beta_1}\cdots{\bf \bar{k}}_{\beta_j}
$$
and integrating against
$$
\int_{-\infty}^{\infty}du d\bar{v} = \frac{(2\omega)^2}{({\bf x}_0\cdot{\bf w})^2
({\bf \bar{x}}_0\cdot{\bf \bar{w}})^2}\,\int_{-\infty}^{\infty}d\lambda d\bar{\lambda}\, .
$$
Here we discuss the simplest case of $j=0$, leaving for clearness of exposition 
the general case to appendix \ref{app1}. Consider then first the integral of the RHS of (\ref{propeq}).
Using the explicit form of the $\delta$--function in the  $\{u,v, {\bf w} \}$ coordinate system,
$$
\delta({\bf x},{\bf \bar{x}}) =
 \frac{2}{ ({\bf x}_0\cdot{\bf w})^2}\,
\delta_{H_{d-1}}({\bf w},{\bf \bar{w}})\,
\delta \left(u-\bar{u} \left(1-\frac{ \bar{u}\bar{v}}{4}\right) \right)\,
\delta \left(v- \bar{v} \left(1-\frac{\bar{u}\bar{v}}{4}\right)^{-1} \right)\ ,
$$
we obtain
\be
\frac{2i}{
(1+v\bar{u}/4)^{2}({\bf x}_0\cdot{\bf w})^{2}}\,
\delta_{H_{d-1}}({\bf w},{\bf \bar{w}})\ .
\label{RHS}
\ee
Next we consider the LHS of (\ref{propeq}). Explicitly parametrizing the metric $d{\bf w}^2$ on $H_{d-1}$ in (\ref{AdSMetric}) as
\begin{equation}
d{\bf w}^2 = \frac{d{\chi}^2}{{\chi}^2-1}+({\chi}^2-1)ds^2(S_{d-2}),
\label{hyperboloidmetric}
\end{equation}
where $\chi=-\mathbf{x}_0\cdot \mathbf{w}$, we have that
\bes
\Box_{\rm AdS} \Pi_\Delta=\left[\Box_{H_{d-1}}+2\,\frac{{\chi}^2-1}{{\chi}}\partial_{\chi} \right]\Pi_\Delta + \partial_u(\cdots)\ .
\ees
We do not show the explicit terms of the form $\partial_u(\cdots)$ since they will vanish once integrated along the two geodesics. Integrating in $du d\bar{v}$ we conclude that (\ref{RHS}) must be equated to
$$
-\frac{2i}{(1+v\bar{u}/4)^{2}} \left[\Box_{H_{d-1}}-\Delta(\Delta-d)
+2\,\frac{{\chi}^2-1}{{\chi}}\partial_{\chi}
\right]
\frac{G({\bf w},{\bf \bar{w}})}{{\chi}\bar{{\chi}}}\ .
$$
Using the fact that
$$
\left[\Box_{H_{d-1}},{\chi}^{-1}\right]=\frac{1}{{\chi}}\left(-2\frac{{\chi}^2-1}{{\chi}}\partial_{\chi}+(3-d)-\frac{2}{{\chi}^2}\right)\ ,
$$
we finally deduce that
$$
\left[ \Box_{H_{d-1}} +1-d -\Delta( \Delta-d ) \right] G(  {\bf w} \cdot
{\bf \bar{w}} )
= -\delta(  {\bf w}, {\bf \bar{w}} )\ .
$$
In appendix \ref{app1} we show that this last equation is also valid for general spin $j$. We conclude that
the function $G$ is the scalar Euclidean propagator in the hyperboloid
$H_{d-1}$
of mass squared  $(\Delta-1)(\Delta-1-d+2)$ and corresponding dimension $\Delta-1$. Denoting this propagator by
$\Pi_{\perp}( {\bf w}, {\bf \bar{w}} )$,
 the eikonal amplitude can be written as
\begin{equation}
\begin{array}{rcl}
A &\simeq&
\displaystyle{(2\omega)^2 \int_{-\infty}^{\infty} dv d\bar{u}  \int_{H_{d-1}} d{\bf w} d{\bf \bar{w}}
 F_1(v,{\bf w}) F_3(v,{\bf w})F_2(\bar{u} , {\bf \bar{w}})F_4(\bar{u} , {\bf \bar{w}})}\\
&&
\ \ \ \ \ \ \ \ \ \ \ \ \ \ \  \ \ \ \ \ \ \ \ \ \ \ \ \ \ \ \ \ \ \ \ \ \ \ \ \ \ \ \ 
\displaystyle{\exp \left( \frac{ig^2  }{2}\,s^{j-1}\,
\Pi_{\perp}({\bf w},{\bf \bar{w}}) \right)}\ ,
\end{array}
\label{eikAdS}
\end{equation}
with $s$ given by (\ref{sAdS}).
In the particular case of gravitational interactions in AdS$_5$, the same result was recently derived in  \cite{BrowerEIK}, 
using different approaches.

\begin{figure}
\begin{center}
\includegraphics[height=6cm]{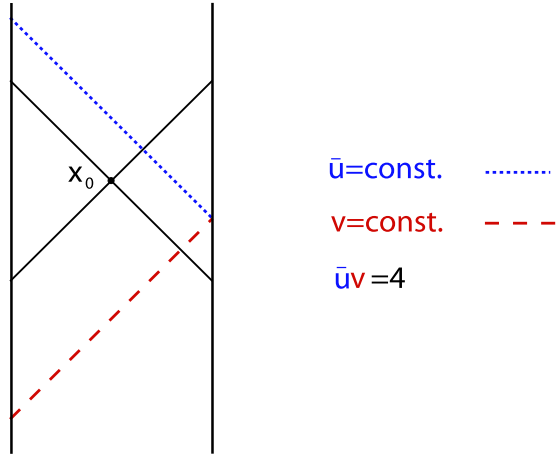}
\caption{  The null geodesics with constant $\bar{u}=-4/v$ are the reflection   in the AdS conformal boundary
of null geodesics with constant $v$.}
\label{reflection}
\end{center}
\end{figure}

\subsection{Localized Wave Functions}

The eikonal amplitude in AdS has a striking similarity with the
standard flat space eikonal amplitude. However, an important
difference is the factor $
\left(1+\frac{v\bar{u}}{4}\right)^{2}$ in the definition (\ref{sAdS}) of $s$, which
makes the exponent in the eikonal amplitude (\ref{eikAdS}) divergent for $v\bar{u}=-4$ and $j=0$. This divergence can be
traced back to the collinearity of the tangent vectors ${\bf k}$
and $ {\bf \bar{k}}$ of the null geodesics labeled by $\{v,{\bf
w} \}$ and $\{\bar{u} , {\bf \bar{w}} \}$ describing classical
trajectories of particle 1 and 2, respectively. When
$v\bar{u}=-4$, one null geodesic can be seen as the reflection of
the other on the AdS boundary (see figure \ref{reflection}). Thus,
the propagator from a point on one geodesic to a point in the
other diverges, since these points are connected by a null
geodesic. This is the physical meaning of the divergence at
$v\bar{u}=-4$. Clearly, we should doubt  the accuracy of the
eikonal approximation in this case of very strong interference. To
avoid this annoying divergence, from now on we shall localize the
external wave functions of particle 1 and 2 around $v=0$ and
$\bar{u}=0$, respectively. More precisely, we shall choose
\begin{equation}
\begin{array}{lll}
&\psi_1({\bf x})\simeq e^{-i\omega v}F(v)F_1({\bf w})\ ,\ \ \ \ \ \ \ \ \ \ \spa{.3}
&\psi_2({\bf \bar{x}})\simeq e^{-i\omega \bar{u}}F(\bar{u}) F_2( {\bf \bar{w}})\ , \\
&\psi_3({\bf x})\simeq e^{\,i\omega v} F^\star(v)F_3({\bf w})\ , \ \ \ \ \ \ \ \ \ \
&\psi_4({\bf \bar{x}})\simeq e^{\,i\omega \bar{u}} F^\star(\bar{u}) F_4( {\bf \bar{w}})\ ,
\end{array}
\label{eikwavefunc}
\end{equation}
where the profile $F(\alpha)$ is localized in the region $|\alpha|<\Lambda \ll 1$ and it
is normalized as
\begin{equation}
\int_{-\infty}^{\infty} d\alpha\,\left| F(\alpha) \right|^2 = \sqrt{2}\ .
\label{norm}
\end{equation}
On the other hand, the smoothness condition $|\partial F|\ll \omega |F|$ requires $\Lambda \gg 1/\omega$.
The two conditions,
$$
1/\omega \ll \Lambda \ll 1 \ ,
$$
are compatible for high energy scattering, when the de Broglie wavelength of the external particles
is much shorter than the radius of AdS ($\ell=1$).
With this choice of external wave functions, the amplitude simplifies to
\begin{equation}
A_{eik}
 \simeq  8\omega^2  \int_{H_{d-1}} d{\bf w} d{\bf \bar{w}}
 F_1({\bf w}) F_3({\bf w})F_2( {\bf \bar{w}})F_4({\bf \bar{w}})
\ \exp \left( \frac{ig^2}{2}\, s^{j-1}\, \Pi_{\perp}({\bf w},{\bf \bar{w}})
\right)\ ,
\label{ampsimp}
\end{equation}
where now 
\begin{equation}
s=\frac{(2\omega)^{2}}{
({\bf x}_0\cdot{\bf w})\, ({\bf x}_0\cdot  {\bf \bar{w}})}\ .
\label{sAdSlocalized}
\end{equation}
The eikonal amplitude (\ref{ampsimp}) is the main result of this chapter which will be
explored in great detail in chapter \ref{ch:eikAdSCFT}.
In the remainder of this chapter we shall extend the eikonal approximation to high energy scattering
in general spacetimes and rederive the  amplitude (\ref{ampsimp}) using gravitational shock waves.

\section{General Spacetime}

The position space derivation of the eikonal amplitude for flat and AdS spacetimes share many
features that suggest a more general description valid for generic backgrounds.
This section is an attempt to achieve such  synthesis. As in the previous sections, we consider
scattering of scalar particles. We also neglect their masses since they are irrelevant in the eikonal regime. 

The starting point is the identification of eikonal wave functions and their associated congruences
of null geodesics describing highly energetic external particles. 
These wave functions have the form
$$
\psi({\bf x})= F({\bf x}) \,e^{-i\omega S({\bf x})}\ ,
$$
where the phase function $S$ obeys
\be
\left( \nabla S \right)^2=0\ , \ \ \ \ \ \ \ \ \ \ \ \ \ \ \ \ \ \ \ 
\Box S=0\ .
\label{eikphase}
\ee
The first condition comes from the leading term at large $\omega$ in the Klein--Gordon equation $\Box \psi=0$.
The second is less obvious but it will be needed in the  derivation of the eikonal approximation.
With these two conditions, any function of $S$ satisfies the massless Klein--Gordon equation.
The semi--classical wave function $\psi$ describes highly energetic particles following null geodesics
with tangent vector 
$$
{\bf U}({\bf x})= -\nabla S({\bf x}) \ .
$$
Indeed, the condition $\left( \nabla S \right)^2=0$ implies that the
 null vector field ${\bf U}({\bf x})$ obeys the geodesic equation
$$
\nabla_{\bf U}\,{\bf U}=0 \ .
$$
Particle's trajectories are then given by the exponential map
$$
{\bf x}(\lambda) = e^{\, \lambda \,\omega {\bf U} }\, {\bf x}(0) \ .
$$
Furthermore, the Klein--Gordon equation at leading order in $1/\omega$, imposes the transversality
$$
\nabla_{\bf U} \, F =0 \ ,
$$
of the modulation function $F$.

In the computation of the eikonal amplitude it is convenient to use a coordinate system built from the
phase function $S$, the affine parameter $\lambda$ of the null geodesics and some other
transverse coordinates $\{ {\rm w}^a \}$. In this coordinate system we have
\be
\delta_\lambda^\mu=\left(\frac{d}{d\lambda}\right)^\mu= \omega {\bf U}^\mu =
-\omega \,g^{\mu\nu}\partial_\nu S = -\omega\,  g^{\mu S}\ .
\label{invmetric}
\ee
In addition, the condition $\Box S=0 $ is equivalent to the closure of the Hodge dual $\star dS $ of the 
the 1--form $dS$. Writing the volume form as
$$
\sqrt{-g} \,dS\wedge d\lambda \wedge d{\rm w}^1 \wedge \cdots \wedge d{\rm w}^{d-1}\ ,
$$
we have
$$
\star dS =\frac{1}{\omega} \sqrt{-g} \,dS \wedge d{\rm w}^1 \wedge \cdots \wedge d{\rm w}^{d-1}\ ,
$$
and the condition $d \star dS=0$ reduces to
\be
\partial_{\lambda}\sqrt{-g}=0\ . \label{measure}
\ee
The results (\ref{invmetric}) and (\ref{measure}) are sufficient to obtain an eikonal approximation
for the propagators of the external particles, similar to (\ref{posprop}) and  (\ref{AdSeikprop}).
At leading order in $1/\omega$, one can replace $\partial_S \simeq  -i\omega$ in the equation
$$
\Box  \Pi \left({\bf x} ,{\bf x'} \right)= i\delta\left({\bf x} ,{\bf x'} \right)\ ,
$$
obtaining
$$
2\partial_{\lambda} \Pi \left( {\bf x} ,{\bf x'} \right)\simeq \frac{1}{\sqrt{-g} } \delta(S-S') \delta(\lambda-\lambda')
\delta( {\rm w}^1 - {\rm w'}^1 ) \cdots \delta( {\rm w}^{d-1} - {\rm w'}^{d-1})
$$
and hence,
\be
 \Pi \left( {\bf x} ,{\bf x'} \right) \simeq \frac{1}{2\sqrt{-g} } \Theta(\lambda-\lambda') \delta(S-S') 
\delta( {\rm w}^1 - {\rm w'}^1 ) \cdots \delta( {\rm w}^{d-1} - {\rm w'}^{d-1})\ . \label{geneikprop}
\ee
This is the essential mark of the eikonal regime: propagation is non--zero only along the classical free
trajectories of the external particles.

The general eikonal amplitude can now be derived just by following the same steps as for Minkowski  and AdS spacetimes.
The scattering process in the eikonal kinematics is characterized by the external wave functions
\begin{align*}
&\psi_1({\bf x})= F_1({\bf x}) \,e^{-i\omega S({\bf x})}\ ,
& &\psi_2({\bf x})= F_2({\bf x}) \,e^{-i\omega \bar{S}({\bf x})}\ ,\\
&\psi_3({\bf x})= F_3({\bf x}) \,e^{\,i\omega S({\bf x})}\ ,
& &\psi_4({\bf x})= F_4({\bf x}) \,e^{\,i\omega \bar{S}({\bf x})}\ .
\end{align*} 
As before, we start from the graph in figure \ref{fig1} describing the exchange of $n$ scalar particles of mass $m$,
\begin{equation*}
\begin{array}{c}
\displaystyle{A_n = (-ig)^{2n} \int d{\bf x}_1 \cdots d{\bf x}_n d{\bf \bar{x}}_1 \cdots d{\bf \bar{x}}_n \,
\psi_3( {\bf x}_n )  \Pi({\bf x}_n,{\bf x}_{n-1})\cdots \Pi({\bf x}_2,{\bf x}_1)\psi_1( {\bf x}_1)}
\spa{0.5}
\\
\displaystyle{\psi_4( {\bf \bar{x}}_n ) \Pi({\bf \bar{x}}_n,{\bf \bar{x}}_{n-1})\cdots
\Pi({\bf \bar{x}}_2,{\bf \bar{x}}_1)\psi_2( {\bf \bar{x}}_1 )
\sum_{{\rm perm} \ \sigma}   \Pi_{m}({\bf x}_1,{\bf \bar{x}}_{\sigma_1})
\cdots \Pi_{m}({\bf x}_n,{\bf \bar{x}}_{\sigma_n})}\ .
\end{array}
\end{equation*}
Using coordinates $\{S,\lambda,{\rm w}^a \}$ for ${\bf x}$ and $\{\bar{S},\bar{\lambda},{\rm \bar{w}}^a \}$ for ${\bf \bar{x}}$,
one can use the eikonal approximation (\ref{geneikprop}) for the propagators of the external particles to simplify the
amplitude $A_n$. Summing over $n$ we arrive at
$$
A_{eik}=(2\omega)^2\int \star dS \, \star d\bar{S}\,  F_1({\bf x}) F_2({\bf \bar{x}}) F_3({\bf x}) F_4({\bf \bar{x}})
\,e^{I/4}\ ,
$$ 
with 
$$
I= (-ig)^2 \int d\lambda d\bar{\lambda}  \,\Pi_{m}\left( {\bf x}(\lambda),{\bf \bar{x}}(\bar{\lambda}) \right)\ .
$$
The generalization to spin $j$ exchanges is straightforward. The only change is in the phase shift determined
by the interaction between two null geodesics,
$$
I= -g^2 \left(-2\omega^2\right)^j\int d\lambda d\bar{\lambda}\,
\frac{ \partial     S  }{\partial {\bf x}^{\alpha_1} }  \cdots 
\frac{ \partial     S  }{\partial {\bf x}^{\alpha_j} }\,
\frac{ \partial  \bar{S} }{\partial{\bf \bar{x}}^{\bar{\alpha}_1}}\cdots 
\frac{ \partial  \bar{S} }{\partial{\bf \bar{x}}^{\bar{\alpha}_j}}\,
\Pi_{m}^{\alpha_1\cdots\alpha_j,\bar{\alpha}_1\cdots\bar{\alpha}_j}\left( {\bf x}(\lambda),{\bf \bar{x}}(\bar{\lambda}) \right)\ .
$$

\vspace{2cm}

\section{Shock Waves}
\label{sec:sw}

As reviewed in the beginning of the chapter, the eikonal approximation in 
non--relativistic Quantum Mechanics is a simple semi--classical approximation where one computes the 
phase shift, induced by the scattering potential, along the free trajectory of the incoming particle.
Similarly,  in relativistic Quantum Field Theory, the eikonal approximation is based
on the interaction between null geodesics, which are the free trajectories of the 
highly energetic incoming particles.
However, in this case, we lack the intuitive picture of a particle accumulating a phase due to 
a background potential, and the derivation of the eikonal approximation relies on a resummation of 
the perturbative series of Feynman diagrams, using simple combinatorics.
Not surprisingly, there is an alternative  derivation  \cite{tHoofteik} of the eikonal amplitude
(\ref{flateik}) based on shock waves in Minkowski spacetime which, as we shall see, are  the direct 
analogue of the scattering potential in Quantum Mechanics.
In this section, we review this derivation and present its generalization to Anti--de Sitter spacetime.

\subsection{Minkowski Spacetime}

We concentrate on the gravitational case where the exchanged particle has spin $j=2$ and mass $m=0$.
This was the original case considered in   \cite{tHoofteik} and it is the most important one since,
as we have seen, it gives the dominant contribution in the eikonal regime.
We shall follow the notation of section \ref{flateikonal}.
The basic idea is to consider the classical energy--momentum tensor of particle 2 as a source
for the gravitational field. 
Since particle 2 moves at the speed of light, it will create a shock
wave in the gravitational field  \cite{Aichelburg,shocktHooft}.
Then, particle 1 scatters in this shock wave background.

First, we study the spacetime geometry produced by particle 2.
In the coordinates (\ref{coord}), the energy--momentum tensor
$$
T^{\mu\nu}({\bf x})=\int d\bar{\lambda} \, \delta\left({\bf x} ,{\bf \bar{x}}(\bar{\lambda} ) \right)
\frac{\partial {\bf \bar{x}}^\mu }{\partial \bar{\lambda}}
\frac{\partial {\bf \bar{x}}^\nu }{\partial \bar{\lambda}}
\ ,
$$
of a massless particle following the null geodesic
${\bf \bar{x}}(\bar{\lambda})={\bf \bar{y}} + \bar{\lambda} {\bf k}_2$, 
has a single non--vanishing component, 
$$
T_{uu}( {\bf x})=\omega\, \delta(u-\bar{u}) \delta({\bf w}-{\bf \bar{w}})\ .
$$
As shown in  \cite{Aichelburg,shocktHooft}, the deformed background is then given by the shock
wave geometry
\begin{equation}
ds^2=d{\bf w}^2 -du\big[ dv-h({\bf w}) \delta(u-\bar{u}) du\big]\ ,
\label{shockmetric}
\end{equation}
where the function $h$ satisfies
$$
\Box_{\mathbb{R}^{d-1}} h({\bf w}) =   - 16\pi G\omega \, \delta({\bf w}-{\bf \bar{w}})\ ,
$$
with $G$ the canonically normalized Newton constant.
Therefore, we encounter again the transverse massless propagator
$$
 h({\bf w}) = 16 \pi G\omega \,\Delta_{\perp}({\bf w}-{\bf \bar{w}})\ .
$$


Secondly, we determine the effect of the shock wave geometry on the wave function $\psi$ of particle 1.
The massless Klein--Gordon equation in the geometry (\ref{shockmetric}) is
$$
\left[-4\partial_u\partial_v  +\Box_{\mathbb{R}^{d-1}} \right]\psi({\bf x} )= 4 \delta(u-\bar{u}) h({\bf w})\psi({\bf x} )\ .
$$
For $u<\bar{u}$ we have the incoming wave function $\psi=\psi_1=e^{-i\omega v}$. Then, solving
$$
\partial_u \psi({\bf x}) = i\omega  \delta(u-\bar{u}) h({\bf w})\psi({\bf x} )
$$
around $u=\bar{u}$, we obtain the wave function just after the shock,
$$
\psi(u=\bar{u}+\epsilon,v,{\bf w}) = e^{-i\omega v} \, e^{\,i\omega  h({\bf w})} 
=e^{-i\omega v} e^{I({\bf w}-{\bf \bar{w}} )/4} \ ,
$$
where $I$ is the interaction between null geodesics introduced in section \ref{flateikonal}.
In fact, to leading order in $1/\omega$,  this is the form of the wave function everywhere
after the gravitational shock wave.
Therefore, after the interaction,  the two particle wave function  reads
$$
\Psi({\bf x},{\bf \bar{x}})\simeq  e^{-i\omega v} e^{-i\omega \bar{u}} e^{I({\bf w}-{\bf \bar{w}} )/4}\ .
$$
Finally, the eikonal amplitude (\ref{flateik}) is recovered  by expanding $\Psi$
in the plane wave basis,
$$
\Psi({\bf x},{\bf \bar{x}})=\frac{1}{2s} \int_{\mathbb{R}^{d-1}} \frac{d {\bf q}}{(2\pi)^{d-1}}
\, \psi_{{\bf q}}({\bf x})\,\bar{\psi}_{-{\bf q}}({\bf \bar{x}})
\,\mathcal{A}( {\bf q} )  \ ,
$$
where
$$
\psi_{{\bf q}}({\bf x})= e^{-i\omega v -i{\bf q}\cdot {\bf w}}\ ,\ \ \ \ \ \ \ \ \ \ \ 
\bar{\psi}_{{\bf q}}({\bf x})= e^{-i\omega u -i{\bf q}\cdot {\bf w}}\ ,
$$
are single particle wave functions.
Indeed, using the Klein--Gordon  scalar products
$$
\left\langle\, \psi_{{\bf q}}  \,|\,\psi_{{\bf q}'} \,\right\rangle = i\int
\star\,\left[ \psi_{{\bf q}}^\star  \,d  \psi_{{\bf q}'} 
- \psi_{{\bf q}'}  \,d \psi_{{\bf q}}^\star \right]
\simeq 2\omega\,(2\pi)^{d-1} \delta( {\bf q}-{\bf q}')\int dv
$$
and
$$
\left\langle\, \bar{\psi}_{{\bf q}}  \,|\,\bar{\psi}_{{\bf q}'} \,\right\rangle 
\simeq 2\omega\,(2\pi)^{d-1} \delta( {\bf q}-{\bf q}')\int du\ ,
$$
we obtain the S--matrix element 
$$
\left\langle\, \psi_{{\bf q}} \, ;  \bar{\psi}_{-{\bf q}}  \, |  \,\Psi \,\right\rangle \simeq V \mathcal{A}( {\bf q} )\ ,
$$
in agreement with the result of section \ref{flateikonal}.


\subsection{Anti--de Sitter Spacetime}

We shall now extend the intuitive shock wave derivation of the eikonal amplitude in flat space
to Anti--de Sitter spacetime. Again, we focus on the gravitational case as the 
dominant example and do not bore the reader with a full rederivation of the general result (\ref{eikAdS}).
We follow the definitions of section \ref{eikonalAdS} and consider the wave functions 
(\ref{eikwavefunc}) as initial and final states of particles 1 and 2.
As explained in section \ref{NCWF}, the initial wave function $\psi_2$ of particle 2 describes highly energetic particles
approximately following the null geodesics
$$
{\bf \bar{x}}(\bar{\lambda})= {\bf \bar{w}} + \bar{\lambda}\,
\frac{{\bf k}_2 }{(-  {\bf x}_0\cdot  {\bf \bar{w}} ) }\ .
$$
We only consider the geodesics in (\ref{pbar}) satisfying ${\bf k}_2 \cdot{\bf \bar{x}} =0$
since  $\psi_2$ is localized around this hypersurface.
The non--vanishing component of the energy--momentum tensor associated with these lightlike geodesics,
$$
T_{uu}( {\bf x})=\omega\,\delta(u) \, \delta_{H_{d-1}}({\bf w},{\bf \bar{w}})\ ,
$$
sources a gravitational shock wave in AdS 
 \cite{Hotta,Podolsky,Sfetsos,Horowitz:1999gf,Arcioni:2001my}.
In the coordinates (\ref{coordAdS}), the metric reads
\begin{equation}
ds^2=d{\bf w}^2 
-({\bf x}_0 \cdot {\bf w})^2 du\big[ dv  - h({\bf w}) \delta(u) du\big]
-\frac{u^2}{4} ({\bf x}_0 \cdot {\bf w})^2 dv^2
\ , \label{AdSshock}
\end{equation}
with the Einstein equations imposing that the function $h$ satisfies
$$
\left[\Box_{H_{d-1}} +2 \frac{\chi^2-1}{\chi}\partial_\chi   \right] h({\bf w})=
-16\pi G \omega \, \frac{\delta_{H_{d-1}}({\bf w},{\bf \bar{w}})}{\chi^2}\ ,
$$
where we recall that $\chi=-{\bf x}_0 \cdot {\bf w} $ in the explicit
parametrization (\ref{hyperboloidmetric}) of the transverse hyperboloid.
Finally, using the techniques of appendix \ref{app1}, we obtain
$$
h({\bf w})= 16\pi G\omega \, \frac{\Pi_\perp ({\bf w},{\bf \bar{w}}) }{ ({\bf x}_0 \cdot {\bf w})( {\bf x}_0 \cdot {\bf \bar{w}}) }\ .
$$

We are now ready to determine the effect of the gravitational shock wave produced by particle 2, on
the wave function $\psi$ of particle 1. As in flat space, the Klein--Gordon equation
governing the propagation of particle 1 in the metric (\ref{AdSshock}), reduces to free propagation in AdS
before and after the shock, plus gluing conditions at the location $u=0$ of the shock.
These are obtained by solving $\left[\Box_{\rm AdS} + \Delta_1(d-\Delta_1)\right]\psi=0$ around $u=0$,
$$
-\partial_u\partial_v \, \psi({\bf x}) = h({\bf w}) \delta(u) \,\partial_v^2 \, \psi({\bf x}) \ .
$$
Since before the shock, particle 1 is characterized by the incoming wave function  
$$
\psi_1({\bf x})= e^{-i\omega v}F(v) F_1({\bf w})\ ,
$$
with $\partial_v \simeq -i\omega $ we obtain
$$
\psi(u=+\epsilon,v,{\bf w}) = e^{\,i\omega  h({\bf w})}\psi_1( {\bf x})
=F(v) F_1({\bf w}) \,e^{-i\omega v} \,e^{I/4} \ ,
$$
where $I$ is the interaction between null geodesics in AdS introduced in section \ref{eikonalAdS}.
We conclude that, at leading order in $1/\omega$,  the final two particle wave function simply receives a phase shift
$$
\Psi({\bf x},{\bf \bar{x}}) \simeq F(v) F_1({\bf w}) e^{-i\omega v} \,
F(\bar{u}) F_2({\bf \bar{w}}) e^{-i\omega \bar{u}}\, e^{I/4}\ .
$$
In the present case, the phase shift is explicitly given by
$$
I= i\,64\pi G\omega^2 \, \frac{\Pi_\perp ({\bf w},{\bf \bar{w}}) }{ ({\bf x}_0 \cdot {\bf w})( {\bf x}_0 \cdot {\bf \bar{w}}) }\ .
$$
As in flat space, we recover the eikonal amplitude  (\ref{ampsimp})
by taking the  Klein--Gordon  scalar product
with the outgoing wave functions $\psi_3^\star$ and $\psi_4^\star$,
$$
A_{eik} = \left\langle\, \psi_3^\star \, ;  \psi_4 ^\star \, |  \,\Psi \,\right\rangle\ ,
$$
and recalling that $g^2=8\pi G$ for the gravitational case $j=2$.

\newpage 

\begin{subappendices}

\section{Spin $j$ Interaction in AdS  \label{app1}}

We wish to extend to result $G({\bf w}\cdot{\bf \bar w})=\Pi_\perp({\bf k},{\bf \bar k})$, derived in section \ref{sectTP},
 to the case of general spin $j$. To this end we use equation (\ref{propeqBIS}), contracting both sides with
$$
(-2)^j\, {\bf k}_{\alpha_1}\cdots{\bf k}_{\alpha_j}\, {\bf \bar{k}}_{\beta_1}\cdots{\bf \bar{k}}_{\beta_j}
$$
and integrating against
$$
\int_{-\infty}^{\infty}du d\bar{v} = \frac{(2\omega)^2}{({\bf x}_0\cdot{\bf w})^2({\bf \bar{x}}_0\cdot{\bf \bar{w}})^2}\,\int_{-\infty}^{\infty}d\lambda d\bar{\lambda}\ .
$$
Using the explicit form of the $\delta$--function in the  $\{u,v, {\bf w} \}$ coordinate system give in section \ref{sectTP},
the RHS reduces to
\be
2i\,(2\omega)^{2j}\, \frac{(1+v\bar{u}/4)^{2j-2}}{
({\bf x}_0\cdot{\bf w})^{2j+2}}\,
\delta_{H_{d-1}}({\bf w},{\bf \bar{w}})\ .
\label{RHSbis}
\ee
Next we consider the LHS of (\ref{propeqBIS}). First we note that the non--vanishing components of the covariant derivatives of ${\bf k}$ are
given by
$$
\nabla_v {\bf k}_v=-\frac{\omega u}{2}\ ,\ \ \ \ \ \ \ \ \ \nabla_v {\bf k}_{\chi}=\nabla_{\chi} {\bf k}_v=\frac{\omega}{{\chi}}\ ,
$$
where we explicitly parametrize the metric on $H_{d-1}$ as in section \ref{sectTP}.
Using these facts, together with the explicit form of the metric and with
$$
\Box_{{\rm AdS}} {\bf k}_\alpha = -d\cdot {\bf k}_\alpha\ , \ \ \ \ \ \ \ \ \ \nabla_\gamma {\bf k}_\alpha
\nabla^\gamma {\bf k}_\beta = \frac{{\chi}^2-1}{{\chi}^2} {\bf k}_\alpha {\bf k}_\beta\, ,
$$
we conclude, after a tedious but straightforward computation, that
\beas
&&(-2)^j\, {\bf k}_{\alpha_1}\cdots{\bf k}_{\alpha_j}\, {\bf \bar{k}}_{\beta_1}\cdots{\bf \bar{k}}_{\beta_j}
\Box_{\rm AdS} \Pi_\Delta^{\alpha_1,\cdots,\alpha_j,\beta_1,\cdots,\beta_j}=\\
&&=\Box_{\rm AdS} \Pi_\Delta^{(j)}+j\,\left[2\frac{{\chi}^2-1}{{\chi}}\partial_{\chi}+(d+j-1)-\frac{j+1}{{\chi}^2}\right]\Pi_\Delta^{(j)} + \partial_u(\cdots)=\\
&&=\left[\Box_{H_{d-1}}+2(j+1)\frac{{\chi}^2-1}{{\chi}}\partial_{\chi}+j(d+j-1)-\frac{j(j+1)}{{\chi}^2}\right]\Pi_\Delta^{(j)} + \partial_u(\cdots)\, ,
\eeas
where we do not show the explicit terms of the form $\partial_u(\cdots)$ since they will vanish once integrated along the two geodesics.
Note that the terms in $\cdots$ contain also other components of the spin--$j$ propagator aside from $\Pi_\Delta^{(j)}$.
We conclude that (\ref{RHSbis}) must be equated to
\begin{align*}
-2i\,(2\omega)^{2j}\,\left(1+\frac{v\bar{u}}{4}\right)^{2j-2} \left[\Box_{H_{d-1}}-(\Delta+j)(\Delta-d-j)
+2(j+1)\frac{{\chi}^2-1}{{\chi}}\partial_{\chi}-\frac{j(j+1)}{{\chi}^2}
\right]
\frac{G({\bf w},{\bf \bar{w}})}{({\chi}\bar{{\chi}})^{j+1}}\ .\\ \spa{.6}
\end{align*}
Using the fact that
$$
[\Box_{H_{d-1}},{\chi}^{-1-j}]=\frac{j+1}{{\chi}^{1+j}}\left(-2\frac{{\chi}^2-1}{{\chi}}\partial_{\chi}+(j-d+3)-\frac{j+2}{{\chi}^2}\right)\ ,
$$
we deduce again that
$$
\left[ \Box_{H_{d-1}} +1-d -\Delta( \Delta-d ) \right] G(  {\bf w} \cdot
{\bf \bar{w}} )
= -\delta(  {\bf w}, {\bf \bar{w}} )\ .
$$
and therefore the function $G$ is given by $\Pi_\perp$.

\end{subappendices}

\chapter{Conformal Partial Waves}
\label{ch:cpw}

This chapter prepares the way for the CFT interpretation of the eikonal approximation 
in AdS, developed in the previous chapter.
We review the  Conformal Partial Wave (CPW) decomposition of CFT 
four point functions  \cite{Ferrara,Osborn, Osborn22}.
In addition, we develop an impact parameter representation for 
CPW suitable for the eikonal kinematical regime  \cite{Paper2,Paper3}.

Given the AdS asymptotic structure, acting effectively as a gravitational box,
the word scattering is just colloquial and there is no S--matrix.
Nevertheless, the word scattering is acceptable for wave functions that
have support at infinity, such as the bulk to boundary propagator.
Hence, the natural AdS analogue of the flat space scattering amplitude $\mathcal{A}(s,t)$  
is the reduced four point function $\mathcal{A}(z,\bar{z})$ of the dual CFT.
Therefore, the partial wave analysis of the AdS eikonal amplitude
is just the  CPW decomposition of CFT four point functions.
Indeed, the CPW are the appropriate partial waves since the conformal group is the isometry
group of AdS spacetime.
In the previous chapter, using the impact parameter 
representation of the partial wave expansion,
we saw that the eikonal approximation in flat space determines the 
phase shift of high spin partial waves.
In the next chapter, we shall  describe the analogous result 
for scattering in AdS spacetime, using the tools here developed.

\section{General Definition}

\begin{figure}
[ptb]
\begin{center}
\includegraphics[width=8cm]{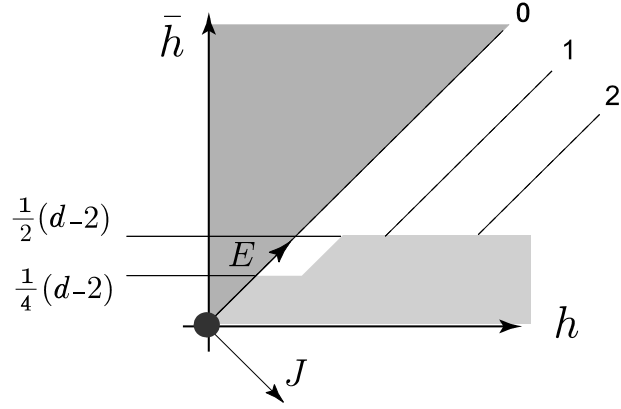} 
 \caption{Intermediate primaries of dimension
$E=h+\bar{h}$ and integer spin $J=h-\bar{h}\ge 0$ can exist in the
white region along the diagonal lines. The light gray region is
excluded due to the unitarity constraint, with the vacuum at
$h=\bar{h}=0$ being the unique exception.}
\label{fig3}
\end{center}
\end{figure}

The CFT four point function,
\[
A\left( {\bf p}_{i} \right) = \left\langle 
\mathcal{O}_{1}\left( {\bf p}_{1}\right)  \mathcal{O}_{2}\left( {\bf p}_{2}\right)
 \mathcal{O}_{1}\left( {\bf p}_{3}\right) \mathcal{O}_{2}\left( {\bf p}_{4}\right) \right\rangle\ ,
\]
of scalar primary operators $\mathcal{O}_{i}$ with  dimensions $\Delta_{1}=\eta+\nu$ and $\Delta_{2}=\eta-\nu$,
is conformal invariant,
$$
\left( {\bf J}_1 + {\bf J}_2+ {\bf J}_3+  {\bf J}_4 \right)_ {\mu\nu} A\left( {\bf p}_{i} \right) = 0\ ,
$$
where 
$$
\left( {\bf J}_j \right)_ {\mu\nu} = i\,\left( {\bf p}_j \right)_{\mu} \frac{\partial}{\partial \left( {\bf p}_j \right)^{\nu}  }
-  i\,\left( {\bf p}_j \right)_{\nu} \frac{\partial}{\partial \left( {\bf p}_j \right)^{\mu}  }\ ,
$$
are the Lorentz generators of  $SO(2,d)$. 
This invariance allows us to express the four point function  $ A\left( {\bf p}_{i} \right) $ 
in terms of the reduced amplitude $\mathcal{A}( z,\bar{z})$
defined in (\ref{CFTamp}).
The amplitude $\mathcal{A}\left(  z,\bar{z}\right)  $ can then be
expanded using the OPE around $z,\bar{z}=0,1,\infty$,
corresponding to the point $\mathbf{p}_{3}$ getting close to
$\mathbf{p}_{1}$, $\mathbf{p}_{2}$ and $\mathbf{p}_{4}$,
respectively. In particular, we will be interested in the
contribution to the amplitude $\mathcal{A}$ coming from the
exchange of a conformal primary operator of dimension $E$ and
integer spin $J\geq0$ in the two channels
\begin{align*}
z,\bar{z}  & \rightarrow
0\ ,\ \ \ \ \ \ \ \ \ \ \ \ \ \ \ \ \ \ \ \ \ \ \text{\emph{T}--channel},\\
z,\bar{z}  & \rightarrow\infty
\ ,\ \ \ \ \ \ \ \ \ \ \ \ \ \ \ \ \ \ \ \ \ \text{\emph{S}--channel},
\end{align*}
together with all of its conformal descendants. 
It will be convenient in the following to use different labels for dimension
and spin $E,J$. We shall use most frequently conformal dimensions
$h,\bar{h}$ defined by
\begin{align}
E  & =h+\bar{h}~,\\
J  & =h-\bar{h}~.
\label{EJhhbar}
\end{align}
We also recall the unitarity bounds $E\geq d-2+J$
for $J\geq1$ and $E\geq\left( d-2\right)  /2$ for $J=0$, with the
single exception of the vacuum with $E=J=0$. This translates into
\begin{align*}
\bar{h}  & \geq\frac{d-2}{2}~,\ \ \ \ \ \ \ \ \ \ \ \ \ \ \ \ \ \ \ \ \left(
J\geq1\right), \\
\bar{h}  & \geq\frac{d-2}{4}~,\ \ \ \ \ \ \ \ \ \ \ \ \ \ \ \ \ \ \ \ \left(
J=0\right),
\end{align*}
again with the exception of the vacuum at $h=\bar{h}=0$. figure \ref{fig3}
summarizes the basic notation regarding the intermediate conformal primaries.

The amplitude $\mathcal{A}\left(  z,\bar{z}\right)  $ can be
expanded in the basis of conformal partial waves in either the
\emph{T} or \emph{S}--channels, whose elements we shall denote by
\[
\mathcal{T}_{h,\bar{h}}~\left(  z,\bar{z}\right)
~,\ \ \ \ \ \ \ \ \ \ \ \ \ \ \ \ \ \ \ \ \mathcal{S}_{h,\bar{h}}~\left(
z,\bar{z}\right)  ~.
\]
Following  \cite{Osborn, Osborn22}, the four point amplitude $A$ associated 
with a single \emph{T}--channel partial wave $\mathcal{A}=\mathcal{T}_{h,\bar{h}}$, satisfies
$$
\frac{1}{4}\left( {\bf J}_2+   {\bf J}_4 \right)^2 A\left( {\bf p}_{i} \right) =c_{h,\bar{h}}  \, A\left( {\bf p}_{i} \right) \ ,
$$
where the constant $c_{h,\bar{h}}$ is the Casimir of the conformal group given by
\[
c_{h,\bar{h}}=h\left(  h-1\right)  +\bar{h}\left(  \bar{h}-d+1\right)\ .
\]
This gives a differential equation for the partial wave $\mathcal{T}_{h,\bar{h}}\left(  z,\bar{z}\right)  $,
namely
$$
D_T\,\mathcal{T}_{h,\bar{h}}=c_{h,\bar{h}}~\mathcal{T}_{h,\bar{h}}\ ,
$$ 
with 
\begin{align}
D_{T}  &=z^{2}\left(  1-z\right)  \partial^{2}-z^{2}\partial+\bar{z}
^{2}\left(  1-\bar{z}\right)  \bar{\partial}^{2}-\bar{z}^{2}\bar{\partial
} +\label{Tchannel}\\
& +\left(  d-2\right)  \frac{z\bar{z}}{z-\bar{z}}\left[  \left(  1-z\right)
\partial-\left(  1-\bar{z}\right)  \bar{\partial}\right]\ . \nonumber
\end{align}
Similarly, using ${\bf J}_1+   {\bf J}_2$ instead of ${\bf J}_2+   {\bf J}_4$, we obtain the differential equation
$$
\left(  z\bar{z}\right)  ^{\eta}D_{S}~\left[  \left(  z\bar{z}\right)
^{-\eta}\mathcal{S}_{h,\bar{h}}\right]  =c_{h,\bar{h}}~\mathcal{S}
_{h,\bar{h}}~, 
$$
for the  \emph{S}--channel partial wave, where
\begin{align}
D_{S}  & =z\left(  z-1\right)  \partial^{2}+\left(  2z-1\right)
\partial+\frac{\nu^{2}}{z}+\nonumber\\
& +\bar{z}\left(  \bar{z}-1\right)  \bar{\partial}^{2}+\left(  2\bar
{z}-1\right)  \bar{\partial}+\frac{\nu^{2}}{\bar{z}}+\label{Schannel}\\
& +\frac{d-2}{z-\bar{z}}\left[  z\left(  z-1\right)  \partial-\bar{z}\left(
\bar{z}-1\right)  \bar{\partial}\right]  ~.\nonumber
\end{align}
Finally, the partial waves $\mathcal{T}_{h,\bar{h}}$
and $\mathcal{S} _{h,\bar{h}}$ must be
symmetric in $z$ and $\bar{z}$ and satisfy the OPE boundary conditions
\begin{equation*}
\lim_{z,\bar{z}\rightarrow0}\mathcal{T}_{h,\bar{h}}   \sim z^{h}~\bar
{z}^{\bar{h}}\ ,\ \ \ \ \ \ \ \ \ \ \ \ \ \ \ \ \ \ \ \ \ \ 
\lim_{z,\bar{z}\rightarrow\infty}\mathcal{S}_{h,\bar{h}}   \sim z^{\eta
-h}~\bar{z}^{\eta-\bar{h}}~,
\end{equation*}
where we choose to take the limit
$\bar{z}\rightarrow0$ or $\infty$ first. The symmetric term with $h$
and $\bar{h}$ interchanged is then sub--leading since
$h\geq\bar{h}$.


\section{Explicit Form in $d=2$}


Explicit expressions for the partial waves
$\mathcal{T}_{h,\bar{h}}$ and $\mathcal{S}_{h,\bar{h}}$ exist for
$d$ even  \cite{Osborn, Osborn22}, and are particularly simple in
$d=2$ where the problem factorizes in left/right equations for $z$
and $\bar{z}$. In this case we have the explicit expressions
\begin{align}
\mathcal{T}_{h,\bar{h}}\left(  z,\bar{z}\right)   & =\mathcal{T}_{h}\left(
z\right)  \mathcal{T}_{\bar{h}}\left(  \bar{z}\right)  +\mathcal{T}_{\bar{h}
}\left(  z\right)  \mathcal{T}_{h}\left(  \bar{z}\right)  ~,\label{eq1000}\\
\mathcal{S}_{h,\bar{h}}\left(  z,\bar{z}\right)   & =\mathcal{S}_{h}\left(
z\right)  \mathcal{S}_{\bar{h}}\left(  \bar{z}\right)  +\mathcal{S}_{\bar{h}
}\left(  z\right)  \mathcal{S}_{h}\left(  \bar{z}\right)  ~,\nonumber
\end{align}
\noindent for $h>\bar{h}$ and
\begin{align*}
\mathcal{T}_{h,h}\left(  z,\bar{z}\right)   & =\mathcal{T}_{h}\left(
z\right)  \mathcal{T}_{h}\left(  \bar{z}\right)  ~, \\
\mathcal{S}_{h,h}\left(  z,\bar{z}\right)   & =\mathcal{S}_{h}\left(
z\right)  \mathcal{S}_{h}\left(  \bar{z}\right)  ~,
\end{align*}
\noindent for $h=\bar{h}$, where
\begin{align}
\mathcal{T}_{h}\left(  z\right)   & =\left(  -z\right)  ^{h}~F \Big(
h,h,2h \Big| z\Big)  ~,\nonumber \\
\mathcal{S}_{h}\left(  z\right)   & =a_{h}~\left(  -z\right)
^{\eta -h}~F\Big(  h+\nu,h-\nu,2h \Big| z^{-1}\Big)
~.\label{2dCPW}
\end{align}

\noindent The specific normalization of the \emph{S}--channel
partial waves
\[
a_{h}=\frac{\Gamma\left(  h+\nu\right)  \Gamma\left(  h-\nu\right)
\Gamma\left(  h+\eta-1\right)  }{\Gamma\left(  \eta+\nu\right)
\Gamma\left(  \eta-\nu\right)  \Gamma\left(  2h-1\right)  \Gamma\left(
h-\eta+1\right)  }
\]
is chosen for later convenience, and it is such that
\begin{equation}
\sum_{h\in\eta+\mathbb{N}_{0}}\mathcal{S}_{h}\left(  z\right)
=1~,\label{ChiralSum}
\end{equation}
\noindent
where $\mathbb{N}_{0}$ is the set of non--negative integers.

It is clear from (\ref{eq1000}) that the $h,\bar{h}$ partial waves
correspond to the exchange of a pair of primary operators of
holomorphic/antiholomorphic dimension $\left(  h,\bar{h}\right)  $
and $\left(  \bar{h},h\right)  $, together with their descendants.


\section{Impact Parameter Representation} \label{impactsection}


We now move back to general dimension $d$, and we consider the
behavior of the \emph{S}--channel partial waves
$\mathcal{S}_{h,\bar{h}}\left(  z,\bar{z}\right)$ for $z,\,
\bar{z} \to 0$. 
More precisely, in strict analogy with the case of flat space studied in section \ref{flatPW}, we
analyze the double limit
\begin{align*}
z,\bar{z}  & \rightarrow0~,\\
h,\bar{h}  & \rightarrow\infty~,
\end{align*}
as in (\ref{limitflat}), with
\[
z \sim \bar{z}\sim h^{-2} \sim \bar{h}^{-2}~.
\]
In this limit, the differential operator $D_{S}$ in (\ref{Schannel}) and the
constant $c_{h,\bar{h}}$ reduce to
\[
-\tilde{D}_{S}=z\partial^{2}+\bar{z}\bar{\partial}^{2}+\partial+\bar{\partial
}-\frac{\nu^{2}}{z}-\frac{\nu^{2}}{\bar{z}}+\frac{d-2}{z-\bar{z}}\left(
z\partial-\bar{z}\bar{\partial}\right)  ~
\]
and
\[
\tilde{c}_{h,\bar{h}}=h^{2}+\bar{h}^{2}~.
\]
We shall denote with $\mathcal{I}_{h,\bar{h}}\left(
z,\bar{z}\right)  $ the approximate \textit{impact parameter}
\emph{S}--channel partial wave, which satisfies
\begin{equation}
\left(  -\tilde{D}_{S}+\tilde{c}_{h,\bar{h}}\right)  \left[  \left(  z\bar
{z}\right)  ^{-\eta}\mathcal{I}_{h,\bar{h}}\right]  =0~.
\label{PDEimpact}
\end{equation}
It is then convenient to write the cross ratios in terms of two points ${\bf p},{\bf \bar{p}}$
in the past Milne wedge $-{\rm M} \subset \mathbb{M}^d$ as
$$
z\bar{z}= {\bf p}^2{\bf \bar{p}}^2 \ ,\ \ \ \ \ \ \ \ \ \ \
z+\bar{z}= 2{\bf p} \cdot {\bf \bar{p}} \ ,
$$
and view the \emph{S}--channel impact parameter
amplitude $\mathcal{I}_{h,\bar{h}}$ as a function of ${\bf p}$ and ${\bf \bar{p}}$.
In analogy with the flat space case (\ref{imprep2}), the function
$\mathcal{I}_{h,\bar{h}}$ admits the following integral
representation over the future Milne wedge $M$
\begin{align}
\mathcal{I}_{h,\bar{h}}  & =\mathcal{N}_{\Delta_{1}}\mathcal{N}_{\Delta_{2}
}~\left(  -{\bf p}^{2}\right)  ^{\Delta_{1}}\left(  -{\bf \bar{p}}^{2}\right)  ^{\Delta_{2}}
\int_{\mathrm{M}}\frac{d{\bf y}}{\left\vert {\bf y}\right\vert ^{d-2\Delta_{1}}}\frac
{d{\bf \bar{y}}}{\left\vert{\bf \bar{y}}\right\vert ^{d-2\Delta_{2}}}~~e^{-2{\bf p}\cdot {\bf y}-2{\bf \bar{p}}\cdot
{\bf \bar{y}}}~\times\nonumber\\
& \times~4h\bar{h}\left(  h^{2}-\bar{h}^{2}\right)  ~\delta\left(  2{\bf y}\cdot
{\bf \bar{y}}~+h^{2}+\bar{h}^{2}\right)  ~\delta\left(  {\bf y}^{2}{\bf \bar{y}}^{2}-h^{2}\bar{h}
^{2}\right)  ~,\label{eq1001}
\end{align}
where
$$
\mathcal{N}_{\Delta}^{-1} = \int_{\mathrm{M}}\frac{d{\bf y}}{\left\vert
y\right\vert ^{d-2\Delta}}~~e^{2{\bf p}\cdot {\bf y}}=\Gamma\left(
2\Delta\right) \int_{\mathrm{H}_{d-1}}\frac{d{\bf y}}{\left(
-2{\bf p}\cdot {\bf y}\right) ^{2\Delta}} =\frac{\pi^{\frac{d}{2}-1}}{2}\Gamma\left(  \Delta\right)
\Gamma\left( \Delta-\frac{d}{2}+1\right)  ~,
$$
with ${\bf p}\in\mathrm{H}_{d-1}$ arbitrary \footnote{In the appendix, we
show that (\ref{eq1001}) is a solution of the differential
equation (\ref{PDEimpact}). On the other hand, we do not have a
complete proof of (\ref{eq1001}), since we cannot distinguish
different cases with the same Casimir $\tilde{c}_{h,\bar{h}}$. 
Nevertheless, the form (\ref{eq1001}) is strongly suggested by
the case $d=2$, where we can check explicitly that (\ref{eq1001})
is the impact parameter approximation to $\mathcal{S}_{h,\bar{h}}$
(see section \ref{d=2Impact}).}. 

Any function $\Sigma(z,\bar{z})$
can be decomposed in the impact parameter partial waves
$\mathcal{I} _{h,\bar{h}}$ and we have chosen the normalization of
$\mathcal{I}_{h,\bar{h} }$ such that
\begin{align}
 \Sigma\left(  z,\bar{z}\right) & =\int_{0}^{\infty}dh\int_{0}^{h}d\bar
{h}~~\sigma\left(  h^{2}\bar{h}^{2},h^{2}+\bar{h}^{2}\right)  ~~\mathcal{I}
_{h,\bar{h}}\left(  z,\bar{z}\right) \label{intrep}\\
&=\mathcal{N}_{\Delta_{1}}\mathcal{N}_{\Delta_{2}}~\left(  -{\bf p}^{2}\right)
^{\Delta_{1}}\left(  -{\bf \bar{p}}^{2}\right)  ^{\Delta_{2}}\times\nonumber\\
&\ \ \  \times\int_{\mathrm{M}}\frac{d{\bf y}}{\left\vert {\bf y}\right\vert ^{d-2\Delta_{1}}
}\frac{d{\bf \bar{y}}}{\left\vert {\bf \bar{y}}\right\vert ^{d-2\Delta_{2}}}~~e^{-2{\bf p}\cdot {\bf y}-2{\bf \bar{p}}\cdot
{\bf \bar{y}}}~\sigma\left(   {\bf y}^{2}{\bf \bar{y}}^{2},-2 {\bf y}\cdot{\bf \bar{y}}\right)  ~.\nonumber
\end{align}
\noindent
In particular, setting $\sigma=1$ we have
\[
\int_{0}^{\infty}dh\int_{0}^{h}d\bar{h}~~~\mathcal{I}_{h,\bar{h}}=1~.
\]
Note that, in (\ref{intrep}), the leading behavior of the function
$\Sigma$ for $z,\bar{z}\rightarrow0$ is controlled by the behavior
of $\sigma$ for $h,\bar{h}\rightarrow\infty$. In fact, when
$\sigma\sim\left( h\bar{h}\right)^{-a}$ for large $h,\bar{h}$ with
$a<2\Delta_1$ and $a<2\Delta_2$, then $\Sigma\sim\left(
z\bar{z}\right)^{a/2}$ for small $z$ and $\bar{z}$.

At this point, the explicit expression (\ref{eq1001}) for the impact parameter partial wave $\mathcal{I}_{h,\bar{h}}$
might seem mysterious, but it will be crucial, as we shall see in the next chapter, 
for the CFT interpretation of the AdS eikonal amplitude derived in the previous chapter.
Moreover, the connection with the AdS eikonal amplitude will show that the name
impact parameter representation is indeed appropriate.


\subsection{Impact Parameter Representation in $d=2$} \label{d=2Impact}


In $d=2$ the constant $\mathcal{N}_{\Delta}$ is given explicitly
by $2/\Gamma\left(  \Delta\right)  ^{2}$. Choosing 
null coordinates $(u,v)$ for $\mathbb{M}^2$ such that
$$
d {\bf y}^2=du dv \ ,
$$
the points ${\bf p}$ and ${\bf \bar{p}}$ in the past Milne wedge can be written as
 ${\bf p}=(-1,z)$ and ${\bf \bar{p}}=(-1,\bar{z})$.
Then, the general expression (\ref{eq1001}) reduces to
\begin{align*}
\mathcal{I}_{h,\bar{h}}  & =\frac{\left(  z\bar{z}\right)  ^{\Delta_{1}}
}{\Gamma\left(  \Delta_{1}\right)  ^{2}\Gamma\left(  \Delta_{2}\right)  ^{2}
}~\int_{0}^{\infty}\frac{dudv}{\left(  uv\right)  ^{1-\Delta_{1}}
}\frac{d\bar{u}d\bar{v}}{\left(  \bar{u}\bar{v}\right)  ^{1-\Delta_{2}}}~~e^{-v +u z
-\bar{v}+\bar{z}\bar{u} }\times\\
& \times4h\bar{h}\left(  h^{2}-\bar{h}^{2}\right)  
~\delta\left(  u \bar{v}+\bar{u}v~-h^{2}-\bar{h}^{2}\right)  ~\delta\left(  uv\bar{u}\bar{v}
-h^{2}\bar{h}^{2}\right)  ~.
\end{align*}
The $v,\bar{v}$ integrals localize at two points, namely at
\[
v=\bar{u}^{-1}\bar{h}^{2},~\ \ \ \ \ \ \ \ \ \ \ \ \ \ \bar{v}=u^{-1} h^{2}~,
\]
and at the point obtained by exchanging $h$ with $\bar{h}$.
Summing the two contributions we obtain
\[
\mathcal{I}_{h,\bar{h}}\left(  z,\bar{z}\right)  =\mathcal{I}_{h}\left(
z\right)  \mathcal{I}_{\bar{h}}\left(  \bar{z}\right)  +\mathcal{I}_{\bar{h}
}\left(  z\right)  \mathcal{I}_{h}\left(  \bar{z}\right)  ~,
\]
where
\begin{align*}
\mathcal{I}_{h}\left(  z\right)   & =2\, \frac{h^{2\Delta_{2}-1}~\left(
-z\right)  ^{\Delta_{1}}}{\Gamma\left(  \Delta_{1}\right)  \Gamma\left(
\Delta_{2}\right)  }\int_{0}^{\infty}\frac{du}{u^{1-2\nu}
}~~e^{uz-h^{2}u^{-1}}\\
& =4\, \frac{h^{2\eta-1}\left(  -z\right)  ^{\eta}}{\Gamma\left(  \Delta
_{1}\right)  \Gamma\left(  \Delta_{2}\right)  }~~K_{2\nu}\left(  2h\sqrt
{-z}\right)  ~.
\end{align*}

\noindent One can check that the function $\mathcal{I}_{h}\left(
z\right)$ is indeed the impact parameter approximation of
$\mathcal{S}_{h}\left(z\right)$ in (\ref{2dCPW}). Moreover it
satisfies
\[
\int_{0}^{\infty}dh~\mathcal{I}_{h}\left(  z\right)  =1~,
\]
which corresponds to (\ref{ChiralSum}).


\section{Free Propagation}      \label{sec:freeprop}


\begin{figure}
\begin{center}
\includegraphics[width=10cm]{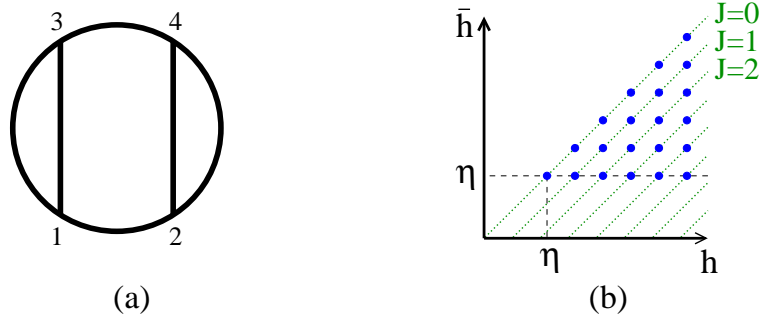} 
\caption{(a) Disconnected Witten diagram describing free propagation in AdS.
The resulting four point function is just the product of two boundary
propagators so that $\mathcal{A}=1$. In the \emph{T}--channel decomposition
only the vacuum contributes, whereas the \emph{S}--channel
decomposition receives contributions from the full lattice (b) of
$\mathcal{O}_{1} \partial  \cdots \partial \mathcal{O}_{2}$ composites
of dimensions $h,\bar {h}\in\eta+\mathbb{N}_{0}$.
}
\label{freeprop}
\end{center}
\end{figure}
It is instructive to consider the conformal partial wave expansion of the 
disconnected contribution to the four point function,
$$
A\left( \mathbf{p}_{1},\cdots ,\mathbf{p}_{4}\right) =\left\langle
\mathcal{O}_{1}\left( \mathbf{p}_{1}\right)
\mathcal{O}_{1}\left( \mathbf{p}_{3}\right) \right\rangle\,\left\langle
 \mathcal{O}_{2}\left( \mathbf{p}_{2}\right) 
\mathcal{O}_{2}\left( \mathbf{p}_{4}\right) \right\rangle \ ,
$$
which corresponds to 
\[
\mathcal{A}=1\ .
\]
In AdS this contribution comes from the Witten diagram in figure \ref{freeprop}(a).
From the graph, it is natural that only
the vacuum state with $h=\bar{h}=0$ contributes to its \emph{T}--channel decomposition.
In fact, with an appropriate normalization for
$\mathcal{T}_{h,\bar{h}}$ we have that
\begin{equation}
\mathcal{A}=\mathcal{T}_{0,0}\ .\label{tfree}
\end{equation}
On the other hand, the \emph{S}--channel decomposition of $\mathcal{A}$ is
more subtle. Indeed, as shown in figure \ref{freeprop}(b), we expect that the
composites  \footnote{
Throughout this thesis we will use this schematic notation to
represent the primary composite operators of spin $J$ and
conformal dimension $E=\Delta _{1}+\Delta _{2}+J+2n$, avoiding the
rather cumbersome exact expression. We
shall also use the simpler notation $\mathcal{O}_{1}\partial \cdots \partial \mathcal{O}_{2}$ whenever possible.}
\[
\mathcal{O}_{1}\partial_{\mu_{1}}\cdots\partial_{\mu_{J}}
\partial^{2n} \mathcal{O}_{2}\, ,
\]
of dimension $E=\Delta_1+\Delta_2+J+2n$ and spin $J$, contribute to the
\emph{S}--channel decomposition and define a lattice of operators of dimension
$h=\eta+J+n$, $\bar{h}=\eta+n$ given by
\[
h,\bar{h}\in\eta+\mathbb{N}_{0}\text{ }
,~\ \ \ \ \ \ \ \ \ \ \ \ \ \ \ \ \ \ \ \eta\leq\bar{h}\leq h~.
\]
Again, with an appropriate normalization for the \emph{S}--channel
partial waves $\mathcal{S}_{h,\bar{h}}$, we have the decomposition
\begin{equation}
\mathcal{A}=\sum_{\eta\leq\bar{h}\leq h}\ \mathcal{S}_{h,\bar{h}
}~,\label{sfree}
\end{equation}
where it is understood that the sum is restricted to
$h,\bar{h}\in \eta+\mathbb{N}_{0}$. Finally, we have the impact
parameter representation corresponding to (\ref{sfree})
\[
\mathcal{A}=\int_{0}^{\infty}dh\int_{0}^{h}d\bar{h}~~~\mathcal{I}
_{h,\bar{h}}~.
\]
Note that the normalizations chosen for the specific case $d=2$ are compatible
with (\ref{tfree}) and (\ref{sfree}).




\newpage

\begin{subappendices}


\section{Impact Parameter Representation}


In this appendix, we show that the integral representation
(\ref{eq1001}) of the impact parameter partial wave
$\mathcal{I}_{h,\bar{h}}$ solves the differential equation
(\ref{PDEimpact}). We start by recalling the relevant kinematics
$$
z\bar{z}= {\bf p}^2{\bf \bar{p}}^2 \ ,\ \ \ \ \ \ \ \ \ \ \
z+\bar{z}= 2{\bf p} \cdot {\bf \bar{p}} \ ,
$$
with ${\bf p}, {\bf \bar{p}}$ in the past Milne wedge of $\mathbb{M}$.
In what follows, we choose once and for all a fixed point
${\bf \bar{p}}\in-\mathrm{H} _{d-1}$, so that ${\bf \bar{p}}^{2}=-1$. We then view the
\emph{S}--channel impact parameter amplitude
$\mathcal{I}_{h,\bar{h}}$ as a function just of ${\bf p}$. Recall that,
in terms of $z$ and $\bar{z}$, the function $\left(  z\bar{z}\right)
^{-\eta}\mathcal{I}_{h,\bar{h}}$ satisfies the following
differential equation
\[
z\partial^{2}+\bar{z}\bar{\partial}^{2}+\partial+\bar{\partial}+\frac
{d-2}{z-\bar{z}}\left(  z\partial-\bar{z}\bar{\partial}\right)  =\frac{\nu
^{2}}{z}+\frac{\nu^{2}}{\bar{z}}-h^{2}-\bar{h}^{2}.
\]
A tedious computation shows that, in terms of ${\bf p}$, the above equation can be
written as
\begin{equation}
\left(  {\bf p}^{\mu} {\bf \bar{p}}^{\nu}-\frac{1}{2}\eta^{\mu\nu}{\bf p} \cdot {\bf \bar{p}}\right)  \frac{\partial
}{\partial {\bf p}^{\mu}}\frac{\partial}{\partial {\bf p}^{\nu}}+\frac{d}{2}{\bf \bar{p}}^{\mu}
\frac{\partial}{\partial {\bf p}^{\mu}}=\nu^{2}\frac{2{\bf \bar{p}}\cdot {\bf p}}{{\bf p}^{2}}+h^{2}+\bar
{h}^{2}~.\label{Aeq1}
\end{equation}
Consider first the following function
\[
f\left(  {\bf y} \right)  =\left\vert {\bf y}\right\vert ^{d}\int_{\mathrm{M}
}d{\bf \bar{y}}~~e^{-2{\bf \bar{p}}\cdot {\bf \bar{y}}}~\delta\left(  2{\bf \bar{y}}\cdot {\bf y}~+h^{2}+\bar{h}^{2}\right)
~\delta\left(  {\bf y}^{2}{\bf \bar{y}}^{2}-h^{2}\bar{h}^{2}\right)  ~,
\]
where we integrate over the future Milne cone $\mathrm{M}\subset\mathbb{M}^{d}$
given by ${\bf \bar{y}}^{2}\leq0,~{\bf \bar{y}}^{0}\geq0$. Changing integration variable to
\[
{\bf z}=-{\bf y}\frac{\left(  {\bf y}\cdot {\bf \bar{y}}\right)  \left(  {\bf y}\cdot {\bf \bar{p}}\right)  -{\bf y}^{2}\left(
{\bf \bar{y}}\cdot {\bf \bar{p}}\right)  }{\left(  {\bf y}\cdot {\bf \bar{p}}\right)  ^{2}+{\bf y}^{2}}+
{\bf \bar{p}}{\bf y}^{2}\frac{\left(
{\bf y}\cdot {\bf \bar{p}}\right)  \left(  {\bf \bar{y}}\cdot {\bf \bar{p}}\right)  +
\left(  {\bf \bar{y}}\cdot {\bf y}\right)  }{\left(
{\bf y}\cdot {\bf \bar{p}}\right)  ^{2}+{\bf y}^{2}},
\]
with ${\bf z}^{2}=-{\bf y}^{2}{\bf \bar{y}}^{2}$, 
${\bf z}\cdot {\bf \bar{p}}=-{\bf \bar{y}}\cdot {\bf y}$, ${\bf z}\cdot {\bf y}=-{\bf y}^{2}{\bf \bar{p}}\cdot {\bf \bar{y}}$
and $d{\bf z}=\left\vert {\bf y}\right\vert ^{d}d{\bf \bar{y}}$, we also have the integral
representation
\[
f\left(  {\bf y}\right)  =\int_{\mathrm{M}}d{\bf z}~~e^{-\frac{2{\bf y}\cdot {\bf z}}{{\bf y}^{2}}}
~\delta\left(  2{\bf z}\cdot {\bf \bar{p}}~-h^{2}-\bar{h}^{2}\right)  ~\delta\left(  {\bf z}^{2}
+h^{2}\bar{h}^{2}\right)  ~,
\]
from which it is clear that the function $f\left( {\bf y}\right)$ satisfies
\begin{equation}
~{\bf \bar{p}}^{\mu}\frac{\partial}{\partial\left(  {\bf y}^{\mu}/{\bf y}^{2}\right)  }f=\left(
{\bf y}^{2}{\bf \bar{p}}^{\nu}-2{\bf \bar{p}}\cdot {\bf y}~{\bf y}^{\nu}\right)  \frac{\partial}{\partial {\bf y}^{\nu}}f=-\left(
h^{2}+\bar{h}^{2}\right)  ~f~.\label{Aeq3}
\end{equation}
We now consider the following function
\begin{equation}
g\left(  {\bf p}\right)  =\left(  -{\bf p}^{2}\right)  ^{\nu}\int_{\mathrm{M}}\frac
{d{\bf y}}{\left\vert {\bf y}\right\vert ^{d-4\nu}}~e^{-2{\bf p}\cdot {\bf y}}~f\left(  {\bf y}\right)
~.\label{Aeq2}
\end{equation}
We claim that $g\left( {\bf p}\right)$ satisfies the differential equation
(\ref{Aeq1}). Replacing
\begin{align*}
\frac{\partial}{\partial {\bf p}^{\mu}}  & \rightarrow2\left(  \nu\frac{{\bf p}_{\mu}}{{\bf p}^{2}
}-{\bf y}_{\mu}\right), \\
\frac{\partial}{\partial {\bf p}^{\mu}}\frac{\partial}{\partial {\bf p}^{\nu}}  &
\rightarrow2\nu\left(  \frac{\eta_{\mu\nu}}{{\bf p}^{2}}-2\frac{{\bf p}_{\mu}{\bf p}_{\nu}}{{\bf p}^{4}
}\right)  +4\left(  \nu\frac{{\bf p}_{\mu}}{{\bf p}^{2}}-{\bf y}_{\mu}\right)  \left(  \nu
\frac{{\bf p}_{\nu}}{{\bf p}^{2}}-{\bf y}_{\nu}\right),
\end{align*}
one can easily show that (\ref{Aeq1}) is equivalent to
\begin{align*}
& \left(  -{\bf p}^{2}\right)  ^{\nu}\int_{\mathrm{M}}\frac{d{\bf y}}{\left\vert
{\bf y}\right\vert ^{d-4\nu}}~e^{-2{\bf p}\cdot {\bf y}}~\left[  4\left(  {\bf p}\cdot {\bf y}\right)
\left(  {\bf \bar{p}}\cdot {\bf y}\right)  -2{\bf y}^{2}\left(  {\bf p}\cdot {\bf \bar{p}}\right)  -\right. \\
& \left.  -\left(  d+4\nu\right)  \left(  {\bf \bar{p}}\cdot {\bf y}\right)  -h^{2}-\bar{h}
^{2}\right]  ~f\left(  {\bf y}\right)  =0~.
\end{align*}
Using (\ref{Aeq3}) the above is equivalent to
\begin{align*}
& \left(  -{\bf p}^{2}\right)  ^{\nu}\int_{\mathrm{M}}\frac{d{\bf y}}{\left\vert
{\bf y}\right\vert ^{d-4\nu}}~e^{-2{\bf p}\cdot {\bf y}}~\left[  4\left(  {\bf p}\cdot {\bf y}\right)
\left(  {\bf \bar{p}}\cdot {\bf y}\right)  -2{\bf y}^{2}\left(  {\bf p}\cdot {\bf \bar{p}}\right)  -\right. \\
& \left.  -\left(  d+4\nu\right)  \left(  {\bf \bar{p}}\cdot {\bf y}\right)  +\left(  {\bf y}^{2}
{\bf \bar{p}}^{\nu}-2{\bf \bar{p}}\cdot {\bf y}~{\bf y}^{\nu}\right)  \frac{\partial}{\partial {\bf y}^{\nu}}\right]
~f\left(  {\bf y}\right)  =0~,
\end{align*}
which in turn is equal to
\[
\int_{\mathrm{M}}d{\bf y}~\frac{\partial}{\partial {\bf y}^{\nu}}\left[  \left(
-{\bf y}^{2}\right)  ^{-\frac{d}{2}+2\nu}\left(  {\bf y}^{2}{\bf \bar{p}}^{\nu}-2{\bf \bar{p}}\cdot {\bf y}~{\bf y}^{\nu}\right)
e^{-2{\bf p}\cdot {\bf y}}~f\left(  {\bf y}\right)  \right]  =0~.
\]
This last equation is true since the boundary value on $\partial\mathrm{M}$, of
the term in the square brackets, vanishes.

We have therefore proved that $\left( -{\bf p}^{2}\right)^{-\eta}\mathcal{I}_{h,\bar{h}}\propto g$. 
Choosing a convenient normalization and going back to a general choice of ${\bf \bar{p}}$ we have arrived at the result (\ref{eq1001}).

\end{subappendices}


\chapter{Eikonal Approximation in AdS/CFT}
\label{ch:eikAdSCFT}


The AdS/CFT correspondence predicts the existence of a dual
CFT$_{d}$ living on the boundary of AdS$_{d+1}$. In particular,
the AdS high energy scattering amplitude we determined in chapter \ref{ch:eikonal} 
is directly related to the CFT four point--function of scalar
primary operators at particular kinematics. We shall now explore this connection to find
specific properties of  CFTs with AdS duals.

As briefly described in the introduction, by the AdS/CFT correspondence, CFT  correlators can be computed using
string theory in Anti--de Sitter spacetime. We shall work in the limit of small string length compared
to the radius of AdS, where the supergravity description is valid.
In this regime, the  four--point correlator
$$
A\left( \mathbf{p}_{1},\mathbf{p}_{2},\mathbf{p}_{3},\mathbf{p}_{4} \right) =
\left\langle \mathcal{O}_1 ( \mathbf{p}_{1} )  \mathcal{O}_2 ( \mathbf{p}_{2} ) 
 \mathcal{O}_1 ( \mathbf{p}_{3} )  \mathcal{O}_2 ( \mathbf{p}_{4} ) \right\rangle \ ,
$$
is given by the sum of
all Feynman--Witten diagrams like the one in figure \ref{fig1}, 
with bulk to boundary propagators $K_{\Delta}({\bf p}, {\bf x} )$ as
external wave functions,
\begin{eqnarray*}
\psi_1({\bf x} )=K_{\Delta_1}({\bf p}_1, {\bf x} )\ ,\ \ \ \ \ \ \ \ \ \ \ \
\psi_2({\bf x} )=K_{\Delta_2}({\bf p}_2, {\bf x} )\ ,\\
\psi_3({\bf x} )=K_{\Delta_1}({\bf p}_3, {\bf x} )\ ,\ \ \ \ \ \ \ \ \ \ \ \
\psi_4({\bf x} )=K_{\Delta_2}({\bf p}_4, {\bf x} )\ .
\end{eqnarray*}
More generally, we can prepare any on--shell wave function  in the bulk by
superposing bulk to boundary propagators from many boundary points. For example,
$$
\psi_1({\bf x} )=\int_{\Sigma} d{\bf p}_1\, \phi_1({\bf p}_1 )\, K_{\Delta_1}({\bf p}_1, {\bf x} )\ ,
$$
where the boundary integration is done along a specific section $\Sigma$ of the light--cone.
Therefore, given boundary wave functions $\phi_i$, such that the corresponding
bulk  wave functions $\psi_i$ are of the eikonal type as defined in chapter \ref{ch:eikonal},
we have
$$
\int_{\Sigma} d{\bf p}_1\cdots  d{\bf p}_4\,
\phi_1({\bf p}_1 )\cdots \phi_4({\bf p}_4 )\,
A\left( \mathbf{p}_{1},\mathbf{p}_{2},\mathbf{p}_{3},\mathbf{p}_{4} \right)
\simeq A_{eik}\ ,
$$
where $A_{eik}$ is given by (\ref{ampsimp}).
The purpose of this chapter is to use the knowledge of $A_{eik}$ to extract information about 
the CFT four point function $A\left( \mathbf{p}_{i} \right)$.

\section{CFT Eikonal Kinematics}
\label{CFTeik}

In order to construct the relevant eikonal wave functions, we shall
need to analyze more carefully the global structure of AdS.  Consider
a point $\mathbf{Q}$, either in AdS or on its boundary. The future and
past light--cones starting from $\mathbf{Q}$ divide global AdS and its
boundary into an infinite sequence of regions, which we label by an
integer.  Given a generic point $\mathbf{Q}^{\prime }$, we introduce
the integral function $n\left( \mathbf{Q}^{\prime },\mathbf{Q}\right)$
which vanishes when $\mathbf{Q}^{\prime }$ is space--like related to
$\mathbf{Q}\,$ and which increases (decreases) by one unit as $\mathbf{Q}^{\prime
}$ moves forward (backward) in global time and crosses the light cone
of $\mathbf{Q}$.  Clearly $n\left( \mathbf{Q},\mathbf{Q}^{\prime
}\right) =-n\left( \mathbf{Q}^{\prime },\mathbf{Q}\right) $.  
As explained in section \ref{AdSdyn},
the boundary and the bulk to boundary propagators
are then given by
\begin{equation}
K_{\Delta}({\bf p}, {\bf p}' )=
\frac{\mathcal{C}_{\Delta}}{|2{\bf p} \cdot {\bf p}'|^{\Delta}}\,i^{\,-2\Delta|n\left( \mathbf{p},\mathbf{p}^{\prime }\right)|}\ ,
\ \ \ \ \ \ \ \ \ \
K_{\Delta}({\bf p}, {\bf x} )=
\frac{\mathcal{C}_{\Delta}}{|2{\bf p} \cdot {\bf x}|^{\Delta}}\,i^{\,-2\Delta|n\left( \mathbf{p},\mathbf{x}\right)|}\ .
\label{propagators}
\end{equation}

\begin{figure}
\begin{center}
\includegraphics[width=9cm]{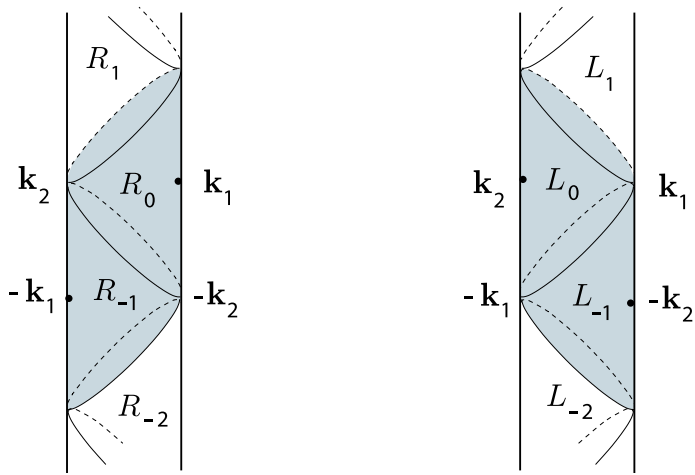}
\caption{ The momenta
$\mathbf{k}_{1},\mathbf{k}_{2}$ divide AdS space in Poincar\'{e}
patches $L_n$ and $R_n$. The boundary wave functions $\phi_1$ and
$\phi_3$ ($\phi_2$ and $\phi_4$) are localized on the boundary of
$R_{-1}$ and $R_{0}$ ($L_{-1}$ and $L_{0}$). } \label{FigPP}
\end{center}
\end{figure}

Recall that the momenta $\mathbf{k}_{1}$ and $\mathbf{k}_{2}$ indicate, respectively, the
outgoing directions of particles $1$ and $2$, whereas $-\mathbf{k}_{1}$ and $-\mathbf{k}_{2}$ 
indicate the incoming ones.
These null vectors are identified with boundary points as in figure \ref{FigPP}.
We therefore expect the boundary wave functions to be localized around
these points. Implicit in the discussion of section \ref{eikonalAdS}
is the assumption that 
$n\left( \mathbf{k}_{1},\mathbf{k}_{2}\right) =n\left( -\mathbf{k}_{1},-\mathbf{k}_{2}\right) =0$,
whereas $n\left( \mathbf{k}_{2},-\mathbf{k}_{1}\right) =n\left( \mathbf{k}_{1},-\mathbf{k}_{2}\right) =1$. 
The momentum $\mathbf{k}_{1}$
divides global AdS$_{d+1}$ space into a set of Poincar\`{e}
patches $L_{n}$ of points $\mathbf{x}$ such that 
$n\left(\mathbf{x,k}_{1}\right) =n$, which are separated by the surface
$\mathbf{x}\cdot \mathbf{k}_{1}=0$, as shown explicitly in figure
\ref{FigPP}.
Similarly, we have the patches $R_{n}$ of points $\mathbf{x}$ with
$n\left( \mathbf{x,k}_{2}\right) =n$, separated by the surface
$\mathbf{x}\cdot \mathbf{k}_{2}=0$. A point $\mathbf{x}$, either
in AdS or on its boundary, with $\mathbf{x\cdot k}_{1}<0$ ($\mathbf{x\cdot k}_{1}>0$), 
will be within a region $L_{n}$ with $n$ even (odd), and
similarly
for the regions $R_{n}$. From our previous construction, 
we see that the interaction takes
place around the hyperboloid $H_{d-1}$ defined by the intersection of the boundary between $R_{0}$ and $R_{-1}$ 
($\mathbf{x\cdot k}_{2}=0 $),
and the boundary between $L_{0}$ and $L_{-1}$ ($\mathbf{x\cdot k}_{1}=0 $). Let us then consider the
incoming wave $\phi _{1}\left( \mathbf{p}_{1}\right) $. In order to achieve
the required eikonal kinematics, we shall localize $\phi _{1}$ on the boundary
of $R_{-1}$, around the point $-\mathbf{k}_{1}$.
We shall show in the next section that, if we choose only
positive frequency modes with respect to the action of time translation in
this patch, which is generated by $\mathbf{T}_{2}$, the corresponding bulk
wave function $\psi _{1}$ will have support only on patches $R_{n}$ with $n\geq -1$. 
Similarly, we shall localize $\phi _{3}$ on $\partial R_{0}$,
around the point $\mathbf{k}_{1}$,
with negative frequency modes only, so that $\psi _{3}$
will have support on $R_{n}$ for $n\leq 0$. The overlap of $\psi_{1}$ and 
$\psi _{3}$ will then be non vanishing only in regions
$R_{-1}$ and $R_{0}$, which are those explicitly parametrized 
by the coordinates $\{u,v,\mathbf{w}\}$. In a symmetric way, we shall localize 
$\phi_{2}$  ($\phi_{4}$) on $\partial L_{-1}$  ($\partial L_{0}$), around the
point  $-\mathbf{k}_{2}$ ($\mathbf{k}_{2}$), with positive
(negative) frequency modes with respect to $\mathbf{T}_{1}$.  The overlap 
of $\psi _{2}$  and $\psi _{4}$ is then localized
in regions $L_{-1}$ and $L_{0}$, parametrized by $\{\bar{u},\bar{v},\mathbf{\bar{w}}\}$. 
Summarizing, the relevant choice of kinematics for the four
points $\mathbf{p}_{i}$ ($i=1,\cdots ,4$) is given by
\begin{equation}
\begin{array}{lllll}
\mathbf{p}_{1} \sim -\mathbf{k}_{1}&\Rightarrow\ &
\mathbf{p}_{1} \,\in\, \partial R_{-1}\ \ \ \ \ \ \ \ \ \ \ \ \ 
&\left( \mathbf{p}_{1}\cdot \mathbf{k}_{2}>0\right) \ ,  \spa{0.2}\\
\mathbf{p}_{2} \sim -\mathbf{k}_{2}&\Rightarrow&
\mathbf{p}_{3} \,\in\, \partial R_{0}
&\left( \mathbf{p}_{3}\cdot \mathbf{k}_{2}<0\right) \ , \spa{0.2} \\
\mathbf{p}_{3} \sim \mathbf{k}_{1}&\Rightarrow&
\mathbf{p}_{2} \,\in\, \partial L_{-1}
&\left( \mathbf{p}_{2}\cdot \mathbf{k}_{1}>0\right) \ ,  \spa{0.2}\\
\mathbf{p}_{4} \sim \mathbf{k}_{2}&\Rightarrow&
\mathbf{p}_{4} \,\in\, \partial L_{0}
&\left( \mathbf{p}_{4}\cdot \mathbf{k}_{1}<0\right) \ ,
\end{array}
\label{boundp}
\end{equation}
so that
\begin{equation*}
n\left( \mathbf{p}_{1},\mathbf{p}_{2}\right)  = 
n\left( \mathbf{p}_{3},\mathbf{p}_{4}\right) =0\ ,  \ \ \ \ \ \ \ \ \ \ \ \ 
n\left( \mathbf{p}_{4},\mathbf{p}_{1}\right)  =
n\left( \mathbf{p}_{3},\mathbf{p}_{2}\right) =1\ .  
\end{equation*}
We shall choose, once and for all, a
specific normalization of the $\mathbf{p}_{i}$ by rescaling the external
points, so that
\[
2\mathbf{p}_{1}\cdot \mathbf{k}_{2}=-2\mathbf{p}_{3}\cdot \mathbf{k}_{2}=
2\mathbf{p}_{2}\cdot \mathbf{k}_{1}=-2\mathbf{p}_{4}\cdot \mathbf{k}_{1}=\left( 2\omega \right) ^{2}~.
\]
This corresponds to choosing specific light cone sections, similar to (\ref{kofy}), to represent the AdS
boundary containing the points $\mathbf{p}_{i}$. 
Notice that these choices precisely cover the required Poincar\'e patches of the AdS boundary.
It is also convenient to parametrize the $\mathbf{p}_{i}$ in terms of Poincar\'{e}
coordinates. Using the splitting $\mathbb{R}^{2,d}\simeq \mathbb{M}^{2}\times \mathbb{M}^{d}$ introduced in section \ref{NCWF}, 
we  write
\begin{equation}
\begin{array}{l}
\displaystyle{\mathbf{p}_{1} = - \mathbf{k}_{1} -  \mathbf{y}_{1}^2 \mathbf{k}_{2} +2\omega\, \mathbf{y}_{1} \ ,} \spa{0.3}\\
\displaystyle{\mathbf{p}_{2} = - \mathbf{k}_{2} -  \mathbf{y}_{2}^2 \mathbf{k}_{1} + 2\omega\, \mathbf{y}_{2} \ ,} \spa{0.3}\\
\displaystyle{\mathbf{p}_{3} =  \mathbf{k}_{1} +  \mathbf{y}_{3}^2 \mathbf{k}_{2} - 2\omega\, \mathbf{y}_{3} \ ,} \spa{0.3}\\
\displaystyle{\mathbf{p}_{4} = \mathbf{k}_{2} +  \mathbf{y}_{4}^2 \mathbf{k}_{1} - 2\omega\, \mathbf{y}_{4} \ ,}
\end{array}
\label{pcoord}
\end{equation}
with the Poincar\'{e} positions $ \mathbf{y} _{i}\in \mathbb{M}^{d}$ orthogonal to $\mathbf{k}_{1}$ and 
$\mathbf{k}_{2}$ and  small, i.e. in components $| \mathbf{y} _{i}^{\,a}|\ll 1$.

\begin{figure}
\begin{center}
\includegraphics[width=5.5cm]{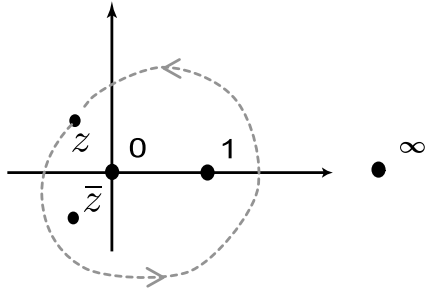}
\caption{ Analytic continuation necessary to obtain
$\hat{\mathcal{A}}$ from the Euclidean amplitude ${\mathcal{A}}$.}
\label{FigCC}
\end{center}
\end{figure}

We shall denote the corresponding Lorentzian CFT amplitude, computed with this
kinematics, by
\begin{equation}
\hat{A}\left( \mathbf{p}_{1},\cdots ,\mathbf{p}_{4}\right) =K_{\Delta
_{1}}\left( \mathbf{p}_{1},\mathbf{p}_{3}\right) K_{\Delta _{2}}\left(
\mathbf{p}_{2},\mathbf{p}_{4}\right)  \hat{\mathcal{A}}\left( z,\bar{z}
\right) ~,  
\label{zzz1}
\end{equation}
where the cross ratios are small and satisfy
\be
z\bar{z} \simeq {\bf p}^{2}{\bf \bar{p}}^{2}~,~\ \ \ \ \ \ \ \ \ \ \ \ \ \ \ \ \ \ \ \ 
z+\bar{z} \simeq 2{\bf p} \cdot {\bf \bar{p}}~,
\label{cba}
\ee
with $ {\bf p}$ and ${\bf \bar{p}}$ points in the $\mathbb{M}^d$ orthogonal to  $\mathbf{k}_{1}$ and 
$\mathbf{k}_{2}$ given by
$$
{\bf p}= \mathbf{y}_3-\mathbf{y}_1\, ,\ \ \ \ \ \ \ \ \ \ \ \ \ \ \ \ \ 
{\bf \bar{p}}=\mathbf{y}_2-\mathbf{y}_4\ .
$$
We shall reserve the labels $A$ and $\mathcal{A}$ for the amplitudes computed
on the principal Euclidean sheet, where $n\left( \mathbf{p}_{i},\mathbf{p}_{j}\right) =0$. 
As we shall discuss in detail in section \ref{sectAC}, the
amplitude $\hat{\mathcal{A}}\left( z,\bar{z}\right) $ is related to 
$\mathcal{A}\left( z,\bar{z}\right) $ by analytic continuation. More
precisely, we shall show that
\be
\hat{\mathcal{A}}\left( z,\bar{z}\right) =\mathcal{A}^{\circlearrowleft
}\left( z,\bar{z}\right) ~,
\label{CCeq}
\ee
where the right--hand side indicates the function obtained by keeping $\bar{z}$ 
fixed and rotating $z$ counter--clockwise  around the branch points  $0$ and $1$, 
as shown in figure \ref{FigCC}.

Let us now discuss the boundary propagators $K_{\Delta }$ in (\ref{zzz1}).
The only subtle issue comes from the appropriate phase factors  \cite{Paper1}. 
More precisely, given the choices in (\ref{boundp}) and the form of the boundary
propagator in (\ref{propagators}), 
we have that $K_{\Delta _{1}}\left( \mathbf{p}_{1},\mathbf{%
p}_{3}\right) $ is given by $\mathcal{C}_{\Delta _{1}}\left\vert 2\mathbf{p}%
_{1}\cdot \mathbf{p}_{3}\right\vert ^{-\Delta _{1}}$ times the following
phases
\begin{eqnarray}
&&1~\ \ \ \ \ \ \ \ \ \ \ \ \ \ \ \ \ \ \ \ \ 
\mathbf{p}_{1},\mathbf{p}_{3}\text{ spacelike separated}  \nonumber \\
&&i^{-2\Delta _{1}}~\ \ \ \ \ \ \ \ \ \ \ \ \ \ \ 
\mathbf{p}_{3}\text{ in the future of }\mathbf{p}_{1}\text{ with }\mathbf{p}_{1}\cdot \mathbf{p}_{3}>0  
\label{phases} \\
&&i^{-4\Delta _{1}}~\ \ \ \ \ \ \ \ \ \ \ \ \ \ \ 
\mathbf{p}_{3}\text{ in the future of }\mathbf{p}_{1}\text{ with }\mathbf{p}_{1}\cdot \mathbf{p}_{3}<0  \nonumber
\end{eqnarray}
A similar statement applies to the propagator $K_{\Delta _{2}}\left( \mathbf{%
p}_{2},\mathbf{p}_{4}\right) $.
The amplitude (\ref{zzz1}) is then given, in terms of ${\bf p} ,{\bf \bar{p}}$ by
\begin{equation}
\hat{A}\left( {\bf p} ,{\bf \bar{p}}\right) = \frac{(2\omega\,i)^{-2\Delta _{1}} 
\mathcal{C}_{\Delta _{1}}}{\left( {\bf p} ^{2}+i\epsilon _{{\bf p} }\right) ^{\Delta _{1}}}\,
\frac{(2\omega\,i)^{-2\Delta _{2}}\mathcal{C}_{\Delta _{2}}}{\left( {\bf \bar{p}}^{2}
-i\epsilon _{{\bf \bar{p}}}\right) ^{\Delta _{2}}} ~\hat{\mathcal{A}}
\left( z,\bar{z}\right)\ ,   \label{zzz2}
\end{equation}
where we have explicitly written the two propagators using
\[
\epsilon _{{\bf p} }=\epsilon \,{\rm sign}\left( -{\bf x}_0\cdot {\bf p} \right) ~,~
\]
which picks the correct branch of the logarithm consistent with the phase
prescription in (\ref{phases}). Notice that ${\bf x}_0$ is any future directed vector in $\mathbb{M}^d$,
chosen to be the reference point introduced in section  \ref{eikonalAdS}.

\section{Boundary Wave Functions}\label{boundfunc}

\begin{figure}
\begin{center}
\includegraphics[height=4.8cm]{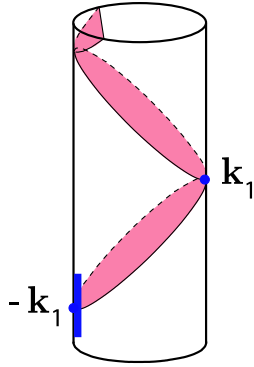}
\caption{
The boundary wave function  $\phi_1$ is localized along a small timelike segment centered in   $-{\bf k}_1$.
The bulk wave function $\psi_1$ is mainly supported around the region ${\bf k}_1\cdot {\bf x}=0 $
in the future of the boundary point $-{\bf k}_1$.}
\label{psi1}
\end{center}
\end{figure}

We shall now describe in detail a particularly convenient choice of boundary wave functions, consistent with the general description
of the previous section, and which correspond to bulk wave functions of the eikonal type.
First recall that in section \ref{NCWF}, ${\bf k}_1$ defined a surface in AdS containing 
the null geodesics that go from the boundary point $-{\bf k}_1$ to  ${\bf k}_1$.
We have then used the AdS isometry generated by $ {\bf T}_2$ to build the congruence of null
geodesics associated to particle 1. 
This isometry is time translation in the Poincar\'e patch $R_{-1}$, with boundary centered
at  $-{\bf k}_1$.
It is then natural
to localize the boundary wave function of $\mathcal{O}_{1}$ along the timelike line
$$
{\bf p}_1(t)=-e^{\,t {\bf T}_2}\,{\bf k}_1=-{\bf k}_1+t\omega \,{\bf x}_0+\frac{t^2}{4}\,{\bf k}_2\ .
$$
In fact, parametrizing ${\bf p}_1(t)$ in Poincar\'e coordinates as in (\ref{pcoord}), we have that
$$
{\bf y}_1(t) = \frac{t}{2}\,{\bf x}_0 \ ,
$$
so, as a function of $t$, we are moving in the future time direction indicated by ${\bf x}_0$.
We then modulate the boundary function with $\omega\,F(t)\, e^{-i\omega t} $,
where the function $F$ is the profile function introduced in (\ref{eikwavefunc}).
The bulk wave function $\psi_1$ is then given by
$$
\psi_1({\bf x} )=\omega
\int dt \,F(t)\, e^{-i\omega t}\,\frac{\mathcal{C}_{\Delta_1} }{\big( -2{\bf p}_1(t) \cdot {\bf x}+i\epsilon \big)^{\Delta_1}}\ ,
$$
where the $i\epsilon$ prescription is correct for all points ${\bf x}$ in region $R_{-1}$. 
Since $F(t)$ is non--vanishing only for $|t|< \Lambda$, the above description is valid 
also in part of region $R_{0}$, as we shall show shortly.
In the coordinate system (\ref{coordAdS}), valid in $R_{-1}$ and $R_{0}$,  we have
$$
-2{\bf p}_1(t) \cdot {\bf x} = -2\omega (t-v)\left(1+\frac{u}{4} (t-v) \right)({\bf x}_0 \cdot {\bf w} ) \ ,
$$
showing that the integrand diverges for $t=v$ and $t=v-4/u$. The first divergence corresponds 
to the future directed signal from point
${\bf p}_1(t)$, whereas the second divergence comes from the reflection at the AdS boundary
 for $u>0$ and from the backward signal from ${\bf p}_1(t)$ for $u<0$. The backwards signal is 
relevant in region $R_{-1}$, where the $i\epsilon$ prescription is valid. For positive $\omega$,
 one may close the $t$ contour avoiding completely the singularity from the backwards signal,
 showing that positive frequencies propagate forward in global time. In region $R_0$, 
on the other hand, the $i\epsilon$ prescription is valid up to the reflected signal at
 the second singularity, more precisely for $u\Lambda<|4-vu|$. In this part of $R_0$ and in region
 $R_{-1}$, for large $\omega$, the integral is dominated by the divergence at $t=v$, and we have that
$$
\psi_1({\bf x} ) \simeq \omega
\,F(v)\, \int dt \, e^{-i\omega t}\,\frac{ \mathcal{C}_{\Delta_1} }{\big(-2\omega (t-v)({\bf x}_0 \cdot {\bf w} )+i\epsilon \big)^{\Delta_1}}\ .
$$
It is then clear that, for large $\omega$, the wave function $\psi_1$ has precisely the required form
$$
\psi_1({\bf x} )\simeq e^{-i\omega v}\,F(v) F_1( {\bf w} )\ ,
$$
with
$$
F_1( {\bf w} )= i^{-\Delta_1} \,\frac{2\pi\,\mathcal{C}_{\Delta_1}}{\Gamma(\Delta_1)}
 \left( -2{\bf x}_0 \cdot {\bf w} \right)^{-\Delta_1}  \ .
$$
Thus, the wave function  $\psi_1$ is supported mainly around the future directed null geodesics starting from
the point $-{\bf k}_1$ of the boundary, as depicted in figure \ref{psi1}.
Similarly, we choose the boundary wave function of $\mathcal{O}_{2}$
localized along the timelike line
$$
{\bf p}_2(t)=-e^{\,t {\bf T}_1}\,{\bf k}_2=-{\bf k}_2+t\omega \,{\bf x}_0+\frac{t^2}{4}\,{\bf k}_1\ ,
$$
which means 
$$
{\bf y}_2(t) = \frac{t}{2}\,{\bf x}_0\ .
$$
The bulk wave function  $\psi_2$ has then the required eikonal form in (\ref{eikwavefunc}) with
$$
F_2( {\bf \bar{w}} )=i^{-\Delta_2} \,\frac{2\pi\,\mathcal{C}_{\Delta_2} }{\Gamma(\Delta_2)}
 \left( -2{\bf x}_0 \cdot {\bf \bar{w}} \right)^{-\Delta_2} \ .
$$

The boundary wave functions $\phi_3$ and $\phi_4$ will be the complex conjugates of $\phi_1$ and $\phi_2$,
but localized along slightly different curves,
$$
{\bf p}_3(t)=e^{\,t {\bf T}_2}\,( {\bf k}_1 + {\bf q}) \ ,\ \ \ \ \ \ \ \ \ \ \ \ \ \ \ \ \
{\bf p}_4(t)=e^{\,t {\bf T}_1}\,( {\bf k}_2 + {\bf \bar{q}}) \ .
$$
In analogy with flat space, the eikonal regime corresponds to ${\bf q}^2, {\bf \bar{q}}^2 \ll \omega^2 $.
The fact that ${\bf p}_3 $ and  ${\bf p}_4$ must be null vectors yields the conditions
$$
{\bf q}^2=- 2{\bf k}_1 \cdot {\bf q}\ ,\ \ \ \ \ \ \ \ \ \ \ \ \ \ \ \ \
{\bf \bar{q}}^2=-2  {\bf k}_2 \cdot {\bf \bar{q}}\ .
$$
The parts of $ {\bf q} $ and ${\bf \bar{q}}$ that are,
respectively,  proportional to ${\bf k}_1 $ and ${\bf k}_2 $ are
irrelevant since we stay in the same null rays. This freedom can
be used to fix
$$
{\bf k}_2 \cdot {\bf q}=0\ ,\ \ \ \ \ \ \ \ \ \ \ \ \ \ \ \ \
{\bf k}_1 \cdot {\bf \bar{q}}=0\ .
$$
We conclude that, to leading order in ${\bf q}/\omega$ and ${\bf \bar{q}}/\omega$,
both $ {\bf q} $ and ${\bf \bar{q}}$ belong to the $\mathbb{M}^d \subset \mathbb{R}^{2,d}$
orthogonal to ${\bf k}_1$ and ${\bf k}_2$. 
Therefore, in the Poincar\'e coordinates (\ref{pcoord}), we have
$$
{\bf y}_3(t) = \frac{t}{2}\,{\bf x}_0 - \frac{1}{2\omega} \,{\bf q}\ ,
\ \ \ \ \ \ \ \ \ \ \ \ \ 
{\bf y}_4(t) = \frac{t}{2}\,{\bf x}_0 -\frac{1}{2\omega} \,{\bf \bar{q}}\, .
$$
Furthermore, we shall choose ${\bf q}$  and ${\bf \bar{q}}$ orthogonal to ${\bf x}_0$.
We then have that
\begin{eqnarray*}
&&{\bf p}_3(t)\cdot {\bf x} = - {\bf p}_1(t)\cdot {\bf x} + {\bf q}\cdot {\bf w}
+\frac{u }{4\omega}\, ({\bf x}_0 \cdot {\bf w} )\,{\bf q}^2
\ ,\\
&&{\bf p}_4(t)\cdot {\bf \bar{x}} =  -{\bf p}_2(t)\cdot {\bf  \bar{x}} + {\bf  \bar{q}}\cdot {\bf  \bar{w}}
+\frac{\bar{v}}{4\omega} \, ( {\bf x}_0 \cdot {\bf \bar{w}}  )\,{\bf \bar{q}}^2
 \ .
\end{eqnarray*}
At large $\omega$, the leading contribution to the bulk wave function $\psi_3$ is given by
\begin{eqnarray*}
\psi_3({\bf x} )
&=&
\omega\int dt \,F^\star(t)\, e^{i\omega t}\,
\frac{\mathcal{C}_{\Delta_1}}{\big( -2{\bf p}_3(t) \cdot {\bf x}+i\epsilon \big)^{\Delta_1}}\\
&\simeq&
e^{i\omega v}\,F^\star(v) F_3( {\bf w} )\ ,
\end{eqnarray*}
where the transverse modulation function  $ F_3( {\bf w} )$ is
\begin{eqnarray*}
F_3( {\bf w} )&=&
\mathcal{C}_{\Delta_1}
\int dl \, e^{il} \big(2({\bf x}_0 \cdot {\bf w})   l - 2{\bf q}\cdot {\bf w}  +i\epsilon \big)^{-\Delta_1}  \\
&=& i^{-\Delta_1}\,\frac{2\pi\,\mathcal{C}_{\Delta_1}}{\Gamma(\Delta_1)}
 \left( -2{\bf x}_0 \cdot {\bf w} \right)^{-\Delta_1}
\exp\left(i\frac{{\bf q}\cdot {\bf w} }{{\bf x}_0 \cdot {\bf w} }\right)\ .
\end{eqnarray*}
Similarly, $\psi_4$ has the form in (\ref{eikwavefunc}) with
\begin{equation*}
F_4( {\bf \bar{w}} )
=i^{-\Delta_2} \,\frac{2\pi\,\mathcal{C}_{\Delta_2}}{\Gamma(\Delta_2)}
 \left( -2{\bf x}_0 \cdot {\bf \bar{w}} \right)^{-\Delta_2}
\exp\left(i\frac{{\bf \bar{q}}\cdot {\bf \bar{w}} }{{\bf x}_0 \cdot {\bf \bar{w}} }\right)\ .
\end{equation*}

With the specific choice of wave functions just described, the AdS eikonal
amplitude (\ref{ampsimp}) becomes
\begin{equation}
\begin{array}{r}
\displaystyle{A_{eik} \simeq 2i^{-2\Delta _{1}}i^{-2\Delta _{2}}
\left( \frac{8\pi^{2}\omega \,\mathcal{C}_{\Delta _{1}}\mathcal{C}_{\Delta _{2}}}{\Gamma
(\Delta _{1})\Gamma (\Delta _{2})}\right) ^{2}\int_{H_{d-1}}d\mathbf{w}d
\mathbf{\bar{w}}\left( -2\mathbf{x}_{0}\cdot \mathbf{w}\right) ^{-2\Delta
_{1}}\left( -2\mathbf{x}_{0}\cdot \mathbf{\bar{w}}\right) ^{-2\Delta _{2}}}\spa{.7}
\\ 
\displaystyle{
\exp \left( i\frac{\mathbf{q}\cdot \mathbf{w}}{\mathbf{x}_{0}\cdot 
\mathbf{w}}+i\frac{\mathbf{\bar{q}}\cdot \mathbf{\bar{w}}}{\mathbf{x}_{0}\cdot
\mathbf{\bar{w}}}+\frac{ig^{2}}{2}\left( 2\omega \right) ^{2j-2}\frac{\Pi
_{\perp }(\mathbf{w},\mathbf{\bar{w}})}{\left( (\mathbf{x}_{0}\cdot \mathbf{w})
(\mathbf{x}_{0}\cdot \mathbf{\bar{w}})\right) ^{j-1}}\right)} \ .
\end{array}
\label{eikEQ}
\end{equation}
By construction, the above expression should approximate, in the limit of
large $\omega $, the CFT correlator $\hat{A}\left({\bf p},{\bf \bar{p}} \right) $ in (\ref{zzz2})
integrated against the corresponding boundary
wave--functions $\phi _{i}\left( \mathbf{p}_{i}\right) $, 
\begin{equation}
A_{eik} \simeq\omega ^{4}\int dt_{1}\cdots dt_{4}\, F( t_{1}) F(t_{2}) F^{\star}(t_{3}) F^{\star}(t_{4})
\,e^{i\omega \left( t_{3}-t_{1}\right) +i\omega \left( t_{4}-t_{2}\right) }
\hat{A}\big( {\bf p}(t_i),{\bf \bar{p}}(t_i) \big) \,,  \label{CFTamplitude}
\end{equation}
with 
\begin{align*}
{\bf p}(t_i)&=\frac{t_3-t_1}{2}\,{\bf x}_0 - \frac{1}{2\omega}\,{\bf q}\ ,\spa{.5}\\
{\bf \bar{p}}(t_i)&=\frac{t_2-t_4}{2}\,{\bf x}_0 +\frac{1}{2\omega}\,{\bf \bar{q}}\ .
\end{align*}
Before deriving the consequences of this result, we must clarify the structure
of the four point correlator $\hat{A}$ in (\ref{CFTamplitude}).
We shall devote the next three sections to this purpose and return to  equations (\ref{eikEQ}) and (\ref{CFTamplitude})
only in section \ref{anomdim}.


\section{Analytic Continuation}

\label{sectAC}

Let us discuss the issue of analytic continuation of
the amplitude $A\left( \mathbf{p}_{i}\right) $, showing in
particular how to derive (\ref{CCeq}). First note that the cross
ratios $z$ and $\bar{z}$, as defined in (\ref{zzbar}), are invariant under
rescalings $\mathbf{p}_{i}\rightarrow \lambda _{i}\mathbf{p}_{i}$,
with $\lambda _{i}$ arbitrary and, in particular, negative.
Moreover, two different boundary points differing by a $2\pi $
translation in AdS global time have the same embedding
representation and therefore also give rise to the same values of
$z$ and $\bar{z}$. On
the other hand, in global AdS, different sets of boundary points $\mathbf{p}%
_{i}$ with the same values of $z$ and $\bar{z}$ have, in general, different
reduced amplitudes $\mathcal{A}\left( z,\bar{z}\right) $ related by analytic
continuation. More precisely, the amplitude $\mathcal{A}$ is a multi--valued
function of $z$ and $\bar{z}$ with branch points at $z,\bar{z}=0,1,\infty $, and
different sets $\{\mathbf{p}_{i}\}$ with the same cross ratios correspond,
in general, to different sheets. The best way to understand this is to start
from the Euclidean reduced four--point amplitude $\mathcal{A}\left( z,\bar{z}%
\right) $ and then Wick rotate to the Lorentzian setting.

\begin{figure}
\begin{center}
\includegraphics[width=14cm]{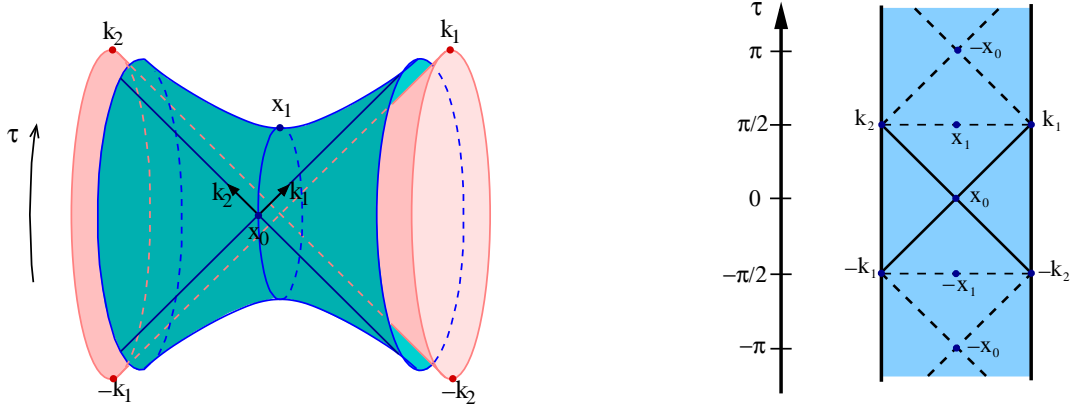}
\caption{Unwrapping the AdS$_2$ global time circle.}
\label{globaltime}
\end{center}
\end{figure}

We start by choosing a global time $\tau $ in AdS. From the
embedding space perspective, global time translations are
rotations in a timelike plane. We choose this to be
the plane generated by the normalized timelike vectors $\mathbf{x}_{0}$ 
and $\mathbf{x}_{1}$, with $2\omega \,\mathbf{x}_{1}=\mathbf{k}_{1}+\mathbf{k}_{2}$ 
(see figure \ref{globaltime}). A generic boundary point $\mathbf{p}$ can then be written as
\[
\mathbf{p}=\lambda \big[\cos (\tau )\,\mathbf{x}_{0}+\sin (\tau )\,\mathbf{x}_{1}+\mathbf{n}\big]\ ,
\]
where the vector $\mathbf{n}$ belongs to the $(d-1)$--dimensional
unit sphere embedded in the space $\mathbb{R}^{d}$ orthogonal to
$\mathbf{x}_{0}$ and $\mathbf{x}_{1}$, and the constant $\lambda>0$ depends on the choice of 
representative $\mathbf{p}$ for each
null ray. We can then consider, for
each of the boundary points under consideration, the standard Wick rotation 
parametrized by $0\leq \theta \leq 1$,
\[
\mathbf{p}=\lambda \left[ \cos \left( -i\tau e^{\frac{i\pi }{2}\theta}\right) 
\,\mathbf{x}_{0}+\sin \left( -i\tau e^{\frac{i\pi }{2}\theta}\right) 
\,\mathbf{x}_{1}+\mathbf{n}\right] \ ,
\]
where $\theta =0$ corresponds to the Euclidean setting and $\theta
=1$ to the Minkowski one. Given the coordinates $\tau _{i}$
and $\mathbf{n}_{i}$ of the four boundary points
$\mathbf{p}_{i}$, the corresponding variables
$z ( \theta )  $, $\bar{z} ( \theta )  $ define
two paths in the complex plane parametrized by $0\leq \theta \leq
1$. The paths $z ( \theta )  $, $\bar{z} ( \theta
) $ are explicitly obtained by replacing
\[
\mathbf{p}_{i}\cdot \mathbf{p}_{j}\rightarrow \mathbf{n}_{i}\cdot \mathbf{n}%
_{j}-\cos \left( -i(\tau _{i}-\tau _{j})e^{\frac{i\pi }{2}\theta }\right) \ ,
\]
in the expressions (\ref{zzbar}). 
The Lorentzian amplitude $\hat{\mathcal{A}}$ is then given by the analytic
continuation of the basic Euclidean amplitude $\mathcal{A}$ following the
paths $z( \theta ) $, $\bar{z}( \theta )$ from $\theta =0$ to $\theta =1$.

\begin{figure}
\begin{center}
\includegraphics[width=16cm]{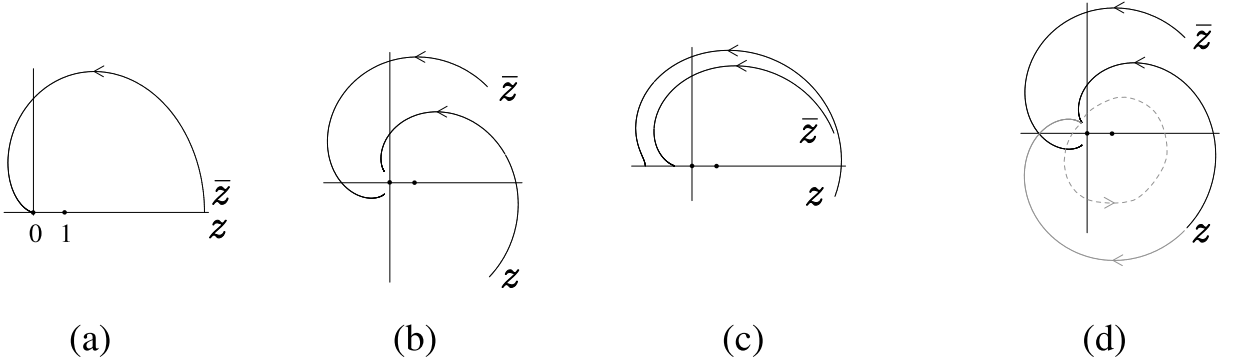}
\end{center}
\caption{ Figures (a), (b) and (c) show the curves $z(\theta)$ and ${\bar z}(\theta)$
starting from the Euclidean setting at $\theta=0$, with $z(0)={\bar z}^{\star}(0)$.
Plot (a) corresponds to the limiting path 
$z(\protect\theta) =\bar{z}( \protect\theta)$
where $t_{i}=0$ and $\mathbf{q}=\mathbf{\bar{q}}=0$. Plots (b) and (c) correspond to general paths.
Figure (d) shows the relevant
analytic continuation relating $\hat{\mathcal{A}}$ to $\mathcal{A}$. Starting from path (b),
the curve $z(\theta)$, shown in black, is equivalent to the path shown in gray, which, in turn, is composed
of two parts. The continuous part, which is the complex conjugate of the curve $\bar z (\theta)$, computes
$\mathcal{A}$ on the principal sheet. The dashed part, also shown in figure \ref{FigCC}, 
rotates $z$ counter--clockwise around the singularities at $0$ and $1$.
Therefore $\hat{\mathcal{A}}=\mathcal{A}^{\circlearrowleft}$.}
\label{complexpaths}
\end{figure}

In our particular case, we have
\begin{eqnarray*}
\tau _{1}\simeq -\frac{\pi }{2}+t_{1}\ ,&\ \ \ \ \ \ \ \ &\mathbf{n}%
_{1}\simeq \frac{1}{2\omega }\left( \mathbf{k}_{2}-\mathbf{k}_{1}\right) \ ,
\\
\tau _{2}\simeq -\frac{\pi }{2}+t_{2}\ ,&\ \ \ \ \ \ \ \ &\mathbf{n}%
_{2}\simeq \frac{1}{2\omega }\left( \mathbf{k}_{1}-\mathbf{k}_{2}\right) \ ,
\\
\tau _{3}\simeq \frac{\pi }{2}+t_{3}\ ,\ \ &\ \ \ \ \ \ \ \ \ \ &\mathbf{n}%
_{3}\simeq \frac{1}{2\omega }\left( \mathbf{k}_{1}-\mathbf{k}_{2}+2\mathbf{q}%
+\frac{\mathbf{q}^{2}}{2\omega ^{2}}\,\mathbf{k}_{2}\right) \ , \\
\tau _{4}\simeq \frac{\pi }{2}+t_{4}\ ,\ \ &\ \ \ \ \ \ \ \ \ \ &\mathbf{n}%
_{4}\simeq \frac{1}{2\omega }\left( \mathbf{k}_{2}-\mathbf{k}_{1}+2\mathbf{%
\bar{q}}+\frac{\mathbf{\bar{q}}^{2}}{2\omega ^{2}}\,\mathbf{k}_{1}\right) \ ,
\end{eqnarray*}%
in the relevant regime of $t_{i}\ll 1$ and $\mathbf{q}^{2},\mathbf{\bar{q}}^{2}\ll \omega ^{2}$. 
Therefore, the complex paths $z( \theta ) ,\bar{z}( \theta )$ will be small deformations of the paths
\[
z( \theta) =\bar{z}( \theta) =\cos ^{2}\left( ie^{\frac{i\pi }{2}\theta }\pi /2\right)
\]
obtained in the special case $t_{i}=0$ and $\mathbf{q}=\mathbf{\bar{q}}=0$.
This limiting path is plotted in figure \ref{complexpaths}(a). We also show,
in figures \ref{complexpaths}(b) and \ref{complexpaths}(c) 
two generic paths, respectively with Lorentzian values $\bar{z}(1) =z^{\star}(1)$ and 
$\im \,z(1)=\im \,\bar{z}( 1) =0$. The equations
governing the generic paths are rather cumbersome and are not
important for our present purpose.
At this point we notice that the paths $z(\theta)$ in  figures \ref{complexpaths}(b) and \ref{complexpaths}(c) can be
continuously deformed, without crossing any branch point, to the path complex
conjugate to $\bar{z}(\theta)$, plus a full counter--clockwise turn around $0$ and $1$, 
as shown in figure \ref{complexpaths}(d). Thus, the Lorentzian amplitude $%
\hat{\mathcal{A}}(z,\bar{z})$ is obtained from the basic Euclidean amplitude
$\mathcal{A}(z,\bar{z})$ after transporting $z$ anti--clockwise around
$0$ and $1$ keeping $\bar{z}$ fixed,
\[
\hat{\mathcal{A}}(z,\bar{z})=\mathcal{A}^{\circlearrowleft}(z,\bar{z})\ .
\]


\section{Anomalous Dimensions as Phase Shift}


As explained in sections \ref{CFTeik} and \ref{boundfunc}, the AdS eikonal regime probes the Lorentzian 
amplitude $\hat{\mathcal{A}}$ for small values of the cross ratios $z$ and $\bar{z}$. 
Here we shall relate the behavior of $\hat{\mathcal{A}}$ in this regime to the 
anomalous dimensions of the composite primary operators,
\[
\mathcal{O}_{1}\partial _{\mu _{1}}\cdots \partial _{\mu _{J}}\partial ^{2n}%
\mathcal{O}_{2}\ ,
\]
of large dimension $E=\Delta_1 + \Delta_2 +J+2n$ and large spin $J$. We shall also use the
conformal dimensions $h\geq \bar{h}\geq 0$ as defined in (\ref{EJhhbar}).

Consider the expansion of the Euclidean amplitude $\mathcal{A}$ in 
\emph{S}--channel conformal partial waves, corresponding to the 
OPE at $z,\bar{z}\rightarrow \infty $ (or $\mathbf{p}_{1}\rightarrow \mathbf{p}_{2}$).
In analogy with flat space, we shall assume  that the \emph{S}--channel 
decomposition of the Euclidean amplitude $\mathcal{A}$ at
large $h,\bar{h}$ is dominated by the $\mathcal{O}_{1}\partial \cdots \partial \mathcal{O}_{2}$
composites already present at zeroth order, as explained in section \ref{sec:freeprop}.
Denoting their anomalous dimensions by $2\Gamma (h,\bar{h})$, we
can write
\begin{equation}
\mathcal{A}(z,\bar{z})\simeq \sum_{h\geq \bar{h}\geq \eta }\left( 1+R(h,%
\bar{h})\right) ~\mathcal{S}_{h+\Gamma (h,\bar{h}),\bar{h}+\Gamma (h,\bar{h}%
)}(z,\bar{z})\ ,  \label{generalex}
\end{equation}%
where the coefficient $R(h,\bar{h})$ encodes the three point coupling
between $\mathcal{O}_{1}$, $\mathcal{O}_{2}$ and the exchanged composite  primary
field with dimensions $h, \bar{h}$. The sum is over the same lattice found in  section \ref{sec:freeprop},
\[
h,\bar{h}\in \eta +\mathbb{N}_{0}\ ,\ \ \ \ \ \ \ \ \ \ \ \ \ \ \ \ \ \ \
\eta \leq \bar{h}\leq h~.
\]%
In section \ref{impactsection} we introduced an impact parameter representation $\mathcal{I}_{h,\bar{h}}$ 
for the \emph{S}--channel partial waves $\mathcal{S}_{h,\bar{h}}$, 
which approximates the latter for small $z$ and $\bar{z}$. Moreover, we
showed that in the regime of small $z$ and $\bar{z}$ one can replace the sum over
\emph{S}--channel partial waves in (\ref{generalex}) by an integral over
their impact parameter representation,
\[
\mathcal{A}(z,\bar{z})\simeq \int dh d\bar{h}\left( 1+R(h,\bar{h})\right) ~%
\mathcal{I}_{h+\Gamma (h,\bar{h}),\bar{h}+\Gamma (h,\bar{h})}(z,\bar{z})\ .
\]
Expanding in powers of $\Gamma $ and dropping the explicit reference to $h,%
\bar{h}$, this equation reads
\begin{eqnarray*}
\mathcal{A}(z,\bar{z}) &\simeq &\int dhd\bar{h}\left( 1+R\right) \left(
1+\Gamma \partial +\frac{1}{2}\Gamma ^{2}\partial ^{2}+\frac{1}{3!}\Gamma
^{3}\partial ^{3}+\cdots \right) ~\mathcal{I}_{h,\bar{h}}(z,\bar{z}) \\
&\simeq &\int dhd\bar{h}\left[ 1-(\partial \Gamma -R)+\partial \big( \Gamma
(\partial \Gamma -R)\big) -\frac{1}{2}\partial ^{2}\left( \Gamma
^{2}(\partial \Gamma -R)\right) +\cdots \right] \mathcal{I}_{h,\bar{h}}(z,%
\bar{z})\ ,
\end{eqnarray*}%
where $\partial $ denotes $\partial _{h}+\partial _{\bar{h}}$, and in the
second equation we have integrated by parts inside the integral over
conformal weights $h,\bar{h}$. On one hand, the standard OPE guarantees that
the Euclidean amplitude $\mathcal{A}$ is regular for small values of $z$ and $\bar{z}$.
Indeed, the leading contribution  behaves as $(z\bar{z})^{(d-2)/4}$ and
corresponds to the exchange of a scalar primary of lowest dimension allowed by the unitarity bound.
Then, from the comments at the end of section \ref{impactsection},
this implies that the coefficients of the above S--channel partial wave
expansion vanish for large $h,\bar{h}$ as fast as $(h\bar{h})^{(2-d)/2}$. 
On the other hand, the coefficients $R$ and the anomalous dimensions $\Gamma $ are computed in perturbation
theory with a leading contribution at order $g^{2}$. Therefore, the
consecutive terms in the last expression have increasing leading order in
the coupling $g^{2}$ and cannot cancel among themselves. We then conclude
that  
 \footnote{More precisely, $R-\partial \Gamma$ has to go to zero, for $h,\bar h \to \infty$,
at least as fast as $(h\bar h)^{(2-d)/2}$.}
\be
R\simeq \partial \Gamma \ , \label{RdelGamma}
\ee
to all orders in the coupling $g^2$.

In order to explore the consequences of the results of the previous
sections, we must analytically continue equation (\ref{generalex}) to find
the partial wave expansion of the Lorentzian amplitude $\hat{\mathcal{A}}=
\mathcal{A}^{\circlearrowleft }$. Using the perturbative form,
\[
\mathcal{A}(z,\bar{z})\simeq \sum \left( 1+\partial \Gamma \right) \left(
1+\Gamma \partial +\frac{1}{2}\Gamma ^{2}\partial ^{2}+\frac{1}{3!}\Gamma
^{3}\partial ^{3}+\cdots \right) \mathcal{S}_{h,\bar{h}}(z,\bar{z})\ ,
\]%
of equation (\ref{generalex}), we just need to compute the analytic
continuation
\[
\left[ \left( \partial _{h}+\partial _{\bar{h}}\right) ^{n}\,\mathcal{S}_{h,%
\bar{h}}(z,\bar{z})\right] ^{\circlearrowleft }\ .
\]%
This can be easily determined using the OPE expansion
\[
\mathcal{S}_{h,\bar{h}}(z,\bar{z})=z^{\eta -h}\,
\bar{z}^{\eta -\bar{h}%
}\sum_{n,\bar{n}\geq 0}z^{-n}\bar{z}^{-\bar{n}}c_{n,\bar{n}}(h,\bar{h}%
)\ \ \ +\ \ \ \left( z\leftrightarrow \bar{z}\right) \ ,
\]%
of the \emph{S}--channel partial waves around $z,\bar{z}\sim
\infty $. The differential operator
\[
\tilde{\partial} =  z^{-h}\bar{z}^{-\bar{h}}\,\partial \,z^{h}\bar{z}^{%
\bar{h}}=\partial +\ln (z\bar{z})\ ,
\]%
acting on $\mathcal{S}_{h,\bar{h}}$ for $h,\bar{h}\in \eta +\mathbb{N}_{0}$%
, is invariant under the analytic continuation $\circlearrowleft $.
Therefore,
\begin{eqnarray*}
\left[ \partial ^{n}\mathcal{S}\right] ^{\circlearrowleft } &=&\left[ \left(
\tilde{\partial}-\ln (z\bar{z})\right) ^{n}\mathcal{S}\right]
^{\circlearrowleft } \\
&=&\left( \tilde{\partial}-\ln (e^{\,2\pi i}z\bar{z})\right) ^{n}\,\mathcal{S%
} \\
&=&\left( \partial -2\pi i\right) ^{n}\,\mathcal{S}\ .
\end{eqnarray*}%
The Lorentzian amplitude $\hat{\mathcal{A}}=\mathcal{A}^{\circlearrowleft }$
is then given by
\[
\hat{\mathcal{A}}(z,\bar{z})\simeq \sum \left( 1+\partial \Gamma \right)
\left( 1+\Gamma (\partial -2\pi i)+\frac{1}{2}\Gamma ^{2}(\partial -2\pi
i)^{2}+\frac{1}{3!}\Gamma ^{3}(\partial -2\pi i)^{3}+\cdots \right) \mathcal{S}_{h,\bar{h}}(z,\bar{z})\ .
\]
Focusing in the small $z$ and $\bar{z}$ regime we can write
\[
\hat{\mathcal{A}}(z,\bar{z})\simeq \int dh d\bar{h}\left( 1-2\pi i\Gamma +
\frac{2\pi i}{2}(2\pi i+\partial )\Gamma ^{2}-\frac{2\pi i}{3!}(2\pi
i+\partial )^{2}\Gamma ^{3}+\cdots \right) \mathcal{I}_{h,\bar{h}}(z,\bar{z})\ ,
\]
where we have integrated by parts inside the integral over conformal
dimensions $h,\bar{h}$. In the large $h,\bar{h}$ limit we can neglect the
derivative $\partial  =  \partial _{h}+\partial _{\bar{h}}$ with respect
to the constant $2\pi i$, obtaining
\begin{equation}
\hat{\mathcal{A}}(z,\bar{z})\simeq \int dh d\bar{h}\,e^{-2\pi i\,\Gamma (h,\bar{h})}\,
\mathcal{I}_{h,\bar{h}}(z,\bar{z})\ .  \label{xyz22}
\end{equation}
Hence, in the impact parameter representation of the reduced Lorentzian amplitude $\hat{\mathcal{A}}$,
the anomalous dimensions $2\Gamma$ play the role of a phase shift.

\section{Impact Parameter Representation}

Now we wish to find an explicit form of the impact parameter representation for the
Lorentzian amplitude $\hat{A}$ in (\ref{zzz2}). First we recall the  result (\ref{eq1001}) derived 
in chapter \ref{ch:cpw}.
For $ {\bf p}  ,{\bf \bar{p} }$ in the past Milne wedge $-\mathrm{M}$, the impact parameter
partial wave $\mathcal{I}_{h,\bar{h}}$ admits the integral representation    \footnote{
The impact parameter representation derived in this section is valid in
general for $ {\bf p} = {\bf y} _{3}$ and $\bar{ {\bf p} }= {\bf y} _{2}$, with $ {\bf y} _{1}= {\bf y} _{4}=0$. The general
case is then related by a conformal transformation, whose precise form is
rather cumbersome, but reduces to $ {\bf p} \simeq  {\bf y} _{3}- {\bf y} _{1}$ and $ {\bf \bar{  p} }\simeq
 {\bf y} _{2}- {\bf y} _{4}$ for the case of interest $\left\vert  {\bf y} _{i}^a\right\vert \ll 1$.}
 over the future Milne wedge $\mathrm{M}$
\begin{align*}
\mathcal{I}_{h,\bar{h}} =\mathcal{N}_{\Delta _{1}}\mathcal{N}_{\Delta_{2}}
\left( - {\bf p}  ^{2}\right) ^{\Delta _{1}}\left( - {\bf \bar{  p} }^{2}\right)
^{\Delta _{2}}\int_{\mathrm{M}}
\frac{d {\bf y}  }{\left\vert  {\bf y}  \right\vert^{d-2\Delta _{1}}}\,
\frac{d {\bf \bar{  y} }}{\left\vert  {\bf \bar{  y} }\right\vert^{d-2\Delta _{2}}}\,
e^{-2 {\bf p}  \cdot  {\bf y}  -2 {\bf \bar{  p} }\cdot  {\bf \bar{  y} }}& \\
4h\bar{h}~\delta \left( 2 {\bf y}  \cdot  {\bf \bar{  y} }+h^{2}+\bar{h}^{2}\right) 
\,\delta \left(  {\bf y}  ^{2} {\bf \bar{  y} }^{2}-h^{2}\bar{h}^{2}\right)& \ ,
\end{align*}
where the cross ratios $z$ and $\bar{z}$ are related to $ {\bf p} , {\bf \bar{  p} }$ as in (\ref{cba}).
Expression (\ref{xyz22}) for the reduced amplitude becomes then
\begin{equation}
\hat{\mathcal{A}} \simeq \mathcal{N}_{\Delta _{1}}\mathcal{N}_{\Delta
_{2}}\left( - {\bf p}  ^{2}\right) ^{\Delta _{1}}\left( - {\bf \bar{  p} }^{2}\right)
^{\Delta _{2}}
\int_{\mathrm{M}}\frac{d {\bf y}  }{\left\vert  {\bf y}  \right\vert^{d-2\Delta _{1}}}\,
\frac{d {\bf \bar{  y} }}{\left\vert  {\bf \bar{  y} }\right\vert^{d-2\Delta _{2}}}
\,e^{-2 {\bf p}  \cdot  {\bf y}  -2 {\bf \bar{  p} }\cdot  {\bf \bar{  y} }}\,
e^{-2\pi i\,\Gamma( h ,\bar{h}) }\ ,
\label{impred}
\end{equation}
where $\Gamma(h,\bar{h})$ depends on $ {\bf y} , {\bf \bar{  y} }$ through
\begin{equation}
h^{2}\bar{h}^{2} =  {\bf y}  ^{2}\, {\bf \bar{  y} }^{2} \ , \ \ \ \ \ \ \ \ \ \ \ \ \ \ \ \ \  
h^{2}+\bar{h}^{2} = -2 {\bf y}  \cdot  {\bf \bar{  y} } \ .
\label{abc1}
\end{equation}
The fact that $\hat{\mathcal{A}}$ is uniquely a function of
the cross--ratios $z$ and $\bar{z}$, translates into the fact that the phase shift 
$\Gamma $ depends only on $ {\bf y}  ^{2}\, {\bf \bar{  y} }^{2}$ and $-2 {\bf y}  \cdot {\bf \bar{  y} }$.

To write the impact parameter representation for the full Lorentzian amplitude $\hat{A}$, 
consider first the boundary propagators in (\ref{zzz2}). For 
$ {\bf p}  , {\bf \bar{  p} }$ in the past Milne wedge $-{\rm M}$ we have
$$
\left(  {\bf p}  ^{2}+i\epsilon _{ {\bf p}  }\right) ^{-\Delta _{1}} = i^{2\Delta _{1}}\left( - {\bf p}  ^{2}\right) ^{-\Delta _{1}} \ ,
\ \ \ \ \ \ \ \ \ \ \ \ \ \ \ \ \ 
\left(  {\bf \bar{  p} }^{2}-i\epsilon _{ {\bf \bar{  p} }}\right) ^{-\Delta _{2}} =i^{-2\Delta _{2}}\left( - {\bf \bar{  p} }^{2}\right)^{-\Delta _{2}}\ ,
$$
where we recall that $\epsilon _{ {\bf p}  }=\epsilon {\rm sign}\left( - {\bf x} _0\cdot  {\bf p}  \right)$
 with $ {\bf x} _0\in\mathrm{M}$.
Rotating the radial part of the $ {\bf y} , {\bf \bar{  y} }$ integrals over the Milne wedges in (\ref{impred}), so that
$ {\bf y}  \to i {\bf y} $ and $ {\bf \bar{  y} }\to -i {\bf \bar{  y} }$,  (\ref{zzz2}) becomes
\begin{equation}
\hat{A}\left(  {\bf p}  , {\bf \bar{  p} }\right) \simeq
\frac{\mathcal{C}_{\Delta _{1}}\mathcal{C}_{\Delta _{2}}\mathcal{N}_{\Delta _{1}}\mathcal{N}_{\Delta _{2}} } 
{(2\omega\, i)^{2\Delta _{1}+2\Delta _{2}} }
\int_{\mathrm{M}}\frac{d {\bf y}  }{\left\vert  {\bf y}  \right\vert
^{d-2\Delta _{1}}}\,\frac{d {\bf \bar{  y} }}{\left\vert  {\bf \bar{  y} }\right\vert
^{d-2\Delta _{2}}}\,e^{2i {\bf p}  \cdot  {\bf y}  -2i {\bf \bar{  p} }\cdot  {\bf \bar{  y} }}\,e^{-2\pi i\,\Gamma ( h ,\bar{h})}\ .
\label{risultato}
\end{equation}
Although this representation was derived assuming $ {\bf p}  , {\bf \bar{  p} }$ in the past Milne wedge,
 we claim it is valid for
generic $ {\bf p}  , {\bf \bar{  p} }\in \mathbb{M}^d$. In fact,
for the $\Gamma=0$  non--interacting amplitude,
we recover the boundary propagators from 
the Fourier transform 
\begin{equation}
\mathcal{N}_{\Delta }\int_{\mathrm{M}}\frac{d {\bf y}  }{\left\vert  {\bf y} 
\right\vert ^{d-2\Delta }}~e^{\pm 2i {\bf p}  \cdot  {\bf y}  }=\frac{1}{\left(  {\bf p} 
^{2}\pm i\epsilon _{ {\bf p}  }\right) ^{\Delta }}\ ,
\label{suca1234}
\end{equation}
which we recall in some detail in appendix  \ref{app3}.

\section{Anomalous Dimensions of Double Trace Operators}     \label{anomdim}

We are now in position to use the AdS/CFT prediction given by
equations (\ref{eikEQ}) and (\ref{CFTamplitude}) to determine the phase
shift in the impact parameter representation (\ref{risultato}), and
therefore to compute the anomalous dimension of double trace primary
operators. First, replace (\ref{risultato}) in (\ref{CFTamplitude})
\begin{eqnarray*}
A_{eik}& \simeq &\omega ^{4}\,\left( 2\omega\, i\right) ^{-2\Delta _{1} -2\Delta _{2}}
\mathcal{C}_{\Delta _{1}}\mathcal{C}_{\Delta _{2}}
\mathcal{N}_{\Delta _{1}}\mathcal{N}_{\Delta _{2}}\\
&&\int dt_{1}\cdots
dt_{4}\,F\left( t_{1}\right) F\left( t_{2}\right) F^{\star }\left(
t_{3}\right) F^{\star }\left( t_{4}\right) \,e^{i\omega \left(
t_{3}-t_{1}\right) +i\omega \left( t_{4}-t_{2}\right) }\\
&&\int_{\mathrm{M}}\frac{d {\bf y}  }{\left\vert  {\bf y}  \right\vert^{d-2\Delta _{1}}}\,
\frac{d {\bf \bar{  y} }}{\left\vert  {\bf \bar{  y} }\right\vert^{d-2\Delta _{2}}}\,
e^{i\left( t_{3}-t_{1}\right) {\bf x}_0\cdot  {\bf y}  +i\left(t_{4}-t_{2}\right) {\bf x}_0\cdot  {\bf \bar{  y} }
-  i {\bf q}\cdot  {\bf y} /\omega - i {\bf \bar{q}}\cdot  {\bf \bar{  y} }/\omega    }\,e^{-2\pi i~\Gamma (h ,\bar{h} ) }\ .
\end{eqnarray*}
At high frequencies $\omega $, we have $t_{1}\sim t_{3}$ and $t_{2}\sim t_{4}$. Therefore, the
integrals over the sums $\frac{1}{2}\int d\left( t_{1}+t_{3}\right) F\left(
t_{1}\right) F^{\star }\left( t_{3}\right) $ and $\frac{1}{2}\int d\left(
t_{2}+t_{4}\right) F\left( t_{2}\right) F^{\star }\left( t_{4}\right) $ give
an overall factor of $2$ from the normalization (\ref{norm}). 
We are then left with the integrals over the differences, which give
\[
\left( 2\pi \right) ^{2}\,\delta \left( {\bf x}_0\cdot  {\bf y}  +\omega \right)
\delta \left( {\bf x}_0\cdot  {\bf \bar{  y} }+\omega \right) ~.
\]
It is easy to see that the integral in $ {\bf y}  $ in the future Milne
wedge $\mathrm{M}$ at fixed time component ${\bf x}_0\cdot  {\bf y}  $ is
equivalent to the integral over points ${\bf w}$ in the hyperboloid $H_{d-1}$,
with the change of coordinates
\begin{eqnarray*}
 {\bf y}   &=&-\frac{ \omega }{{\bf x}_0\cdot {\bf w}} \,{\bf w}   \ ,\\
\int_{\mathrm{M}}d {\bf y}  ~\delta \left( {\bf x}_0\cdot  {\bf y}  +\omega \right)
&=&2^{d}\,\omega ^{d-1}\int_{H_{d-1}}\frac{d{\bf w}}{\left( -2{\bf x}_0\cdot {\bf w}\right)^{d}}\ .
\end{eqnarray*}
We then get
\begin{eqnarray}
A_{eik}& \simeq &2\left( 2\pi \omega \right) ^{2}\,i^{-2\Delta _{1}-2\Delta _{2}}\mathcal{C}
_{\Delta _{1}}\mathcal{C}_{\Delta _{2}}\mathcal{N}_{\Delta _{1}}\mathcal{N}
_{\Delta _{2}} \label{lastaeik}   \\
&&\int_{H_{d-1}}\frac{d{\bf w}}{\left( -2{\bf x}_0\cdot {\bf w}\right) ^{2\Delta _{1}}}
\,\frac{d\bar{{\bf w}}}{\left( -2{\bf x}_0\cdot \bar{{\bf w}}\right) ^{2\Delta _{2}}}
\,\exp\left( i  \,\frac{{\bf q}\cdot {\bf w}}{{\bf x}_0\cdot {\bf w}}+i  \,\frac{{\bf \bar{q}}\cdot \bar{{\bf w}}}{
{\bf x}_0\cdot \bar{{\bf w}}}-2\pi i~\Gamma (h ,\bar{h}) \right) \ , \nonumber
\end{eqnarray}
where $\Gamma(h,\bar{h})$ depends on ${\bf w},\bar{{\bf w}}$ through
\begin{equation}
4h\bar{h} = \frac{\left( 2\omega \right) ^{2}}{\left( {\bf x}_0\cdot {\bf w}
\right) \left( {\bf x}_0\cdot \bar{{\bf w}}\right) } \ ,
\ \ \ \ \ \ \ \ \ \ \ \ \ \ \ 
\frac{\bar{h}}{h}+\frac{h}{\bar{h}} = -2 {\bf w}\cdot {\bf \bar{w}} \ .
\label{fromhhbartogeod}
\end{equation}
Finally, comparing equation (\ref{lastaeik})
with (\ref{eikEQ}), 
we obtain a prediction for the large $h,\bar{h}$ behavior of the anomalous
dimensions due to the AdS exchange of a spin $j$ particle of
dimension $\Delta $,
\begin{equation}
2\Gamma \left( h,\bar{h}\right) \simeq-\frac{g^{2}}{2\pi }\left( 4h\bar{h}\right)
^{j-1}~\Pi _{\perp }\left( h/\bar{h}\right) \ .
\label{adimresult}
\end{equation}
The transverse propagator $\Pi _{\perp }$  is the Euclidean scalar propagator on $H_{d-1}$ with dimension $\Delta -1$.
Its explicit form in terms of the hypergeometric 
function is
\begin{eqnarray*}
\Pi _{\perp }(h,\bar{h}) &=& \frac{1}{2\pi^{\frac{d}{2}-1}}\,\frac{\Gamma
\left( \Delta -1\right) }{\Gamma\left( \Delta -\frac{d}{2} +1\right) }
\,\left( \frac{\left( h-\bar{h}\right) ^{2}}{h\bar{h}}\right) ^{1-\Delta }\  \\
&&F\left( \Delta -1,\frac{2\Delta -d+1}{2},2\Delta -d+1,~-\frac{4h\bar{h}}{%
\left( h-\bar{h}\right) ^{2}}\right) \ .
\end{eqnarray*}%
In particular, in dimensions $d=2$ and $d=4$ the above expression simplifies
to%
\begin{eqnarray*}
\Pi _{\perp }(h,\bar{h}) &=&\frac{1}{2\left( \Delta -1\right) }
\left( \frac{h}{\bar{h}}\right) ^{1-\Delta}\ \ \ \ \ \ \ \ \ \ \ \ \ \ \ \left(d=2\right)\ ,  \\
&=&\frac{1}{2\pi }\frac{{h}^{2}}{h^{2}-\bar{h}^{2}}
\left( \frac{h}{\bar{h}}\right) ^{1-\Delta}\ \ \ \ \ \ \ \ \ \ \ \ \ \ \left( d=4\right)\ .
\end{eqnarray*}
We remark that, although we have considered contributions to the four point amplitude from all orders in the 
AdS coupling $g$, the anomalous dimensions just obtained are determined
by the tree level result alone.
In other words, the loop corrections to the
anomalous dimensions of primary operators with large $h,\bar{h}$
are subleading with respect to the tree level contribution. This
is analogous to the flat space statement that the loop
corrections to the phase shift of large spin partial waves are
subleading with respect to the tree level contribution.

Expressions (\ref{fromhhbartogeod}) are very suggestive of the physical process we are describing.
From the AdS/CFT correspondence, a composite primary operator 
$\mathcal{O}_{1}\partial \cdots \partial \mathcal{O}_{2}$ is dual to a two particle state in AdS,
whose energy and spin are given, respectively, by the  full dimension $E+2\Gamma$ and spin $J$ of the CFT dual operator.
In general, AdS states are hard to describe. However, it is natural to expect that a classical description emerges
in the limit of large charges $E$ and $J$, which we are considering.
Indeed, from (\ref{fromhhbartogeod}), we conclude that the AdS state in question simply corresponds
to two highly energetic particles approximately following two null geodesics as in figure \ref{nullgeodincylinder}
and bouncing repeatedly on the AdS boundary.
The null geodesics are precisely given  by (\ref{p}) and (\ref{pbar}) with $v=0$ and $\bar{u}=0$, i.e.
$$
{\bf x}= {\bf w} + \lambda \,{\bf k}\ ,\ \ \ \ \ \ \ \ \ \ \ \ \ 
{\bf \bar{x}}= {\bf \bar{w}} + \bar{\lambda}\, {\bf\bar{ k}}\ ,
$$
where the null momenta satisfies
$$
s  = -2{\bf k} \cdot  {\bf\bar{ k}} =4h\bar{h}=E^{2}-J^{2}
$$
and the  transverse impact parameter $r$ is determined by
\begin{equation}
2\cosh r = -2 {\bf w}\cdot {\bf \bar{w}} 
=\frac{\bar{h}}{h} + \frac{h}{\bar{h}}=\frac{E^2+J^2}{E^2-J^2}\ .
\label{imppar}
\end{equation}
It is easy to verify that these two null geodesics carry energy $E$ 
with respect to the global time $\tau$ defined by rotations in the timelike plane
containing ${\bf x}_0$ and ${\bf x}_1=( {\bf k}_1 + {\bf k}_2)/(2\omega)$.
Also, they have total angular momentum $J$ along a spacelike surface of constant $\tau$.
This picture naturally leads to the heuristic explanation, given in the Outline of the Thesis,
of the relation between the eikonal phase shift in AdS and the anomalous dimensions
of double trace operators with large $h,\bar{h}$.

We emphasize that, for large $h,\bar h$, the anomalous dimensions (\ref{adimresult}) are
dominated by the AdS particles with highest spin.
Moreover, when $h\gg\bar h$ the lightest particle of maximal spin
determines $\Gamma$, since in this limit the propagator decays as $\Pi_\perp \sim (h/\bar h)^{1-\Delta}$.
In theories with a gravitational description, this particle is 
\emph{the graviton}. This yields a universal prediction for CFT's
with AdS duals in the gravity limit 
\be 
2\Gamma (h,\bar{h})\simeq -G\, \frac{4\,\Gamma\left( \frac{d-1}{2}\right) }{d\,\pi^{\frac{d-1}{2} }}\,
 \frac{ (2\bar{h})^d}{h^{d-2}}\ \ \ \ \ \ \ \ \ \ 
\left( h\sim\bar{h}\rightarrow \infty \ ,\ \ \ \ h\gg\bar{h} \right)\ .
\label{sucabis}
\ee 
This reduces to the result (\ref{resSYM4}) stated in the Outline for $\mathcal{N}=4$ SYM in 4 dimensions.

\vspace{2cm}

\section{\emph{T}--channel Decomposition} \label{sec:tchannel}

\begin{figure}
\begin{center}
\includegraphics[width=4cm]{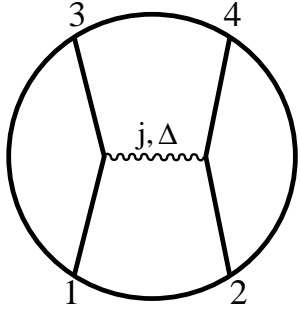}
\caption{Witten diagram representing the \emph{T}--channel exchange of an AdS particle with spin $j$ and 
dimension $\Delta$.}
\label{AdStchannel}
\end{center}
\end{figure}

We have found that the eikonal approximation in AdS determines the small $z$ and $\bar{z}$ 
behavior of the reduced Lorentzian amplitude $\hat{\mathcal{A}}$.
In the previous sections we have explored this result using  the  \emph{S}--channel partial 
wave expansion.
We shall now study the \emph{T}--channel partial wave decomposition of the
tree--level diagram in figure \ref{AdStchannel}.
The corresponding  Euclidean amplitude $\mathcal{A}_{1}$ can be expanded in 
\emph{T}--channel partial waves,
\begin{equation}
\mathcal{A}_{1}=\sum\ \mu_{h,\bar{h}}~\mathcal{T}_{h,\bar{h}}\ .
\label{AinT}
\end{equation}
On the other hand, from the term of order $g^2$ in   (\ref{impred}), we have
\begin{equation*}
\mathcal{A}_{1}^{\circlearrowleft}\simeq i\,g^2\,2^{2j-3}\,\mathcal{N}_{\Delta _{1}}\mathcal{N}_{\Delta
_{2}}\left( - {\bf p}  ^{2}\right) ^{\Delta _{1}}\left( - {\bf \bar{  p} }^{2}\right)
^{\Delta _{2}}
\int_{\mathrm{M}}\frac{d {\bf y}\, d {\bf \bar{  y} } \,e^{-2 {\bf p}  \cdot  {\bf y}  -2 {\bf \bar{  p} }\cdot  {\bf \bar{  y} }}  }
{\left\vert  {\bf y}  \right\vert^{d-2\Delta _{1} +1-j}    \,
\left\vert  {\bf \bar{  y} }\right\vert^{d-2\Delta _{2} +1-j}       }\,
\Pi_{\perp}\left(\frac{{\bf y} }{\left\vert  {\bf y}  \right\vert }   , 
\frac{{\bf \bar{ y}} }{\left\vert  {\bf\bar{  y}}  \right\vert } \right)\ ,
\end{equation*}
where we recall the expressions $z\bar{z} \simeq {\bf p}^{2}{\bf \bar{p}}^{2}$ and
$z+\bar{z} \simeq 2{\bf p} \cdot {\bf \bar{p}}$ relating the cross ratios to the points
${\bf p}$ and ${\bf \bar{p}}$ in the past Milne wedge $-{\rm M}$.
Performing the radial integrals over $\left\vert  {\bf y}  \right\vert$ and 
$\left\vert  {\bf \bar{  y} }\right\vert$ one obtains
\begin{equation}
\mathcal{A}_{1}^{\circlearrowleft}\simeq  i\,g^2\,\mathcal{K}
\left( z\bar{z}\right)^{(1-j)/2}
\int_{H_{d-1}} d {\bf w}\, d {\bf \bar{ w}} \,\frac{ \Pi_{\perp}\left( {\bf w} , {\bf \bar{ w}} \right)}
{ \left(-2 {\bf e}  \cdot  {\bf w}\right)^{2\Delta _{1} -1+j} 
\left( -2 {\bf \bar{  e} }\cdot  {\bf \bar{  w} }\right)^{2\Delta _{2} -1+j }  }\ ,
\label{treeleveleikonal}
\end{equation}
with  the constant $\mathcal{K}$ given by
\begin{equation}
\mathcal{K}=2^{2j-3}\,\mathcal{N}_{\Delta _{1}}\mathcal{N}_{\Delta_{2}}
\,\Gamma(2\Delta _{1} -1+j )\,\Gamma(2\Delta _{2} -1+j )     \label{constantK}
\end{equation}
and $ {\bf e},{\bf \bar{  e} } \in H_{d-1}$ defined by
$$
{\bf e} = -\frac{ {\bf p} }{|{\bf p}|} \ ,\ \ \ \ \ \ \ \ \
{\bf \bar{  e} } =- \frac{  {\bf \bar{  p} }  }{  |{\bf \bar{  p} }| } \ ,
$$
so that 
$$
-2 {\bf e} \cdot {\bf \bar{  e} } = \sqrt{\frac{z }{\bar{z}  }} +   \sqrt{\frac{\bar{z}  }{z }}\ .
$$
In order to understand the expansion of (\ref{treeleveleikonal}) in 
\emph{T}--channel partial waves,
\begin{equation}
\mathcal{A}_{1}^{\circlearrowleft}=\sum\ \mu_{h,\bar{h}}~\mathcal{T}_{h,\bar{h}}^{\circlearrowleft} \ ,
\label{LorAinT}
\end{equation}
we must first analyze in detail the behavior of the functions
$\mathcal{T}_{h,\bar{h}}\left(  z,\bar{z}\right)  $ as we rotate the point $z$
anti--clockwise around $0,1$, keeping $\bar{z}$ fixed. Since the
eikonal result (\ref{treeleveleikonal}) holds around $z,\bar{z}\sim0$, we shall only need
the leading behavior of the Lorentzian \emph{T}--channel partial waves
$\mathcal{T}_{h,\bar{h}}^{\circlearrowleft}$
around the origin.

Consider first the behavior of $\mathcal{T}_{h,\bar{h}}$ in the limit $\bar
{z}\rightarrow0$, with $z$ fixed. The operator $D_{T}$ in (\ref{Tchannel}) reduces to
\[
z^{2}\left(  1-z\right)  \partial^{2}-z^{2}\partial+\bar{z}^{2}\bar{\partial
}^{2}-\left(  d-2\right)  \bar{z}\bar{\partial}~.
\]
Using the boundary condition $\mathcal{T}_{h,\bar{h}}\sim\left(  -z\right)
^{h}\left(  -\bar{z}\right)  ^{\bar{h}}$ around the origin, we conclude that
\[
\mathcal{T}_{h,\bar{h}}\sim\left(  -\bar{z}\right)  ^{\bar{h}}\left(
-z\right)  ^{h}F\left(  h,h,2h, z\right)  ~.
\]
Since
$$
\left(  -z \right)  ^{h}F \left(  h,h,2h, z \right)    =\frac{\Gamma\left(
2h\right)  }{\Gamma\left(  h\right)  ^{2}}\sum_{n\geq0}\frac{\left(  h\right)
_{n}\left(  1-h\right)  _{n}}{\left(  n!\right)  ^{2} \,z^n}
\Big[  \ln\left(  -z\right)  +2\psi\left(  n+1\right)  -\psi\left(
h+n\right)  -\psi\left(  h-n\right)  \Big]
$$
we obtain that
\[
\mathcal{T}_{h,\bar{h}}^{\circlearrowleft} \sim 2\pi i \frac{\Gamma\left(
2h\right)  }{\Gamma\left(  h\right)  ^{2}}\left(  -\bar{z}\right)
^{\bar{h} }F\left(  h,1-h,1 , \frac{1}{z} \right) ~.
\]
In the limit of small $z$ the leading behavior is
\begin{equation}
\mathcal{T}_{h,\bar{h}}^{\circlearrowleft} \sim 2\pi i\frac{\Gamma\left(
2h\right)  \Gamma\left(  2h-1\right)  }{\Gamma\left(  h\right)  ^{4}}~\left(
-z\right)  ^{1-h}\left(  -\bar{z}\right)  ^{\bar{h}}~.
\label{bcDiscT}
\end{equation}
Recall that we derived this result in the limit $\bar{z}\rightarrow0$. To
understand the general behavior around $z,\bar{z}\sim0$ of
$\mathcal{T}_{h,\bar{h}}^{\circlearrowleft}$, we expand $\mathcal{T}
_{h,\bar{h}}$ in powers of $\bar{z}$ as
\begin{equation}
\mathcal{T}_{h,\bar{h}}\sim\sum_{n\geq0}\left(  -\bar{z}\right)  ^{\bar{h}
+n}g_{n}\left(  z\right)  ~.\label{eqTex}
\end{equation}
We have just determined that $g_{0}^{\circlearrowleft} \sim z^{1-h}$ for
small $z$. The other functions $g_{n}$ are determined recursively by expanding
the differential equation $D_{T}=c_{h,\bar{h}}$ in powers of $\bar{z}$. A
rather cumbersome but straightforward computation shows that
$g_{n}^{\circlearrowleft} \sim z^{1-h-n}$ for small $z$. Therefore, we
conclude that in general
\begin{equation}
\mathcal{T}_{h,\bar{h}}^{\circlearrowleft}\simeq \left( -z\right)
^{1-h}\left(  -\bar{z}\right)  ^{\bar{h}}~G_{h,\bar{h}}\left( \frac{\bar{z}
}{z}\right) \label{discT}
\end{equation}
around $z,\bar{z}\sim0$. The function $G_{h,\bar{h}}\left(  w\right)  $ is
regular around $w=0$ and, using (\ref{bcDiscT}), 
satisfies $G_{h,\bar{h}}\left(  0\right) = 2\pi i\Gamma\left(  2h\right)  \Gamma\left(  2h-1\right)  /\Gamma\left(  h\right)^{4}$.
 The careful reader will have noticed that in equation (\ref{eqTex}) 
we have implicitly neglected to symmetrize in $z\leftrightarrow\bar{z}$. 
Had we not, the function $\mathcal{T}_{h,\bar{h}}$ would have had an extra
contribution of the form $\sum_{n\geq0}\left(  -\bar{z}\right)  ^{h+n}
f_{n}\left(  z\right)  $ with $f_{n}^{\circlearrowleft}\sim
z^{1-\bar{h}-n}$. These terms then give sub--leading contributions to
(\ref{discT}).

To compute explicitly the function $G_{h,\bar{h}}$, we consider the operator
$D_{T}$ in (\ref{Tchannel}) near $z,\bar{z}\sim0$, which reduces to
\[
z^{2}\partial^{2}+\bar{z}^{2}\bar{\partial}^{2}+\left(  d-2\right)
\frac{z\bar{z}}{z-\bar{z}}\left(  \partial-\bar{\partial}\right)  ~.
\]
Acting on (\ref{discT}) the differential equation $D_{T}=c_{h,\bar{h}}$
becomes of the hypergeometric form
\[
2w\left(  1-w\right)  G^{\prime\prime}+\left[  2\left(  h+\bar{h}\right)
\left(  1-w\right)  -\left(  d-2\right)  \left(  1+w\right)  \right]
G^{\prime}=\left(  d-2\right)  \left(  h+\bar{h}-1\right)  G~,
\]
in terms of $w=\bar{z}/z$. We then arrive at the result
\begin{equation}
G_{h,\bar{h}}= 2\pi i\frac{\Gamma\left(  2h\right)  \Gamma\left(  2h-1\right)
}{\Gamma\left(  h\right)  ^{4}}~F \left(  \frac{d}{2}-1,h+\bar{h}-1,h+\bar {h}+1-\frac{d}{2} ,
 \frac{\bar{z}}{z}\right)  ~. \label{Ghhbar}
\end{equation}
It is interesting to notice that using the identity
$$
F(a,b,b-a+1,w)=\left(1-\sqrt{w}\right)^{-2b}
F\left( b,b-a+\frac{1}{2},2b-2a+1,-\frac{ 4\sqrt{w} }{ \left( 1-\sqrt{w} \right)^2 }\right) \ ,
$$
we obtain  the scalar propagator $\Pi_{\perp,E-1}$ of dimension $E-1$ in the transverse hyperboloid $H_{d-1}$,
\begin{equation}
\mathcal{T}_{h,\bar{h}}^{\circlearrowleft}\simeq  
 i\,\mathcal{D}\, \left(z\bar{z} \right)^{(1-J)/2}\,
\Pi_{\perp,E-1}({\bf e},{\bf \bar{  e} }  )\ ,
 \label{Trotated}
\end{equation}
where the constant $\mathcal{D}$ is given by
$$
\mathcal{D}= \frac{4\pi^{\frac{d}{2}} \Gamma\left(E-\frac{d}{2}+1\right)\Gamma\left(  E+J\right) 
 \Gamma\left(  E+J-1\right)
}{  \Gamma\left(  E-1\right) \Gamma\left( ( E+J)/2\right)  ^{4}}\ .
$$

We are now in position to determine the implications of the eikonal result
(\ref{treeleveleikonal}) for the \emph{T}--channel expansion coefficients $\mu_{h,\bar
{h}}$ in (\ref{AinT}).
\begin{figure}
[ptb]
\begin{center}
\includegraphics[width=6cm]{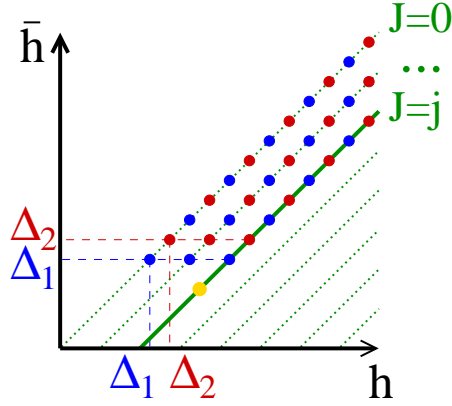} 
\caption{\emph{T}--channel decomposition of the spin
$j$ exchange graph in figure \ref{AdStchannel}. Only partial waves
with spin $J\leq j$ contribute. The eikonal amplitude determines
the contributions with $J=j$. 
Along this line of spin $J=j$, there is the operator dual
to the exchanged AdS particle, with dimension $\Delta$,
and the composites
$\mathcal{O} _{1}\partial_{\mu_1}\cdots \partial_{\mu_j} \partial^{2n}\mathcal{O}_{1}$ and
$\mathcal{O}_{2} \partial_{\mu_1}\cdots \partial_{\mu_j} \partial^{2n} \mathcal{O}_{2}$, with
dimensions $2\Delta _{1}+j+2n$ and
$2\Delta_{2}+j+2n$, respectively.}
\label{tlattice}
\end{center}
\end{figure}
We immediately conclude from (\ref{Trotated}) that, in the decomposition
(\ref{AinT}), only partial waves with
\[
J=h-\bar{h}\leq j
\]
can appear, as shown in figure \ref{tlattice}. Moreover, the coefficients
$\mu_{h,\bar{h}}$ for $J=h-\bar{h}=j$, which we denote by $\mu_E$, are determined directly from
the expansion 
\begin{equation}
\sum_{E}\mu_{E}\,\mathcal{D}\,\Pi_{\perp,E-1}({\bf e},{\bf \bar{  e} }  )=
 g^2\,\mathcal{K}
\int_{H_{d-1}} d {\bf w}\, d {\bf \bar{ w}} \,\frac{ \Pi_{\perp,\Delta-1}\left( {\bf w} , {\bf \bar{ w}} \right)}
{ \left(-2 {\bf e}  \cdot  {\bf w}\right)^{2\Delta _{1} -1+j} 
\left( -2 {\bf \bar{  e} }\cdot  {\bf \bar{  w} }\right)^{2\Delta _{2} -1+j }  }\ .
\label{mucoef}
\end{equation}

\begin{figure}
[ptb]
\begin{center}
\includegraphics[width=12cm]{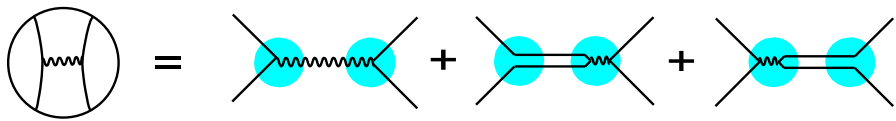}
\caption{\emph{T}--channel decomposition of the graph in figure  \ref{AdStchannel}.
The contributions come from the operator dual to the exchanged particle as well
as from the $\mathcal{O}_{1} \partial_{\mu_1}\cdots \partial_{\mu_j} \partial^{2n} \mathcal{O}_{1}$ 
and $\mathcal{O}_{2} \partial_{\mu_1}\cdots \partial_{\mu_j} \partial^{2n} \mathcal{O}_{2}$ composites.}
\label{texchanges}
\end{center}
\end{figure}

We recall that both sides of the last equation, 
are, by construction, only a function of the invariant chordal distance $ ({\bf e} - {\bf \bar{  e} })^2$.
It is then natural to consider their expansion in harmonic functions on $H_{d-1}$ which we
review in Appendix \ref{harmony}.
We shall consider the regular  eigenfunctions $\Omega_{\nu}({\bf e},{\bf \bar{  e} })$ of the
Laplacian operator,
$$
\Box_{H_{d-1}} \Omega_{\nu}({\bf e},{\bf \bar{  e} }) =
 -\left( \nu^2 + \frac{(d-2)^2}{4} \right) \Omega_{\nu}({\bf e},{\bf \bar{  e} }) \ ,
$$
which form a basis labeled by $\nu\in \mathbb{R}$.
Any  \footnote{Square integrable function on $H_{d-1}$.} function $F({\bf e},{\bf \bar{  e} })$ 
of the invariant $ ({\bf e} - {\bf \bar{  e} })^2$ can then be expanded
in this basis,
$$
F({\bf e},{\bf \bar{  e} })=\int_{-\infty}^{\infty}d\nu \,f(\nu)\, \Omega_{\nu}({\bf e},{\bf \bar{  e} }) \ ,
$$
with the transform $f(\nu)$ given by
$$
f(\nu)=\frac{1}{ \Omega_{\nu}({\bf e},{\bf  e }) } \int_{H_{d-1}} d {\bf  \bar{  e}}
\,F({\bf e},{\bf \bar{  e} })\,\Omega_{\nu}({\bf e},{\bf \bar{  e} })\ .
$$
In particular, the scalar propagator $\Pi_{E-1}$ has the simple decomposition
$$
\Pi_{\perp,E-1}({\bf e},{\bf \bar{  e} })= \int_{-\infty}^{\infty}d\nu
 \frac{ \Omega_{\nu}({\bf e},{\bf \bar{  e} }) }{\nu^2+\left(E- d/2 \right)^2}\ .
$$
Moreover, the right hand side of (\ref{mucoef}) is a convolution of three functions on $H_{d-1}$ and therefore
its transform is just the product of the transforms of these functions.
The transform of equation (\ref{mucoef}) is then 
\begin{equation}
\sum_{E}\frac{ \mu_{E} \, \mathcal{D}}{\nu^2+\left(E- d/2 \right)^2}=
 g^2\,\mathcal{K}\,
\frac{ \varphi_{2\Delta_1+j-1}(\nu) \,\varphi_{2\Delta_2+j-1}(\nu)  }{ \nu^2+\left(\Delta- d/2 \right)^2 }\ ,
\label{transfmucoef}
\end{equation}
where
$$
\varphi_{a}(\nu)= \frac{1}{ \Omega_{\nu}({\bf e},{\bf  e }) } \int_{H_{d-1}} d {\bf  w}
\,\Omega_{\nu}({\bf e},{\bf w })\, \left(-2 {\bf e}  \cdot  {\bf w}\right)^{-a}\ .
$$
Therefore, each pair of poles at $\nu=\pm i (E-d/2)$ in the right hand side of (\ref{transfmucoef}),
corresponds to a \emph{T}--channel exchange of a primary of spin $j$ and dimension $E$.
In the $d=2$ case, $\varphi_{a}(\nu)$  is given by a simple Fourier transform
\begin{equation*}
\varphi_{a}(\nu)=2\int_0^{\infty}dr \cos(\nu r) \left( 2\cosh r\right)^{-a}
= \frac{\Gamma\left( \frac{a+i\nu}{2}\right)\Gamma\left( \frac{a-i\nu}{2}\right) }{ 2\, \Gamma(a)}\ .
\end{equation*}
In  Appendix \ref{harmony} we show that, for generic $d$, the only singularities of the function $\varphi_a(\nu)$ 
are  simple poles at $\nu=\pm i(a+1-d/2+2n)$ with $n=0,1,2,\cdots$.
We conclude that the \emph{T}--channel decomposition of the 
tree--level exchange of an AdS particle with spin $j$ and dimension $\Delta$,
includes several primary operators of spin $J=j$. Namely,
the dual operator to the AdS exchanged particle, with dimension $\Delta$,
and the composites $\mathcal{O} _{1}\partial_{\mu_1}\cdots \partial_{\mu_j} \partial^{2n}\mathcal{O}_{1}$ and
$\mathcal{O}_{2} \partial_{\mu_1}\cdots \partial_{\mu_j} \partial^{2n} \mathcal{O}_{2}$, for $n=0,1,2,\cdots$, with
dimensions $2\Delta _{1}+j+2n$ and
$2\Delta_{2}+j+2n$, respectively.
The expansion coefficients $ \mu_{E} $ are determined by matching the residues of the
$\nu$ poles in equation (\ref{transfmucoef}).
These results are summarized in figures \ref{tlattice} and \ref{texchanges}.

\section{An Example in $d=2$}         \label{SectExample}

To be more explicit we include in this chapter a simple example where we
can check our results. We shall consider the case $d=2$, $j=0$ and $\Delta=2$
corresponding to a massless scalar exchange in AdS$_3$.
The basic tree level amplitude $\mathcal{A}_{1}$ is given by
\[
\frac{\mathcal{C}_{\Delta_{1}}\mathcal{C}_{\Delta_{2}}}{\mathbf{p}
_{13}^{\Delta_{1}}\mathbf{p}_{24}^{\Delta_{2}}}\mathcal{A}_{1}=-ig\int
_{\mathrm{AdS}} d\mathbf{x}\frac{\mathcal{C}_{\Delta_{1}}
^{\,2}}{\left(  -2\mathbf{x\cdot p}_{1}\right)  ^{\Delta_{1}}\left(
-2\mathbf{x\cdot p}_{3}\right)  ^{\Delta_{1}}}~\mathbf{h}\left(
\mathbf{x}\right)  ~,
\]
where
\[
\mathbf{h}\left(  \mathbf{x}\right)  =-ig \int_{\mathrm{AdS}
}d\mathbf{\bar{x}}~\Pi \left(  \mathbf{x,\bar{x}}\right) \
\frac{\mathcal{C}_{\Delta_{2}}^{\,2}}{\left(  -2\mathbf{\bar{x}\cdot
p}_{2}\right)
^{\Delta_{2}}\left(  -2\mathbf{\bar{x}\cdot p}_{4}\right)  ^{\Delta_{2}}}
\]
and $\Pi\left( \mathbf{x,\bar{x}}\right)$ is the massless propagator in $\mathrm{AdS}_3$.
We shall concentrate, in particular, on the simple case $\Delta_{2}=2$, 
so that the scalar field dual to the operator $\mathcal{O}_{2}$ is massless in $\mathrm{AdS}_{3}$. 
In this case we can use the general technique in  \cite{Rastelli2} and easily compute
\[
\mathbf{h}\left( \mathbf{x}\right) = -\frac{g}{16 \pi ^{2}
}\frac{1}{~\mathbf{p}_{24}\left( -2\mathbf{x\cdot
p}_{2}\right)  \left( -2\mathbf{x}\cdot\mathbf{p}_{4}\right)  }~,
\]
where we have used $\mathcal{C}_{\Delta}=1/\left( 2\pi\right)$ for $d=2$. In terms of the standard D--functions
\[
D_{\Delta_{i}}^{d}\left(  \mathbf{p}_{i}\right)  =\int_{H_{d+1}
}\mathbf{~}~\frac{\mathbf{d\mathbf{x}}}{
{\textstyle\prod\nolimits_{i}}
\left(  -2\mathbf{x\cdot p}_{i}\right)  ^{\Delta_{i}}}~,~
\]
reviewed in appendix \ref{Dfunctions}, we conclude, after Wick rotation
\[
\int_{\mathrm{AdS}_{d+1}} d\mathbf{x}\rightarrow-i\int
_{H_{d+1}}d\mathbf{x}~,
\]
that the exact tree level amplitude is given by
\begin{equation}
\mathcal{A}_{1}=\frac{g^2}{16 \pi ^{2} }~\mathbf{p}_{13}^{\Delta_{1}}\mathbf{p}
_{24}~D_{\Delta_{1},\Delta_{1},1,1}^{2}\left(  \mathbf{p}_{1},\mathbf{p}
_{3},\mathbf{p}_{2},\mathbf{p}_{4}\right)  \ . \label{ampex}
\end{equation}

On the other hand, we may also explicitly compute the integral 
in (\ref{treeleveleikonal}), which we claim to control the leading behavior as $z,\bar{z}\rightarrow0$
of the Lorentzian amplitude $\hat{\mathcal{A}}_{1}= \mathcal{A}_{1}^{\circlearrowleft}$. 
For $\Delta_{2}=2$, $j=0$ we can use again the methods of  \cite{Rastelli2} to explicitly perform 
the ${\bf \bar{w}}$--integral in (\ref{treeleveleikonal}). We then arrive at the result
\[
\mathcal{A}_{1}^{\circlearrowleft}\simeq i\,g^2\,\frac{\Gamma\left(  2\Delta_{1}-1\right)  }
{8 \Gamma\left(  \Delta_{1}\right)  ^{2}}\,
\sqrt{z\bar{z}}
\,D_{2\Delta_{1}-1,1}^{0}\left({\bf e},{\bf \bar{e}}\right)  ~.
\]
Using the explicit form of $D_{2\Delta_{1}-1,1}^{0}$ given in appendix \ref{Dfunctions}, 
we conclude that 
\begin{equation}
\mathcal{A}_{1}^{\circlearrowleft}\simeq - \frac{i \,g^2}{16(2\Delta_{1}-1)}\,\bar{z}\,
 F \left(  1,\Delta_{1}
,2\Delta_{1} ,\frac{z-\bar{z}}{z}\right) , \label{sss1}
\end{equation}
where $F$ is the standard hypergeometric function.

Now we verify that the eikonal limit of the exact tree level amplitude (\ref{ampex}) is indeed
our result (\ref{sss1}).
We shall restrict our attention to the special case $\Delta_{1}=2$, 
where the amplitude (\ref{ampex}) can be explicitly computed  \cite{Bianchi}
\begin{equation}
\mathcal{A}_{1}\left(  z,\bar{z}\right)
=-\frac{g^2}{16\pi} \frac{z^{2}\bar{z}^{2}}{\left(
\bar{z}-z\right)  }\left[  \frac{1}{1-\bar{z}}\partial-\frac{1}{1-z}
\bar{\partial}\right]  a\left(  z,\bar{z}\right)  ~, \label{explicitamp}
\end{equation}
with
\[
a\left(  z,\bar{z}\right)  =\frac{\left(  1-z\right)  \left(
1-\bar {z}\right)  }{\left(  z-\bar{z}\right)  }\left[
\mathrm{Li}_{2}\left( z\right)  -\mathrm{Li}_{2}\left(
\bar{z}\right)  +\frac{1}{2}\mathrm{\ln }\left(  z\bar{z}\right)
\mathrm{\ln}\left(  \frac{1-z}{1-\bar{z}}\right) \right]  ~
\]
and  $\mathrm{Li}_{2}\left(  z\right)  $  the standard dilogarithm. 
Using the symmetry $a\left( z,\bar{z}\right) = a\left( z^{-1},\bar{z}^{-1}\right)$, 
we quickly deduce that
\[
a^{\circlearrowleft}\left( z,\bar{z}\right) =i\pi
\frac{\left( 1-z\right) \left( 1-\bar{z}\right)  }{\left(
z-\bar
{z}\right) }\;\mathrm{\ln}\left( \frac{1-z}{1-\bar{z}}\cdot\frac{\bar{z}}
{z}\right)  \ .
\]
Applying to $a^{\circlearrowleft}\left( z,\bar{z}\right)$ the differential operator relating 
$a\left( z,\bar{z}\right)$ with $\mathcal{A}_{1}\left( z,\bar{z}\right)$, 
we obtain an exact expression for the Lorentzian amplitute
\begin{equation*}
\mathcal{A}_{1}^{\circlearrowleft} =-\frac{ig^2}{16}\frac{z\bar{z}}{\left( z-\bar{z}\right)
^{3}}\left[
z^{2}-\bar{z}^{2}+\ln\left( \frac{1-z}{1-\bar{z}}\cdot\frac{\bar{z}}
{z}\right) \left( 2z\bar{z}-z\bar{z}^{2}-z^{2}\bar{z}\right)
\right]  \ .
\end{equation*}
For small $z$ and $\bar{z}$ the above expression simplifies to 
\[
\mathcal{A}_{1}^{\circlearrowleft} \simeq -\frac{ig^2}{16}\frac{z\bar{z}}{\left( z-\bar{z}\right)
^{3}}\left[ z^{2}-\bar{z}^{2}+2z\bar{z} \ln\left(\bar{z}/z\right) \right]  \ ,
\]
which is exactly (\ref{sss1}) for $\Delta_{1}=2$.

Finally let us consider the expansion of $\mathcal{A}_{1}\,$ in the
$S$--channel.  Denoting by $\sigma_{h,\bar{h}}$ and
$\rho_{h,\bar{h}}$ the contribution of the tree--level graph to
$\Gamma(h,\bar{h})$ and $R(h,\bar{h})$ in equation (\ref{generalex}), we can write
\begin{equation*}
\mathcal{A}_{1}=\sum_{\eta\leq\bar{h}\leq h}\ \sigma_{h,\bar{h}}\left(
\frac{\partial}{\partial h}+\frac{\partial}{\partial\bar{h}}\right)
\mathcal{S}_{h,\bar{h}}+\sum_{\eta\leq\bar{h}\leq h}\ \rho_{h,\bar{h}
}~\mathcal{S}_{h,\bar{h}}\ . 
\end{equation*}
Using a symbolic manipulation program we can check this decomposition
of the amplitude (\ref{explicitamp}) up to very high order, with
\begin{align*}
\sigma_{h,\bar{h}}  & =-\frac{g^2}{32\,h\left(  h-1\right)  }~,\\
\rho_{h,\bar{h}}  & =\left(  \frac{\partial}{\partial h}+\frac{\partial
}{\partial\bar{h}}\right)  \sigma_{h,\bar{h}}+\frac{g^2}{32\,h\left(  h-1\right)
\bar{h}\left(  \bar{h}-1\right)  }~.
\end{align*}
In the limit of large dimensions $h,\bar{h} \rightarrow\infty$, 
we have the anomalous dimensions $2\sigma_{h,\bar{h}}\sim-g^2/(16\, h^{2})$, as
predicted from the general formula (\ref{adimresult}). Also, in the same
limit, we have $\rho_{h,\bar{h}}\simeq\partial
\sigma_{h,\bar{h}}$, in agreement with (\ref{RdelGamma}).

\section{Graviton Dominance}

\begin{figure}
\begin{center}
\includegraphics[width=12cm]{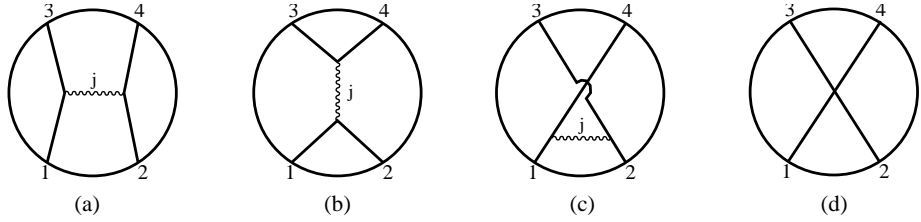}
\caption{Some possible interactions at tree--level in AdS.
When decomposing the amplitude in the \emph{S}--channel, only graph (a), 
with maximal spin $j=2$, dominates the dynamics at large intermediate spin and dimension.} 
\label{AdSdiagrams}
\end{center}
\end{figure}

We have analyzed in great detail the tree--level exchange of a
spin $j$ particle in the \emph{T}--channel, given by graph \ref{AdSdiagrams}(a).
In section \ref{anomdim}, we have determined the \emph{S}--channel partial wave decomposition
controlling the small $z$ and $\bar{z}$ behavior of the corresponding Lorentzian amplitude,
\begin{equation}
\hat{\mathcal{A}}_1(z,\bar{z})\simeq -2\pi i \sum_{\eta\leq\bar{h}\leq h} \Gamma (h,\bar{h}) \,
\mathcal{S}_{h,\bar{h}}(z,\bar{z})\ , \label{TinS}
\end{equation}
with the anomalous dimensions $2\Gamma$ given by (\ref{anomdim}).
In addition, we found that the dominant contribution to the anomalous dimensions 
at large $h,\bar{h}$ comes from the maximal value for the spin $j$ of the
exchanged AdS particle.
In gravitational theories in AdS,
this particle is the graviton, whose exchange dominates the
interaction and determines the tree--level anomalous dimensions of
the double trace $\mathcal{O}_{1}\mathcal{O}_{2}$ composites for large
$h,\bar{h} $. 
On the other hand, the full gravitational theory in
AdS will have more interactions at tree--level, like \emph{S}
and \emph{U}--channel exchanges, as well as contact and non--minimal
interactions. Just as in flat space, though, all these other
interactions are subdominant in the large spin and energy limit,
and can be neglected in first approximation. We will not give a
complete proof of this fact, but we shall rather concentrate on
some specific significant examples. In particular, we will analyze
the graphs in figure \ref{AdSdiagrams}, and we will concentrate on the case
$\Delta_{1}=\Delta_{2}=\eta$ for simplicity.

Letting $\mathcal{A}_{1}(z,\bar{z})$ be the amplitude for graph
\ref{AdSdiagrams}(a),  the amplitudes for graphs \ref{AdSdiagrams}(b) and
\ref{AdSdiagrams}(c) are simply obtained by permuting the external
particles and are given explicitly by
\begin{align*}
\left( z\bar{z}\right)^{\eta}\mathcal{A}_{1}\left( \frac{1}{z}, \frac{1}{\bar{z}} \right),
\text{ \ \ \ \ \ \ \ \ \ \ \ \ \ \ \ \ \ \ \ \ \ \ \ \ \ }  & \text{graph
\ref{AdSdiagrams}(b)},\\
\left(  \frac{z\bar{z}}{\left(  1-z\right)  \left(
1-\bar{z}\right) }\right)  ^{\eta}\mathcal{A}_{1}\left(
1-z,1-\bar{z}\right),  \text{ \ \ \ \ \ \ \ \ \ \ \ \ \ \ \ \ \ \
\ \ \ }  & \text{graph \ref{AdSdiagrams}(c)}.
\end{align*}
The  \emph{S}--channel decomposition of graph \ref{AdSdiagrams}(b)
can be trivially deduced from the results of section
\ref{sec:tchannel}, which considers the mirror \emph{T}--channel
decomposition of graph \ref{AdSdiagrams}(a). Without any further
analysis, we conclude that \ref{AdSdiagrams}(b) contributes only to
\emph{S}--channel partial waves of spin $J\leq j$, and is therefore
local in spin as in flat space. 
Thus,  graph \ref{AdSdiagrams}(b) does not contribute to the 
anomalous dimensions of the large spin composite operators we considered.


To analyze the $S$--channel decomposition of graph \ref{AdSdiagrams}(c), let
us first note that the differential operator $D_{S}$ in (\ref{Schannel}) is
invariant under $z\rightarrow1-z$ and $\bar{z}\rightarrow1-\bar{z}$, whenever
$2\nu=\Delta_1-\Delta_2=0$. We therefore conclude that
\[
\left(  \frac{z\bar{z}}{\left(  1-z\right)  \left(  1-\bar{z}\right)
}\right)  ^{\eta}\mathcal{S}_{h,\bar{h}}\left(  1-z,1-\bar{z}\right)
=\left(  -\right)  ^{h-\bar{h}}\mathcal{S}_{h,\bar{h}}\left(  z,\bar
{z}\right)  ~,
\]
where the normalization is fixed by recalling the leading behavior
of $\mathcal{S}_{h,\bar{h}}\sim
z^{\eta-h}\bar{z}^{\eta-\bar{h}}$ when
$z,\bar{z}\rightarrow\infty$. In fact, if we choose $z=i\lambda$,
$\bar {z}=-i\lambda$ to avoid branch cuts on the real axis,
 we have that $\mathcal{S}_{h,\bar{h}}\left(  z,\bar{z}\right)
\sim\left(  i\lambda\right)  ^{\eta-h}\left(  -i\lambda\right)
^{\eta-\bar{h}}$ and that $\mathcal{S}_{h,\bar{h}}\left(  1-z,1-\bar
{z}\right)  \sim\left(  -i\lambda\right)  ^{\eta-h}\left(  i\lambda\right)
^{\eta-\bar{h}}$, thus fixing the relative normalization to $\left(
-\right)  ^{h-\bar{h}}$. We then conclude that, if the $S$--channel expansion
of $\hat{\mathcal{A}}_{1}$ is given by (\ref{TinS}), then the Lorentzian amplitude
associated to graph \ref{AdSdiagrams}(c) is given by
\[
-2\pi i \sum_{\eta\leq\bar{h}\leq h}\ \left(  -\right)  ^{h-\bar{h}}~\Gamma\left(h,\bar{h}\right)
\mathcal{S}_{h,\bar{h}}\ .
\]
Therefore, for instance, the extra contribution to the anomalous dimension is
given by
\[
\left(  -\right)  ^{h-\bar{h}}~2\Gamma\left(h,\bar{h}\right)\ ,
\]
which oscillates as a function of spin. In a coarse--grained
picture, in which we consider large and
continuous spins, these oscillations average to a 
subleading contribution, just as in the flat space case considered in section \ref{flatPW}.

Finally, let us analyze the contact interaction of graph
\ref{AdSdiagrams}(d). Using the techniques developed in chapter \ref{ch:eikonal},
one can easily establish that the Lorentzian amplitude corresponding to
graph \ref{AdSdiagrams}(d) is proportional to
\begin{equation}
\sqrt{z\bar{z}}
\int_{H_{d-1}} d {\bf w} \,\frac{ 1}
{ \left(-2 {\bf e}  \cdot  {\bf w}\right)^{2\eta -1} 
\left( -2 {\bf \bar{  e} }\cdot  {\bf w }\right)^{2\eta -1 }  }\ .\label{contactint}
\end{equation}
This expression is essentially equation (\ref{treeleveleikonal})
with $j=0$ and the propagator $\Pi_{\perp}\left( {\bf w}, {\bf \bar{w}}\right)$ replaced by
the delta--function $\delta\left( {\bf w}, {\bf \bar{w}}\right)$ on the transverse
space $\mathrm{H}_{d-1}$. It is therefore easy to interpret it in the \emph{S} and in the
\emph{T}--channel partial wave expansions. On one hand, the
 \emph{S}--channel decomposition of the Lorentzian amplitude using the impact parameter representation
gives high spin and energy anomalous dimensions, which are now
proportional to
\[
\frac{1}{s}\delta\left(  {\bf w}, {\bf \bar{w}} \right)  ~,
\]
where we recall that $s=4h\bar{h}$ and $-2 {\bf w}\cdot {\bf \bar{w}}
=\bar{h}/h+h/\bar{h}$. Therefore the delta--function fixes
$h=\bar{h}$, \textit{i.e.}, spin $J=0$. On the other hand, since
the  Lorentzian amplitude (\ref{contactint}) goes like $\sqrt{z\bar{z}}$ for
small $z$ and $\bar{z}$, we have only spin $J=0$ partial waves appearing
in the \emph{T}--channel decomposition. Expression
(\ref{contactint}) also generates the relative weights of all
these spin zero contributions. In both channels, as expected,
graph \ref{AdSdiagrams}(d) only contains spin zero intermediate
primaries, and therefore does not effect the anomalous dimensions
of large spin operators,
which are only controlled by graph \ref{AdSdiagrams}(a).

\newpage

\begin{subappendices}


\section{Some Relevant Fourier Transforms}\label{app3}


Start by recalling  the standard generalized Feynman propagator%
\[
\frac{1}{\pi ^{d}}\int_{\mathbb{M}^d } \frac{d{\bf p} }{\left( {\bf p} ^{2}\mp i\epsilon \right)
^{\Delta }}~e^{2i{\bf y} \cdot {\bf p} }=\pm ~\frac{\pi ^{-\frac{d}{2}}\Gamma
\left( \frac{d}{2}-\Delta \right) }{\Gamma \left( \Delta \right) }~\frac{i}{%
\left( {\bf y} ^{2}\pm i\epsilon \right) ^{\frac{d}{2}-\Delta }}~.
\]%
We now wish to consider the Fourier transform of interest%
\[
f\left( {\bf y} \right) =\frac{1}{\pi ^{d}}\int_{\mathbb{M}^d } ~\frac{d{\bf p} }{\left( {\bf p}
^{2}-i\epsilon _{{\bf p} }\right) ^{\Delta }}e^{2i{\bf y} \cdot {\bf p} }
\]%
We consider first the case $ y ^{0} =  -{\bf y} \cdot
{\bf x}_0<0$. In this case $f\left( {\bf y} \right) $ vanishes since we
can deform the ${ p} ^{0}$ contour in the upper complex plane
$\im \, { p} ^{0}>0$. By Lorentz invariance, $f\left( {\bf y}
\right) $ also vanishes whenever ${\bf y} $ is spacelike, and
$f\left( {\bf y} \right) $ is therefore supported only in the
future Milne wedge $\mathrm{M}$, where it is proportional to
$\left\vert {\bf y} \right\vert ^{2\Delta -d}$. To find the
constant of proportionality, we note that, when ${ y} ^{0}>0$ we
may deform the ${ p} ^{0}$ contours in the lower complex plane and
show that
\[
f\left( {\bf y} \right) =\frac{1}{\pi ^{d}}\int_{\mathbb{M}^d } \left[ \frac{d{\bf p} }{\left(
{\bf p} ^{2}+i\epsilon \right) ^{\Delta }}+\frac{d{\bf p} }{\left( {\bf p}
^{2}-i\epsilon \right) ^{\Delta }}\right] \, e^{2i{\bf y} \cdot {\bf p} }~.~\ \ \
\ \ \ \ \ \ \ \ \ \left( { y} ^{0}>0\right)
\]%
We then deduce that%
\begin{eqnarray*}
f\left( {\bf y} \right)  &=&-i\,\frac{\pi ^{-\frac{d}{2}}\Gamma \left( \frac{d}{%
2}-\Delta \right) }{\Gamma \left( \Delta \right) }\left( i^{2\Delta
}-i^{-2\Delta }\right) \left\vert {\bf y} \right\vert ^{2\Delta -d} \\
&=&\frac{2\pi ^{1-\frac{d}{2}}}{\Gamma \left( \Delta \right) \Gamma
\left( 1+\Delta -\frac{d}{2}\right) }\left\vert {\bf y} \right\vert ^{2\Delta
-d}~\ \ \ \ \ \ \ \ \ \ \ \ \left( {\bf y} \in \mathrm{M}\right)
\end{eqnarray*}%
and $f\left( {\bf y} \right) =0$ for ${\bf y} \notin \mathrm{M}$, thus proving
equation (\ref{suca1234}).

\section{Tree--level Eikonal from Shock Waves}
\label{BtoBSec}

In this section, we present an alternative derivation of the tree--level eikonal amplitude (\ref{treeleveleikonal})
based on the direct computation of the Witten diagram describing
the exchange of an AdS particle of dimension $\Delta$ and spin $j$,
 with a specific choice of external wave functions.
Indeed, we start from the general expression
\begin{equation}
\mathcal{E}_{1}=-g^2\,2^j\,\int
_{\mathrm{AdS}} d\mathbf{x} d\mathbf{\bar{x}}\, T_{\alpha_1 \cdots \alpha_j} (\mathbf{x})
\,\Pi_\Delta^{\alpha_1 \cdots \alpha_j,\beta_1\cdots\beta_j}(\mathbf{x}, \mathbf{\bar{x}})
\,\bar{T}_{\beta_1\cdots\beta_j}( \mathbf{\bar{x}}) \ ,
\label{treeWittenj}
\end{equation}
with the sources given by
\begin{eqnarray*}
 T_{\alpha_1 \cdots \alpha_j}(\mathbf{x}) &=&\psi_1(\mathbf{x})  \nabla_{\alpha_1} \cdots  \nabla_{\alpha_j} \psi_3(\mathbf{x}) 
+ \cdots \ ,
\\
\bar{T}_{\beta_1\cdots\beta_j}( \mathbf{x}) &=& \psi_2(\mathbf{x})
  \nabla_{\beta_1} \cdots  \nabla_{\beta_j} \psi_4(\mathbf{x})+ \cdots \ ,
\end{eqnarray*}
where the $\cdots$ stand for terms that can be dropped inside the integral (\ref{treeWittenj}),
because the propagator $\Pi_\Delta^{\alpha_1 \cdots \alpha_j,\beta_1\cdots\beta_j}(\mathbf{x}, \mathbf{\bar{x}})$
is traceless and divergenceless.
For now, assume that one can find external wave functions $\psi_2$ and $\psi_4$ such that
the source $\bar{T}$ is localized along the null hypersurface ${\bf k}_2 \cdot {\bf x}=0$.
This hypersurface corresponds to the condition  $u=0$ in the coordinates (\ref{coordAdS}) introduced in chapter \ref{ch:eikonal}.
Moreover, we demand that the only non--vanishing component of $\bar{T}$ has the form
\begin{equation}
\bar{T}_{u \cdots u}(\mathbf{x})= \delta(u) \bar{T}({\bf w}) \ ,    \label{Tbar}
\end{equation}
again in  the coordinates (\ref{coordAdS}).
In this case, the integral over $\mathbf{\bar{x}}$ in equation (\ref{treeWittenj}) greatly simplifies.
Defining,
$$
h^{\alpha_1 \cdots \alpha_j} (\mathbf{x})=
\int_{\mathrm{AdS}}d\mathbf{\bar{x}}
\,\Pi_\Delta^{\alpha_1 \cdots \alpha_j,\beta_1\cdots\beta_j}(\mathbf{x}, \mathbf{\bar{x}})
\,\bar{T}_{\beta_1\cdots\beta_j}( \mathbf{\bar{x}}) \ ,
$$
we have
\begin{equation}
\Big[ \Box_{{\rm AdS}} -\Delta(\Delta-d)+j\Big] h_{\alpha_1 \cdots \alpha_j}(\mathbf{x}) =
i\, \bar{T}_{\alpha_1 \cdots \alpha_j}({\bf x}) \ .   \label{hsource}
\end{equation}
When the source $\bar{T}$ has form (\ref{Tbar}),
the field $h$ defines a shock wave localized at $u=0$.
Writing the ansatz
$$
h_{u \cdots u}(\mathbf{x})= -i\,\delta( u )\,(-{\bf x}_0\cdot {\bf w} )^{j-1}\, h ( {\bf w})\ ,
$$
one can check, either by a tedious direct computation or by using the methods of appendix \ref{app1}, 
that equation (\ref{hsource}) reduces to
$$
\Big[  \Box_{H_{d-1}} -\Delta(\Delta-d) -d+1 \Big]h (\mathbf{w}) =-
(-{\bf x}_0\cdot {\bf w} )^{1-j}\, \bar{T}({\bf w})\ ,
$$
which has solution,
$$
h ( {\bf w})=\int_{H_{d-1}} d {\bf \bar{w}} 
\, \Pi_\perp( {\bf w} ,{\bf \bar{w}} ) \,(-{\bf x}_0\cdot {\bf \bar{w}} )^{1-j}\, \bar{T}({\bf \bar{w}})\ .
$$
Therefore, equation (\ref{treeWittenj}) becomes
$$
\mathcal{E}_{1}=i\,g^2\,2^{j-1}\, \int
_{H_{d-1}} d\mathbf{w} \,h ( {\bf w})\,(-{\bf x}_0\cdot {\bf w} )^{j+1} \int_{-\infty}^\infty dv\, T^{u \cdots u} (u=0,v,\mathbf{w})\ .
$$
We now choose wave functions
$$
\psi_1 ({\bf x})= K_{\Delta_1}({\bf p}_1,{\bf x})\ ,\ \ \ \ \ \ \  \ \ \ \ \  \ \ \ \
\psi_3({\bf x})= K_{\Delta_1}({\bf p}_3, {\bf x})\ ,
$$
with the boundary points
$$
{\bf p}_1=-{\bf k}_1 -{\bf y}_1^2{\bf k}_2+2\omega    {\bf y}_1 \ ,\ \ \ \ \ \ \ \ \ \ \ \ \ \
{\bf p}_3={\bf k}_1+ {\bf y}_3^2{\bf k}_2-2\omega    {\bf y}_3 \ ,
$$
where ${\bf y}_1$ and ${\bf y}_3$ are orthogonal to ${\bf k}_1$ and ${\bf k}_2$, as in (\ref{pcoord}).
In the coordinates (\ref{coordAdS}) we obtain
\begin{eqnarray*}
\psi_1(u=0,v,{\bf w})&=&\mathcal{C}_{\Delta_1} \Big(2\omega ({\bf x}_0\cdot {\bf w} )v -4\omega ({\bf y}_1\cdot {\bf w} )
 +i\epsilon \Big)^{-\Delta_1}\\
&=& \frac{\mathcal{C}_{\Delta_1} }{(2\omega i)^{\Delta_1} \Gamma(\Delta_1) } 
\int_0^\infty ds\,s^{\Delta_1-1}\,e^{is  ({\bf x}_0\cdot {\bf w} )v - 2is({\bf y}_1\cdot {\bf w} ) }
\end{eqnarray*}
and, similarly,
$$
\psi_3(u=0,v,{\bf w})=
\frac{\mathcal{C}_{\Delta_1} }{(2\omega i)^{\Delta_1} \Gamma(\Delta_1) } 
\int_0^\infty ds\,s^{\Delta_1-1}\,e^{-is  ({\bf x}_0\cdot {\bf w} )v + 2is({\bf y}_3\cdot {\bf w} ) }\ .
$$
With this choice one can compute the source,
\begin{align*}
\int_{-\infty}^\infty dv\, T^{u \cdots u} (u=0,v,\mathbf{w})=
\frac{\mathcal{C}^2_{\Delta_1} }{(2\omega i)^{2\Delta_1} \Gamma^2(\Delta_1) }
\int_0^\infty ds_1 ds_3\,s_1^{\Delta_1-1}\,s_3^{\Delta_1-1} \,e^{-2is_1  ({\bf y}_1\cdot {\bf w} ) + 2is_3({\bf y}_3\cdot {\bf w} ) }
&\\
\int_{-\infty}^\infty dv\,
e^{is_1  ({\bf x}_0\cdot {\bf w} )v} \left( -\frac{2}{ ({\bf x}_0\cdot {\bf w} )^2 } \frac{\partial}{\partial v}\right)^j
e^{-is_3  ({\bf x}_0\cdot {\bf w} )v} & \ .
\end{align*}
The integral over $v$ forces $s_1=s_3$ and the expression simplifies to
\begin{align*}
& \frac{2\pi (-2i)^j\mathcal{C}^2_{\Delta_1} }{(2\omega i)^{2\Delta_1} \Gamma^2(\Delta_1) }  (-{\bf x}_0\cdot {\bf w} )^{-1-j}
\int_0^\infty ds\, s^{2\Delta_1+j-2} \,e^{2is  ({\bf y}_3- {\bf y}_1)\cdot {\bf w}  }=  \spa{.5} \\
&=
-i\,2^{j+1}\frac{\pi \,\mathcal{C}^2_{\Delta_1}  \Gamma(2\Delta_1+j-1)}{(2\omega )^{2\Delta_1} \Gamma^2(\Delta_1) } 
\frac{(-{\bf x}_0\cdot {\bf w} )^{-1-j}}{\left( 2{\bf p}\cdot {\bf w} +i\epsilon \right)^{2\Delta_1+j-1}} \ .
\end{align*}
where ${\bf p}={\bf y}_3- {\bf y}_1 $ as before.
We conclude that if we can find wave functions $\psi_2$ and $\psi_4$ such that their source $\bar{T}$
is of the form (\ref{Tbar}), then the corresponding Witten diagram is given by
\begin{equation}
\mathcal{E}_{1}=g^2\,\frac{\pi\, 4^j\,\mathcal{C}^2_{\Delta_1}  \Gamma(2\Delta_1+j-1)}
{(2\omega )^{2\Delta_1} \Gamma^2(\Delta_1) }  
\int_{H_{d-1}} d\mathbf{w} d\mathbf{\bar{w}} 
\frac{ \Pi_\perp( {\bf w} ,{\bf \bar{w}} ) \,(-{\bf x}_0\cdot {\bf \bar{w}} )^{1-j}\, \bar{T}({\bf \bar{w}}) }
{\left( 2{\bf p}\cdot {\bf w} +i\epsilon \right)^{2\Delta_1+j-1} }\ .      \label{shockfourpoint}
\end{equation}


\subsection{Creating the Shock Wave Geometry\label{CreateShock}}


In this section, we construct the wave functions $\psi_{2}$ and $\psi_{4}$ such that the corresponding source $\bar{T}$
is of the form (\ref{Tbar}).
In other words, the wave functions $\psi_2$ and $\psi_4$ will be chosen such that 
they source a shock wave in the exchanged spin $j$ field.
This will be achieved by choosing $\psi_{2}$ and $\psi_{4}$ 
to be a particular linear combinations of bulk to boundary propagators.
More precisely, the 
fields $\psi_{2}$ and $\psi_{4}$ will respectively vanish after and before the shock, 
so that the source $\bar{T}\sim \psi_{2}\partial^{j}\psi_{4}$ is 
supported only at $u=0$. Moreover, near $u\sim 0$, the functions $\psi_{2}$ and $\psi_{4}$ 
will be respectively chosen to behave in the light cone directions $u$ and $\bar{v}$ as $u^{\Delta_{2}+j-1}$ 
and $u^{-\Delta_{2}}$, so that their overlap $\psi_{2}\partial_{u }^{j} \psi_{4}$ goes as 
$1/u \sim\delta\left( u \right)$. 

\begin{figure}
[ptb]
\begin{center}
\includegraphics[keepaspectratio,height=6cm]{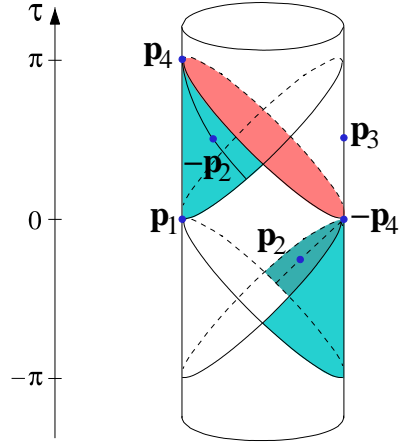} 
\caption{Construction of the external wave
functions $\psi_{2}$ and $\psi_{4}$ starting from the bulk to
boundary propagators $K_{\Delta_2}(\pm\mathbf{p}_{2},{\bf x})$ and $K_{\Delta_2}(\pm\mathbf{p}_{4},{\bf x})$.
The points $\pm\mathbf{p}_{4}=\pm\mathbf{k}_{2}$ are fixed, whereas the points $\pm\mathbf{p}_{2}$ are free to move
in the past Milne wedges (shaded regions of the boundary). In
particular, in (\ref{Gline}), the points $\pm\mathbf{p}_{2}$ lie
along a ray from the origin, as shown. The source points
$\mathbf{p}_{1}$, $\mathbf{p}_{3}$ are also shown.}
\label{fig12}
\end{center}
\end{figure}

Following figure \ref{fig12}, we start by choosing as external
state $\psi_{4}$ the linear combination
\begin{eqnarray}
\psi_{4} ( {\bf x} ) & =&i^{2\Delta_{2}}\,K_{\Delta_{2} }(\mathbf{p}_{4},{\bf x})
 - K_{\Delta_{2} }(-\mathbf{p}_{4}, {\bf x} )  \nonumber
\\
&=& \frac{\mathcal{C}_{\Delta_2} }{
 (-2\omega\, {\bf x}_0\cdot {\bf w} )^{\Delta_{2} } }
\left[  (-u - i\epsilon)^{-\Delta_{2}} - (-u + i\epsilon )^{-\Delta_{2}} \right]  \ ,
\label{psi4}
\end{eqnarray}
where we have chosen 
$$
\mathbf{p}_{4} =\mathbf{k}_{2}\ .
$$
The wave function
$\psi_4$ clearly vanishes before the shock for $u<0$.
Similarly, the function $\psi_{2}$ will be given by the general
linear combination
\begin{equation}
\psi_{2}( {\bf x} )=\int_{\mathbb{M}^d}\frac{d{\bf \bar{p}} }{\left(  2\pi\right)  ^{d}}
G\left(  {\bf \bar{p}} \right)
\left[ K_{\Delta_{2} } (-\mathbf{p}_{2} ,{\bf x}) -i^{2\Delta_{2}} \,K_{\Delta_{2} }(\mathbf{p}_{2},{\bf x}) \right] \ ,
\label{psi2}
\end{equation}
where we write
$$
{\bf p}_2=-{\bf k}_2 -{\bf \bar{p}}^2{\bf k}_1+2\omega    {\bf \bar{p}} \ ,
$$
with ${\bf \bar{p}}\in \mathbb{M}^d$ orthogonal to ${\bf k}_1$ and ${\bf k}_2$, similarly to  (\ref{pcoord}).
 The integrand in
(\ref{psi2}) vanishes for $\mathbf{x\cdot p}_2<0$, so that, for
$G( {\bf \bar{p}} )$ supported in the past Milne wedge $-{\rm M}$, the wave function
$\psi_2$ vanishes after the shock for $u>0$. Recall that we are
interested in the overlap $\psi_4\partial_u^j\psi_2$, so that we
need only the behavior of $\psi_2$ for $u \sim 0$. This in turn
is controlled only by $G({\bf \bar{p}})$ for ${\bf \bar{p}}\sim 0$. 
To show this, we shall
for a moment assume that $G({\bf \bar{p}})$ in (\ref{psi2}) is homogeneous in
${\bf \bar{p}}$ as $G(\lambda {\bf \bar{p}})=\lambda^c \,G({\bf \bar{p}})$. Then,
from the explicit form of the bulk to boundary propagator in the coordinates (\ref{coordAdS}),
$$
K_{\Delta_{2} }( \mathbf{p}_{2},{\bf x} )=\mathcal{C}_{\Delta_2} \left( 2\omega\right)^{-\Delta_2 }
\left[-2 {\bf \bar{p}}\cdot{\bf w} +({\bf x}_0 \cdot{\bf w})
\left(u +-uv  {\bf \bar{p}}\cdot{\bf x}_0  +v\left(1 -\frac{uv}{4} \right){\bf \bar{p}}^2  \right) +i\epsilon
\right]^{-\Delta_2 }\ ,
$$
it is clear that $\psi_2$ is an
eigenfunction of $u\partial_u-v\partial_v$ as follows
\begin{equation*}
\psi_{2}(  \lambda u,\lambda^{-1} v,{\bf w}) =\lambda^{c-\Delta
_{2}+d}\psi_{2}(u,v,{\bf w} )  \ .
\end{equation*}
Therefore, close to $u\sim 0$, it behaves as
$$
\psi_2(u,v,{\bf w} )\simeq (-u)^{c-\Delta_2+d}\psi_2(-1,0,{\bf w} )\ .
$$
In general, we shall take, for reasons which
shall become clear shortly,
\begin{align*}
G\left({\bf \bar{p}}\right) &= G_0 \left({\bf \bar{p}}\right) +\cdots\, ,\\
G_0\left( \lambda {\bf \bar{p}}\right) &= \lambda^{2\Delta_{2}+j-d-1} G_0
\left( {\bf \bar{p}}\right)\, ,
\end{align*}
where the dots denote sub--leading terms for ${\bf \bar{p}}\sim 0$. The above
discussion then immediately implies that the behavior of
$\psi_{2}$ just before the shock is given by
\[
\psi_{2}\left(  u,v ,{\bf w}  \right)  \simeq
(-u)^{\Delta _{2}+j-1}g\left({\bf w} \right)+\cdots  ~,
\]
where $g\left({\bf w} \right)$ is determined uniquely by $G_0({\bf \bar{p}})$.

We are now in a position to explicitly compute the source term
$\bar{T}$ in (\ref{treeWittenj}). We first recall the following
representation of the delta--function
\begin{align*}
& \Gamma\left(  \alpha\right)  \Gamma\left(  \beta\right)  \Big[(u
-i\epsilon)^{-\alpha}-(u+i\epsilon)^{-\alpha}\Big] \Big[(-u-i\epsilon)^{-\beta
}-(-u+i\epsilon)^{-\beta}\Big]=\\
& =\left\{
\begin{array}[c]{ll} 
-2\pi^{2}\delta (  u )  ~\ ,~\ \ \ \ \ \ \ \ \ \ \ 
& \left(\alpha+\beta=1\right)\ , \\
0~\ ,~\ \ \ \ \ \ \ \ \ \ \ 
&\left( \alpha+\beta<1\right)\ .
\end{array}
\right.
\end{align*}
Writing the leading behavior of $\psi_{2}$ as
\[
\frac{\Gamma\left(  \Delta_{2}+j\right)  \Gamma\left(
1-\Delta_{2}-j\right) }{2\pi i} \Big[
(u-i\epsilon)^{\Delta_{2}+j-1}-(u+i\epsilon
)^{\Delta_{2}+j-1} \Big]  g\left({\bf w} \right)
\]
and using the above representation of $\delta\left( u\right)$
we conclude that the source function $\bar{T}$ has precisely the form (\ref{Tbar})
with
$$
\bar{T}\left( {\bf w} \right)   =
\frac{i\pi\,\mathcal{C}_{ \Delta_{2}} \Gamma\left( \Delta_{2}+j\right)  }{
\left(2\omega\right)^{\Delta_{2}} \Gamma\left(\Delta_{2}\right)  }
\left(- {\bf x}_0 \cdot{\bf w} \right)^{-\Delta_{2}}
g\left({\bf w}\right)  \ .
$$

To explicitly compute the function $g\left({\bf w}\right)$ in terms of
the weight function $G_0\left({\bf \bar{p}}\right)$, we must simply evaluate
(\ref{psi2}) at $\{ u= -1,v=0,{\bf w}\}$, with $G$ replaced by $G_0$.
The first term in (\ref{psi2})  gives
\begin{align*}
 &\mathcal{C}_{\Delta_{2}} \left( 2\omega\right)^{-\Delta_2 }\int_{\mathbb{M}}\frac{d{\bf \bar{p}}}{\left(  2\pi\right)  ^{d}}
\frac{G_0\left( {\bf \bar{p}}\right)  }{\left( {\bf x}_0 \cdot{\bf w} +2{\bf \bar{p}}\cdot {\bf w}+i\epsilon\right)  ^{\Delta_{2}}}
\spa{0.5} \\&=\frac{\mathcal{C}_{\Delta_{2}}~i^{-\Delta_{2}}}{  \left( 2\omega\right)^{\Delta_2 } \Gamma\left(
\Delta _{2}\right)
}\int_{0}^{\infty}ds\,s^{\Delta_{2}-1}~\int_{\mathbb{M}}\frac{d{\bf \bar{p}}}{\left(
2\pi\right)  ^{d}}G_0\left(  {\bf \bar{p}}\right)  ~e^{is\left( {\bf x}_0 \cdot{\bf w} 
+2{\bf \bar{p}}\cdot {\bf w} \right)  } \spa{0.5}\\
&=(i/2)^{j-1} \frac
{\mathcal{C}_{\Delta_{2}}\Gamma\left(  1-\Delta_{2}-j\right)  ~}{ \left( 8\omega\right)^{\Delta_2 } \Gamma\left(
\Delta_{2}\right) } 
( -{\bf x}_0 \cdot{\bf w} )^{ \Delta_{2}+j-1}
\hat{G}_0\left( {\bf w}\right)  \ ,
\end{align*}
where we denote with $\hat{G}_0\left( {\bf y} \right)
=\left( 2\pi\right)^{-d}\int d{\bf \bar{p}} \,e^{i {\bf \bar{p}} \cdot {\bf y} }~G_0\left( {\bf \bar{p}}\right)$
the Fourier transform of $G_0\left({\bf \bar{p}}\right)$. The second term in
(\ref{psi2}) is similarly given by
\begin{align*}
&\frac{i^{2\Delta_{2}}\mathcal{C}_{\Delta_{2}}}{
 \left( 2\omega\right)^{\Delta_2 } }
\int\frac{d{\bf \bar{p}}}{\left(
2\pi\right) ^{d}}\frac{G_0\left( {\bf \bar{p}}\right)  }
{\left( -{\bf x}_0 \cdot{\bf w}-2{\bf \bar{p}}\cdot {\bf w}+i\epsilon\right)
^{\Delta_{2}}} \spa{0.5} \\
&=
(2i)^{1-j}
\frac{\mathcal{C}_{\Delta_{2}}\Gamma\left(  1-\Delta_{2}-j\right)  ~}{ \left( 8\omega\right)^{\Delta_2 } \Gamma\left(
\Delta_{2}\right) } 
( -{\bf x}_0 \cdot{\bf w} )^{ \Delta_{2}+j-1}
\hat{G}_0\left( -{\bf w}\right)
\end{align*}
Note that, in this case, the $i\epsilon$ prescription is correct
since $G_0\left({\bf \bar{p}}\right)$ is supported only in the past Milne
wedge $-\mathrm{M}$. We finally conclude that the $T$--channel
exchange Witten diagram in figure \ref{AdSdiagrams}(a) with external wave
functions $\psi_4$ and $\psi_2$, respectively as in (\ref{psi4})
and (\ref{psi2}), is given by (\ref{shockfourpoint}) with
\begin{equation}
\bar{T}\left(  {\bf w}\right)
=(-)^j \frac{ \pi\,\mathcal{C}_{\Delta_{2}}^{2} \Gamma\left( 1-\Delta_{2}\right) }{
2^{2\Delta_{2}+j-1}  \left( 2\omega\right)^{2\Delta_2} \Gamma\left(\Delta _{2}\right) } 
( -{\bf x}_0 \cdot{\bf w} )^{ j-1}
\Big[  i^{j}\hat{G}_0\left(  {\bf w}\right)
+i^{-j}\hat{G}_0\left( -{\bf w}\right) \Big]  \ .\label{GtoT}
\end{equation}
Denoting
with $A_{1}^{\pm\pm}$ the tree level correlator associated to
graph \ref{AdSdiagrams}(a) when the external points are at
$\mathbf{p}_{1}$, $\mathbf{p}_{3}$ and $\pm\mathbf{p}_{2}$,
$\pm\mathbf{p}_{4}$, the same Witten diagram can be written as
\begin{equation}
\mathcal{E}_{1} =
\int\frac{d{\bf \bar{p}}}{\left(
2\pi\right)  ^{d}}G\left(  {\bf \bar{p}}\right)  \left(  i^{2\Delta_{2}}A_{1}
^{+-}+i^{2\Delta_{2}}A_{1}^{-+}-A_{1}^{--}-i^{4\Delta_{2}}A_{1}^{++}\right)~.
\label{Int4pt}
\end{equation}

It is particularly convenient to choose a weight function $G\left(
{\bf \bar{p}}\right)$ supported along a straight line as shown in figure \ref{fig12},
\begin{equation}
G\left(  {\bf \bar{p}}\right)  =\int_0^a dt~\ t^{2\Delta_{2}+j-2}~\left(
2\pi\right) ^{d}\delta\left( {\bf \bar{p}}+t\,{\bf \bar{e}}\right)
~,\label{Gline}
\end{equation}
with ${\bf \bar{e}}\in H_{d-1}$ a unit vector. Note that the behavior of
$G({\bf \bar{p}})$ for ${\bf \bar{p}}\sim 0$ is independent of the upper limit $a$, and the leading
behavior $G_0({\bf \bar{p}})$ is obtained by setting $a=\infty$ in
(\ref{Gline}). We then have
\[
\hat{G}_0\left( {\bf y} \right)  =i^{2\Delta_{2}+j-1}\frac{\Gamma\left(
2\Delta _{2}+j-1\right)  }{\left( - {\bf \bar{e}}\cdot{\bf y}+i\epsilon \right)
^{2\Delta_{2} +j-1}}\ ,
\]
and finally, for ${\bf w} \in H_{d-1}$,
\[
\bar{T}\left( {\bf w}\right)  =
\frac{(2\pi)^2\, \mathcal{C}^2_{\Delta_{2}}\Gamma\left(  2\Delta_{2}+j-1\right)}
{2  \left( 2\omega\right)^{2\Delta_2} \Gamma^2\left(\Delta _{2}\right)}
\frac{ \left( - {\bf x}_0\cdot{\bf w}\right)^{j-1}   }{\left( - 2{\bf \bar{e}}\cdot{\bf w}\right)^{2\Delta_{2} +j-1} }\ .
\]
For this particular source the four point function
(\ref{shockfourpoint}) becomes
\begin{equation}
\mathcal{E}_{1}=g^2\, \mathcal{K}\,\frac{\pi\,\mathcal{C}_{\Delta_{1}}\mathcal{C}_{\Delta_{2}} }
{(2\omega )^{2\Delta_1+2\Delta_2}}
\, \int_{H_{d-1}} d\mathbf{w} d\mathbf{\bar{w}} 
\frac{ \Pi_\perp( {\bf w} ,{\bf \bar{w}} ) }
{\left( 2{\bf p}\cdot {\bf w}\right)^{2\Delta_1+j-1}\left( - 2{\bf \bar{e}}\cdot{\bf \bar{w}}\right)^{2\Delta_{2} +j-1} }\ ,
 \label{partE1}
\end{equation}
where the constant $ \mathcal{K}$ is given (\ref{constantK}) and we 
recall that both ${\bf p}\cdot {\bf w} $ and $-{\bf \bar{e}}\cdot{\bf \bar{w}}$ are
positive if ${\bf p} $ is in the past Milne wedge $-\mathrm{M}$.


\subsection{Relation to the Dual CFT Four--Point Function\label{secsix}}


In this section, we shall express the Lorentzian four point
correlators in (\ref{Int4pt}) in terms of the Euclidean
four point function by means of analytic continuation. We will
denote with $\mathcal{A}_{1}^{\pm\pm}\left( z,\bar{z}\right)$ the
Lorentzian amplitudes corresponding to the tree level correlators
$A_{1}^{\pm\pm}$. More precisely, we have
\[
A_{1}^{\pm\pm}=
K_{\Delta_{1}}\left({\bf p}_1, {\bf p}_3\right)
K_{\Delta_{2}}\left(\pm{\bf p}_2, \pm {\bf p}_4\right)
\,\mathcal{A}_{1}^{\pm\pm}\left(z,\bar{z}\right)\ .
\]
As in the previous chapter,
the functions $\mathcal{A}_{1}^{\pm\pm }\left(z,\bar{z}\right)$ are given by specific analytic continuations of
the basic Euclidean four point amplitude $\mathcal{A}_{1}\left(z,\bar{z}\right)$.

From now on we fix $ {\bf p} \in -\mathrm{M}$. Recalling that $G\left(  {\bf \bar{p}}\right)$ is
non--vanishing only for ${\bf \bar{p}} \in-\mathrm{M}$, we have that the
phases of the boundary propagators are determined from 
$$
n\left({\bf p}_1, {\bf p}_3\right)=0\ ,\ \ \ \ \ \ \ 
n\left({\bf p}_2, {\bf p}_4\right)=2 \ ,\ \ \ \ \ \ \ 
n\left(-{\bf p}_2, -{\bf p}_4\right)=0\ 
$$
and
$$
n\left(-{\bf p}_2, {\bf p}_4\right)=n\left({\bf p}_2, -{\bf p}_4\right)=1\ .
$$ 
Therefore, we can write (\ref{Int4pt})
as
\[
\mathcal{E}_{1}=
\frac{\mathcal{C}_{\Delta_{1}}\mathcal{C}_{\Delta_{2}}}
{(2\omega )^{2\Delta_1+2\Delta_2} }
\int\frac
{d{\bf \bar{p}}}{\left(  2\pi\right)  ^{d}}\frac{G\left(  {\bf \bar{p}}\right)  }{\left(
-{\bf p}^{2}\right)  ^{\Delta_{1}}\left(  -{\bf \bar{p}}^{2}\right)
^{\Delta_{2}}}\left(
\mathcal{A}_{1}^{+-}+\mathcal{A}_{1}^{-+}-\mathcal{A}_{1}^{++}-\mathcal{A}
_{1}^{--}\right)  ~,
\]
where we recall that $z$ and $\bar{z}$ are implicitly defined by
\[
z\bar{z}={\bf p}^{2}{\bf \bar{p}}^{2}~,~\ \ \ \ \ \ \ \ \ \ z+\bar{z}=2{\bf p}\cdot {\bf \bar{p}}~.
\]
In particular, choosing $-{\bf p}={\bf e}\in H_{d-1}$ of unit norm and
$G\left( {\bf \bar{p}}\right)$ as in (\ref{Gline}), we obtain the expression
\begin{equation}
\mathcal{E}_{1}=
\frac{\mathcal{C}_{\Delta_{1}}\mathcal{C}_{\Delta_{2}}}
{(2\omega )^{2\Delta_1+2\Delta_2} }
\int_{0}^{a}dt~\ t^{~j-2}\left(  \mathcal{A}_{1}^{+-}+\mathcal{A}_{1}^{-+}
-\mathcal{A}_{1}^{++}-\mathcal{A}_{1}^{--}\right)  ~,\label{cutEQ}
\end{equation}
where now
\begin{align}
z &  =-tw^{-1/2},~~\ \ \ \ \ \ \bar{z}=-tw^{1/2}~,\label{ss2}\\
w^{1/2}+w^{-1/2} &  =-2{\bf e}\cdot {\bf \bar{e}}\ .\nonumber
\end{align}
Notice that for ${\bf e}, {\bf \bar{e}}\in H_{d-1}$ we have that 
$-{\bf e}\cdot {\bf \bar{e}} \geq 1$ and therefore both $z$ and $\bar{z}$ are
real and negative.

\begin{figure}
[ptb]
\begin{center}
\includegraphics[height=6cm]{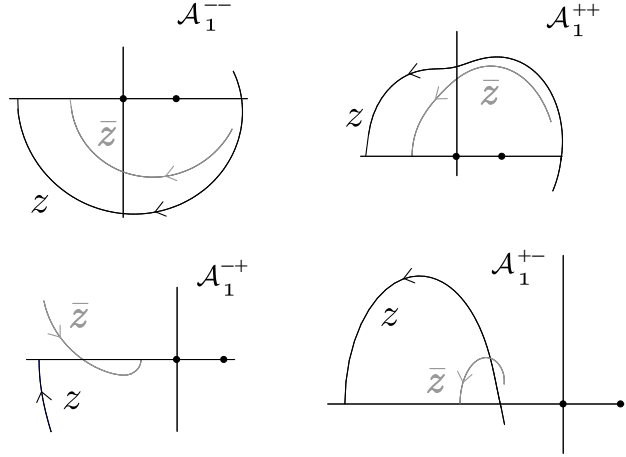} 
\caption{Wick rotation of $z$,$\bar{z}$ when the
external points are at $\mathbf{p}_{1}$, $\mathbf{p}_{3}$,
$\pm\mathbf{p}_{2}$, $\pm\mathbf{p}_{4}$. On the Euclidean
principal sheet of the amplitude we have $\bar{z}=z^{\star}$
(initial points of the curves), while after Wick rotating to the
Lorentzian domain $z$ and $\bar{z}$ are real and negative (final
points).}
\label{4wickrotations}
\end{center}
\end{figure}

We now must consider more carefully the issue of the analytic
continuation as we did in section \ref{sectAC}. 
Given the four
points $\mathbf{p}_{1}$, $\mathbf{p}_{3}$, $\pm\mathbf{p}_{2}$,
$\pm\mathbf{p}_{4}$, we may follow the cross--ratios $z\left(
\theta\right)$, $\bar{z}\left( \theta\right)$ as a function of
the parameter $\theta$ of the Wick rotation. 
The plots of $z\left( \theta\right)$, $\bar{z}\left(
\theta\right)$ in the four cases $\mathcal{A}_{1}^{\pm\pm}$ are
shown in figure \ref{4wickrotations}.
Note that, in the Euclidean limit, we
have that
\[
\bar{z}\left(  0\right)  =\left(  z\left(  0\right)  \right)
^{\star}\ ,
\]
as expected. On the other hand, when $\theta=1$, the cross--ratios
$z\left( 1\right), \bar{z}\left( 1\right)$ are given by (\ref{ss2}).
From figure \ref{4wickrotations}, we deduce that
\begin{align*}
\mathcal{A}_{1}^{--}\left(  z,\bar{z}\right)   &  =\mathcal{A}_{1}
^{\circlearrowright}\left(  z-i\epsilon,\bar{z}-i\epsilon\right)  ~,\\
\mathcal{A}_{1}^{-+}\left(  z,\bar{z}\right)   &
=\mathcal{A}_{1}\left(
z-i\epsilon,\bar{z}-i\epsilon\right)  ~,\\
\mathcal{A}_{1}^{+-}\left(  z,\bar{z}\right)   &
=\mathcal{A}_{1}\left(
z+i\epsilon,\bar{z}+i\epsilon\right)  ~,\\
\mathcal{A}_{1}^{++}\left(  z,\bar{z}\right)   &  =\mathcal{A}_{1}
^{\circlearrowleft}\left(  z-i\epsilon,\bar{z}+i\epsilon\right) ~,
\end{align*}
where $\mathcal{A}_{1}^{\circlearrowright}$
($\mathcal{A}_{1}^{\circlearrowleft}$) is the analytic
continuation of the Euclidean amplitude $\mathcal{A}_{1}$ obtained
by keeping $\bar{z}$ fixed and by transporting $z$ clockwise
(counterclockwise) around $0,1$. 
In general we have that the Euclidean amplitude is real, in the
sense that
\[
\mathcal{A}_{1}\left( z,\bar{z}\right)  =\mathcal{A}_{1}\left(
\bar {z},z\right)  ~,~\ \ \ \ \ \ \ \ \ \ \
\mathcal{A}_{1}^{\star}\left( z,\bar{z}\right)
=\mathcal{A}_{1}\left( z^{\star},\bar{z}^{\star}\right)  ~.
\]
Therefore
\[
\mathcal{A}_{1}^{\circlearrowleft}\left( z,\bar{z}\right) =\left[
\mathcal{A}_{1}^{\circlearrowright}\left(
z^{\star},\bar{z}^{\star}\right) \right]  ^{\star}~.
\]
We then conclude that
\[
\mathcal{A}_{1}^{+-}+\mathcal{A}_{1}^{-+}-\mathcal{A}_{1}^{++}-\mathcal{A}_{1}^{--}=
2\,\re \left[ \mathcal{A}_{1}\left(z-i\epsilon,\bar {z}-i\epsilon\right)  
-\mathcal{A}_{1}^{\circlearrowleft}\left(z-i\epsilon,\bar {z}-i\epsilon\right) \right]\ ,
\]
so that (\ref{cutEQ}) becomes
\begin{equation}
\mathcal{E}_1=-
\frac{2\,\mathcal{C}_{\Delta_{1}}\mathcal{C}_{\Delta_{2}}}
{(2\omega )^{2\Delta_1+2\Delta_2} }
\int_{0}^{a}dt\, t^{\,j-2}\,
\re \left[ \mathcal{A}_{1}^{\circlearrowleft}\left(z-i\epsilon,\bar {z}-i\epsilon\right)  
-\mathcal{A}_{1}\left(z-i\epsilon,\bar {z}-i\epsilon\right) \right]\ ,
\label{cut2}
\end{equation}
with $z$ and $\bar{z}$ depending on $t$ trough (\ref{ss2}).
Recall from the discussion in the previous section that the above
integral is independent of $a$. Therefore, the integrand is
supported at $t=0$, and 
$$
\re \left[ \mathcal{A}_{1}^{\circlearrowleft}\left(z,\bar {z}\right)  \right]=
\re \left[ \mathcal{A}_{1}\left(z,\bar {z}\right) \right] = 
\mathcal{A}_{1}\left(z,\bar {z}\right) \ ,
$$
for $z$ and $\bar{z}$ along the negative real axis.
On the other hand, a non--vanishing integral requires
the leading behavior
\begin{equation}
\mathcal{A}_{1}^{\circlearrowleft}\left(z-i\epsilon,\bar {z}-i\epsilon\right)  
\simeq i\left(t+i\epsilon \right)^{1-j} M(w)\ , \label{itislate}
\end{equation}
with $M^{\star}(w)=M(w^{\star})$. Note, in particular, that the
residue function $M\left(  w\right)  $ must be real in order for
the integrand in (\ref{cut2}) to be localized at $t=0$, which
follows from the independence of the integral on the upper limit
of integration $a$. Then (\ref{cut2}) becomes
\[
\mathcal{E}_1=
\frac{2\,\mathcal{C}_{\Delta_{1}}\mathcal{C}_{\Delta_{2}}}
{(2\omega )^{2\Delta_1+2\Delta_2} }
M(w)  \int_{0}^{a}dt\,t^{\,j-2} \,\im \frac{1}{(t+i\epsilon)^{j-1}}
=-
\frac{\pi\,\mathcal{C}_{\Delta_{1}}\mathcal{C}_{\Delta_{2}}}
{(2\omega )^{2\Delta_1+2\Delta_2} }
M(w)\ .
\]
The two--point function $\mathcal{E}_{1}$ is, on the other hand,
given by (\ref{partE1}) with $-{\bf p}={\bf e}\in H_{d-1}$. This gives
then an integral representation for $M\left( w\right)$ given by
\begin{equation}
M(w)=- g^2\, \mathcal{K}\,
 \int_{H_{d-1}} d\mathbf{w} d\mathbf{\bar{w}} 
\frac{ \Pi_\perp( {\bf w} ,{\bf \bar{w}} ) }
{\left(- 2{\bf e}\cdot {\bf w}\right)^{2\Delta_1+j-1}\left( - 2{\bf \bar{e}}\cdot{\bf \bar{w}}\right)^{2\Delta_{2} +j-1} }\ ,
\label{resultM}
\end{equation}
where we recall that $w^{1/2}+w^{-1/2}=-2 {\bf e} \cdot{\bf \bar{e}} $.
Clearly we have that
\[
M(w) = M \left( 1/ w \right).
\]
Finally, we recognize that equations (\ref{itislate}) and (\ref{resultM}) reproduce the tree--level eikonal 
amplitude (\ref{treeleveleikonal}). 

\section{Harmonic Analysis on $H_{d-1}$ \label{harmony}}
Consider functions $F({\bf e},{\bf \bar{  e} })$ with ${\bf e},{\bf \bar{  e} }\in H_{d-1} $,
which depend only on the geodesic distance $r$ between  ${\bf e}$ and ${\bf \bar{  e} }$,
given by
$$
\cosh r = - {\bf e} \cdot {\bf \bar{  e} }
$$
and which are square integrable
$$
 \int_{H_{d-1}} d {\bf  e}\,| F({\bf e},{\bf \bar{  e} })|^2 =
\frac{2\pi^{\frac{d-1}{2}}}{\Gamma\left( \frac{d-1}{2} \right)}
\int_0^\infty dr \, \sinh^{d-2} r \, |F(r)|^2
<\infty\ .
$$
The Euclidean propagator $\Pi_{\perp,\Delta-1}({\bf e},{\bf \bar{  e} })$ belongs to this class
for $\Delta>d/2$.

We shall be interested in expanding the functions $F$ in a basis of regular  eigenfunctions 
of the Laplacian operator. We define the latter by
\be
\Omega_{\nu}({\bf e},{\bf \bar{  e} })= \frac{i\nu}{2\pi} \left[
\Pi_{\perp,i\nu+d/2 -1}({\bf e},{\bf \bar{  e} }) - \Pi_{\perp,-i\nu +d/2-1}({\bf e},{\bf \bar{  e} })
\right]\ ,    \label{defOmega}
\ee
so that 
$$
\Box_{H_{d-1}} \Omega_{\nu}({\bf e},{\bf \bar{  e} }) =
 -\left( \nu^2 + \frac{(d-2)^2}{4} \right) \Omega_{\nu}({\bf e},{\bf \bar{  e} }) \ 
$$
and $\Omega_{\nu}({\bf e},{\bf \bar{  e} })$ is regular when ${\bf e} \to {\bf \bar{  e} }$.
From the explicit expression of the propagator
\begin{align*}
\Pi _{\perp,i\nu+d/2 -1}({\bf e},{\bf \bar{  e} })  &= \frac{\Gamma
\left( i\nu+d/2 -1\right) }{2\pi^{\frac{d}{2}-1}\,\Gamma\left( i\nu +1\right) }
\,\big[ 2\sinh (r/2)\big] ^{2-d-2i\nu }
\\  &\ \ \ \ 
F\left( i\nu+\frac{d}{2} -1,i\nu+\frac{1}{2},2i\nu+1,-\sinh^{-2} (r/2) \right) \\
&=\frac{\Gamma
\left( i\nu+d/2 -1\right) }{2\pi^{\frac{d}{2}-1}\,\Gamma\left( i\nu +1\right) }\,
e^{-r(i\nu+d/2-1)}\,F\left( i\nu+\frac{d}{2} -1,\frac{d}{2}-1,i\nu+1,e^{-2r} \right)\ ,
\end{align*}
it is easy to see that its only singularities in the $\nu$ complex plane are poles at $\nu=i(n+d/2-1)$ with $n=0,1,2,\cdots$.
Therefore, we conclude that 
$$
\Pi_{\perp,E-1}({\bf e},{\bf \bar{  e} })= \int_{-\infty}^{\infty}d\nu\,
 \frac{ \Omega_{\nu}({\bf e},{\bf \bar{  e} }) }{\nu^2+\left(E- d/2 \right)^2}\ ,
$$
just by using the definition (\ref{defOmega}) and closing the $\nu$--contour integral  appropriately.
Furthermore, applying the operator $\big[ \Box-(E-1)(E-d+1)\big]$ to the last equation one obtains
\be
\delta_{H_{d-1}}({\bf e},{\bf \bar{  e} })= \int_{-\infty}^{\infty}d\nu\,
\Omega_{\nu}({\bf e},{\bf \bar{  e} }) \ .      \label{deltaident}
\ee
Finally, we can write the explicit form of the harmonic functions $\Omega_{\nu}$ in terms
of the hypergeometric function
$$
\Omega_{\nu}({\bf e},{\bf \bar{  e} })=
\frac{\nu \sinh \pi\nu \,\Gamma\left( \frac{d}{2} -1-i\nu \right) \Gamma\left( \frac{d}{2} -1+i\nu \right) }
{ 2^{d-1}\,\pi^{\frac{d+1}{2}}\, \Gamma\left( \frac{d-1}{2}\right) }
\,F\left( \frac{d}{2} -1-i\nu ,  \frac{d}{2} -1+i\nu, \frac{d-1}{2}, -\sinh^{2} \frac{r}{2}  \right)\ .
$$

The functions  $\Omega_{\nu}$ with  $\nu\in \mathbb{R}$ form a basis,
$$
F({\bf e},{\bf \bar{  e} })=\int_{-\infty}^{\infty}d\nu \,f(\nu)\, \Omega_{\nu}({\bf e},{\bf \bar{  e} }) \ ,
$$
with the transform $f(\nu)=f(-\nu)$ since $\Omega_{\nu}$ is an even function of $\nu$.
In order to invert this transform, we consider the convolution
$$
C({\bf e},{\bf \bar{  e} })= \int_{H_{d-1}} d {\bf w} \, \Omega_{\nu}({\bf e},{\bf w })\,\Omega_{\bar{\nu}}({\bf w},{\bf \bar{  e} })\ ,
$$
which is clearly a function of ${\bf e} \cdot {\bf \bar{  e} } $ and therefore
invariant under the permutation ${\bf e} \leftrightarrow {\bf \bar{  e} } $.
Hence 
$$
\Big[\Box_{{\bf e} } - \Box_{{\bf \bar{  e} }} \Big]  C({\bf e},{\bf \bar{  e} })= 
\left(\bar{\nu}^2-\nu^2 \right)C({\bf e},{\bf \bar{  e} })=0
$$
and the convolution can only be non--zero if $\bar{\nu}=\pm \nu $.
In fact,
\be
\int_{H_{d-1}} d {\bf w} \, \Omega_{\nu}({\bf e},{\bf w })\,\Omega_{\bar{\nu}}({\bf w},{\bf \bar{  e} })=
\frac{1}{2}\left[ \delta(\nu- \bar{\nu}) +  \delta(\nu+\bar{\nu}) \right]\Omega_{\nu}({\bf e},{\bf \bar{  e} })\ ,
\label{convoluident}
\ee
where we have fixed the normalization using  (\ref{deltaident}).
From the identity (\ref{convoluident}), we have
$$
\int_{H_{d-1}} d {\bf w} \, \Omega_{\nu}({\bf e},{\bf w })\,F({\bf w},{\bf \bar{  e} })=
f(\nu)\, \Omega_{\nu}({\bf e},{\bf \bar{  e} })
$$
and, choosing ${\bf e}={\bf \bar{  e} }$, we obtain the inverse transform,
$$
f(\nu)=\frac{1}{ \Omega_{\nu}({\bf e},{\bf  e }) } \int_{H_{d-1}} d {\bf  w}\,\Omega_{\nu}({\bf e},{\bf w })
\,F({\bf w},{\bf e})\ .
$$
In addition, we note that  (\ref{convoluident}) implies that the transform of the convolution of two radial functions on $H_{d-1}$ 
is just the product of the transforms of the two functions.

In section \ref{sec:tchannel}, the transform
$$
\varphi_{a}(\nu)= \frac{1}{ \Omega_{\nu}({\bf e},{\bf  e }) } \int_{H_{d-1}} d {\bf  w}
\,\Omega_{\nu}({\bf e},{\bf w })\, \left(-2 {\bf e}  \cdot  {\bf w}\right)^{-a}
$$
was required.
Using the explicit form of the harmonic functions  $\Omega_{\nu}$, we can write this transform as a radial integral,
$$
\varphi_{a}(\nu)=  \frac{2\pi^{\frac{d-1}{2}}}{\Gamma\left( \frac{d-1}{2}\right)}
\int_0^{\infty} dr\,\sinh^{d-2}r \, \left( 2\cosh r \right)^{-a}
\,F\left( \frac{d}{2} -1-i\nu ,  \frac{d}{2} -1+i\nu, \frac{d-1}{2}, -\sinh^{2} \frac{r}{2}  \right)\ ,
$$
whose integrand is an analytic function of $\nu$. Thus, the poles of $\varphi_{a}(\nu)$ must result from
divergences of the integral at $r=\infty$.
To understand the structure of the integral for large $r$ it is convenient to rewrite it in the following way
\begin{align*}
\varphi_{a}(\nu)=  (4\pi)^{\frac{d-2}{2}} 
\int_0^{\infty} dr\,&\sinh^{d-2}r \, \left( 2\cosh r \right)^{-a} e^{r\left(1-\frac{d}{2}\right)}\\
&\left[ e^{i\nu r} \frac{\Gamma\left( i\nu \right) }{\Gamma\left( \frac{d}{2} -1 +i\nu \right) }
\,F\left( \frac{d}{2} -1-i\nu ,  \frac{d-2}{2}, 1-i\nu, e^{-2r} \right) + 
(\nu \leftrightarrow -\nu) \right]\ ,
\end{align*}
showing that, at large $r$, we have
$$
\varphi_{a}(\nu)= \int_0^{\infty} dr\, e^{r\left( i\nu -a +\frac{d}{2}-1 \right)}\,\sum_{n=0}^{\infty}c_n e^{-2nr}
\ +\  \big(\nu \leftrightarrow -\nu\big)\ ,
$$
for some expansion coefficients $c_n$. Hence, we conclude that the function $\varphi_{a}(\nu)$ has poles
precisely at  $\nu=\pm i(a+1-d/2+2n)$ with $n=0,1,2,\cdots$, as stated in  section \ref{sec:tchannel}.
We remark that for some dimensions $d$ the integral can be done explicitly and the poles are found
exactly as expected.
For example, in the important cases of $d=2$ and $d=4$ we have
\begin{equation*}
\varphi_{a}(\nu)=2\int_0^{\infty}dr \cos(\nu r) \left( 2\cosh r\right)^{-a}
= \frac{\Gamma\left( \frac{a+i\nu}{2}\right)\Gamma\left( \frac{a-i\nu}{2}\right) }{ 2\, \Gamma(a)}
\end{equation*}
and
\begin{equation*}
\varphi_{a}(\nu)=\frac{4\pi}{\nu}\int_0^{\infty}dr \sinh r \sin(\nu r) \left( 2\cosh r\right)^{-a}
= \frac{\pi\, \Gamma\left( \frac{a-1+i\nu}{2}\right)\Gamma\left( \frac{a-1-i\nu}{2}\right) }{ 2\, \Gamma(a)}\ ,
\end{equation*}
respectively.


\section{D--functions} \label{Dfunctions}


The contact $n$--point functions are defined by
\[
D_{\Delta_{i}}^{d}\left(  \mathbf{Q}_{i}\right)  =\int_{H_{d+1}} d\mathbf{x}
\,{\textstyle\prod\nolimits_{i}}
\,\left(  -2\mathbf{x\cdot Q}_{i}\right)^{-\Delta_{i}}\ ,
\]
where the points $\mathbf{Q}_{i}$  are future directed timelike or null vectors of the embedding space $\mathbb{M}^{d+2}$.
Introducing $n$ auxiliary integrals one obtains
\be
D_{\Delta_{i}}^{d}\left(  \mathbf{Q}_{i}\right)  =
\frac{1}{{\textstyle\prod\nolimits_{i}}
\Gamma\left(  \Delta_{i}\right)  }
\int_0^\infty  {\textstyle\prod\nolimits_{i}}\,dt_{i}\,t_{i}^{\Delta_{i}-1}
\int_{H_{d+1}} d\mathbf{x}\,
e^{2\mathbf{x}\cdot \mathbf{Q}}\ ,  \label{tiDfunc}
\ee
with $\mathbf{Q}=\sum_i \,t_i\, \mathbf{Q}_{i}$ a generic  future directed vector in $\mathbb{M}^{d+2}$.
The integral over $\mathbf{x}$ can then be performed by choosing coordinates for $H_{d+1}$ centered around 
the unit vector $  \mathbf{Q}/| \mathbf{Q} |$,
$$
\int_{H_{d+1}} d\mathbf{x}\,
e^{2\mathbf{x}\cdot \mathbf{Q}}=
\frac{2\pi^{\frac{d+1}{2}}}{\Gamma\left( \frac{d+1}{2} \right)}
\int_0^\infty dr \, \sinh^d r \,e^{-2| \mathbf{Q} |\cosh r}\ .
$$
More conveniently, one can compute the $\mathbf{x}$ integral by parametrizing the hyperboloid $H_{d+1}$ with Poincar\'e coordinates,
as given in equation (\ref{Poincare}) for AdS$_{d+1}$. The sole difference in the present Euclidean
case  is that ${\bf y}\in \mathbb{R}^d$ instead of $\mathbb{M}^d$.
Then, choosing the time direction ${\bf k}+{\bf \bar{k}}$ of the embedding space $\mathbb{M}^{d+2}$ such that
${\bf Q} = |{\bf Q} |( {\bf k}+{\bf \bar{k}}) $, we arrive at
$$
\int_{H_{d+1}} d\mathbf{x}\,
e^{2\mathbf{x}\cdot \mathbf{Q}}=
\int_0^{\infty}dy\,y^{-d-1}\int_{ \mathbb{R}^d }d{\bf y} \,
e^{-| \mathbf{Q} | \left(1+y^2 + {\bf y}^2\right)/y }\ .
$$
One can now evaluate the Gaussian integral over ${\bf y}$ and change to the variable $s= | \mathbf{Q} |\,  y$ to obtain
$$
\int_{H_{d+1}} d\mathbf{x}\,
e^{2\mathbf{x}\cdot \mathbf{Q}}=
\pi^{\frac{d}{2}}\int_0^\infty ds \,s^{-\frac{d}{2}-1}\,e^{-s+\mathbf{Q}^2 /s}\ .
$$
Returning to (\ref{tiDfunc}) and rescaling $t_i\to t_i\sqrt{s}$, we have
\begin{align*}
D_{\Delta_{i}}^{d}\left(  \mathbf{Q}_{i}\right)  &=
\frac{\pi^{\frac{d}{2}}}{{\textstyle\prod\nolimits_{i}}
\Gamma\left(  \Delta_{i}\right)  }
\int_0^\infty  {\textstyle\prod\nolimits_{i}}\,dt_{i}\,t_{i}^{\Delta_{i}-1}
e^{\mathbf{Q}^2}
\int_0^\infty ds \,s^{\frac{\Delta-d}{2}-1}\,e^{-s}\\
&=
\frac{\pi^{\frac{d}{2}}  \Gamma\left(  \frac{\Delta-d}{2}\right)   }{{\textstyle\prod\nolimits_{i}}
\Gamma\left(  \Delta_{i}\right)  }
\int_0^\infty  {\textstyle\prod\nolimits_{i}}\,dt_{i}\,t_{i}^{\Delta_{i}-1}
e^{\mathbf{Q}^2}\ ,
\end{align*}
where $\Delta={\textstyle\sum\nolimits_{i}}\Delta_{i}$.
Finally, recalling that  $\mathbf{Q}=\sum_i \,t_i\, \mathbf{Q}_{i}$ we arrive at the 
following the integral representation
\begin{align*}
D_{\Delta_{i}}^{d}\left(  \mathbf{Q}_{i}\right)   & =\pi^{\frac{d}{2}}
\frac{\Gamma\left(  \frac{\Delta-d}{2}\right)  }{
{\textstyle\prod\nolimits_{i}}
\Gamma\left(  \Delta_{i}\right)  }\int_0^\infty
{\textstyle\prod\nolimits_{i}}\,
dt_{i}\,t_{i}^{\Delta_{i}-1}~e^{\mathbf{-}\frac{1}{2}\sum_{i,j}t_{i}
t_{j}~\mathbf{Q}_{ij}}\\
& =\frac{\pi^{\frac{d}{2}}}{2}\frac{\Gamma\left(
\frac{\Delta-d}{2}\right)
\Gamma\left(  \frac{\Delta}{2}\right)  }{
{\textstyle\prod\nolimits_{i}}
\Gamma\left(  \Delta_{i}\right)  }\int_0^\infty
{\textstyle\prod\nolimits_{i}}\,
dt_{i}\,t_{i}^{\Delta_{i}-1}\ \frac{\delta\left(
{\textstyle\sum\nolimits_{i}}
t_{i}-1\right)  ~}{\left(  \frac{1}{2}\sum_{i,j}t_{i}t_{j}~\mathbf{Q}
_{ij}\right)  ^{\frac{\Delta}{2}}}~,
\end{align*}
with $\mathbf{Q}_{ij}=-2\mathbf{Q}_{i}\cdot\mathbf{Q}_{j}\ge 0$. 

The two--point function integral can be explicitly evaluated,
\[
D_{\Delta_{1},\Delta_{2}}^{d} \left( \mathbf{Q}_{1}, \mathbf{Q}_{2} \right) =
\left( -\mathbf{Q}_{1}^{2}\right)^{-\Delta_{1}/ 2} \left( -\mathbf{Q}_{2}^{2}\right)^{-\Delta_{2}/2}\,
\frac{\pi^{\frac{d}{2}}\,\Gamma\left( \frac{\Delta-d}{2}\right) \Gamma\left( \frac{\Delta }{2}\right)}
{2\,\Gamma\left( \Delta\right)}\,
\alpha^{\Delta_{2}}\,
F  \left( \Delta_{2},\frac{\Delta}{2},\Delta , 1-\alpha^{2}\right)  \ ,
\]
where
\[
\alpha+\frac{1}{\alpha}=\frac{-2\mathbf{Q}_{1}\cdot\mathbf{Q}_{2}}{\left(
-\mathbf{Q}_{1}^{2}\right)  ^{\frac{1}{2}}\left(
-\mathbf{Q}_{2}^{2}\right) ^{\frac{1}{2}}}~.
\]
Finally, as shown in  \cite{Bianchi}, the four--point function $D_{2,2,1,1}^{2}$ can be 
explicitly computed in terms of standard one--loop box integrals, as given in section \ref{SectExample}.

\end{subappendices}








\chapter{Conclusions \& Open Questions}
\label{ch:conc}

The AdS/CFT correspondence is one of the most remarkable achievements of String Theory.
It is the latest great advance in the long quest for the a string description of Yang--Mills theories.
Although we  still seem far from discovering the string theory dual to real QCD,
the subject has already accumulated a huge amount of results as can be judged by the vast literature.
However, the exploration of the conjectured AdS/CFT duality has been, so far, essentially limited to the planar limit,
which corresponds to zero string coupling, as explained in the Introduction. 
This thesis summarizes the first steps of a program to study string interaction effects in AdS/CFT and
therefore test the correspondence outside the planar limit.
From the gauge theory point of view, these correspond to non--planar contributions in the $1/N$ expansion.
In string theory, these effects are very important since they are responsible for decay processes of excited strings
into smaller ones, and for non--trivial scattering of strings.

The understanding of the worldsheet theory describing free strings in AdS is still incomplete, despite
the many recent advances. Thus, direct  computations of string scattering amplitudes as worldsheet correlation functions
are presently out of reach.
In this thesis, we overcome this obstacle by focusing on the particle limit $\ell_s \to 0$, where  string
theory reduces to a gravitational theory, and on high energy scattering processes, where 
the gravitational loop expansion can be resummed using eikonal methods.
Here we shall very briefly summarize our main findings.

Our first step was the generalization of the eikonal approximation in flat space to AdS.
This was achieved with a very clear physical picture of semi--classical interaction between
null geodesics.
Since AdS has a timelike boundary, it is effectively a gravitational box and the word scattering is just colloquial.
In fact, the eikonal phase shift in AdS simply determines the interaction energy between the two
colliding high energetic strings. Therefore, using eikonal methods, we were able to determine
the energy shift of specific two string states in AdS, which, by the AdS/CFT correspondence,
is the anomalous dimension of the dual double trace primary operators.
Although the predicted anomalous dimension depends on the type of interaction in AdS,
for high energy strings scattering at large impact parameters,
the graviton contribution dominates and the result yields a universal prediction for CFT's with gravitational AdS duals.

The investigations of this thesis open many interesting possibilities for future studies of non--planar effects
in the AdS/CFT correspondence. We conclude with a list of some natural next steps in this program.

\begin{itemize}
\item Study the leading string corrections to the eikonal amplitude in AdS.
In flat space, the leading string effects in high energy scattering can be understood as
the exchange of particles of all spins lying in the leading Regge trajectory,
resulting in an effective reggeon interaction of spin
approximately $2$ for large string tension  \cite{ACV,GG}. 
As recalled in  \cite{GG},
in flat space the leading corrections to the pure gravity result occur
due to tidal forces which
excite internal modes of the scattering strings. These effects start to be
relevant at impact parameters of the order of $\ell _{\mathrm{Planck}}\left(
\mathcal{E}\ell _{s}\right) ^{2/(d-1)}$, where $\ell _{\mathrm{Planck}}$ is
the Planck length in the $\left( d+1\right) $--dimensional spacetime, and
where $\mathcal{E}$ is the energy of the process. Translating into AdS
variables and recalling from (\ref{imppar}) that the impact parameter distance $r$ 
is given by $r=\ell \,\mathrm{ln}(h/\bar{h})$, we then expect tidal string excitations to play a role for
$r\lesssim \left (G h\bar{h}\, \ell ^{\,d-3} \ell _{s}^{\,2}\right)^{1/(d-1)}$, i.e. for
$$
\Big[ \ln \left( h /\bar{h} \right) \Big]^{\, d-1}  \lesssim G \, \frac{\ell _{s}^{\,2}}{ \ell^{\,2}}\,  h\bar{h} \ ,
$$
where $G$ is the dimensionless
Newton constant in terms of the AdS radius $\ell$.
The appropriate treatment of these string effects requires an extension of
Regge theory to conformal field theories which is quite natural in
our formalism  \cite{Paper4}, with results which reproduce and extend those of
 \cite{Polchinski,Polfriends}.

\item At small enough impact parameters, there is the possibility of the incoming strings 
change their internal state due to the tidal forces of the interaction field produced by them.
This type of inelastic scattering in flat space was described  in  \cite{ACV,AmatiSW,VegaSW}, by
promoting the eikonal phase shift to an operator acting on the internal states of the incoming strings.
The phase shift $\Gamma $ in AdS will then
become an operator acting on two--string states, which will include both an
orbital part as well as a contribution from the internal excitation of the
two scattering strings. 
In AdS, one could try to determine such an eikonal phase shift operator
acting on the space of  states of two small strings moving along null geodesics  \cite{BMN},
by studying string propagation in the gravitational shock wave background of section \ref{sec:sw},
in the spirit of   \cite{AmatiSW,VegaSW}.
Given the relation of phase shift and anomalous dimension,
$2\Gamma $ would be a
generalization, to double trace operators, of the dilatation operator  \cite{Beisert} 
which has played a crucial role in analyzing the spectrum of
single trace states in $\mathcal{N}=4$ SYM theory.

\item Up to now we have always considered large 't Hooft coupling $\lambda$. 
It is natural to ask if one can determine the anomalous dimensions of large dimension and spin composite 
operators using perturbative techniques on the field theory side of the duality.
Indeed, it is possible  \cite{Paper4} to compute the four point function in the eikonal kinematical regime
at weak 't Hooft coupling using BFKL techniques   \cite{BFKL1, BFKL2, BFKL3, BFKL4, LipatovSol, LipatovRev}.  
Following the initial results of  \cite{Polfriends}, we shall
relate our formalism to that of BFKL,
describing hard pomeron exchange at weak coupling, including the
non--trivial transverse dependence relevant at non--vanishing momentum
transfer. 
Then, using the methods of this thesis, one can determine the anomalous dimensions of high spin and dimension
composite operators at small $\lambda$.
Recently, Maldacena and Alday  \cite{MaldaAlday} argued that the large spin ($J=h-\bar{h}$) and
low twist ($2\bar{h}=E-J$) double trace operators of $\mathcal{N}=4$ SYM, 
have $1/N$ corrections to their anomalous dimension of the form
\begin{equation}
\frac{g(\lambda,\bar{h})}{N^2\,J^2}  \ ,\label{goflambda}
\end{equation}
corresponding to the field theory exchange of the energy--momentum tensor.
The relevance of these low twist double trace operators for deep inelastic scattering at strong 't Hooft coupling
was emphasized in  \cite{PolchinskiDIS}.
Although our results were derived in a different regime, namely large $h$ and $\bar{h}$ with fixed ratio  $h/\bar{h}$,
the eikonal approximation in AdS for $\bar{h}\ll h$ reproduces the behavior (\ref{goflambda}), 
determining the large $\lambda$ limit
$$
\lim_{\lambda \to \infty} g(\lambda,\bar{h}) = -4 \,\bar{h}^4\ .
$$
Using the BFKL approach we plan to investigate the small $\lambda$ behavior of  $g(\lambda,\bar{h})$.

\item A direct application to the prototypical example of $\mathcal{N}=4$ SYM would also be very interesting.
However, for impact parameters of the order of the $S^5$ radius the transverse sphere should become important.
Corrections to (\ref{sucabis}), due to massive KK modes of the graviton,
will start to be relevant at $r\sim \ell $. These corrections are computable
with an extension of the methods of this thesis, which includes the sphere $S^{5}$ 
in the transverse space.

\end{itemize}

\newpage

\appendix
\chapter{List of PhD  Publications}

This thesis summarizes the content of the following three publications.\newline \newline
{\bf Eikonal approximation in AdS/CFT: Resumming the gravitational loop expansion.}\newline
Lorenzo Cornalba, Miguel S. Costa, Jo\~ao Penedones. \newline
To appear in JHEP. arXiv:0707.0120 [hep-th]\newline
\emph{Abstract}: We derive an eikonal approximation to high energy interactions in
Anti--de Sitter spacetime, by generalizing a position space derivation of the
eikonal amplitude in flat space. We are able to resum, in terms of a
generalized phase shift, ladder and cross ladder graphs associated to the
exchange of a spin $j$ field, to all orders in the coupling constant.
Using the AdS/CFT correspondence, the resulting amplitude determines the
behavior of the dual conformal field theory four--point function $%
\left\langle \mathcal{O}_{1}\mathcal{O}_{2}\mathcal{O}_{1}\mathcal{O}%
_{2}\right\rangle $ for small values of the cross ratios, in a
Lorentzian regime. Finally we show that the phase shift is
dominated by graviton exchange and computes, in the dual CFT, the
anomalous dimension of the double trace primary operators
$\mathcal{O}_{1}\partial \cdots \partial \mathcal{O}_{2}$ of large dimension and
spin, corresponding to the relative motion of the two interacting
particles. The results are valid at strong t'Hooft coupling and 
are exact in the $1/N$ expansion.
\newline \newline \newline
{\bf Eikonal Approximation in AdS/CFT: Conformal Partial Waves and Finite N Four-Point Functions.}\newline
Lorenzo Cornalba, Miguel S. Costa, Jo\~ao Penedones, Ricardo Schiappa.\newline
Published in Nucl.Phys.B767:327-351,2007. hep-th/0611123\newline 
\emph{Abstract}: We introduce the impact parameter representation for
conformal field theory correlators of the form $\mathcal{A}\sim
\left\langle\mathcal{O}_{1}\mathcal{O}_{2}\mathcal{O}_{1}\mathcal{O}_{2}\right\rangle$.
This representation is appropriate in the eikonal kinematical
regime, and approximates the conformal partial wave decomposition
in the limit of large spin and dimension of the exchanged primary.
Using recent results on the two--point function $\left\langle
\mathcal{O}_{1}\mathcal{O}_{1}\right\rangle_{\textrm{shock}}$ in
the presence of a shock wave in Anti--de Sitter, and its relation
to the discontinuity of the four--point amplitude $\mathcal{A}$
across a kinematical branch cut, we find  the high spin and
dimension conformal partial wave decomposition of all tree--level
Anti--de Sitter Witten diagrams. We show that, as in flat space,
the eikonal kinematical regime is dominated by the
\emph{T}--channel exchange of the massless particle with highest
spin ({\em graviton dominance}). We also compute the anomalous
dimensions of the high spin $\mathcal{O}_{1}\mathcal{O}_{2}$
composites. Finally, we conjecture a formula re--summing
crossed--ladder Witten diagrams to \textit{all} orders in the
gravitational coupling.
\newline \newline \newline
{\bf  Eikonal Approximation in AdS/CFT: From Shock Waves to Four-Point Functions.}\newline
Lorenzo Cornalba, Miguel S. Costa, Jo\~ao Penedones, Ricardo Schiappa. \newline
To appear in JHEP. hep-th/0611122 \newline 
\emph{Abstract}: We initiate a program to generalize the standard eikonal approximation to compute amplitudes in
Anti--de Sitter spacetimes. Inspired by the shock wave derivation of the eikonal amplitude in flat space,
we study the two--point function $\mathcal{E} \sim \left\langle \mathcal{O}_{1} \mathcal{O}_{1} \right\rangle_{\mathrm{shock}}$
in the presence of a shock wave in Anti--de Sitter, where $\mathcal{O}_{1}$ is a scalar primary operator in the dual
conformal field theory. At tree level in the gravitational coupling, we relate the shock two--point
function $\mathcal{E}$ to the discontinuity across a kinematical branch cut of the conformal field theory four--point
function $\mathcal{A} \sim \left\langle \mathcal{O}_{1} \mathcal{O}_{2} \mathcal{O}_{1} \mathcal{O}_{2} \right\rangle$,
where $\mathcal{O}_2$ creates the shock geometry in Anti--de Sitter. Finally, we extend the above results by computing
$\mathcal{E}$ in the presence of shock waves along the horizon of Schwarzschild BTZ black holes. This work gives new tools
for the study of Planckian physics in Anti--de Sitter spacetimes.
\newline \newline \newline

My research activities during the PhD were not restricted to the specific subject reflected in this thesis and 
included  other areas of String Theory.
The visible outcome of these other studies is the following list of publications.\newline \newline
{\bf  From Fundamental Strings to Small Black Holes.}\newline
Lorenzo, Miguel S. Costa, Jo\~ao Penedones, Pedro Vieira.\newline
Published in JHEP 0612:023,2006. hep-th/0607083\newline 
\emph{Abstract}: We give evidence in favour of a string/black hole transition in the case of BPS fundamental string states of the
Heterotic string. Our analysis
goes beyond the counting of degrees of freedom and considers the evolution
of dynamical quantities in the process. As the coupling increases, the string states decrease their size 
up to the string scale when a small black hole is formed. We compute the
absorption cross section for several fields in both the black hole and the perturbative string phases. 
At zero frequency, these cross sections can be seen as order parameters
for the transition. In particular, for the scalars fixed at the horizon the cross section evolves 
to zero when the black hole is formed.
\newline \newline 
{\bf  Hagedorn transition and chronology protection in string theory.}\newline
Miguel S. Costa, Carlos A.R. Herdeiro, J. Penedones, N. Sousa. \newline
Published in Nucl.Phys.B728:148-178,2005. hep-th/0504102\newline 
\emph{Abstract}: We conjecture  chronology is protected in string theory due to the condensation of light winding strings near closed
null curves. This condensation triggers a Hagedorn phase transition, whose end-point target space geometry should be
chronological. Contrary to conventional arguments, chronology is protected by an infrared effect. We support this 
conjecture by studying strings in the O-plane orbifold, where we show that some winding string states are unstable and
condense in the non-causal region of spacetime. The one-loop string partition function has infrared divergences associated
to the condensation of these states.
\newline \newline \newline
{\bf  Brane nucleation as decay of the tachyon false vacuum.}\newline
Lorenzo Cornalba, Miguel S. Costa, Jo\~ao Penedones.\newline
Published in Phys.Rev.D72:046002,2005. hep-th/0501151 \newline
\emph{Abstract}: It is well known that spherical D--branes are nucleated in the presence of an external RR electric field. 
Using the description of D--branes as solitons of the tachyon field
on non--BPS D--branes, we show that the brane nucleation process can be seen as the 
decay of the tachyon false vacuum. This process can describe the decay of flux--branes
in string theory or the decay of quintessence potentials arising in flux compactifications.
\newline \newline

\newpage

\addcontentsline{toc}{chapter}{Bibliography}

\bibliographystyle{hieeetr}
\bibliography{tese}

\end{fmffile}
\end{document}